\documentclass[opre,nonblindrev]{informs1update}              % for a regular run
\usepackage{bm,dsfont}
\usepackage{physics}
\usepackage{amsmath}
\usepackage{tikz}
\usepackage{mathdots}
\usepackage{yhmath}
\usepackage{cancel}
\usepackage{color}
\usepackage{siunitx}
\usepackage{array}
\usepackage{multirow}
\usepackage{amssymb}
\usepackage{gensymb}
\usepackage{tabularx}
\usepackage{extarrows}
\usepackage{booktabs}
\usetikzlibrary{fadings}
\usetikzlibrary{patterns}
\usetikzlibrary{shadows.blur}
\usetikzlibrary{shapes}
\usepackage{natbib}
 \NatBibNumeric
 \def\bibfont{\small}%
 \bibpunct[, ]{[}{]}{,}{n}{}{,}%

\usepackage[colorlinks=true,breaklinks=true,bookmarks=true,urlcolor=blue,
     citecolor=blue,linkcolor=blue,bookmarksopen=false,draft=false]{hyperref}

\def\EMAIL#1{\href{mailto:#1}{#1}}% When hyperref is used, otherwise outcomment 
\def\URL#1{\href{#1}{#1}}         % When hyperref is used, otherwise outcomment 

\TheoremsNumberedThrough     % Preferred (Theorem 1, Lemma 1, Theorem 2)
\EquationsNumberedThrough    % Default: (1), (2), ...
\begin{document}
%%%%%%%%%%%%%%%%

%\setcounter{page}{1}
%\DOI{10.1287/xxxx.0000.0000}%

\RUNAUTHOR{Bayrak et al.}

\RUNTITLE{Distributionally Robust Optimal Allocation with Costly Verification}

% \TITLE{Bundling Information Goods of Decreasing Value}
\TITLE{Distributionally Robust Optimal Allocation with Costly Verification}

\ARTICLEAUTHORS{%
\AUTHOR{Halil \.{I}brahim Bayrak}
\AFF{Department of Industrial Engineering, Bilkent University, Turkey, \EMAIL{halil.bayrak@bilkent.edu.tr}, \URL{}}
\AUTHOR{\c{C}a\u{g}{\i}l Ko\c{c}yi\u{g}it}
\AFF{Luxembourg Centre for Logistics and Supply Chain Management, University of Luxembourg, Luxembourg, \EMAIL{cagil.kocyigit@uni.lu}, \URL{}}
\AUTHOR{Daniel Kuhn}
\AFF{Risk Analytics and Optimization Chair, \'Ecole Polytechnique F\'ed\'erale de Lausanne, Switzerland, \EMAIL{daniel.kuhn@epfl.ch}, \URL{}}
\AUTHOR{Mustafa \c{C}elebi P{\i}nar}
\AFF{Department of Industrial Engineering, Bilkent University, Turkey, \EMAIL{mustafap@bilkent.edu.tr}, \URL{}}
} % end of the block

\ABSTRACT{%
We consider the mechanism design problem of a principal allocating a single good to one of several agents without monetary transfers. Each agent desires the good and uses it to create value for the principal. We designate this value as the agent's private type. Even though the principal does not know the agents' types, she can verify them at a cost. The allocation of the good thus depends on the agents' self-declared types and the results of any verification performed, and the principal's payoff matches her value of the allocation minus the costs of verification. It is known that if the agents' types are independent, then a favored-agent mechanism maximizes her expected payoff. However, this result relies on the unrealistic assumptions that the agents' types follow known independent probability distributions. In contrast, we assume here that the agents' types are governed by an ambiguous joint probability distribution belonging to a commonly known ambiguity set and that the principal maximizes her worst-case expected payoff. We study support-only ambiguity sets, which contain all distributions supported on a rectangle, Markov ambiguity sets, which contain all distributions in a support-only ambiguity set satisfying some first-order moment bounds, and Markov ambiguity sets with independent types, which contain all distributions in a Markov ambiguity set under which the agents' types are mutually independent. In all cases, we construct explicit favored-agent mechanisms that are not only optimal but also Pareto robustly optimal.
}%

%\KEYWORDS{deterministic inventory theory; infinite linear programming duality; 
%  existence of optimal policies; semi-Markov decision process; cyclic schedule}
%\MSCCLASS{Primary: 90B05; secondary: 90C40, 90C90}
%\ORMSCLASS{Primary: Inventory/production: deterministic multi-item;
%  secondary: dynamic programming/optimal control: deterministic 
%  semi-Markov; programming: infinite dimensional}
%\HISTORY{Received November 20, 2003; revised March 8, 2004, and March 26, 2004.}

%\MSCCLASS{}
%\ORMSCLASS{Primary: ; secondary: }

\maketitle
%%%%%%%%%%%%%%%%%%%%%%%%%%%%%%%%%%%%%%%%%%%%%%%%%%%%%%%%%%%%%%%%%%%%%%
% Samples of sectioning (and labeling) in MOOR.
% NOTE: (1) all section levels end with a period,
%       (2) capitalization is as shown (sentence style, not title style).
%
%\section{Introduction.}\label{intro} %%1.
%\subsection{Duality and the classical EOQ problem.}\label{class-EOQ} %% 1.1.
%\subsection{Outline.}\label{outline1} %% 1.2.
%\subsubsection{Cyclic schedules for the general deterministic SMDP.}
%  \label{cyclic-schedules} %% 1.2.1
%\section{Problem description.}\label{problemdescription} %% 2.
\section{Introduction}

Consider a principal (``she'') who allocates a good to one of several agents without using monetary transfers. Each agent (``he'') derives strictly positive utility from owning the good and has a private type, which reflects the value he creates for the principal if receiving the good. %In other words, each agent privately knows the value he generates for the principal if he is allocated the good. 
The principal is unaware of the agents' types but can verify any of them at a cost. Any verification will perfectly reveal the corresponding agent's type to the principal. The good is allocated based on the agents' self-declared types as well as the results of any verification performed. The principal aims to design an allocation mechanism that maximizes her payoff, $i.e.$, the value of allocation minus any costs of verification.

This generic mechanism design problem arises in many different contexts. %The examples of such an environment are vast. 
For example, the rector of a university may have funding for a new faculty position and needs to allocate it to one of the school's departments, the ministry of health may need to decide in which town to open up a new hospital, a venture capitalist may need to select a start-up business that should receive seed funding, the procurement manager of a manufacturing company may need to choose one of several suppliers, or a consulting company may need to identify a team that leads a new project. In all of these examples, the principal wishes to put the good into use where it best contributes to her organization or the society as a whole. Each agent desires the good and is likely to be well-informed about the value he will generate for the principal if he receives the good. In addition, monetary transfers may be inappropriate in all of the described situations, but the principal can collect information through costly investigation or audit.

Mechanism design problems of the above type are usually referred to as `allocation with costly verification.' \citet{ben2014optimal} describe the first formal model for their analysis and introduce the class of favored-agent mechanisms, which are attractive because of their simplicity and interpretability. As in most of the literature on mechanism design, \cite{ben2014optimal} model the agents' types as independent random variables governed by a commonly known probability distribution, which allows them to prove that any mechanism that maximizes the principal's expected payoff is a randomization over favored-agent mechanisms.
%a favored-agent mechanism maximizes the principal's expected payoff. 
Any favored-agent mechanism is characterized by a favored agent and a threshold value,
%Under a favored-agent mechanism, the agents are asked to report their types. 
and it assigns the good to the favored agent without verification whenever the reported types of all other agents---adjusted for the costs of verification---fall below the given threshold. Otherwise, it allocates the good to any agent for which the reported type minus the cost of verification is maximal and verifies his reported type. 
%If no agent other than the favored agent reports a type such that the value of his type minus his verification cost falls above the threshold, then the principal gives the good to the favored agent without verification. Otherwise, an agent whose reported type maximizes the value of type minus... across all agents receives the good after his reported type is verified. 
%The choice of the favored agent is predicated on the principal's prior beliefs about the agents' types. The favored agent receives the good without verification if no other agent reports a sufficiently high type. Otherwise, the principal verifies the maximum adjusted type reported by any agent and allocates the good to the respective agent. 
This mechanism is incentive compatible, that is, no agent has an incentive to misreport his true type; see Section~\ref{sec: Problem Statement and Preliminaries} for more details.

The vast majority of the literature on allocation with costly verification (see, $e.g.$, \cite{li2020mechanism,mylovanov2017optimal} and the references therein) sustains the modeling assumptions of \cite{ben2014optimal}, thus assuming that the agents' types are independent random variables and that their distribution is common knowledge. In reality, however, it is often difficult to justify the precise knowledge of such a distribution. 
%For example, a rector simply does not have the time to constantly monitor all departments' activities to form reliable beliefs about the payoff of allocating a job slot to each department. 
This prompts us to study allocation problems with costly verification under the more realistic assumption that the principal has only partial information about the distribution of the agents' types. Specifically, we assume that the distribution of the agents' types is unknown but belongs to a commonly known ambiguity set ($i.e.$, a family of multiple---perhaps infinitely many---distributions). In addition, we assume that the principal is ambiguity averse in the sense that she wishes to maximize her worst-case expected payoff in view of all distributions in the ambiguity set. Under these assumptions, the mechanism design problem at hand can be cast as a zero-sum game between the principal, who chooses a mechanism to allocate the good, and some fictitious adversary, who chooses the distribution of the agents' types from the ambiguity set in order to inflict maximum damage to the principal.
%We formulate this mechanism design problem as a distributionally robust optimization problem (see, $e.g.$, \cite{delage2010distributionally,wiesemann2014distributionally}). \noteck{Can we write this using `zero-sum game'?} 
Using techniques from distributionally robust optimization (see, $e.g.$, \cite{delage2010distributionally,wiesemann2014distributionally}), we characterize optimal and Pareto robustly optimal mechanisms for well-known classes of ambiguity sets: (i) support-only ambiguity sets containing all distributions supported on a rectangle, (ii) Markov ambiguity sets containing all distributions in a support-only ambiguity set whose mean values fall within another (smaller) rectangle, and (iii) Markov ambiguity sets with independent types containing all distributions in a Markov ambiguity set under which the agents' types are mutually independent. 
{ Ambiguity sets of these three classes are parsimonious, that is, they require only limited prior distributional information such as knowledge of the minimal, maximal and most likely (or expected) types of all agents. Indeed, as the allocation problems studied in this paper are expected to occur infrequently, these indicators are the only properties of the unknown type distribution that can reasonable be extracted from scarce data, expert knowledge or common sense reasoning. Support-only and Markov ambiguity sets are widely studied in contexts where data is scarce (see, $e.g.$, \cite{ben2009robust, delage2010distributionally}). In particular, they are successfully used in other mechanism design problems (see, $e.g.$, 
\cite{bandi2014optimal, carrasco2018optimal, koccyiugit2020distributionally, suzdaltsev2020optimal, wang2020minimax}).} %We emphasize that support-only as well as Markov ambiguity sets contain distributions under which the agents' types are mutually dependent.

Pareto robust optimality is an important solution concept in robust optimization \citep{iancu2014pareto}. In the distributionally robust context considered here, a mechanism is called Pareto robustly optimal if it is not Pareto robustly dominated, that is, if there is no other mechanism that generates a non-inferior expected payoff under every distribution and a strictly higher expected payoff under at least one distribution in the ambiguity set. Every Pareto robustly optimal mechanism is also robustly optimal, but the converse implication is not true in general. Mechanisms that fail to be Pareto robustly optimal would not be used by any rational agent. 

The concept of Pareto robust optimality was invented because robustly optimal solutions are often highly degenerate, that is, typical (distributionally) robust optimization problems admit a multitude of robustly optimal solutions that all attain the  same worst-case expected payoff. However, most of these solutions underperform when average (non-worst-case) conditions prevail~\citep{iancu2014pareto}. This phenomenon is particularly pronounced in {\em adjustable} robust optimization~\citep{bertsimas2020paro}. We emphasize that the mechanism design problem studied in this paper can be viewed as an adjustable (distributionally) robust optimization problem because the allocation probabilities encoding different mechanisms represent functions of the agents' unknown types. { While adjustable robust optimization problems are generically NP-hard, we will show that the mechanism design problem at hand can be solved in closed form. However, we will also see} that this problem suffers from massive solution degeneracy. We emphasize that this degeneracy cannot be avoided if the goal is to optimize worst-case performance in the face of distributional ambiguity.

The Pareto robustly optimal mechanisms form a small subset of the family of all robustly optimal mechanisms, and they are the only robustly optimal mechanisms of practical value. Indeed, any other robustly optimal mechanism unnecessarily sacrifices performance under at least one distribution in the ambiguity set. Seeking Pareto robustly optimal mechanisms is therefore a natural goal. By identifying Pareto robustly optimal {\em favored-agent} mechanisms, we also establish a bridge to the classical theory of allocation with costly verification \citep{ben2014optimal}. On the methodological front, we { establish a new sufficient condition for Pareto robust optimality, which does not depend on any specifics of the mechanism design problem at hand and may therefore be useful for other applications, %---It is worthwhile to emphasize that this sufficient condition does not rely on any structural properties of the ambiguity set and any specifics of the mechanism design problem and can readily be extended to other distributionally robust optimization problems under arbitrary ambiguity sets---and} 
and we develop a new {\em spatial induction} technique for checking this condition. Spatial induction} exploits the lack of locality of the incentive compatibility constraints in our robust mechanism design problem. %{ in order to prove inductively that our new sufficient condition for Pareto robust optimality holds}. 
In fact, while the constraints of standard adjustable robust optimization problems are {\em local} in the sense that they are separable with respect to different uncertainty realizations, the incentive compatibility constraints of our robust mechanism design problem are {\em non-local} in the sense that they couple the allocation probabilities of any feasible mechanism across different types. {  We exploit this non-locality to prove that a judiciously chosen favorite-agent mechanism satisfies our new sufficient condition for Pareto robust optimality.

On a high level, our proof strategy can be explained as follows.} For the sake of contradiction, we first assume that there exists another feasible mechanism that Pareto robustly dominates the chosen favorite agent mechanism. Second, we use spatial induction to prove that both mechanisms must generate the {\em same} payoff in {\em every} scenario in the type space. To this end, we partition the type space into disjoint subsets that are tailored to the problem data and---in particular---to the ambiguity set at hand. We then use elementary arguments to prove that the two mechanisms generate the same payoff in every scenario in a first (particularly benign) subset; this is the {\em base step} of our spatial induction. Next, we use the non-locality of the incentive compatibility constraints to relate any scenario in the second subset to a scenario in the first subset of the type space. Exploiting our knowledge that the two mechanisms generate the same payoff throughout the first subset, we can then prove that they also generate the same payoff throughout the second subset. This is the first {\em induction step}. We then iterate through the remaining subsets of the type space one by one and apply each time a similar induction step that exploits the non-locality of the incentive compatibility constraints. This eventually proves that both mechanisms generate the same payoff throughout the {\em entire} type space, which contradicts the initial assumption that the prescribed favorite-agent mechanism is Pareto robustly dominated by some other mechanism. Hence, it is Pareto robustly optimal. Increasingly involved versions of this conceptual proof strategy based on spatial induction will be used to analyze support-only ambiguity sets as well as Markov ambiguity sets with and without independent types.
 
The main contributions of this paper can now be summarized as follows.
\begin{itemize}
	\item[(i)] For support-only ambiguity sets, we first show that not every robustly optimal mechanism represents a randomization over favored-agent mechanisms. This result is unexpected in view of the classical theory on stochastic mechanism design \citep{ben2014optimal}.
	We then construct an explicit favored-agent mechanism that is not only robustly optimal but also Pareto robustly optimal. This mechanism selects the favored agent from among those whose types have the highest possible lower bound, and it sets the threshold to this lower bound.
	\item[(ii)] For Markov ambiguity sets, we also construct an explicit favored-agent mechanism that is both robustly optimal as well as Pareto robustly optimal. This mechanism selects the favored agent from among those whose {\em expected} types have the highest possible lower bound, and it sets the threshold to the highest possible {\em actual} (not {\em expected}) type of the favored agent.
	\item[(iii)] For Markov ambiguity sets with independent types, we identify again a favored-agent mechanism that is robustly optimal as well as Pareto robustly optimal. Here, the favored agent is chosen exactly as under an ordinary Markov ambiguity set, but the threshold is set to the lowest possible {\em expected} (not {\em actual}) type of the favored agent. 
   \item[(iv)] {  We establish a new sufficient condition for Pareto robust optimality, which may be useful beyond distributionally robust mechanism design. We also develop a new spatial induction technique for proving that the above favored-agent mechanisms satisfy our sufficient condition and are therefore Pareto robustly optimal. This technique crucially exploits the non-locality of the incentive compatibility constraints of the distributionally robust mechanism design problem.}
    %\item[(v)] We develop a new spatial induction technique to prove the Pareto robust optimality of the above favored-agent mechanisms. This technique crucially exploits the non-locality of the incentive compatibility constraints of our mechanism design problem { and leverages the sufficient condition for Pareto robust optimality}.
\end{itemize}
Our results show that favored-agent mechanisms continue to play an important role in allocation with costly verification even if the unrealistic assumption of a commonly known type distribution is abandoned. In addition, they suggest that robust optimality alone is not a sufficiently distinctive criterion to single out practically useful mechanisms under distributional ambiguity. However, our results also show that among possibly infinitely many robustly optimal mechanisms, one can always find a simple and interpretable Pareto robustly optimal favored-agent mechanism. Unlike in the classical theory that assumes the type distribution to be known~\citep{ben2014optimal}, the favored agent as well as the threshold of our Pareto robustly optimal mechanisms are {\em in}dependent of the verification costs. 

%\noteck{Could we find a sentence about why this is good?} \notehib{This is because any optimal mechanism verifies no report under the worst-case distributions and always assigns the good to the agent with the maximum worst-case type. In other words, both ambiguity sets contain a worst-case distribution under which verification is sub-optimal. Hence, an optimal mechanism can be designed solely by looking at the ambiguity sets.}

\textit{Literature Review.} The first treatise of allocation with costly verification is due to \citet{townsend1979optimal}, who studies a principal-agent model with monetary transfers involving a single agent. \citet{ben2014optimal} extend this model to multiple agents but rule out the possibility of monetary transfers. Their seminal work has inspired considerable follow-up research in economics. For example, \citet{mylovanov2017optimal} study a variant of the problem where verification is costless, but the principal can impose only limited penalties and only partially recover the good when agents misreport their types. \citet{li2020mechanism} accounts both for costly verification and for limited penalties, thereby unifying the models in~\cite{ben2014optimal} and~\cite{mylovanov2017optimal}. \citet{chua2019optimal} further extend the model in~\cite{ben2014optimal} to multiple homogeneous goods, assuming that each agent can receive at most one good. \citet{bayrak2017optimal} spearhead the study of allocation with costly verification under distributional ambiguity. However, for reasons of computational tractability, they focus on ambiguity sets that contain only two discrete distributions. In this paper, we investigate ambiguity sets that contain infinitely many (not necessarily discrete) type distributions characterized by support and moment constraints, and we derive robustly as well as Pareto robustly optimal mechanisms in closed form.

This paper also contributes to the growing literature on (distributionally) robust mechanism design. Note that any mechanism design problem is inherently affected by uncertainty due to the private information held by different agents. The vast majority of the extant mechanism design literature models uncertainty through random variables that are governed by a commonly known probability distribution. The robust mechanism design literature, on the other hand, explicitly accounts for (non-stochastic) distributional uncertainty and seeks mechanisms that maximize the worst-case payoff, minimize the worst-case regret or minimize the worst-case cost in view of all distributions consistent with the information available. Robust mechanism design problems have recently emerged in different contexts such as pricing (see, $e.g.$, \cite{balseiro2021futility,carrasco2018optimal,chen2021screening,koccyiugit2021robust,pinar2017robust,wang2020minimax}), auction design (see, $e.g.$, \cite{anunrojwong2022robustness,bandi2014optimal,he2022correlation,koccyiugit2020distributionally,suzdaltsev2020optimal}) or contracting (see, $e.g.$, \cite{walton2022general}). This literature is too vast to be discussed in detail. %We refer interested readers to the papers cited and the references therein.
To our best knowledge, however, we are the first to derive closed-form optimal and Pareto robustly optimal mechanisms for the allocation problem with costly verification under distributional ambiguity. 
%{ Our paper is most closely related to the independent concurrent work by Chen et al.~\cite {chen2022information}, who also study allocation problems with costly verification under distributional uncertainty. They assume that the agents have only access to a noisy signal that correlates with their (unknown) types and that the principal has only access to limited information about the signal distribution. They identify the worst-case and best-case signal distributions for the principal and the agents. They also study a distributionally robust mechanism design problem over a (what we call a) Markov ambiguity set, where the signals have known means. However, Chen et~al.~\cite{chen2022information} do not address the multiplicity of robustly optimal mechanisms, and consequently they do not identify Pareto robustly optimal mechanisms.} 
Our paper is most closely related to the independent concurrent work by \citet{chen2022information}, who also study allocation problems with costly verification under distributional uncertainty. They assume that the agents have only access to a signal that correlates with their (unknown) types and that the principal has only access to the signal distribution, which is selected by a fictitious information designer.
They identify the worst- and best-case signal distributions for the principal and the best-case signal distributions for the agents. They also study a distributionally robust mechanism design problem over a (what we call a) Markov ambiguity set, where the agents' types have known means. However, \cite{chen2022information} do not address the multiplicity of robustly optimal mechanisms, and consequently they do not identify Pareto robustly optimal mechanisms.

The remainder of this paper is structured as follows. Section~\ref{sec: Problem Statement and Preliminaries} introduces our model and establishes preliminary results. Sections~\ref{sec: support only},~\ref{sec: Markov Ambiguity Sets} and~\ref{sec: Markov Ambiguity Sets with Independent Types} solve the proposed mechanism design problem for support-only ambiguity sets, Markov ambiguity sets, and Markov ambiguity sets with independent types, respectively. { Section \ref{sec: Numerical Illustration} assesses the performance of the proposed mechanisms numerically.} Conclusions are drawn in Section~\ref{sec:conclusions}, and all proofs are relegated to the online appendix.

\textit{Notation.} For any $\bm{t} \in \mathbb{R}^I$, we denote by $t_i$ the $i^{\text{th}}$ component and by $\bm{t}_{-i}$ the subvector of~$\bm t$ without~$t_i$. The indicator function of a logical expression~$E$ is defined as~$\mathds{1}_{E} = 1$ if~$E$ is true and as~$\mathds{1}_{E} = 0$ otherwise. For any Borel sets~$\mathcal{S} \subseteq \mathbb{R}^n$ and~$\mathcal{D} \subseteq \mathbb{R}^m$, we use~$\mathcal{P}_0(\mathcal{S})$ and~$\mathcal{L}(\mathcal{S}, \mathcal{D})$ to denote the family of all probability distributions on $\mathcal{S}$ and the set of all bounded Borel-measurable functions from~$\mathcal{S}$ to~$\mathcal{D}$, respectively. Random variables are designated by symbols with tildes ($e.g.$, $\tilde{\bm{t}}$), and their realizations are denoted by the same symbols without tildes ($e.g.$,~$\bm{t}$).

\section{Problem Statement and Preliminaries}\label{sec: Problem Statement and Preliminaries}

A principal aims to allocate a single good to one of~$I \geq 2$ agents. Each agent $i\in\mathcal{I}= \{1, 2,  \dots, I\}$ derives a strictly positive deterministic benefit from receiving the good and uses it to generate a value $t_i \in \mathcal{T}_i = [\underline{t}_i, \overline{t}_i]$ for the principal, where $0 \leq \underline{t}_i < \overline{t}_i < \infty$. We henceforth refer to $t_i$ as agent~$i$'s type, and we assume that~$t_i$ is privately known to agent~$i$ but unknown to the principal and the other agents. Thus, the principal perceives the vector $\tilde{\bm{t}} = (\tilde{t}_1, \tilde{t}_2 \dots, \tilde{t}_I )$ of all agents' types as a random vector governed by some probability distribution $\mathbb{P}_0$ on the type space~$\mathcal{T} = \prod_{i \in \mathcal{I}} \mathcal{T}_i$. However, the principal can inspect agent~$i$'s type at cost $c_i > 0$, and the inspection perfectly reveals~$t_i$.  In contrast to much of the existing literature on mechanism design, we assume here that neither the principal nor the agents know~$\mathbb{P}_0$. Instead, they are only aware that~$\mathbb{P}_0$ belongs to some commonly known ambiguity set $\mathcal{P} \subseteq \mathcal{P}_0(\mathcal{T})$. 
%Even though the probability distribution $\mathbb{P}_0$ is unknown to the agents and the principal, the ambiguity set $\mathcal{P}$ is commonly known. 
On this basis, the principal aims to design a mechanism for allocating the good. A mechanism is an extensive-form game between the principal and the agents, where the principal commits in advance to her strategy (for a formal definition of extensive-form games, see, $e.g.$, \cite{fudenberg1991game}). %The agents' actions are cheap talk statements
Such a mechanism may involve multiple stages of cheap talk statements by the agents, while the principal's actions include the decisions on whether to inspect certain agents and how to allocate the good. Monetary transfers are not allowed, $i.e.$, the agents and the principal cannot exchange money at any time.

%In principle, a mechanism can be very complex, $e.g.$, it could have a dynamic structure where the agents and the principal act sequentially. 

%It could ``have  multiple stages of cheap talk statements by the agents and checking by the principal, where who can speak and which agents are checked depend on past statements and the results from past checks, finally culminating in the allocation of the good, perhaps to no one.'' 

%Any mechanism of this nature can be thought of as an extensive form game between the agents and the principal, where the principal is committed in advance to her strategy. An extensive form game contains the information: set $\mathcal{I} \cup \{\text{principal}\}$ of players, the order of moves represented by a game tree $T$ that is a finite collection of ordered nodes $x \in \mathcal{X}$, 

Given any mechanism represented as an extensive form game, we denote by $\mathcal{H}_i$ the family of all information sets of agent $i$ and by $\mathcal{A}(h_i)$ the actions available to agent $i$ at the nodes in the information set $h_i\in \mathcal{H}_i$. 
All agents select their actions strategically in view of their individual preferences and the available information. In particular, agent~$i$'s actions depend on his type~$t_i$. Thus, we model any (mixed) strategy of agent~$i$ as a function $s_i \in \mathcal{L}(\mathcal{T}_i, \prod_{h_i \in \mathcal{H}_i} \mathcal{P}_0 (\mathcal{A}(h_i)))$ that maps each of his possible types to a complete contingency plan~$a_i \in \prod_{h_i \in \mathcal{H}_i} \mathcal{P}_0 (\mathcal{A}(h_i))$, which represents a probability distribution over the actions available to agent~$i$ for all information sets~$h_i\in\mathcal{H}_i$. 
% for every~$i\in\mathcal{I}$.
%Consider any such mechanism and denote by $\mathcal{A}_i$ the set of available action plans to agent $i$, $i \in \mathcal{I}$. An action plan $a_i \in \mathcal{A}_i$ is a complete contingency plan specifying the action to be taken at each information set. Given a mechanism, the agents select their action plans strategically to induce the most desirable outcome in view of their individual preferences. In particular, agent~$i$ selects an action plan depending on his type $t_i$. Thus, his (possibly mixed) strategy must be modeled as a function $s_i : \mathcal{T}_i \rightarrow \mathcal{P}_0(\mathcal{A}_i)$ that maps each of his possible types to a probability distribution over the set $\mathcal{A}_i$ of actions. 
In the following, we denote by $\text{prob}_i(a_i; \bm{t}, \bm{a}_{-i})$ the probability that agent $i \in \mathcal{I}$ receives the good under the principal's mechanism if the agents have types $\bm{t}$ and play the contingency plans~$\bm{a} = (a_1, a_2, \dots, a_I)$. We also restrict attention to mechanisms that admit an ex-post Nash equilibrium.

%Note that the probability $prob_i(a_i; \bm{t}, \bm{a}_{-i})$ depends on the mechanism fixed by the principal. 
\begin{definition}[Ex-Post Nash Equilibrium]
An $I$-tuple $\bm{s} = (s_1, s_2, \dots, s_I)$ of mixed strategies $s_i \in \mathcal{L}(\mathcal{T}_i, \prod_{h_i \in \mathcal{H}_i} \mathcal{P}_0 (\mathcal{A}(h_i)))$, $i\in\mathcal{I}$, is called an {\em ex-post Nash equilibrium} if 
\begin{equation*}
\begin{aligned}
\text{\rm prob}_i(s_i(t_i); \bm{t}, \bm{s}_{-i}(\bm{t}_{-i})) \geq \text{\rm prob}_i(a_i; \bm{t}, \bm{s}_{-i}(\bm{t}_{-i})) \quad \forall i \in \mathcal{I},\, \forall \bm{t} \in \mathcal{T},\, \forall a_i \in {\textstyle \prod}_{h_i \in \mathcal{H}_i} \mathcal{P}_0 (\mathcal{A}(h_i)).
\end{aligned}
\end{equation*}
\end{definition}

{  Recall that all agents assign a strictly positive deterministic value to the good, and therefore the expected utility of agent~$i$ conditional on~$\tilde{\bm{t}}=\bm{t}$ must grow proportionally to $\text{prob}_i(a_i; \bm{t}, \bm{a}_{-i})$ under {\em any} non-decreasing utility function.}
%The definition of an ex-post Nash equilibrium thus captures rational behavior of the agents under any reasonable decision criteria.} %---We emphasize that our only assumption about the expected utility of agent~$i$ conditional on~$\tilde{\bm{t}}=\bm{t}$ is that it is proportional to the probability of receiving the good, which is reasonable to expect for any rational agent---}. 
In an ex-post Nash equilibrium, each agent~$i$ maximizes this probability simultaneously for all type scenarios~$\bm{t}\in\mathcal{T}$. Hence, it is clear that insisting on the existence of an ex-post Nash equilibrium restricts the family of mechanisms to be considered. Note that \citet{ben2014optimal} study the larger class of mechanisms that admit a Bayesian Nash equilibrium. However, these mechanisms generically depend on the type distribution~$\mathbb{P}_0$ and can therefore not be implemented by a principal who lacks knowledge of~$\mathbb{P}_0$. %, however, the principal is unable to predict the outcome of a Bayesian Nash equilibrium. 
It is therefore natural to restrict attention to mechanisms that admit ex-post Nash equilibria, which remain well-defined in the face of distributional ambiguity. { 
In Section~\ref{sec:Robustness Against the Agents’ Attitude Towards Ambiguity}, we show, by restricting attention to mechanisms that admit an ex-post Nash equilibrium, that the principal also hedges against uncertainty about the agents’ attitude towards ambiguity.} We further assume from now on that the principal is ambiguity averse in the sense that she wishes to maximize her worst-case expected payoff in view of all distributions in the ambiguity set~$\mathcal{P}$. 

The class of all mechanisms that admit an ex-post Nash equilibrium is vast. An important subclass is the family of all truthful direct mechanisms. A direct mechanism~$(\bm{p}, \bm{q})$ consists of two $I$-tuples $\bm{p} = (p_1, p_2, \dots, p_I)$ and $\bm{q} = (q_1, q_2, \dots, q_I)$ of allocation functions $p_i, q_i \in \mathcal{L}(\mathcal{T}, [0,1])$, $i \in \mathcal{I}$. Any direct mechanism $(\bm{p}, \bm{q})$ is implemented as follows. First, the principal announces~$\bm{p}$ and $\bm{q}$, and then she collects a bid $t'_i \in \mathcal{T}_i$ from each agent~$i\in\mathcal{I}$. Next, the principal implements randomized allocation and inspection decisions. Specifically, $p_i(\bm{t}')$ represents the total probability that agent~$i$ receives the good, while~$q_i(\bm{t}')$ represents the probability that agent~$i$ receives the good {\em and} is inspected. If the inspection reveals that agent $i$ has misreported his type, the principal penalizes the agent by repossessing the good. %In summary, if the bids are given by~$\bm{t}'$, then $p_i(\bm{t}')$ denotes the probability that agent~$i$ receives the resource, and $q_i(\bm{t}')$ denotes the probability that agent $i$ is inspected. 
Any direct mechanism $(\bm{p}, \bm{q})$ must satisfy the feasibility conditions
\begin{equation}\label{eq:FC} \tag{FC}
\begin{aligned}
q_i(\bm{t}') \leq p_i(\bm{t}') \;\;\forall i \in \mathcal{I} \quad \text{and}\quad \sum_{i \in \mathcal{I}} p_i(\bm{t}') \leq 1 \;\;\forall \bm{t}' \in \mathcal{T}.
\end{aligned}
\end{equation}
The first inequality in \eqref{eq:FC} holds because only agents who receive the good may undergo an inspection. The second inequality in \eqref{eq:FC} ensures that the principal allocates the good at most once. A direct mechanism~$(\bm{p}, \bm{q})$ is called truthful if it is optimal for each agent~$i$ to report his true type~$t'_i=t_i$. Thus, $(\bm{p}, \bm{q})$ is truthful if and only if it satisfies the incentive compatibility constraints
\begin{equation}\label{eq:IC} \tag{IC}
\begin{aligned}
p_i(\bm{t}) \geq p_i(t'_i, \bm{t}_{-i}) - q_i(t'_i, \bm{t}_{-i}) &&\forall i \in \mathcal{I},\, \forall t'_i \in \mathcal{T}_i,\, \forall \bm{t} \in \mathcal{T},
\end{aligned}
\end{equation} 
which ensure that if all other agents report their true types $\bm{t}_{-i}$, then the probability $p_i(\bm{t})$ of agent~$i$ receiving the good if he reports his true type $t_i$ exceeds the probability $p_i(t'_i, \bm{t}_{-i}) - q_i(t'_i, \bm{t}_{-i})$ of agent $i$ receiving the good if he misreports his type as~$t'_i\neq t_i$. %As every agent strictly prefers receiving the resource to not receiving it, the incentive compatibility constraints can be expressed only in terms of the probabilities agents receive the resource; see \cite{ben2014optimal} for details.
By leveraging a variant of the Revelation Principle detailed in \cite{ben2014optimal}, one can show that for any mechanism that admits an ex-post Nash equilibrium, there exists an equivalent truthful direct mechanism that duplicates or improves the principal's worst-case expected payoff; see the online appendix of \cite{ben2014optimal} for details. Without loss of generality, the principal may thus focus on truthful direct mechanisms, which greatly simplifies the problem of finding an optimal mechanism.
Consequently, the principal's mechanism design problem can be formalized as the following distributionally robust optimization problem.
\begin{equation}\label{eq:MDP} \tag{MDP}
\begin{aligned}
z^\star \;=\; &\sup_{\bm{p}, \bm{q}} &&\inf_{\mathbb{P} \in \mathcal{P}} \; \mathbb{E}_{\mathbb{P}}\left[\sum_{i \in \mathcal{I}} (p_i(\tilde{\bm{t}}) \tilde{t}_i - q_i(\tilde{\bm{t}}) c_i) \right]\\
&\;\text{s.t.} && p_i, q_i \in \mathcal{L}(\mathcal{T}, [0,1]) \;\;\forall i \in \mathcal{I}\\
&&&\eqref{eq:IC}, \; \eqref{eq:FC}
\end{aligned}
\end{equation}
From now on, we will use the shorthand $\mathcal X$ to denote the set of all $(\bm{p},\bm{q})$ feasible in \eqref{eq:MDP}. Note that the feasible set $\mathcal X$ does not depend on the ambiguity set~$\mathcal P$. { Recall also that a maximization (minimization) problem is called {\em solvable} if its supremum (infimum) is attained by a feasible solution. Thus, problem~\eqref{eq:MDP} is solvable if there is~$(\bm p^\star,\bm q^\star)\in\mathcal X$ with worst-case expected payoff~$z^\star$.}
%Cagil:This is not really necessary but I add it to address a comment by AE.
%, $i.e.$, $(\bm{p},\bm{q}) \in \mathcal X$ if $p_i,q_i \in \mathcal{L}(\mathcal{T}, [0,1])$ for $i \in \mathcal{I}$ and they satisfy the constraints \eqref{eq:IC} and \eqref{eq:FC}.

In the remainder, we will demonstrate that~\eqref{eq:MDP} often admits multiple optimal solutions. While different optimal mechanisms generate the same expected profit in the worst case, they may offer dramatically different expected profits under generic non-worst-case distributions. This observation prompts us to seek mechanisms that are not only optimal but perform also well under {\em all} type distributions in the ambiguity set~$\mathcal{P}$. More precisely, we hope to identify an optimal mechanism for which there exists no other feasible mechanism that generates a non-inferior expected payoff under {\em every} distribution in~$\mathcal{P}$ and a higher expected payoff under {\em at least one} distribution in~$\mathcal{P}$. A mechanism with this property is called {\it Pareto robustly optimal}. This terminology is borrowed from the theory of Pareto efficiency in classical robust optimization~\citep{iancu2014pareto}.

\begin{definition}[Pareto Robust Optimality]
We say that a mechanism $(\bm{p}', \bm{q}')$ that is feasible in~\eqref{eq:MDP} \textit{weakly Pareto robustly dominates} another feasible mechanism $(\bm{p}, \bm{q})$ if
%A mechanism $(\bm{p}, \bm{q})$ that is feasible in \eqref{eq:MDP} is called Pareto robustly optimal if there exists no other feasible mechanism $(\bm{p}', \bm{q}')$ with
\begin{equation}\label{eq:dominance}
    \mathbb{E}_{\mathbb{P}}\left[\sum_{i \in \mathcal{I}} (p'_i(\tilde{\bm{t}}) \tilde{t}_i - q'_i(\tilde{\bm{t}}) c_i) \right] \geq  \mathbb{E}_{\mathbb{P}}\left[\sum_{i \in \mathcal{I}} (p_i(\tilde{\bm{t}}) \tilde{t}_i - q_i(\tilde{\bm{t}}) c_i) \right] \quad\forall \mathbb{P} \in \mathcal{P}.
\end{equation}
If the inequality~\eqref{eq:dominance} holds for all $\mathbb P\in\mathcal P$ and is strict for at least one $\mathbb P\in\mathcal P$, we say that~$(\bm{p}', \bm{q}')$ Pareto robustly dominates $(\bm{p}, \bm{q})$. A mechanism $(\bm{p}, \bm{q})$ that is optimal in \eqref{eq:MDP} is called Pareto robustly optimal if there exists no other feasible mechanism $(\bm{p}', \bm{q}')$ that Pareto robustly dominates~$(\bm{p}, \bm{q})$.
%where the inequality is strict for at least one distribution $\mathbb{P} \in \mathcal{P}$.
\end{definition}

%If~\eqref{eq:dominance} holds for every~$\mathbb{P}\in\mathcal{P}$, we say that $(\bm{p}', \bm{q}')$ \textit{weakly} Pareto robustly dominates $(\bm{p}, \bm{q})$. If, in addition, \eqref{eq:dominance} is strictly satisfied for some $\mathbb{P} \in \mathcal{P}$, we say that $(\bm{p}', \bm{q}')$ Pareto robustly dominates~$(\bm{p}, \bm{q})$. 
Note that any mechanism that weakly Pareto robustly dominates an optimal mechanism is also optimal in \eqref{eq:MDP}. Moreover, a Pareto robustly optimal mechanism typically exists. However, there may not exist any mechanism that Pareto robustly dominates all other feasible mechanisms.

%\subsection{Favored-Agent Mechanisms}

We now define the notion of a favored-agent mechanism, which was first introduced in \cite{ben2014optimal}.

%In the context of allocation with costly verification, the most well-known mechanisms are the . 
%{\color{red}Definition... Feasibility... Tell about optimality results in non-ambiguous setting...}

\begin{definition}[Favored-Agent Mechanism]\label{def:fav-agent-mechanism}
	A mechanism $(\bm{p}, \bm{q})$ is a {\em favored-agent mechanism} if there is a favored agent $i^\star \in \mathcal{I}$ and a threshold value $\nu^\star \in \mathbb{R}$ such that the following~hold.
	\begin{itemize}
	\item[(i)] If $\max_{i \neq i^\star} t_i-c_i < \nu^\star$, then $p_{i^\star}(\bm{t})=1$, $q_{i^\star}(\bm{t})=0$ and $p_i(\bm{t}) = q_i(\bm{t})=0$ for all $i \neq i^\star$.
	\item[(ii)] If $\max_{i \neq i^\star} t_i-c_i > \nu^\star$, then $p_{i'}(\bm{t}) = q_{i'}(\bm{t})=1$ for some $i' \in \arg \max_{i \in \mathcal{I}} (t_i-c_i) $ and $p_i(\bm{t}) = q_i(\bm{t})=0$ for all $i \neq i'$.
	\end{itemize}
If $\max_{i\neq i^\star} t_i-c_i =\nu^\star$, then we are free to define~$(\bm{p}(\bm{t}), \bm{q}(\bm{t}))$ either as in~(i) or as in~(ii).
\end{definition}

%Figure Favored-agent mechanism definition
% Pattern Info
 
\tikzset{
pattern size/.store in=\mcSize, 
pattern size = 5pt,
pattern thickness/.store in=\mcThickness, 
pattern thickness = 0.3pt,
pattern radius/.store in=\mcRadius, 
pattern radius = 1pt}
\makeatletter
\pgfutil@ifundefined{pgf@pattern@name@_ifjv4t7dw}{
\pgfdeclarepatternformonly[\mcThickness,\mcSize]{_ifjv4t7dw}
{\pgfqpoint{0pt}{-\mcThickness}}
{\pgfpoint{\mcSize}{\mcSize}}
{\pgfpoint{\mcSize}{\mcSize}}
{
\pgfsetcolor{\tikz@pattern@color}
\pgfsetlinewidth{\mcThickness}
\pgfpathmoveto{\pgfqpoint{0pt}{\mcSize}}
\pgfpathlineto{\pgfpoint{\mcSize+\mcThickness}{-\mcThickness}}
\pgfusepath{stroke}
}}
\makeatother

% Pattern Info
 
\tikzset{
pattern size/.store in=\mcSize, 
pattern size = 5pt,
pattern thickness/.store in=\mcThickness, 
pattern thickness = 0.3pt,
pattern radius/.store in=\mcRadius, 
pattern radius = 1pt}
\makeatletter
\pgfutil@ifundefined{pgf@pattern@name@_vx41q1cae}{
\pgfdeclarepatternformonly[\mcThickness,\mcSize]{_vx41q1cae}
{\pgfqpoint{0pt}{0pt}}
{\pgfpoint{\mcSize+\mcThickness}{\mcSize+\mcThickness}}
{\pgfpoint{\mcSize}{\mcSize}}
{
\pgfsetcolor{\tikz@pattern@color}
\pgfsetlinewidth{\mcThickness}
\pgfpathmoveto{\pgfqpoint{0pt}{0pt}}
\pgfpathlineto{\pgfpoint{\mcSize+\mcThickness}{\mcSize+\mcThickness}}
\pgfusepath{stroke}
}}
\makeatother

% Pattern Info
 
\tikzset{
pattern size/.store in=\mcSize, 
pattern size = 5pt,
pattern thickness/.store in=\mcThickness, 
pattern thickness = 0.3pt,
pattern radius/.store in=\mcRadius, 
pattern radius = 1pt}
\makeatletter
\pgfutil@ifundefined{pgf@pattern@name@_qhgl2nnj1}{
\pgfdeclarepatternformonly[\mcThickness,\mcSize]{_qhgl2nnj1}
{\pgfqpoint{-\mcThickness}{-\mcThickness}}
{\pgfpoint{\mcSize}{\mcSize}}
{\pgfpoint{\mcSize}{\mcSize}}
{
\pgfsetcolor{\tikz@pattern@color}
\pgfsetlinewidth{\mcThickness}
\pgfpathmoveto{\pgfpointorigin}
\pgfpathlineto{\pgfpoint{0}{\mcSize}}
\pgfusepath{stroke}
}}
\makeatother
\tikzset{every picture/.style={line width=0.75pt}} %set default line width to 0.75pt        

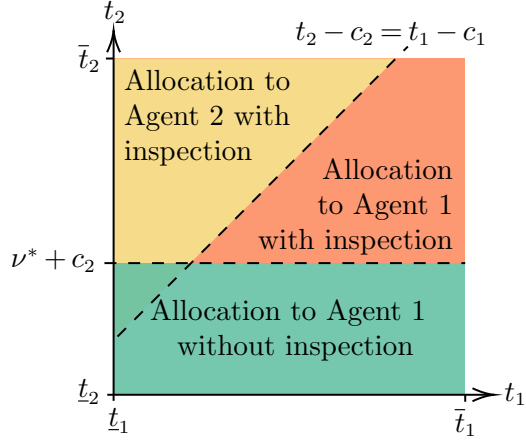
\begin{figure}
    \centering

\begin{tikzpicture}[x=0.75pt,y=0.75pt,yscale=-1,xscale=1]

%Shape: Rectangle [id:dp42648192421898856] 
\draw  [draw opacity=0][fill={rgb, 255:red, 252; green, 141; blue, 98 }  ,fill opacity=0.9 ] (252.25,40.72) -- (429.59,40.72) -- (429.59,143.87) -- (252.25,143.87) -- cycle ;
%Shape: Rectangle [id:dp5526954899105041] 
\draw  [draw opacity=0][pattern=_ifjv4t7dw,pattern size=8.774999999999999pt,pattern thickness=0.75pt,pattern radius=0pt, pattern color={rgb, 255:red, 255; green, 255; blue, 255}] (252.25,39.86) -- (429.59,39.86) -- (429.59,142.86) -- (252.25,142.86) -- cycle ;
%Shape: Right Triangle [id:dp9674540945739405] 
\draw  [draw opacity=0][fill={rgb, 255:red, 246; green, 220; blue, 138 }  ,fill opacity=1 ] (252.25,182.48) -- (394.65,40.72) -- (252.25,40.72) -- cycle ;
%Shape: Right Triangle [id:dp16718568141189905] 
\draw  [draw opacity=0][pattern=_vx41q1cae,pattern size=8.55pt,pattern thickness=0.75pt,pattern radius=0pt, pattern color={rgb, 255:red, 255; green, 255; blue, 255}] (252.25,182.48) -- (396.18,39.86) -- (252.25,39.86) -- cycle ;
%Straight Lines [id:da15767309707066413] 
\draw    (248.36,40.72) -- (252.25,40.72) ;
%Straight Lines [id:da39211708326032935] 
\draw    (429.59,213.96) -- (429.59,210.12) ;
%Shape: Rectangle [id:dp6481937948297329] 
\draw  [draw opacity=0] (252.25,94.74) -- (429.59,94.74) -- (429.59,210.12) -- (252.25,210.12) -- cycle ;
%Shape: Rectangle [id:dp9302020864830929] 
\draw  [draw opacity=0][fill={rgb, 255:red, 102; green, 194; blue, 165 }  ,fill opacity=0.9 ] (252.25,143.87) -- (429.59,143.87) -- (429.59,210.12) -- (252.25,210.12) -- cycle ;
%Shape: Rectangle [id:dp40078216748053563] 
\draw  [draw opacity=0][pattern=_qhgl2nnj1,pattern size=6.8999999999999995pt,pattern thickness=1.1pt,pattern radius=0pt, pattern color={rgb, 255:red, 255; green, 255; blue, 255}] (251.48,143.87) -- (428.82,143.87) -- (428.82,213.96) -- (251.48,213.96) -- cycle ;
%Straight Lines [id:da09330406232710797] 
\draw  [dash pattern={on 4.5pt off 4.5pt}]  (247.64,143.98) -- (422.42,143.98) -- (429.59,143.98) ;
%Straight Lines [id:da6351934800668948] 
\draw  [dash pattern={on 4.5pt off 4.5pt}]  (254.25,181.05) -- (400.65,34.66) ;
%Straight Lines [id:da07080944225532093] 
\draw    (249.07,210.43) -- (442.18,210.43) ;
\draw [shift={(444.18,210.43)}, rotate = 180] [color={rgb, 255:red, 0; green, 0; blue, 0 }  ][line width=0.75]    (10.93,-3.29) .. controls (6.95,-1.4) and (3.31,-0.3) .. (0,0) .. controls (3.31,0.3) and (6.95,1.4) .. (10.93,3.29)   ;
%Straight Lines [id:da9973655119240143] 
\draw    (252.25,213.5) -- (252.25,28.61) ;
\draw [shift={(252.25,26.61)}, rotate = 90] [color={rgb, 255:red, 0; green, 0; blue, 0 }  ][line width=0.75]    (10.93,-3.29) .. controls (6.95,-1.4) and (3.31,-0.3) .. (0,0) .. controls (3.31,0.3) and (6.95,1.4) .. (10.93,3.29)   ;

% Text Node
\draw (245.85,10) node [anchor=north west][inner sep=0.75pt]    {$t_{2}$};
% Text Node
\draw (447.97,203) node [anchor=north west][inner sep=0.75pt]    {$t_{1}$};
% Text Node
\draw (342.5,21) node [anchor=north west][inner sep=0.75pt]    {$t_{2} -c_{2} =t_{1} -c_{1}$};
% Text Node
\draw (199,135) node [anchor=north west][inner sep=0.75pt]    {$\nu ^{*} +c_{2}$};
% Text Node
\draw (422.77,215.58) node [anchor=north west][inner sep=0.75pt]    {$\overline{t}_{1}$};
% Text Node
\draw (248.15,215.46) node [anchor=north west][inner sep=0.75pt]    {$\underline{t}_{1}$};
% Text Node
\draw (232.87,201) node [anchor=north west][inner sep=0.75pt]    {$\underline{t}_{2}$};
% Text Node
\draw (233,31.88) node [anchor=north west][inner sep=0.75pt]    {$\overline{t}_{2}$};
% Text Node
\draw (344,178) node [align=left][inner sep=0.75pt] {\begin{minipage}[lt]{110pt}\setlength\topsep{0pt}
Allocation to Agent 1
\begin{center}
    without inspection
\end{center}
\end{minipage}};
% Text Node
\draw (302,90) node [anchor=north west][inner sep=0.75pt]   [align=left] {\begin{minipage}[lt]{90pt}\setlength\topsep{0pt}
\begin{flushright}
Allocation\\ to Agent 1\\ with inspection
\end{flushright}

\end{minipage}};
% Text Node
\draw (256,45) node [anchor=north west][inner sep=0.75pt]   [align=left] {\begin{minipage}[lt]{90pt}\setlength\topsep{0pt}
\begin{flushleft}
Allocation to\\ Agent 2 with\\ inspection
\end{flushleft}

\end{minipage}};

\end{tikzpicture}

\caption{{ A favored-agent mechanism for two agents with favored agent $i^\star = 1$ and threshold value $\nu^\star$.}\label{favored-agent-figure}}
\end{figure}

%{\color{red}Type 1 favored-agent mechanism ($\leq$ in (i)) and type 2 favored-agent mechanism ($\geq$ in (ii))}

%For ease of exposition, we will henceforth use the lexicographic tie-breaker that sets $p_{i'}(\bm{t}) = q_{i'}(\bm{t})=1$ for $i' = \min \arg \max_{i \in \mathcal{I}} (t_i-c_i)$ in item $(ii)$. We emphasize that our results will not depend on this particular choice.
Intuitively, if $t_i$ is smaller than the adjusted cost of inspection $c_i+ \nu^\star$ for every agent $i \neq i^\star$, then we are in case~(i), and the favored-agent mechanism allocates the good to the favored agent~$i^\star$ without inspection. %In this case, the principal inspects none of the agents. 
If there exists an agent $i \neq i^\star$ whose type $t_{i}$ exceeds the adjusted cost of inspection $c_{i} + \nu^\star$, then we are in case~(ii), and the favored-agent mechanism allocates the good to an agent~$i'$ with highest net payoff~$t_{i'}-c_{i'}$, 
%$t_{i'}-c_{i'}\geq t_i-c_i$ for all~$i'\in\mathcal{I}$, 
and this agent is inspected. 
%{\em with} inspection.
Note that in case~(ii) the good can also be allocated to the favored agent. { Figure~\ref{favored-agent-figure} illustrates the allocations of a favored-agent mechanism in the special case when there are only two agents.}

A favored-agent mechanism is uniquely determined by a favored agent $i^\star$, a threshold value~$\nu^\star$, and two tie-breaking rules. The first tie-breaking rule determines the winning agent in case~(ii) when $\arg \max_{i \in \mathcal{I}} (t_i-c_i)$ is not a singleton. From now on we will always use the lexicographic tie-breaking rule in this case, which sets $i' = \min \arg \max_{i \in \mathcal{I}} (t_i-c_i)$. The second tie-breaking rule determines whether $(\bm{p}(\bm{t}), \bm{q}(\bm{t}))$ should be constructed as in case~(i) or as in case~(ii) when $\max_{i\neq i^\star} t_i-c_i =\nu^\star$. From now on we say that a favored-agent mechanism is of type~(i) if $(\bm{p}(\bm{t}), \bm{q}(\bm{t}))$ is always defined as in~(i) and that it is of type~(ii) if $(\bm{p}(\bm{t}), \bm{q}(\bm{t}))$ is always defined as in~(ii) in case of a tie. Note that both tie-breaking rules are irrelevant in the Bayesian setting considered in \citet{ben2014optimal}, but they are relevant for us because the ambiguity sets~$\mathcal P$ to be studied below contain discrete distributions, under which ties have a strictly positive probability.

%that determines who receives the good in case of ties, that is, when the set $\arg \max_{i \in \mathcal{I}} (t_i-c_i)$ is not a singleton, {\color{red}and the definition of~$(\bm{p}(\bm{t}), \bm{q}(\bm{t}))$ when $\max_{i\neq i^\star} t_i-c_i =\nu^\star$, that is, whether to define~$(\bm{p}(\bm{t}), \bm{q}(\bm{t}))$ as in~(i) or as in~(ii). From now on, we say that a favored-agent mechanism is of {\it type 1} if $(\bm{p}(\bm{t}), \bm{q}(\bm{t}))$ is defined as in~(i) and is of {\it type 2} if $(\bm{p}(\bm{t}), \bm{q}(\bm{t}))$ is defined as in~(ii) when $\max_{i\neq i^\star} t_i-c_i =\nu^\star$. Note that  Ben-Porath et al. \cite{ben2014optimal} do not define $(\bm{p}(\bm{t}), \bm{q}(\bm{t}))$ when $\max_{i \neq i^\star} t_i-c_i = \nu^\star$ as the authors assume $\mathbb P_0$ to be continuous under which case the probability of $\tilde{\bm{t}} = \bm t$ is zero. On the other hand, we assume that $\mathbb{P}_0$ belongs to an ambiguity set $\mathcal P$ that contains discrete distributions. Thus, the definition of~$(\bm{p}(\bm{t}), \bm{q}(\bm{t}))$ when $\max_{i \neq i^\star} t_i-c_i = \nu^\star$ plays a role in our analyses as the probability of $\tilde{\bm{t}} = \bm t$ is strictly positive under some $\mathbb P \in \mathcal P$.}

All favored-agent mechanisms are feasible in \eqref{eq:MDP}, see Remark~1 in \cite{ben2014optimal}. In particular, they are incentive compatible, that is, the agents have no incentive to misreport their types. To see this, recall that under a favored-agent mechanism the winning agent receives the good with probability one, and the losing agents receive the good with probability zero. Thus, if an agent wins by truthful bidding, he cannot increase his chances of receiving the good by lying about his type. If an agent loses by truthful bidding, on the other hand, he has certainly no incentive to lower his bid $t_i$ because the chances of receiving the good are non-decreasing in~$t_i$. Increasing his bid~$t_i$ may earn him the good provided that $t_{i} - c_i$ attains the maximum of $t_{i'}-c_{i'}$ over $i'\in\mathcal{I}$. However, in this case, the agent's type is inspected with probability one. Hence, the lie will be detected and the good will be repossessed. This shows that no agent benefits from lying under a favored-agent mechanism.

If $\mathcal{P} = \{\mathbb{P}_0\}$ is a singleton, the agents' types are independent under~$\mathbb{P}_0$, and~$\mathbb{P}_0$ has an everywhere positive density on~$\mathcal{T}$, then problem \eqref{eq:MDP} is solved by a favored-agent mechanism~\citep[Theorem~1]{ben2014optimal}. The favored-agent mechanism with favored agent~$i$ and threshold~$\nu_i$ generates an expected payoff~of
\begin{equation*}
\begin{aligned}
    &\mathbb{E}_{\mathbb{P}_0}\left[ \tilde t_i \mathds{1}_{\tilde y_i\leq\nu_i} +\max\left\{\tilde t_i-c_i, \tilde y_i \right\} \mathds{1}_{\tilde y_i\geq \nu_i}\right]\\[1ex]
    &= \int_{-\infty}^{\nu_i} \mathbb{E}_{\mathbb{P}_0}\left[ \tilde t_i\right] \rho_i(y_i){\rm d}y_i+ \int_{\nu_i}^\infty \mathbb{E}_{\mathbb{P}_0}\left[ \max\left\{\tilde t_i-c_i, y_i \right\} \right] \rho_i(y_i){\rm d}y_i,
     \end{aligned}
\end{equation*}
where the random variable~$\tilde y_i= \max_{j\neq i} \tilde t_j-c_j$ with probability density function~$\rho_i(y_i)$ is independent of~$\tilde t_i$ under~$\mathbb P_0$. The threshold value~$\nu_i^\star$ that maximizes this expression thus solves the first-order optimality condition
%In addition, an optimal favored-agent mechanism can be constructed as follows (see~\cite[Theorems~2 and~3]{ben2014optimal}). Assign each agent~$i\in\mathcal{I}$ a threshold~$\nu_i^\star$ defined as the unique solution of the equation
\begin{align}
\label{eq:Ben-Porath interpretation}
\mathbb{E}_{\mathbb{P}_0} \left[\tilde t_i\right]=\mathbb{E}_{\mathbb{P}_0}\left[\max\left\{\tilde t_i-c_i, \nu_i\right\}\right].
\end{align}
Note that $\nu_i^\star$ is unique because the right-hand side of~\eqref{eq:Ben-Porath interpretation} strictly increases in~$\nu_i$ on the domain of interest; see~\cite[Theorem~2]{ben2014optimal} for additional details. One can further prove that within the finite class of favored-agent mechanisms with optimal thresholds, the ones with the highest threshold are optimal. More specifically, any favored-agent mechanism with favored agent~$i^\star \in \arg \max_{i \in \mathcal{I}} \nu_i^\star$ and threshold~$\nu^\star = \max_{i \in \mathcal{I}} \nu_i^\star$ is optimal within the class of favored-agent mechanisms \cite[Theorem~3]{ben2014optimal}. Hence, any such mechanism must be optimal in~\eqref{eq:MDP}.
%Thus, the favored agent's threshold coincides with the threshold $\nu^\star$ used in the mechanism. 
%For the subsequent discussion we define $i'\in \arg\max_{i\neq i'}t_i-c_i$ as any agent with second highest agent-wise threshold
%One can show that if~$\max_{j \neq i} t_j-c_j=\nu^\star_i$, then the principal is indifferent between allocating the good to agent~$i$ without checking (resulting in an expected profit equal to the left-hand side of~\eqref{eq:Ben-Porath interpretation}) or to an agent~$j$ with the highest net payoff~$t_j-c_j$ after a check. If~$i$ was the favored agent, $\nu^\star_i$ would thus naturally represent the threshold that the net payoffs of the agents~$j\neq i$ should meet to prevent the principal from allocating the good to agent~$i$ without inspection. The optimal favored agent~$i^\star$ maximizes this threshold. 
Finally, one can also show that for mutually distinct cost coefficients~$c_i$, $i\in\mathcal{I}$, the optimal favored-agent mechanism is unique.

In the remainder of the paper, we will address instances of the mechanism design problem~\eqref{eq:MDP} where~$\mathcal{P}$ is {\em not} a singleton, and we will prove that favored-agent mechanisms remain optimal. Under distributional ambiguity, however, the construction of~$i^\star$ and $\nu^\star$ described above is no longer well-defined because it depends on a particular choice of the probability distribution of~$\tilde{\bm t}$. We will show that if $\mathcal{P}$ is not a singleton, then there may be infinitely many optimal favored-agent mechanisms with different thresholds~$\nu^\star$. In this situation, it is expedient to look for Pareto robustly optimal favored-agent mechanisms. { Before studying specific ambiguity sets, we formally introduce the basics of spatial induction, which is our main tool for proving Pareto robust optimality. To facilitate a concise presentation, we first define the concept of unilateral reachability.

\begin{definition}[Unilateral Reachability]
 For any fixed agent~$i \in \mathcal I$, a scenario $\bm t' \in \mathcal T$ is called \emph{$i$-unilaterally reachable} from another scenario $\bm t \in \mathcal T$ if $\bm t'_{-i} = \bm t_{-i}$.
\end{definition}
Note that $\bm t' \in \mathcal T$ is $i$-unilaterally reachable from $\bm t \in \mathcal T$ if $\bm t'$ is obtained by changing the type~$t_i$ of agent~$i$ to $t'_i$ while keeping the types~$\bm t_{-i}$ of all other agents fixed.

\begin{lemma}
\label{lem:deviationForFavored}
    The following hold for any~$(\bm p, \bm q) \in \mathcal X$, $i \in \mathcal I$ and~$\bm t \in \mathcal T$.
    \begin{itemize}
        \item[(i)] If $p_i(\bm t)-q_i(\bm t)=1$, then the mechanism~$(\bm p, \bm q)$ allocates the good to agent~$i$ with probability~$1$ in any scenario that is $i$-unilaterally reachable from $\bm t$, $i.e.$, $p_i(t_i',\bm t_{-i})=1$ for all $t'_i \in \mathcal T_i$.
        \item[(ii)] If $p_i(\bm t)=0$, then the mechanism~$(\bm p, \bm q)$ inspects agent~$i$ whenever he wins the good in any scenario that is $i$-unilaterally reachable from $\bm t$, $i.e.$, $p_i(t_i',\bm t_{-i})=q_i(t_i',\bm t_{-i})$ for all $t_i' \in \mathcal T_i$.
    \end{itemize}
\end{lemma}
The proof of Lemma~\ref{lem:deviationForFavored} follows directly from the~\eqref{eq:IC} and~\eqref{eq:FC} constraints. 
Note that, by~\eqref{eq:FC}, the condition $p_i(\bm{t}) - q_i(\bm{t}) = 1$ in Lemma \ref{lem:deviationForFavored}(i) is satisfied if and only if $p_i(\bm{t}) = 1$ and $q_i(\bm{t}) = 0$. Thus, Lemma \ref{lem:deviationForFavored}(i) asserts that if we allocate the good to agent~$i$ without inspection in scenario~$\bm{t}$, then we should also allocate the good to agent~$i$ in any $i$-unilaterally reachable scenario~$\bm{t}'$. However, we may or may not inspect agent~$i$ in scenario $\bm{t}'$. Lemma~\ref{lem:deviationForFavored}(ii) asserts that if agent~$i$ does not receive the good in scenario~$\bm{t}$ (which implies via~\eqref{eq:FC} that agent~$i$ is {\em not} inspected in scenario~$\bm{t}$), then we must inspect agent~$i$ in any $i$-unilaterally reachable scenario~$\bm{t}'$ in which he receives the good.

The spatial induction technique to be developed in this paper critically relies on the following proposition, which introduces a sufficient condition for the Pareto robust optimality of a mechanism. 
\begin{proposition}
\label{prop:unifyingPRO}
    An optimal mechanism $(\bm p^\star, \bm q^\star) \in \mathcal{X}$ for problem~\eqref{eq:MDP} is Pareto robustly optimal if there exists a partition $\mathcal{S}_1, \ldots, \mathcal{S}_m$ of the type space $\mathcal T$ such that the following conditions hold for any index $k \in \{1,\ldots,m\}$ and for any scenario $\bm t \in \mathcal{S}_k$.
    \begin{itemize}
        \item[(i)] There exists a probability distribution $\mathbb P \in\mathcal P\cap \mathcal{P}_0(\cup_{l=1}^k \mathcal{S}_l)$ with $\mathbb P(\tilde{\bm t}= \bm t)>0$.
        \item[(ii)] The mechanism $(\bm p^\star, \bm q^\star)$ solves the following auxiliary scenario problem.
            \begin{equation}\label{eq:MDPk(t)} \tag{$\text{SP}_{k}(\bm t)$}
            \begin{array}{cl}
                \displaystyle \max_{(\bm{p}, \bm{q}) \in \mathcal{X}} & \displaystyle \sum_{i \in \mathcal{I}} (p_i({\bm{t}}) {t}_i - q_i({\bm{t}}) c_i) \\[3ex]
                \operatorname{s.t.} & \displaystyle  \sum_{i \in \mathcal{I}} (p_i({\bm{t}'}) {t}'_i - q_i({\bm{t}}') c_i) = \sum_{i \in \mathcal{I}} (p_i^\star({\bm{t}}') {t}'_i - q_i^\star({\bm{t}'}) c_i) \quad \forall {\bm t}' \in \cup_{l=1}^{k-1} \mathcal{S}_l
            \end{array}
            \end{equation}
    \end{itemize}
    %{\color{orange} Furthermore, if the above conditions hold true, then any mechanism $(\bm p,\bm q) \in \mathcal{X}$ that weakly Pareto robustly dominates $(\bm p^\star,\bm q^\star)$ generates the same payoff as $(\bm p^\star,\bm q^\star)$ in every scenario $\bm t \in \mathcal{T}$.}
\end{proposition}

Note that problem~\ref{eq:MDPk(t)} not only depends on~$k$ and~$\bm t$ but also on the prescribed optimal mechanism~$(\bm p^\star, \bm q^\star)$. However, this dependence is notationally suppressed to avoid clutter. 

Proposition~\ref{prop:unifyingPRO} establishes a new sufficient condition for Pareto robust optimality that does not exploit any structural properties of the ambiguity set~$\mathcal P$. We also emphasize that the proof of Proposition~\ref{prop:unifyingPRO} does not exploit any specifics of the mechanism design problem~\eqref{eq:MDP} and thus readily extends to generic distributionally robust optimization problems. Thanks to Proposition~\ref{prop:unifyingPRO}, the Pareto robust optimality of a given optimal mechanism can be proved by identifying a partition $\mathcal{S}_1, \ldots, \mathcal{S}_m$ of~$\mathcal T$ that satisfies two simple conditions. First, for any fixed $k\in\{1,\ldots,m\}$ and~$\bm t\in\mathcal{S}_k$, there must exist an admissible distribution that assigns~$\bm t$ a strictly positive probability, while assigning zero probability to~$\mathcal{S}_l$ for every~$l > k$.  Second, the given mechanism should solve the scenario problems~\ref{eq:MDPk(t)} simultaneously for all $k\in\{1,\ldots,m\}$ and $\bm t \in \mathcal{S}_k$. %, maximizing the payoff under each scenario. 
%Each \ref{eq:MDPk(t)} has a restricted feasibility set so that any mechanism that does not exactly match the payoff of the given Pareto robustly optimal mechanism under the scenarios of the subsets preceding $\mathcal{S}_k$ is excluded. 
%Conversely, Proposition~\ref{prop:unifyingPRO} motivates an iterative procedure for constructing a Pareto robustly optimal mechanism by solving all scenario problems successively for $k=1,\ldots,m$. 
%indicating that the payoff cannot be improved under a scenario without sacrificing it elsewhere.
Conditions~(i) and~(ii) imply that the given mechanism cannot be Pareto robustly dominated by any other mechanism.
We emphasize, however, that a mechanism satisfying the conditions~(i) and~(ii) of Proposition~\ref{prop:unifyingPRO} is not necessarily robustly optimal. Thus, the conditions~(i) and~(ii) are not sufficient to guarantee Pareto robust optimality unless we restrict attention to robustly optimal mechanisms.
%Solving this series of scenario problems determines a mechanism on the Pareto frontier, indicating that the payoff cannot be improved under a scenario without sacrificing it elsewhere. It is important to note that a mechanism on the Pareto frontier does not necessarily have to be robustly optimal; therefore, the conditions (i) and (ii) in Proposition \ref{prop:unifyingPRO} are not sufficient for Pareto robust optimality without verifying robust optimality separately.

In each of the subsequent sections, we will leverage Proposition~\ref{prop:unifyingPRO} to show that a given candidate mechanism is Pareto robustly optimal for a particular ambiguity set. Specifically, we will choose a partition $\mathcal{S}_1, \dots, \mathcal{S}_m$ of the type space tailored to the given ambiguity set such that condition~(i) is satisfied. We will then apply the following spatial induction technique. First, we will show that the given candidate mechanism solves the scenario problems $\text{SP}_1(\bm{t})$ for all~$\bm t\in \mathcal{S}_1$. Next, we will exploit the non-locality of the incentive compatibility constraints---as manifested through Lemma~\ref{lem:deviationForFavored}---to relate any scenario in~$\mathcal{S}_2$ to a scenario in~$\mathcal{S}_1$. This will allow us to prove that the given candidate mechanism solves the scenario problems $\text{SP}_2(\bm{t})$ for all~$\bm t\in \mathcal{S}_2$. We then iterate through the remaining subsets of the type space one by one and apply each time a similar induction step.

The following corollary shows that the sufficient condition of Proposition~\ref{prop:unifyingPRO} implies a payoff equivalence principle, which will be useful to prove the main results of this paper. The proof of this corollary follows immediately from that of Proposition~\ref{prop:unifyingPRO} and is thus omitted.
\begin{corollary}[Payoff Equivalence]\label{Corollary of sufficiency proposition}
Assume that $(\bm p^\star, \bm q^\star) \in \mathcal{X}$ is an optimal mechanism satisfying the conditions~(i) and~(ii) of Proposition~\ref{prop:unifyingPRO}. If any other mechanism $(\bm p,\bm q) \in \mathcal{X}$ weakly Pareto robustly dominates $(\bm p^\star,\bm q^\star)$, then it generates the same payoff as~$(\bm p^\star,\bm q^\star)$ in every scenario~$\bm t \in \mathcal{T}$. 
%Given an optimal mechanism $(\bm p^\star, \bm q^\star) \in \mathcal{X}$ for problem~\eqref{eq:MDP}, if there exists a partition $\mathcal{S}_1, \ldots, \mathcal{S}_m$ of the type space $\mathcal T$ such that conditions (i) and (ii) hold for any index $k \in \{1,\ldots,m\}$ and for any scenario $\bm t \in \mathcal{S}_k$, then any mechanism $(\bm p,\bm q) \in \mathcal{X}$ that weakly Pareto robustly dominates $(\bm p^\star,\bm q^\star)$ generates the same payoff as $(\bm p^\star,\bm q^\star)$ in every scenario $\bm t \in \mathcal{T}$.
\end{corollary}
Recall that if $(\bm p, \bm q)$ weakly Pareto robustly dominates $(\bm p^\star, \bm q^\star)$, then the {\em expected} payoff of $(\bm p, \bm q)$ is at least as large as that of~$(\bm p^\star, \bm q^\star)$ under any distribution $\mathbb P\in\mathcal P$. Corollary~\ref{Corollary of sufficiency proposition} additionally asserts that, if~$(\bm p^\star, \bm q^\star)$ satisfies the sufficient condition of Proposition~\ref{prop:unifyingPRO}, then the {\em actual} payoff of~$(\bm p, \bm q)$ is at least as large as (in fact, exactly equal to) that of~$(\bm p^\star, \bm q^\star)$ under any scenario~$\bm t\in\mathcal T$.}

\section{Support-Only Ambiguity Sets}\label{sec: support only}

%{\color{red}Reduction to pure robust problem $\&$ results...}

We now investigate the mechanism design problem~\eqref{eq:MDP} under the assumption that $\mathcal{P} = \mathcal{P}_0(\mathcal{T})$ is the support-only ambiguity set that contains all distributions supported on the type space~$\mathcal{T}$. As~$\mathcal{P}$ contains all Dirac point distributions concentrating unit mass at any~$\bm {t}\in\mathcal{T}$, the worst-case expected payoff over all distributions~$\mathbb{P}\in\mathcal{P}$ simplifies to the worst-case payoff overall type profiles~$\bm{t}\in\mathcal{T}$, and thus it is easy to verify that problem~\eqref{eq:MDP} simplifies to
	\begin{equation} \label{eq:RMDP} 
	\begin{aligned}
	z^\star = ~&\sup_{\bm{p}, \bm{q}}  &&\inf_{\bm{t} \in \mathcal{T}} \; \sum_{i \in \mathcal{I}} (p_i(\bm{t}) t_i - q_i(\bm{t}) c_i)  \\
	&\;\text{s.t.} &&%p_i : \mathcal{T} \rightarrow [0,1] \;\text{and}\; q_i : \mathcal{T} \rightarrow [0,1] \;\;\forall i \in \mathcal{I}
	p_i, q_i \in \mathcal{L}(\mathcal{T}, [0,1]) \;\;\forall i \in \mathcal{I}\\
	&&&\eqref{eq:IC}, \; \eqref{eq:FC}.
	\end{aligned}
	\end{equation}
% \begin{proposition}\label{prop:RMDP}
% 	If $\mathcal{P} = \mathcal{P}_0(\mathcal{T})$, then \eqref{eq:MDP} simplifies to
% 	\begin{equation} \label{eq:RMDP} 
% 	\begin{aligned}
% 	z^\star = &\sup_{\bm{p}, \bm{q}}  &&\inf_{\bm{t} \in \mathcal{T}} \; \sum_{i \in \mathcal{I}} (p_i(\bm{t}) t_i - q_i(\bm{t}) c_i)  \\
% 	&\;\text{s.t.} &&p_i : \mathcal{T} \rightarrow [0,1] \;\text{and}\; q_i : \mathcal{T} \rightarrow [0,1] \;\;\forall i \in \mathcal{I}\\
% 	&&&\eqref{eq:IC}, \; \eqref{eq:FC}.
% 	\end{aligned}
% 	\end{equation}
% \end{proposition}
Similarly, it is easy to verify that an optimal mechanism $(\bm{p}^\star, \bm{q}^\star)$ for problem~\eqref{eq:RMDP} is Pareto robustly optimal if there exists no other feasible mechanism $(\bm{p}, \bm{q})$ with
\begin{equation*}\label{eq:dominance support-only}
\begin{aligned}
    \sum_{i \in \mathcal{I}} (p_i({\bm{t}}) {t}_i - q_i({\bm{t}}) c_i) \geq  \sum_{i \in \mathcal{I}} (p^\star_i({\bm{t}}) {t}_i - q^\star_i({\bm{t}}) c_i) &&\forall \bm t \in \mathcal{T},
\end{aligned}
\end{equation*}
where the inequality is strict for at least one type profile $\bm t \in \mathcal{T}$.
% \begin{proposition}\label{prop:RMDP Pareto Optimality}
% If $\mathcal{P} = \mathcal{P}_0(\mathcal{T})$, then a mechanism $(\bm{p}^\star, \bm{q}^\star)$ that is optimal in \eqref{eq:MDP} (and, consequently, in \eqref{eq:RMDP}) is Pareto robustly optimal if there exists no other feasible mechanism $(\bm{p}, \bm{q})$ with
% \begin{equation}\label{eq:dominance support-only}
% \begin{aligned}
%  &\sum_{i \in \mathcal{I}} (p_i({\bm{t}}) {t}_i - q_i({\bm{t}}) c_i) \geq  \sum_{i \in \mathcal{I}} (p^\star_i({\bm{t}}) {t}_i - q^\star_i({\bm{t}}) c_i) &&\forall \bm t \in \mathcal{T},
% \end{aligned}
% \end{equation}
% where the inequality is strict for at least one type profile $\bm t \in \mathcal{T}$.
% \end{proposition}
%
%The proofs of Propositions \ref{prop:RMDP} and \ref{prop:RMDP Pareto Optimality} follow from elementary arguments and are thus omitted. 
%Take any mechanism $(\bm{p},\bm{q})$ that is optimal in (\ref{eq:MDP}). It should also be optimal in (\ref{eq:RMDP}) since otherwise there would exist another mechanism $(\bm{p}',\bm{q}')$ that yields a strictly higher worst-case payoff for the principle. Then, one can show that the worst-case expected payoff for the principal under $(\bm{p}',\bm{q}')$ is strictly higher than $(\bm{p},\bm{q})$, which contradicts the optimality of $(\bm{p},\bm{q})$.
If the principal knew the agents' types ex ante, she could simply allocate the good to the agent with the highest type and would not have to spend money on inspecting anyone. One can therefore show that the optimal value~$z^\star$ of problem~\eqref{eq:RMDP} is upper bounded by $\inf_{\bm{t} \in \mathcal{T}} \, \max_{i \in \mathcal{I}} {t}_i = \max_{i \in \mathcal{I}} \underline{t}_i$. The following proposition reveals that this upper bound is attained by an admissible mechanism. % and must therefore coincide with~$z^\star$.

\begin{proposition}	\label{prop:optimal-z-support}	
Problem \eqref{eq:RMDP} is solvable, and its optimal value is given by $z^\star = \max_{i \in \mathcal{I}} \underline{t}_i$.
	%The optimal value of \eqref{eq:RMDP} is given by $z^\star = \max_{i \in \mathcal{I}} \underline{t}_i$.
\end{proposition}

%The optimal value of the robust mechanism design problem~\eqref{eq:RMDP} identified in Proposition~\ref{prop:optimal-z-support} is also attained by the favored-agent mechanism with favored agent $i^\star \in \arg \max_{i \in \mathcal{I}} \underline{t}_i$  and threshold value $\nu^\star \geq \max_{i \in \mathcal{I}} (\overline{t}_i - c_i)$. 
The next theorem shows that there are infinitely many optimal favored-agent mechanisms that attain the optimal value $z^\star = \max_{i \in \mathcal{I}} \underline{t}_i$ of problem~\eqref{eq:RMDP}.

\begin{theorem}\label{thr:opt-fav-support-only}
Any favored-agent mechanism with favored agent $i^\star \in {\arg \max}_{i \in \mathcal{I}} \underline{t}_i$ and threshold value $\nu^\star \geq \max_{i \in \mathcal{I}} \underline{t}_i$ is optimal in problem~\eqref{eq:RMDP}.    
\end{theorem}

\begin{remark}
Theorem~\ref{thr:opt-fav-support-only} is sharp in the sense that there are problem instances for which any favored-agent mechanism with favored agent $i^\star \in {\arg \max}_{i \in \mathcal{I}} \underline{t}_i$ and threshold value $\nu < \max_{i \in \mathcal{I}} \underline{t}_i$ is strictly suboptimal in~\eqref{eq:RMDP}. To see this, consider an instance with~$I=2$ agents, where~$\mathcal{T}_1 = [2,8]$, $\mathcal{T}_2 = [0,10]$ and~$c_1=c_2=1$. By Proposition~\ref{prop:optimal-z-support}, the supremum of~\eqref{eq:RMDP} is given by $ \max_{i \in \mathcal{I}} \underline{t}_i=2$. Consider now any favored agent mechanism with favored agent $1\in {\arg \max}_{i \in \mathcal{I}} \underline{t}_i$ and threshold value $\nu<\underline{t}_1=2$. This mechanism is strictly suboptimal. To see this, assume first that~$\nu< 1$. If~$\bm t=(2,2)$, then the mechanism allocates the good to agent~1 or agent~2 with verification and earns $t_1-c_1 =t_2-c_2 =1$. Thus, the worst-case payoff overall $\bm t\in\mathcal T$ cannot exceed~$1$, which is strictly smaller than the optimal worst-case payoff. Assume next that~$\nu \in [1,2)$. If $\bm t=(2,2+\nu/2) \in \mathcal{T}$, then the mechanism allocates the good to agent~$2$ with verification %because $t_2-c_2 = (3-\varepsilon/2) - 1 > \nu$ and $t_2-c_2 >t_1-c_1=1$. The principal's payoff thus amounts to 
and earns $1+\nu/2$. Thus, the worst-case payoff overall $\bm t\in\mathcal T$ cannot exceed~$1+\nu/2$, which is strictly smaller than the optimal worst-case payoff. In summary, the mechanism is strictly suboptimal for all $\nu<2$. \hfill$\square$
%In Theorem~\ref{thr:opt-fav-support-only}, any choice of threshold smaller than ${\max}_{i \in \mathcal{I}} \underline{t}_i$ is not optimal in general. To see this, consider an example with two agents, where $\mathcal{T}_1 = [2,8]$, $\mathcal{T}_2 = [0,10]$, and the verification cost for each agent is 1. Consider the favored agent mechanism with favored agent $1$ and threshold value $\nu<2=\underline{t}_1$. If $\nu< 1$, in scenario $(2,2)$, this mechanism allocates the good to agent 1 or 2 with verification and earns a payoff of 1 as $t_1-c_1 =t_2-c_2 =1> \nu$. As the worst-case payoff of the aforementioned favored-agent mechanism cannot exceed its payoff in scenario $(2,2)$ and $\underline{t}_1 =2$, we conclude that this mechanism is sub-optimal in view of Proposition \ref{prop:optimal-z-support}. If $\nu \in [1,2)$, let $\varepsilon = 2- \nu$ that belongs to the interval $(0,1]$ by definition. Consider the scenario $(2,3-\varepsilon/2) \in \mathcal{T}$. In this scenario, the aforementioned favored-agent mechanism allocates the good to agent $2$ and inspects his type as $t_2-c_2 = (3-\varepsilon/2) - 1 > \nu$ and $t_2-c_2 >t_1-c_1=1$. The principal's payoff thus amounts to $2-\varepsilon/2$ and therefore is smaller than $\underline{t}_1=2$. This mechanism thus cannot be optimal in view of Proposition \ref{prop:optimal-z-support}. \square
\end{remark}

%We emphasize that Theorem~\ref{thr:opt-fav-support-only} holds irrespective of whether the favored-agent mechanism is of type 1 or of type 2. 
As the mechanism design problem~\eqref{eq:RMDP} constitutes a convex program, any convex combination of %randomization over 
optimal favored-agent mechanisms gives rise to yet another optimal mechanism. However, problem~\eqref{eq:RMDP} also admits optimal mechanisms that can neither be interpreted as favored-agent mechanisms nor as convex combinations of %randomizations over 
favored-agent mechanisms. To see this, consider any favored-agent mechanism $(\bm{p}, \bm{q})$ with favored agent $i^\star \in {\arg \max}_{i \in \mathcal{I}} \underline{t}_i$ and threshold value~$\nu^\star\in\mathbb R$ satisfying $\nu^\star \geq \max_{i \in \mathcal{I}} \underline{t}_i$ and $\nu^\star > \max_{i \in \mathcal{I}} \overline{t}_i-c_i$. By Theorem~\ref{thr:opt-fav-support-only}, this mechanism is optimal in problem~\eqref{eq:RMDP}. The second condition on~$\nu^\star$ implies that this mechanism allocates the good to the favored agent without inspection for every~$\bm t\in\mathcal T$ (case~(i) always prevails). Next, construct~$\hat{\bm{t}}\in\mathcal{T}$ through~$\hat{t}_i = \underline{t}_i$ for all~$i \neq i^\star$ and $\hat{t}_{i^\star} = \overline{t}_{i^\star}$, and note that $\hat{\bm{t}} \neq \underline{\bm t}$ because $\underline{t}_{i^\star}<\overline{t}_{i^\star}$. Finally, introduce another mechanism $(\bm{p}, \bm{q}')$, where~$\bm{q}'$ is defined through $q'_i (\bm{t})= q_i(\bm{t})$ for all~$\bm{t}\in\mathcal{T}$ and~$i \neq i^\star$ and
\begin{equation*}
    \begin{aligned}
        q_{i^\star}'(\bm{t}) = \begin{cases}
        \min\{1,(\overline{t}_{i^\star}-\underline{t}_{i^\star})/c_{i^\star}\} &\text{if } \bm t = \hat{\bm t},\\
        q_{i^\star}(\bm{t}) &\text{if }\bm t\in\mathcal T\setminus\{\hat{\bm{t}}\}.
        \end{cases}
    \end{aligned}
\end{equation*}
%The allocation rules $\bm{p}'$ and $\bm{p}$ thus coincide for all $\bm t \in \mathcal T$, while $\bm{q}'$ and $\bm{q}$ coincide for all $\bm t \in \mathcal T \setminus \{\hat{\bm t}\}$.
%
%By construction, the favored-agent mechanism $(\bm{p}, \bm{q})$ satisfies $p_{i^\star}(\hat{\bm{t}})=1$, $q_{i^\star}(\hat{\bm{t}})=0$ and $\sum_{i \in \mathcal{I}} (p_i(\hat{\bm{t}}) \hat{t}_i - q_i(\hat{\bm{t}}) c_i) = \hat{t}_{i^\star} > \underline{t}_{i^\star} = \max_{i \in \mathcal{I}} \underline{t}_i$.
One readily verifies that $(\bm{p}, \bm{q}')$ is feasible in~\eqref{eq:RMDP}. Indeed, as $(\bm{p}, \bm{q}')$ differs from $(\bm{p}, \bm{q})$ only in scenario~$\hat{\bm{t}}$, and as $(\bm{p}, \bm{q})$ is feasible, it suffices to check the feasibility of $(\bm{p}, \bm{q}')$ in scenario~$\hat{\bm{t}}$. Indeed, the modified allocation rule $\bm q'$ is valued in~$[0,1]^I$, and $(\bm{p}, \bm{q}')$ satisfies \eqref{eq:FC} because $0\leq q_{i^\star}'(\hat{\bm{t}}) \leq 1 = p_{i^\star}(\hat{\bm{t}})$, where the equality holds because the favored-agent mechanism~$(\bm{p}, \bm{q})$ allocates the good to agent~$i^\star$ with certainty.
%The allocation probability $q_{i^\star}'(\hat{\bm{t}})$ is non-negative and satisfies $q_{i^\star}'(\hat{\bm{t}}) =\min\{1,(\overline{t}_{i^\star}-\underline{t}_{i^\star})/c_i\} \leq 1 = p_{i^\star}'(\hat{\bm{t}})$. 
Similarly, the modified mechanism $(\bm{p}, \bm{q}')$ satisfies~\eqref{eq:IC} because 
\[
    p_{i^\star}(t_{i^\star}, \hat{\bm{t}}_{-{i^\star}})=1 \geq p_{i^\star}(\hat{\bm{t}}) - q_{i^\star}'(\hat{\bm{t}}) \quad \forall t_{i^\star} \in \mathcal{T}_{i^\star}.
\]
In summary, we have thus shown that the mechanism $(\bm{p}, \bm{q}')$ is feasible in~\eqref{eq:RMDP}. To show that it is also optimal, recall that $(\bm{p}, \bm{q})$ is optimal with worst-case payoff~$\max_{i \in \mathcal{I}} \underline{t}_i$ and that $(\bm{p}, \bm{q}')$ differs from $(\bm{p}, \bm{q})$ only in scenario~$\hat{\bm{t}}$. The principal's payoff in scenario $\hat{\bm{t}}$ amounts to 
\[
    p_{i^\star}(\hat{\bm{t}})\hat{t}_{i^\star}-q_{i^\star}' (\hat{\bm{t}})c_{i^\star}=\hat{t}_{i^\star}-q_{i^\star}'(\hat{\bm{t}})c_{i^\star} \geq \hat{t}_{i^\star}-\frac{\hat{t}_{i^\star}-\underline{t}_{i^\star}}{c_{i^\star}}c_{i^\star} = \underline{t}_{i^\star} =\max_{i \in \mathcal{I}} \underline{t}_i,
\]
where the inequality follows from the definition of~$q_{i^\star}'(\hat{\bm{t}})$. Thus, the worst-case payoff of~$(\bm{p}, \bm{q}')$ amounts to~$\max_{i \in \mathcal{I}} \underline{t}_i$, and~$(\bm{p}, \bm{q}')$ is indeed optimal in~\eqref{eq:RMDP}. However, $(\bm{p}, \bm{q}')$ is {\em not} a favored-agent mechanism for otherwise~$q_{i^\star}'(\hat{\bm{t}})$ would have to vanish; see Definition~\ref{def:fav-agent-mechanism}. In addition, note that $p_{i^\star}(\hat{\bm{t}})-q'_{i^\star}(\hat{\bm{t}})<1$ whereas $p_{i^\star}(t_{i^\star},\hat{\bm{t}}_{-i^\star})-q'_{i^\star}(t_{i^\star},\hat{\bm{t}}_{-i^\star})=1$ for all $t_{i^\star} \neq \hat{t}_{i^\star}$. This implies via Lemma~\ref{lem:randomization} below that $(\bm{p}, \bm{q}')$ is also {\em not} a convex combination of %randomization over 
favored-agent mechanisms.

\begin{lemma}
    \label{lem:randomization}
    If a mechanism $(\bm{p},\bm{q})$ is a convex combination of %randomization over 
    favored-agent mechanisms, then the function $p_i(t_i,\bm{t}_{-i})-q_i(t_i,\bm{t}_{-i})$ is constant in $t_i \in \mathcal{T}_i$ for any fixed $i \in \mathcal{I}$ and $\bm{t}_{-i} \in \mathcal{T}_{-i}$.
\end{lemma}

In summary, we have shown that the robust mechanism design problem~\eqref{eq:RMDP} admits infinitely many optimal solutions. Some of these solutions represent favored-agent mechanisms, while others  represent %randomizations over 
convex combinations of favored-agent mechanisms. However, some optimal mechanisms are neither crisp favored-agent mechanisms nor convex combinations of favored-agent mechanisms. While all robustly optimal mechanisms generate the same payoff in the worst case, however, their payoffs may differ significantly in non-worst-case scenarios. This observation suggests that robust optimality alone is not a sufficient differentiator to distinguish desirable from undesirable mechanisms. Note also that the optimal mechanism constructed above by altering the inspection probabilities of an optimal favored-agent mechanism is Pareto robustly dominated by its underlying favored-agent mechanism. This observation prompts us to seek Pareto robustly optimal mechanisms for problem~\eqref{eq:RMDP}. The next theorem shows that among all robustly optimal favored-agent mechanisms identified in Theorem~\ref{thr:opt-fav-support-only} there is always one that is also Pareto robustly optimal.

\begin{theorem}\label{thr:PRO-fav-agents}
	Any favored-agent mechanism of type~(i) with favored agent
	${i^\star} \in {\arg \max}_{i \in \mathcal{I}} \underline{t}_i$ and threshold value $\nu^\star = \max_{i \in \mathcal{I}} \underline{t}_i= \underline t_{i^\star}$ is Pareto robustly optimal in problem~\eqref{eq:RMDP}.
\end{theorem}

We sketch the proof idea in the special case when there are only two agents. To convey the key ideas without tedious case distinctions, we assume that $\underline{t}_1>\underline{t}_2$ so that ${\arg \max}_{i \in \mathcal{I}} \underline{t}_i = \{1\}$ is a singleton, and we assume that $\overline{t}_2>c_2+\underline{t}_1$ and $\overline{t}_1>c_2+\underline{t}_1$. Throughout the subsequent discussion, we will use the following partition of the type space~$\mathcal{T}$.
{ 
\begin{equation}\label{eq: partition support-only}
\begin{aligned}
\mathcal{S}_{1} &= \{\bm{t} \in  \mathcal{T} \,:\, t_2-c_2 \leq \underline{t}_1 \; \text{and} \; t_1 = \overline{t}_1 \}\\
\mathcal{S}_{2} &=  \{\bm{t} \in  \mathcal{T} \,:\, t_2-c_2 \leq \underline{t}_1  \; \text{and} \;  t_1 < \overline{t}_1\}\\
\mathcal{S}_{3} &= \{\bm{t} \in  \mathcal{T} \,:\, t_2-c_2 >\underline{t}_1  \; \text{and} \; t_1= \underline t_1\}\\
\mathcal{S}_{4} &= \{\bm{t} \in  \mathcal{T} \,:\, t_2-c_2 >\underline{t}_1  \; \text{and} \; t_1 > \underline t_1\}
\end{aligned}
\end{equation}}
The sets $\mathcal{S}_{1}$, $\mathcal{S}_{2}$, $\mathcal{S}_{3}$ and $\mathcal{S}_{4}$ are visualized in Figure \ref{support-only-figure}. Note that all of them are nonempty thanks to the above assumptions about~$\underline t_1$, $\underline t_2$ and~$c_2$.
%$\mathcal{T}_{II}$, $\mathcal{T}_{III}$ and $\mathcal{T}_{IV}$ can be empty if $\underline{t}_1$ or $c_2$ are sufficiently large, while $\mathcal{T}_{I}$ is necessarily nonempty. To convey the key ideas without being repetitive, we assume that $\overline{t}_2>c_2+\underline{t}_1$, which ensures that all four subsets are nonempty and that $\overline{t}_1>c_2+\underline{t}_1$. 
We emphasize, however, that all simplifying assumptions as well as the restriction to two agents are relaxed in the formal proof of Theorem~\ref{thr:PRO-fav-agents}. %  {
%Moreover, in the formal proof, we partition the type set into $\mathcal{T}_I$, $\mathcal{T}_{II}$, and $\mathcal{T}_{III}$, and then further partition $\mathcal{T}_I$ and $\mathcal{T}_{III}$ into smaller sets. The set $\mathcal{T}_{III}$ here is equivalent to $\mathcal{T}_{III}^1$ of the formal proof, and $\mathcal{T}_{IV}$ is equivalent to $\mathcal{T}_{III}^2$.
%The simplifying assumptions here allow us to work with four subsets, whereas the formal proof requires consideration of $2I+1$ subsets in total.} 

  {Denote by~$(\bm{p}^\star,\bm{q}^\star)$ the favored-agent mechanism of type~(i) with favored agent~$1$ and threshold value~$\nu^\star = \underline{t}_1$. By definition, the principal's payoff in scenario~$\bm t$ under~$(\bm{p}^\star,\bm{q}^\star)$ thus amounts to~$t_1$ when $t_2-c_2 \leq \underline{t}_1$ ($i.e.$, when $\bm{t} \in \mathcal{S}_{1} \cup \mathcal{S}_{2}$) and to~$\max_{i \in \mathcal{I}} t_i-c_i$ when $t_2-c_2 > \underline{t}_1$ ($i.e.$, when $\bm{t} \in \mathcal{S}_{3} \cup \mathcal{S}_{4}$). In the following, we will prove that this mechanism is Pareto robustly optimal in problem~\eqref{eq:RMDP}. To this end, we will leverage Proposition~\ref{prop:unifyingPRO}, which provides a sufficient condition for the Pareto robust optimality of robustly optimal mechanisms. From Theorem~\ref{thr:opt-fav-support-only} we already know that~$(\bm{p}^\star,\bm{q}^\star)$  is robustly optimal. To show that~$(\bm{p}^\star,\bm{q}^\star)$ is Pareto robustly optimal, it thus suffices to verify the conditions~(i) and~(ii) in Proposition~\ref{prop:unifyingPRO} for the  partition~\eqref{eq: partition support-only}.
Note first that condition~(i) trivially holds because the support-only ambiguity set contains all Dirac point distributions concentrating unit mass at some scenario~$\bm t \in \mathcal T$. It thus remains to verify condition~(ii). To this end, we will show by induction on~$k$ that $(\bm p^\star, \bm q^\star)$ solves~\ref{eq:MDPk(t)} for all~$\bm t \in \mathcal S_k$. The induction steps exploit the non-locality of the incentive compatibility constraint~\eqref{eq:IC}, which led to Lemma~\ref{lem:deviationForFavored}.}

%Figure PRO for support-only
% Pattern Info
 
\tikzset{
pattern size/.store in=\mcSize, 
pattern size = 5pt,
pattern thickness/.store in=\mcThickness, 
pattern thickness = 0.3pt,
pattern radius/.store in=\mcRadius, 
pattern radius = 1pt}
\makeatletter
\pgfutil@ifundefined{pgf@pattern@name@_1voo82hgq}{
\pgfdeclarepatternformonly[\mcThickness,\mcSize]{_1voo82hgq}
{\pgfqpoint{-\mcThickness}{-\mcThickness}}
{\pgfpoint{\mcSize}{\mcSize}}
{\pgfpoint{\mcSize}{\mcSize}}
{
\pgfsetcolor{\tikz@pattern@color}
\pgfsetlinewidth{\mcThickness}
\pgfpathmoveto{\pgfpointorigin}
\pgfpathlineto{\pgfpoint{0}{\mcSize}}
\pgfusepath{stroke}
}}
\makeatother

% Pattern Info
 
\tikzset{
pattern size/.store in=\mcSize, 
pattern size = 5pt,
pattern thickness/.store in=\mcThickness, 
pattern thickness = 0.3pt,
pattern radius/.store in=\mcRadius, 
pattern radius = 1pt}
\makeatletter
\pgfutil@ifundefined{pgf@pattern@name@_uoqu8884c}{
\pgfdeclarepatternformonly[\mcThickness,\mcSize]{_uoqu8884c}
{\pgfqpoint{0pt}{-\mcThickness}}
{\pgfpoint{\mcSize}{\mcSize}}
{\pgfpoint{\mcSize}{\mcSize}}
{
\pgfsetcolor{\tikz@pattern@color}
\pgfsetlinewidth{\mcThickness}
\pgfpathmoveto{\pgfqpoint{0pt}{\mcSize}}
\pgfpathlineto{\pgfpoint{\mcSize+\mcThickness}{-\mcThickness}}
\pgfusepath{stroke}
}}
\makeatother

% Pattern Info
 
\tikzset{
pattern size/.store in=\mcSize, 
pattern size = 5pt,
pattern thickness/.store in=\mcThickness, 
pattern thickness = 0.3pt,
pattern radius/.store in=\mcRadius, 
pattern radius = 1pt}
\makeatletter
\pgfutil@ifundefined{pgf@pattern@name@_lozk1vl61 lines}{
\pgfdeclarepatternformonly[\mcThickness,\mcSize]{_lozk1vl61}
{\pgfqpoint{0pt}{0pt}}
{\pgfpoint{\mcSize+\mcThickness}{\mcSize+\mcThickness}}
{\pgfpoint{\mcSize}{\mcSize}}
{\pgfsetcolor{\tikz@pattern@color}
\pgfsetlinewidth{\mcThickness}
\pgfpathmoveto{\pgfpointorigin}
\pgfpathlineto{\pgfpoint{\mcSize}{0}}
\pgfusepath{stroke}}}
\makeatother

% Pattern Info
 
\tikzset{
pattern size/.store in=\mcSize, 
pattern size = 5pt,
pattern thickness/.store in=\mcThickness, 
pattern thickness = 0.3pt,
pattern radius/.store in=\mcRadius, 
pattern radius = 1pt}
\makeatletter
\pgfutil@ifundefined{pgf@pattern@name@_f85x3oa8e}{
\pgfdeclarepatternformonly[\mcThickness,\mcSize]{_f85x3oa8e}
{\pgfqpoint{0pt}{0pt}}
{\pgfpoint{\mcSize+\mcThickness}{\mcSize+\mcThickness}}
{\pgfpoint{\mcSize}{\mcSize}}
{
\pgfsetcolor{\tikz@pattern@color}
\pgfsetlinewidth{\mcThickness}
\pgfpathmoveto{\pgfqpoint{0pt}{0pt}}
\pgfpathlineto{\pgfpoint{\mcSize+\mcThickness}{\mcSize+\mcThickness}}
\pgfusepath{stroke}
}}
\makeatother
\tikzset{every picture/.style={line width=0.75pt}} %set default line width to 0.75pt        

\begin{figure}
    \centering

\begin{tikzpicture}[x=0.75pt,y=0.75pt,yscale=-1,xscale=1]

%Shape: Rectangle [id:dp4812135912384887] 
\draw  [draw opacity=0][fill={rgb, 255:red, 102; green, 194; blue, 165 }  ,fill opacity=0.9 ] (249.63,121.84) -- (426.09,121.84) -- (426.09,213.34) -- (249.63,213.34) -- cycle ;
%Shape: Rectangle [id:dp5197845752032535] 
\draw  [draw opacity=0][fill={rgb, 255:red, 141; green, 160; blue, 203 }  ,fill opacity=0.9 ] (249.19,41.17) -- (253.09,41.17) -- (253.09,122.22) -- (249.19,122.22) -- cycle ;
%Straight Lines [id:da5731569368051517] 
\draw    (244.9,41.17) -- (248.85,41.17) ;
%Straight Lines [id:da4364226707663972] 
\draw    (429.16,217.31) -- (429.16,213.41) ;
%Shape: Rectangle [id:dp992511427652162] 
\draw  [draw opacity=0] (248.85,96.1) -- (429.16,96.1) -- (429.16,213.41) -- (248.85,213.41) -- cycle ;
%Shape: Rectangle [id:dp48443727144916093] 
\draw  [draw opacity=0][fill={rgb, 255:red, 252; green, 141; blue, 98 }  ,fill opacity=0.9 ] (253.09,41.17) -- (428.38,41.17) -- (428.38,122.22) -- (253.09,122.22) -- cycle ;
%Shape: Rectangle [id:dp2588134942762801] 
\draw  [draw opacity=0][fill={rgb, 255:red, 231; green, 138; blue, 195 }  ,fill opacity=0.9 ] (425.64,122.22) -- (428.77,122.22) -- (428.77,212.63) -- (425.64,212.63) -- cycle ;
%Shape: Rectangle [id:dp9719978077027427] 
\draw  [draw opacity=0][pattern=_1voo82hgq,pattern size=7.125pt,pattern thickness=1.1pt,pattern radius=0pt, pattern color={rgb, 255:red, 255; green, 255; blue, 255}] (249.63,122.22) -- (426.42,122.22) -- (426.42,211.81) -- (249.63,211.81) -- cycle ;
%Shape: Rectangle [id:dp15546083193561167] 
\draw  [draw opacity=0][pattern=_uoqu8884c,pattern size=8.625pt,pattern thickness=0.75pt,pattern radius=0pt, pattern color={rgb, 255:red, 255; green, 255; blue, 255}] (253.09,41.17) -- (428.77,41.17) -- (428.77,122.22) -- (253.09,122.22) -- cycle ;
%Straight Lines [id:da5356738524580724] 
\draw  [dash pattern={on 4.5pt off 4.5pt}]  (244.17,122.74) -- (421.87,122.74) -- (429.16,122.74) ;
%Straight Lines [id:da04031365786359831] 
\draw    (245.61,213.72) -- (441.99,213.72) ;
\draw [shift={(443.99,213.72)}, rotate = 180] [color={rgb, 255:red, 0; green, 0; blue, 0 }  ][line width=0.75]    (10.93,-3.29) .. controls (6.95,-1.4) and (3.31,-0.3) .. (0,0) .. controls (3.31,0.3) and (6.95,1.4) .. (10.93,3.29)   ;
%Straight Lines [id:da5213435242675668] 
\draw    (248.85,216.84) -- (248.85,28.83) ;
\draw [shift={(248.85,26.83)}, rotate = 90] [color={rgb, 255:red, 0; green, 0; blue, 0 }  ][line width=0.75]    (10.93,-3.29) .. controls (6.95,-1.4) and (3.31,-0.3) .. (0,0) .. controls (3.31,0.3) and (6.95,1.4) .. (10.93,3.29)   ;
%Shape: Rectangle [id:dp9148340427939352] 
\draw  [draw opacity=0][pattern=_lozk1vl61,pattern size=7.125pt,pattern thickness=1.1pt,pattern radius=0pt, pattern color={rgb, 255:red, 255; green, 255; blue, 255}] (431.11,122.22) -- (424.86,122.22) -- (424.86,213.37) -- (431.11,213.37) -- cycle ;
%Shape: Rectangle [id:dp06245524540940184] 
\draw  [draw opacity=0][pattern=_f85x3oa8e,pattern size=8.625pt,pattern thickness=0.75pt,pattern radius=0pt, pattern color={rgb, 255:red, 255; green, 255; blue, 255}] (248.85,41.17) -- (253.09,41.17) -- (253.09,121.84) -- (248.85,121.84) -- cycle ;

% Text Node
\draw (242.47,10) node [anchor=north west][inner sep=0.75pt]    {$t_{2}$};
% Text Node
\draw (445.91,206) node [anchor=north west][inner sep=0.75pt]    {$t_{1}$};
% Text Node
\draw (415.05,158) node [anchor=north west][inner sep=0.75pt]  [font=\Large,color={rgb, 255:red, 0; green, 0; blue, 0 }  ,opacity=1 ] [align=left] {$\displaystyle \mathcal{S}_{1}$};
% Text Node
\draw (238.65,71) node [anchor=north west][inner sep=0.75pt]  [font=\Large,color={rgb, 255:red, 0; green, 0; blue, 0 }  ,opacity=1 ] [align=left] {$\displaystyle \mathcal{S}_{3}$};
% Text Node
\draw (327.6,71) node [anchor=north west][inner sep=0.75pt]  [font=\Large,color={rgb, 255:red, 0; green, 0; blue, 0 }  ,opacity=1 ] [align=left] {$\displaystyle \mathcal{S}_{4}$};
% Text Node
\draw (201,115) node [anchor=north west][inner sep=0.75pt]    {$c_{2} +\underline{t}_{1}$};
% Text Node
\draw (422.39,218.05) node [anchor=north west][inner sep=0.75pt]    {$\overline{t}_{1}$};
% Text Node
\draw (244.81,216.5) node [anchor=north west][inner sep=0.75pt]    {$\underline{t}_{1}$};
% Text Node
\draw (230.32,205) node [anchor=north west][inner sep=0.75pt]    {$\underline{t}_{2}$};
% Text Node
\draw (227,32.35) node [anchor=north west][inner sep=0.75pt]    {$\overline{t}_{2}$};
% Text Node
\draw (328.38,158) node [anchor=north west][inner sep=0.75pt]  [font=\Large,color={rgb, 255:red, 0; green, 0; blue, 0 }  ,opacity=1 ] [align=left] {$\displaystyle \mathcal{S}_{2}$};

\end{tikzpicture}

    \caption{Partition~\eqref{eq: partition support-only} of the type space~$\mathcal T$. \label{support-only-figure}}
\end{figure}
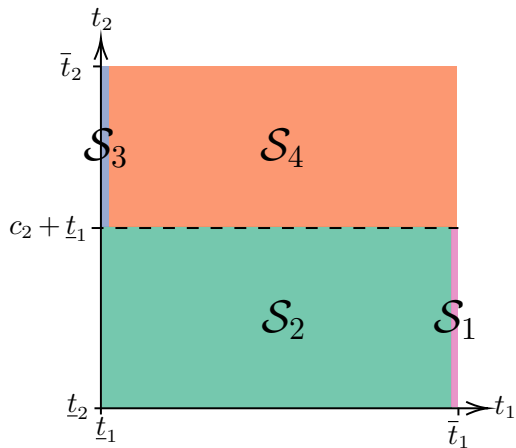

  {As for the base step corresponding to~$k=1$, note that any~$(\bm{p},\bm{q})\in\mathcal{X}$ is feasible in ${\text{SP}}$$_{1}(\bm t)$ for any~$\bm{t} \in \mathcal{S}_{1}$. The objective function value of $(\bm p, \bm q)$ is dominated by that of~$(\bm p^\star, \bm q^\star)$ in~${\text{SP}}$$_{1}(\bm t)$ because
$$
    \sum_{i \in \mathcal{I}} (p_i(\bm t)t_i- q_i(\bm t)c_i) \leq \sum_{i \in \mathcal{I}} p_i(\bm t)t_i \leq t_1 = \sum_{i \in \mathcal{I}} (p_i^\star(\bm t)t_i- q_i^\star(\bm t)c_i),
$$
where the first inequality holds because $c_i > 0$ for all $i \in \mathcal{I}$, while the second inequality follows from our simplifying assumption that $\overline{t}_1>c_2+\underline{t}_1$ and the definition of $\mathcal S_1$, which imply that $t_2 \leq \underline t_1 +c_2 < \overline{t}_1 = t_1$. Thus,~$(\bm{p}^\star,\bm{q}^\star)$ solves ${\text{SP}}$$_{1}(\bm t)$. As a byproduct, we have shown that any mechanism $(\bm{p}, \bm{q})$ feasible in ${\text{SP}}$$_{1}(\bm t)$ can match the payoff $t_1$ of~$(\bm{p}^\star,\bm{q}^\star)$ only if $p_1(\bm{t}) = 1$ and $q_1(\bm{t}) = 0$ because $t_2 < t_1$, $c_i > 0$ and $(\bm{p}, \bm{q})$ satisfies the~\eqref{eq:FC} constraints $\sum_{i \in \mathcal{I}} p_i(\bm{t}) \leq 1$ and $q_i(\bm{t}) \geq 0$.}

{  As for the first induction step corresponding to~$k=2$, consider any~$\bm{t} \in \mathcal{S}_{2}$ and $(\bm p, \bm q)$ feasible in~${\text{SP}}$$_{2}(\bm t)$. From the base step we know that $(\bm p^\star, \bm q^\star)$ solves ${\text{SP}}$$_{1}(\bm t')$ for all $\bm t' \in \mathcal S_{1}$. The constraints of ${\text{SP}}$$_{2}(\bm t)$ thus ensure that $(\bm p, \bm q)$ solves~${\text{SP}}$$_{1}(\bm t')$ for all $\bm t' \in \mathcal{S}_1$, too. The reasoning in the base step further implies that $p_1(\bm t')=1$ and $q_1(\bm t')=0$ for all $\bm t' \in \mathcal{S}_1$. As any scenario~$\bm{t} \in \mathcal{S}_{2}$ is $1$-unilaterally reachable from $(\overline t_1, t_2) \in \mathcal{S}_1$, and as $p_1(\overline t_1, t_2) -q_1(\overline t_1, t_2)=1$, Lemma~\ref{lem:deviationForFavored}(i) implies that $p_1(\bm{t}) = 1$. The objective function value of~$(\bm p, \bm q)$ in~${\text{SP}}$$_{2}(\bm t)$ thus amounts to $t_1 - q_1(\bm{t})c_1$ and is bounded above by $t_1 = \sum_{i \in \mathcal{I}} (p_i^\star(\bm t)t_i- q_i^\star(\bm t)c_i)$ because~$c_1 \geq 0$. Therefore, $(\bm{p}^\star,\bm{q}^\star)$ solves~${\text{SP}}$$_{2}(\bm t)$. In addition, as~$c_1 >0$, a mechanism $(\bm p , \bm q)$ feasible in~${\text{SP}}$$_{2}(\bm t)$ can attain the optimal payoff~$t_1$ only if $p_1(\bm t)=1$ and $q_1(\bm t)=0$.}

{  Next, set $k=3$, and consider any~$\bm{t} \in \mathcal{S}_{3}$ and~$(\bm p, \bm q)$ feasible in~${\text{SP}}$$_{3}(\bm t)$. The constraints of~${\text{SP}}$$_{3}(\bm t)$ ensure that $(\bm p, \bm q)$ solves~${\text{SP}}$$_{1}(\bm t')$ for all $\bm t' \in \mathcal{S}_1$ as well as~${\text{SP}}$$_{2}(\bm t')$ for all $\bm t' \in \mathcal{S}_{2}$. Our earlier reasoning also implies that $p_1(\bm t')=1$ and $q_1(\bm t')=0$ for all~$\bm t' \in \mathcal{S}_{1} \cup \mathcal{S}_{2}$. As any $\bm t \in \mathcal{S}_{3}$ is $2$-unilaterally reachable from $(\underline t_1, \underline t_2) \in \mathcal{S}_{2}$, and as~$p_1(\underline t_1, \underline t_2)=1$ and $p_2(\underline t_1, \underline t_2)=0$, Lemma~\ref{lem:deviationForFavored}(ii) ensures that $p_2(\bm t)=q_2(\bm t)$. Thus, the objective function value of $(\bm p, \bm q)$ in ${\text{SP}}$$_{3}(\bm t)$ is bounded above by that of $(\bm p^\star, \bm q^\star)$ because
$$
\sum_{i \in \mathcal{I}} (p_i(\bm t)t_i- q_i(\bm t)c_i) \leq p_2(\bm t)(t_2-c_2) + p_1(\bm t)t_1 \leq t_2-c_2 = \sum_{i \in \mathcal{I}} (p_i^\star(\bm t)t_i- q_i^\star(\bm t)c_i),
$$
where the first inequality holds because~$c_1 \geq 0$. The second inequality exploits~\eqref{eq:FC} and the definition of~$\mathcal{S}_{3}$, which implies that $t_2-c_2 > \underline t_1 = t_1$. Thus, $(\bm{p}^\star,\bm{q}^\star)$ solves ${\text{SP}}$$_{3}(\bm t)$. As $t_2-c_2 > \underline t_1 = t_1$, 
the mechanism $(\bm p, \bm q)$ can attain the optimal payoff $t_2 - c_2$ of $(\bm{p}^\star,\bm{q}^\star)$ only if~$p_2(\bm t)=q_2(\bm t)=1$.}

{  Finally, set $k=4$, and consider any~$\bm{t} \in \mathcal{S}_{4}$ and $(\bm p, \bm q)$ feasible in ${\text{SP}}$$_{4}(\bm t)$. The constraints of~${\text{SP}}$$_{4}(\bm t)$ ensure that~$(\bm p, \bm q)$ solves~${\text{SP}}$$_{l}(\bm t')$ for all $\bm t' \in \mathcal{S}_l$ and~$l= 1,2,3$. Our earlier reasoning also implies that $p_1(\bm t')=1$ and $q_1(\bm t')=0$ for all $\bm t' \in \mathcal{S}_{1} \cup \mathcal{S}_{2}$ and that $p_2(\bm t')=q_2(\bm t')=1$ for all $\bm t' \in \mathcal{S}_{3}$. As $p_1(\bm t')=1$ and thus $p_2(\bm t')=0$ for all $\bm t' \in \mathcal{S}_{1} \cup \mathcal{S}_{2}$, we can use Lemma~\ref{lem:deviationForFavored}(ii) and similar arguments as before to conclude that $p_2(\bm t)=q_2(\bm t)$. Also, as any scenario $\bm t \in \mathcal{S}_{4}$ is $1$-unilaterally reachable from $(\underline t_1, t_2) \in \mathcal{S}_{3}$, and as $p_2(\underline t_1, t_2)=1$, which implies that $p_1(\underline t_1, t_2)=0$, Lemma~\ref{lem:deviationForFavored}(ii) ensures that $p_1(\bm t)=q_1(\bm t)$. Thus, the objective function value of $(\bm p, \bm q)$ in ${\text{SP}}$$_{4}(\bm t)$ is bounded above by that of $(\bm p^\star, \bm q^\star)$ because
$$
\sum_{i \in \mathcal{I}} (p_i(\bm t)t_i- q_i(\bm t)c_i) \leq \sum_{i \in \mathcal{I}} p_i(\bm t)(t_i-c_i) \leq \max_{i \in \mathcal{I}} t_i-c_i = \sum_{i \in \mathcal{I}} (p_i^\star(\bm t)t_i- q_i^\star(\bm t)c_i).
$$
Hence, $(\bm{p}^\star,\bm{q}^\star)$ solves ${\text{SP}}$$_{4}(\bm t)$. In summary, we have thus verified condition~(ii) of Proposition~\ref{prop:unifyingPRO} by using spatial induction. This establishes the claim that $(\bm{p}^\star,\bm{q}^\star)$ is Pareto robustly optimal.}

\section{Markov Ambiguity Sets}\label{sec: Markov Ambiguity Sets}

Although simple and adequate for situations in which there is no distributional information at all, support-only ambiguity sets may be perceived as conservative in practice. In the following, we thus investigate the mechanism design problem~\eqref{eq:MDP} under the assumption that distributional uncertainty is captured by a Markov ambiguity set of the form
\begin{equation}\label{eq: Markov ambiguity set}
\begin{aligned}
\mathcal{P} = \left\{ \mathbb{P} \in \mathcal{P}_0(\mathcal{T}) \,:\,  \mathbb{E}_{\mathbb{P}}[\tilde{t}_i] \in [\underline{\mu}_i,\overline{\mu}_i] \;\; \forall i \in \mathcal{I} \right\},
\end{aligned}
\end{equation}
where $\underline{\mu}_i$ and $\overline{\mu}_i$ denote lower and upper bounds on the expected type $\mathbb{E}_{\mathbb{P}}[\tilde{t}_i]$ of agent $i \in \mathcal{I}$. We assume without much loss of generality that $\underline{t}_i < \underline{\mu}_i < \overline{\mu}_i < \overline{t}_i$ for all $i \in \mathcal{I}$. Under a Markov ambiguity set, the principal has information about the support as well as the mean of the agents' types.

%%SOME DRAFTS IF WE WANT TO EXPLAIN THE ASSUMPTION FURTHER
%{\color{red} We make this technical assumption because it facilitates the Pareto robust optimality result of this section. On the other hand, this assumption also practically makes sense because ...}
%{  We make this technical assumption to avoid extreme cases that complicate the proofs of this section without adding valuable insights.} 

Recall that if the principal knew the agents' types ex ante, then she could simply allocate the good to the agent with the highest type without inspection. Therefore, the optimal value $z^\star$ of problem~\eqref{eq:MDP} cannot exceed $\inf_{\mathbb{P} \in \mathcal{P}} \mathbb{E}_{\mathbb{P}} [\max_{i \in \mathcal{I}} \tilde{t}_i]$. The next proposition shows that if $\mathcal{P}$ is a Markov ambiguity set, then this upper bound coincides with $\max_{i \in \mathcal{I}} \underline{\mu}_i$ and is attained by an admissible mechanism. 
% (Comment Cagil: We are not talking about a particular mechanism here just yet) Hence, this mechanism must be optimal.

\begin{proposition}\label{prop:optimal-z-markov}
If $\mathcal{P}$ is a Markov ambiguity set of the form \eqref{eq: Markov ambiguity set}, then problem \eqref{eq:MDP} is solvable, and $z^\star = \max_{i \in \mathcal{I}} \underline{\mu}_i$.
\end{proposition}

Contrasting Proposition~\ref{prop:optimal-z-markov} with Proposition~\ref{prop:optimal-z-support} shows that the principal can increase her optimal worst-case expected payoff from ${\max}_{i \in \mathcal{I}} \underline{t}_i$ to ${\max}_{i \in \mathcal{I}} \underline{\mu}_i$ by acquiring information about the mean values of the agents' types. The next theorem characterizes a class of favored-agent mechanisms that attain the optimal value $z^\star = \max_{i \in \mathcal{I}} \underline{\mu}_i$ of problem~\eqref{eq:MDP} under a Markov ambiguity set.

\begin{theorem}\label{thr:opt-fav-markov}
If $\mathcal{P}$ is a Markov ambiguity set of the form \eqref{eq: Markov ambiguity set}, then any favored-agent mechanism with favored agent $i^\star \in {\arg \max}_{i \in \mathcal{I}} \underline{\mu}_i$ and threshold value $\nu^\star \geq \overline{t}_{i^\star}$ is optimal in \eqref{eq:MDP}.  
\end{theorem}

\begin{remark}\label{rem: sub-optimal thresholds}
Theorem~\ref{thr:opt-fav-markov} is sharp in the sense that there are problem instances for which any favored-agent mechanism with favored agent $i^\star \in \arg \max_{i \in \mathcal{I}} \underline{\mu}_i$ and threshold value $\nu < \overline{t}_{i^\star}$ is strictly suboptimal. To see this, consider an instance of problem~\eqref{eq:MDP} with $I=2$ agents, where $\mathcal{T}_1 =[1,6]$, $\mathcal{T}_2 =[0,10]$, $[\underline \mu_1,\overline\mu_1]=[4,5]$, $[\underline \mu_2,\overline\mu_2]=[3,7]$ and $c_1=c_2=2$. By Proposition~\ref{prop:optimal-z-markov}, the optimal value of problem~\eqref{eq:MDP} is thus given by $\max_{i \in \mathcal{I}} \underline{\mu}_i=4$. Consider now any favored agent mechanism with favored agent $1 \in \arg \max_{i \in \mathcal{I}} \underline{\mu}_i$ and threshold value $\nu<\overline{t}_1=6$. In the following, we prove that this mechanism is suboptimal. To this end, assume first that $\nu<1$. If $\bm{t}=\underline{\bm{\mu}}=(4,3)$, then the mechanism allocates the good to agent 1 with verification and earns $t_1-c_1=2$.
%as $\underline{{\mu}}_1 - c_1 = 2 > \underline{{\mu}}_2 - c_2 = 1$ and $\underline{{\mu}}_2 - c_2 =1 > \nu$. 
As the discrete distribution that assigns unit probability to the scenario $\underline{\bm{\mu}}$ belongs to the Markov ambiguity set~\eqref{eq: Markov ambiguity set}, the worst-case expected payoff across all admissible distributions cannot exceed~$2$, which is strictly smaller than the optimal worst-case expected payoff. Assume next that $\nu \in [1,6)$, and 
%let $\varepsilon=6-\nu$ that is strictly positive by definition.
consider the discrete distribution $\mathbb P$ that assigns probability $\frac{1}{2}$ to the scenarios
%$(6,8-\varepsilon/4)$
$(6,6.5+\nu/4)$ and $(2,0)$ each. One readily verifies that $\mathbb P$ belongs to the Markov ambiguity set~\eqref{eq: Markov ambiguity set}. In scenario $(6,6.5+\nu/4)$, the mechanism allocates the good to agent~2 with verification because $t_2-c_2 = (6.5+\nu/4) - 2 > \nu$ and $t_2-c_2 > 4 = t_1-c_1$. In scenario $(2,0)$, on the other hand, the mechanism allocates the good to agent 1 without verification. Thus, the expected payoff of the mechanism under the distribution~$\mathbb P$ amounts to $\frac{1}{2}(4.5+\nu/4)+ \frac{1}{2}2 = 3.25+\nu/8$, and the worst-case expected payoff over all admissible distributions cannot exceed $3.25+\nu/8$, which is strictly smaller than the optimal worst-case expected payoff. In summary, the mechanism is strictly suboptimal for all $\nu<6$.
\hfill$\square$
\end{remark}

%\begin{wrapfigure}{r}{0.45\textwidth}
%\centering
%\includegraphics[width=0.45\textwidth]{remark-markov.png}
%\caption{A sub-optimal favored agent mechanism under \eqref{eq: Markov ambiguity set}. \label{figure:suboptimal-markov}}
%\end{wrapfigure}
In the remainder of this section, we will show that among all optimal favored-agent
mechanism identified in Theorem~\ref{thr:opt-fav-markov} there is always one that is Pareto robustly optimal; see Theorem~\ref{thr:PRO-fav-agents-markov} below. The proof of this main result requires {  two preliminary lemmas.}
%built upon the assumption that $\arg \max_{i \in \mathcal{I}} \underline{\mu}_i$ is a singleton (meaning that the favored agent is uniquely determined). The first one helps us prove the condition~(i) of Proposition \ref{prop:unifyingPRO}, whereas the second one will formalise a partition of the type set under which both conditions of Proposition \ref{prop:unifyingPRO} are satisfied for our Pareto robust optimal mechanism.
We stress that, even though { these lemmas} rely on the assumption that the set $\arg \max_{i \in \mathcal{I}} \underline{\mu}_i$ is a singleton (meaning that the favored agent is uniquely determined), Theorem~\ref{thr:PRO-fav-agents-markov} will {\em not} depend on this restrictive assumption. 

We first show that any type profile $\bm t \in \mathcal T$ has a strictly positive probability under some two-point distribution in the Markov ambiguity set.

\begin{lemma}\label{lem:findP}
If $\mathcal{P}$ is a Markov ambiguity set of the form \eqref{eq: Markov ambiguity set} and $\arg \max_{i \in \mathcal{I}} \underline{\mu}_i = \{i^\star\}$ is a singleton,  then, for any type profile $\bm{t} \in \mathcal{T}$ there exist a scenario $\hat{\bm{t}} \in \mathcal{T}$ with $\max_{i\neq i^\star} \hat t_i< {  \underline{\mu}_{i^\star}}$
%\hat t_{i^\star}$ 
and a discrete distribution $\mathbb{P} \in \mathcal{P}$ such that
(i) $\mathbb{E}_{\mathbb{P}}[\tilde{t}_i] =\underline{\mu}_i$ for all $i \in \mathcal{I}$, (ii) $\mathbb{P}(\tilde{\bm{t}} \in \{\bm{t},\hat{\bm{t}} \})=1$, and
(iii) $\mathbb{P}(\tilde{\bm{t}} =\bm{t})>0$.
\end{lemma}

{ In the next lemma we show that if $\arg \max_{i \in \mathcal{I}} \underline{\mu}_i = \{i^\star\}$ is a singleton, then problem~\eqref{eq:MDP} admits a Pareto robustly optimal favored-agent mechanism~$(\bm{p}^\star,\bm{q}^\star)$. Specifically, we leverage Lemmas~\ref{lem:deviationForFavored} and~\ref{lem:findP} as well as the payoff equivalence principle of Corollary~\ref{Corollary of sufficiency proposition} to show that if a feasible mechanism~$(\bm{p},\bm{q})$ generates the same or a higher {\em expected} payoff than~$(\bm{p}^\star,\bm{q}^\star)$ under every {\em distribution} $\mathbb P\in \mathcal P$, then it must generate the same payoff as $(\bm{p}^\star,\bm{q}^\star)$ in every {\em scenario} $\bm t\in\mathcal T$. %This readily implies that $(\bm{p}^\star,\bm{q}^\star)$ is Pareto robustly optimal in~\eqref{eq:MDP}.
}
\begin{lemma}\label{lem:PRO-revenue equivalence}
    Assume that $\mathcal{P}$ is a Markov ambiguity set of the form \eqref{eq: Markov ambiguity set} and $\arg \max_{i \in \mathcal{I}} \underline{\mu}_i = \{i^\star\}$ is a singleton, and let $(\bm{p}^\star,\bm{q}^\star)$ be the type~(ii) favored-agent mechanism with favored agent ${i^\star}$and threshold value $\nu^\star = \overline{t}_{i^\star}$. Then, any mechanism~$(\bm{p},\bm{q}) \in \mathcal X$ that weakly Pareto robustly dominates $(\bm{p}^\star,\bm{q}^\star)$ must generate the same payoff as $(\bm{p}^\star,\bm{q}^\star)$ in every scenario~$\bm t\in\mathcal T$, that is, we have
    \begin{equation*}\label{eq:revenue-equivalence}
	\begin{aligned}
	   &\sum_{i \in \mathcal{I}} (p_i(\bm{t}) t_i - q_i(\bm{t}) c_i)  =  \sum_{i \in \mathcal{I}} (p_i^\star(\bm{t}) t_i - q_i^\star(\bm{t}) c_i)  &&\forall \bm{t} \in \mathcal{T}.
	\end{aligned}
    \end{equation*}
\end{lemma}
{  The assumption that $\arg \max_{i \in \mathcal{I}} \underline{\mu}_i = \{i^\star\}$ is a singleton will be relaxed in Theorem~\ref{thr:PRO-fav-agents-markov} below. To gain some intuition, we first} sketch the proof of Lemma~\ref{lem:PRO-revenue equivalence} when there are only two agents with $\underline{\mu}_2<\underline{\mu}_1$ such that ${\arg \max}_{i \in \mathcal{I}} \underline{\mu}_i = \{1\}$ is a singleton. In the subsequent discussion, we assume that $\overline{t}_2>c_2+\overline{t}_1$, and we use the following partition of the type space $\mathcal{T}$, which is visualized in Figure~\ref{markov-figure}. 
{ 
\begin{equation}\label{eq: partition}
\begin{aligned}
\mathcal{S}_1 &= \{\bm{t} \in  \mathcal{T} \,:\, t_2-c_2 < \overline t_1, \; t_2<\underline \mu_1 \; \text{and} \; t_1 = \underline{\mu}_1\}\\
\mathcal{S}_2 &=  \{\bm{t} \in  \mathcal{T} \,:\, t_2-c_2 < \overline t_1, \; t_2 < \underline \mu_1 \; \text{and} \; t_1 \neq \underline{\mu}_1\}\\
\mathcal{S}_3 &= \{\bm{t} \in  \mathcal{T} \,:\, t_2 -c_2 < \overline{t}_1, \; t_2 \geq \underline \mu_1 \;\text{and}\; t_1 = \overline t_1 \}\\
\mathcal{S}_4 &= \{\bm{t} \in  \mathcal{T} \,:\, t_2 - c_2 < \overline{t}_1, \; t_2 \geq \underline \mu_1 \; \text{and} \; t_1 < \overline{t}_1 \}\\
\mathcal{S}_5 &= \{\bm{t} \in  \mathcal{T} \,:\, t_2-c_2 \geq \overline{t}_1\}
\end{aligned}
\end{equation}}
Note that some inequalities in the definitions of these sets are redundant and only included for better readability. Using our simplifying assumptions on $\underline{\mu}_1$, $\underline{\mu}_2$, $\overline{t}_1$, $\overline{t}_2$ and $c_2$, one can show that all of the above sets are nonempty. 
%% SEE SCENARIOS IN EACH SET HERE -- Particularly, we have $\underline{\bm{\mu}} \in \mathcal{T}_I$, $(\underline{\mu}_1,\underline{\mu}_1) \in \mathcal{T}_{II}$, $(\overline{t}_1,\overline{t}_1) \in \mathcal{T}_{III}$, $(\hat{t}+c_2,\hat{t}) \in \mathcal{T}_{IV}$ where $\hat{t}= \max\{\underline{t}_1,\overline{t}_1-c_2\}$, and finally $\overline{\bm{t}} \in \mathcal{T}_V$. 
We emphasize, however, that these simplifying assumptions as well as the restriction to two agents are relaxed in the formal proof of Lemma~\ref{lem:PRO-revenue equivalence}.

{ Denote now by $(\bm{p}^\star,\bm{q}^\star)$ the type~(ii) favored-agent mechanism with favored agent $1$ and threshold value $\nu^\star = \overline{t}_1$. By construction, the principal's payoff in scenario~$\bm t$ under~$(\bm{p}^\star,\bm{q}^\star)$ amounts to~$t_1$ when $t_2-c_2 < \overline{t}_1$ ($i.e.$, when $\bm{t} \in \mathcal{S}_{1} \cup \mathcal{S}_{2} \cup \mathcal{S}_{3} \cup \mathcal{S}_{4}$) and to~$\max_{i \in \mathcal{I}} t_i-c_i$ when $t_2-c_2 \geq \overline{t}_1$ ($i.e.$, when $\bm{t} \in \mathcal{S}_{5}$). To prove Lemma~\ref{lem:PRO-revenue equivalence}, it suffices to show that the conditions~(i) and~(ii) of Proposition~\ref{prop:unifyingPRO} are satisfied for the partition~\eqref{eq: partition}. The claim then follows from Corollary~\ref{Corollary of sufficiency proposition}.
%In particular, to show that any mechanism~$(\bm{p},\bm{q}) \in \mathcal X$ that weakly Pareto robustly dominates $(\bm{p}^\star,\bm{q}^\star)$ must generate the same payoff as $(\bm{p}^\star,\bm{q}^\star)$ in every scenario~$\bm t\in\mathcal T$, it suffices to verify the conditions~(i) and~(ii) in Proposition~\ref{prop:unifyingPRO} for the  partition~\eqref{eq: partition}.
We will exploit Lemma~\ref{lem:findP} to verify condition~(i). To verify condition~(ii), we will use induction on~$k$ to show that $(\bm p^\star, \bm q^\star)$ solves the scenario problem~\ref{eq:MDPk(t)} for all~$\bm t \in \mathcal S_k$. The induction step exploits Lemma~\ref{lem:deviationForFavored}.

We first show that condition~(i) is satisfied, that is, for any $\bm t \in \mathcal S_k$ and $k \in \{1, \dots, 5\}$ there exists $\mathbb P \in\mathcal P$ supported on $\cup_{l=1}^k \mathcal{S}_l$ with $\mathbb P(\tilde{\bm t}= \bm t)>0$. Set first $k = 1$, and fix any $\bm t \in \mathcal{S}_1$. By Lemma~\ref{lem:findP-independent}, there exists a scenario $\hat{\bm{t}} \in \mathcal{T}$ with $\max_{i\neq i^\star} \hat t_i< {  \underline{\mu}_{i^\star}}$
%\hat t_{i^\star}$ 
and a discrete distribution $\mathbb{P} \in \mathcal{P}$ such that
$\mathbb{E}_{\mathbb{P}}[\tilde{t}_i] =\underline{\mu}_i$ for all $i \in \mathcal{I}$, $\mathbb{P}(\tilde{\bm{t}} \in \{\bm{t},\hat{\bm{t}} \})=1$, and
$\mathbb{P}(\tilde{\bm{t}} =\bm{t})>0$. By the definition of $\mathcal S_1$, we have $t_{i^\star} = \underline \mu_{i^\star}$, and thus $\mathbb{E}_{\mathbb{P}}[\tilde{t}_{i^\star}] =\underline{\mu}_{i^\star}$ can hold only if $\hat t_{i^\star} = \underline \mu_{i^\star}$. As $\max_{i\neq i^\star} \hat t_i< \underline{\mu}_{i^\star}$, this means that $\hat{\bm t} \in \mathcal S_1$ and $\mathbb P \in \mathcal{P}_0(\mathcal{S}_1)$. Condition~(i) thus holds for $k=1$ and any $\bm t \in \mathcal S_1$. For any $k \in \{2,\ldots,5\}$ and $\bm t \in \mathcal{S}_k$, condition~(i) can easily be verified by using Lemma~\ref{lem:findP} and by noting that $\hat{\bm t} \in \mathcal{S}_1 \cup \mathcal{S}_2 =\{\bm t \in \mathcal{T}: \; \max_{i \neq i^\star} t_i < \underline \mu_{i^\star}\}$, which implies that the discrete distribution $\mathbb P$ is supported on $\mathcal{S}_1 \cup \mathcal{S}_2 \cup \mathcal S_k$. } 

\tikzset{
pattern size/.store in=\mcSize, 
pattern size = 5pt,
pattern thickness/.store in=\mcThickness, 
pattern thickness = 0.3pt,
pattern radius/.store in=\mcRadius, 
pattern radius = 1pt}
\makeatletter
\pgfutil@ifundefined{pgf@pattern@name@_u95qnt6jt}{
\pgfdeclarepatternformonly[\mcThickness,\mcSize]{_u95qnt6jt}
{\pgfqpoint{0pt}{-\mcThickness}}
{\pgfpoint{\mcSize}{\mcSize}}
{\pgfpoint{\mcSize}{\mcSize}}
{
\pgfsetcolor{\tikz@pattern@color}
\pgfsetlinewidth{\mcThickness}
\pgfpathmoveto{\pgfqpoint{0pt}{\mcSize}}
\pgfpathlineto{\pgfpoint{\mcSize+\mcThickness}{-\mcThickness}}
\pgfusepath{stroke}
}}
\makeatother

% Pattern Info
 
\tikzset{
pattern size/.store in=\mcSize, 
pattern size = 5pt,
pattern thickness/.store in=\mcThickness, 
pattern thickness = 0.3pt,
pattern radius/.store in=\mcRadius, 
pattern radius = 1pt}
\makeatletter
\pgfutil@ifundefined{pgf@pattern@name@_sknu39ly0}{
\pgfdeclarepatternformonly[\mcThickness,\mcSize]{_sknu39ly0}
{\pgfqpoint{-\mcThickness}{-\mcThickness}}
{\pgfpoint{\mcSize}{\mcSize}}
{\pgfpoint{\mcSize}{\mcSize}}
{
\pgfsetcolor{\tikz@pattern@color}
\pgfsetlinewidth{\mcThickness}
\pgfpathmoveto{\pgfpointorigin}
\pgfpathlineto{\pgfpoint{0}{\mcSize}}
\pgfusepath{stroke}
}}
\makeatother

% Pattern Info
 
\tikzset{
pattern size/.store in=\mcSize, 
pattern size = 5pt,
pattern thickness/.store in=\mcThickness, 
pattern thickness = 0.3pt,
pattern radius/.store in=\mcRadius, 
pattern radius = 1pt}
\makeatletter
\pgfutil@ifundefined{pgf@pattern@name@_lk8bccha7}{
\pgfdeclarepatternformonly[\mcThickness,\mcSize]{_lk8bccha7}
{\pgfqpoint{0pt}{0pt}}
{\pgfpoint{\mcSize+\mcThickness}{\mcSize+\mcThickness}}
{\pgfpoint{\mcSize}{\mcSize}}
{
\pgfsetcolor{\tikz@pattern@color}
\pgfsetlinewidth{\mcThickness}
\pgfpathmoveto{\pgfqpoint{0pt}{0pt}}
\pgfpathlineto{\pgfpoint{\mcSize+\mcThickness}{\mcSize+\mcThickness}}
\pgfusepath{stroke}
}}
\makeatother

% Pattern Info
 
\tikzset{
pattern size/.store in=\mcSize, 
pattern size = 5pt,
pattern thickness/.store in=\mcThickness, 
pattern thickness = 0.3pt,
pattern radius/.store in=\mcRadius, 
pattern radius = 1pt}
\makeatletter
\pgfutil@ifundefined{pgf@pattern@name@_na8poby7a lines}{
\pgfdeclarepatternformonly[\mcThickness,\mcSize]{_na8poby7a}
{\pgfqpoint{0pt}{0pt}}
{\pgfpoint{\mcSize+\mcThickness}{\mcSize+\mcThickness}}
{\pgfpoint{\mcSize}{\mcSize}}
{\pgfsetcolor{\tikz@pattern@color}
\pgfsetlinewidth{\mcThickness}
\pgfpathmoveto{\pgfpointorigin}
\pgfpathlineto{\pgfpoint{\mcSize}{0}}
\pgfusepath{stroke}}}
\makeatother
\tikzset{every picture/.style={line width=0.75pt}} %set default line width to 0.75pt        

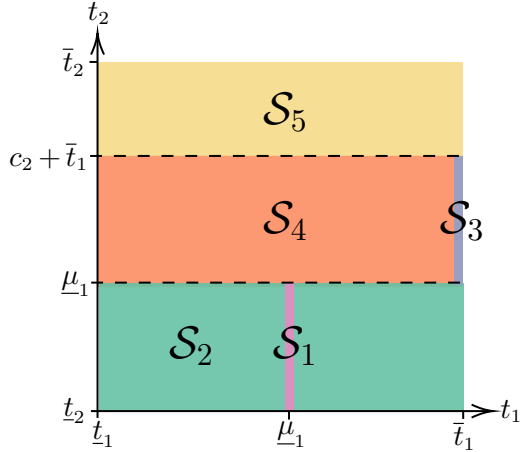
\begin{figure}
\centering             

\begin{tikzpicture}[x=0.75pt,y=0.75pt,yscale=-1,xscale=1]

%Shape: Rectangle [id:dp8726994378435775] 
\draw  [draw opacity=0][fill={rgb, 255:red, 252; green, 141; blue, 98 }  ,fill opacity=0.9 ] (253.03,85.38) -- (437.45,85.38) -- (437.45,151.56) -- (253.03,151.56) -- cycle ;
%Shape: Rectangle [id:dp5008096374782514] 
\draw  [draw opacity=0][pattern=_u95qnt6jt,pattern size=8.625pt,pattern thickness=0.75pt,pattern radius=0pt, pattern color={rgb, 255:red, 255; green, 255; blue, 255}] (253.62,86.18) -- (431.98,86.18) -- (431.98,150.12) -- (253.62,150.12) -- cycle ;
%Shape: Rectangle [id:dp6341416181352075] 
\draw  [draw opacity=0][fill={rgb, 255:red, 102; green, 194; blue, 165 }  ,fill opacity=0.9 ] (253.77,149.8) -- (437.45,149.8) -- (437.45,213.96) -- (253.77,213.96) -- cycle ;
%Shape: Rectangle [id:dp47957965123559676] 
\draw  [draw opacity=0][pattern=_sknu39ly0,pattern size=6.300000000000001pt,pattern thickness=1.1pt,pattern radius=0pt, pattern color={rgb, 255:red, 255; green, 255; blue, 255}] (253.51,149.28) -- (437.19,149.28) -- (437.19,213.17) -- (253.51,213.17) -- cycle ;
%Shape: Rectangle [id:dp4411059767753531] 
\draw  [draw opacity=0][fill={rgb, 255:red, 141; green, 160; blue, 203 }  ,fill opacity=0.9 ] (432.98,85.44) -- (437.45,85.44) -- (437.45,150.07) -- (432.98,150.07) -- cycle ;
%Shape: Rectangle [id:dp24937151659623025] 
\draw  [draw opacity=0][pattern=_lk8bccha7,pattern size=8.625pt,pattern thickness=0.75pt,pattern radius=0pt, pattern color={rgb, 255:red, 255; green, 255; blue, 255}] (432.98,83.84) -- (437.45,83.84) -- (437.45,148.47) -- (432.98,148.47) -- cycle ;
%Shape: Rectangle [id:dp4574083656870669] 
\draw  [draw opacity=0][fill={rgb, 255:red, 231; green, 138; blue, 195 }  ,fill opacity=0.9 ] (347.53,149.32) -- (352,149.32) -- (352,213.96) -- (347.53,213.96) -- cycle ;
%Straight Lines [id:da5291563712691492] 
\draw    (248.98,37.94) -- (253.03,37.94) ;
%Straight Lines [id:da8092673098524588] 
\draw    (437.45,218.11) -- (437.45,214.12) ;
%Straight Lines [id:da5903530420528216] 
\draw  [dash pattern={on 4.5pt off 4.5pt}]  (249.83,149.32) -- (431.59,149.32) -- (439.05,149.32) ;
%Shape: Rectangle [id:dp44159837868731944] 
\draw  [draw opacity=0][fill={rgb, 255:red, 246; green, 220; blue, 138 }  ,fill opacity=0.9 ] (253.03,37.94) -- (437.45,37.94) -- (437.45,85.38) -- (253.03,85.38) -- cycle ;
%Straight Lines [id:da36867648160885325] 
\draw    (249.71,214.17) -- (450.62,214.17) ;
\draw [shift={(452.62,214.17)}, rotate = 180] [color={rgb, 255:red, 0; green, 0; blue, 0 }  ][line width=0.75]    (10.93,-3.29) .. controls (6.95,-1.4) and (3.31,-0.3) .. (0,0) .. controls (3.31,0.3) and (6.95,1.4) .. (10.93,3.29)   ;
%Straight Lines [id:da19476656787521462] 
\draw  [dash pattern={on 4.5pt off 4.5pt}]  (249.3,85.38) -- (431.06,85.38) -- (438.52,85.38) ;
%Straight Lines [id:da002568934481421481] 
\draw    (253.03,217.63) -- (253.03,25.27) ;
\draw [shift={(253.03,23.27)}, rotate = 90] [color={rgb, 255:red, 0; green, 0; blue, 0 }  ][line width=0.75]    (10.93,-3.29) .. controls (6.95,-1.4) and (3.31,-0.3) .. (0,0) .. controls (3.31,0.3) and (6.95,1.4) .. (10.93,3.29)   ;
%Shape: Rectangle [id:dp11895883509483718] 
\draw  [draw opacity=0][pattern=_na8poby7a,pattern size=7.5pt,pattern thickness=1.1pt,pattern radius=0pt, pattern color={rgb, 255:red, 255; green, 255; blue, 255}] (347.53,149.12) -- (352,149.12) -- (352,213.76) -- (347.53,213.76) -- cycle ;
%Straight Lines [id:da2505520426083696] 
\draw    (349.6,218.43) -- (349.6,214.44) ;

% Text Node
\draw (246.67,6) node [anchor=north west][inner sep=0.75pt]    {$t_{2}$};
% Text Node
\draw (454.71,207) node [anchor=north west][inner sep=0.75pt]    {$t_{1}$};
% Text Node
\draw (340,170) node [anchor=north west][inner sep=0.75pt]  [font=\Large,color={rgb, 255:red, 0; green, 0; blue, 0 }  ,opacity=1 ] [align=left] {$\displaystyle \mathcal{S}_{1}$};
% Text Node
\draw (207,77.06) node [anchor=north west][inner sep=0.75pt]    {$c_{2} +\overline{t}_{1}$};
% Text Node
\draw (431,218.02) node [anchor=north west][inner sep=0.75pt]    {$\overline{t}_{1}$};
% Text Node
\draw (249.07,217.92) node [anchor=north west][inner sep=0.75pt]    {$\underline{t}_{1}$};
% Text Node
\draw (234.29,206) node [anchor=north west][inner sep=0.75pt]    {$\underline{t}_{2}$};
% Text Node
\draw (233.29,142) node [anchor=north west][inner sep=0.75pt]    {$\underline{\mu }_{1}$};
% Text Node
\draw (234,29.14) node [anchor=north west][inner sep=0.75pt]    {$\overline{t}_{2}$};
% Text Node
\draw (287.59,170) node [anchor=north west][inner sep=0.75pt]  [font=\Large,color={rgb, 255:red, 0; green, 0; blue, 0 }  ,opacity=1 ] [align=left] {$\displaystyle \mathcal{S}_{2}$};
% Text Node
\draw (424.15,107) node [anchor=north west][inner sep=0.75pt]  [font=\Large,color={rgb, 255:red, 0; green, 0; blue, 0 }  ,opacity=1 ] [align=left] {$\displaystyle \mathcal{S}_{3}$};
% Text Node
\draw (335,107) node [anchor=north west][inner sep=0.75pt]  [font=\Large,color={rgb, 255:red, 0; green, 0; blue, 0 }  ,opacity=1 ] [align=left] {$\displaystyle \mathcal{S}_{4}$};
% Text Node
\draw (335,52) node [anchor=north west][inner sep=0.75pt]  [font=\Large,color={rgb, 255:red, 0; green, 0; blue, 0 }  ,opacity=1 ] [align=left] {$\displaystyle \mathcal{S}_{5}$};
% Text Node
\draw (342.1,218) node [anchor=north west][inner sep=0.75pt]    {$\underline{\mu }_{1}$};

\end{tikzpicture}

\caption{Partition \eqref{eq: partition} of the type space~$\mathcal T$. \label{markov-figure}}
\end{figure}

%~~~~~~~~~~~~~~~~~~Figure2~~~~~~~~~~~~~~~~~~

% \begin{wrapfigure}{R}{0.45\textwidth}
% \centering
% \includegraphics[width=0.45\textwidth]{proof-markov.png}
% \caption{Partition of the type space~$\mathcal T$. \label{markov-figure}}
% \end{wrapfigure}

{ Next, we prove condition~(ii) by induction on~$k$. As for the base step corresponding to $k=1$, note that any $(\bm{p},\bm{q}) \in \mathcal{X}$ is feasible in $\text{SP}_1(\bm t)$ for any $\bm t \in \mathcal{S}_1$. As $t_2 <t_1$ and $c_i >0$, the objective function value of $(\bm{p},\bm{q})$ in $\text{SP}_1(\bm t)$ is bounded above by $t_1 = \sum_{i \in \mathcal{I}} (p_i^\star(\bm t)t_i- q_i^\star(\bm t)c_i)$. Therefore, $(\bm{p}^\star,\bm{q}^\star)$ solves~$\text{SP}_1(\bm t)$. Also, a mechanism $(\bm{p},\bm{q})$ can attain this bound only if $p_1(\bm{t})=1$ and $q_1(\bm{t})=0$ because $t_2 < t_1$, $c_i > 0$ and $(\bm{p}, \bm{q})$ satisfies the~\eqref{eq:FC} constraints $\sum_{i \in \mathcal{I}} p_i(\bm{t}) \leq 1$ and $q_i(\bm{t}) \geq 0$. 

As for the first induction step corresponding to~$k=2$, consider any $\bm t \in \mathcal{S}_2$ and $(\bm{p},\bm{q})$ feasible in~$\text{SP}_2(\bm t)$. From the base step we know that~$(\bm p^\star, \bm q^\star)$ solves~${\text{SP}}$$_{1}(\bm t')$ for all $\bm t' \in \mathcal S_{1}$. The constraints of~${\text{SP}}$$_{2}(\bm t)$ thus ensure that $(\bm p, \bm q)$ solves~${\text{SP}}$$_{1}(\bm t')$ for all $\bm t' \in \mathcal{S}_1$, too. From the proof of the base step we further know that $p_1(\bm t')=1$ and $q_1(\bm t')=0$ for all $\bm t' \in \mathcal{S}_1$. As any scenario~$\bm{t} \in \mathcal{S}_{2}$ is $1$-unilaterally reachable from $(\underline \mu_1, t_2) \in \mathcal{S}_1$, and as $p_1(\underline \mu_1, t_2) -q_1(\underline \mu_1, t_2)=1$, Lemma~\ref{lem:deviationForFavored}(i) implies that $p_1(\bm{t}) = 1$. The objective function value of~$(\bm p, \bm q)$ in~${\text{SP}}$$_{2}(\bm t)$ thus amounts to $t_1 - q_1(\bm{t})c_1$ and is bounded above by $t_1$ because~$c_1 > 0$. Thus,~$(\bm{p}^\star,\bm{q}^\star)$ solves~${\text{SP}}$$_{2}(\bm t)$. In addition, as~$c_1 >0$, a mechanism $(\bm p , \bm q)$ feasible in~${\text{SP}}$$_{2}(\bm t)$ can attain the optimal payoff~$t_1$ only if~$p_1(\bm t)=1$ and $q_1(\bm t)=0$.

Next, set $k=3$, and consider any~$\bm{t} \in \mathcal{S}_{3}$ and~$(\bm p, \bm q)$ feasible in~${\text{SP}}$$_{3}(\bm t)$. The constraints of~${\text{SP}}$$_{3}(\bm t)$ ensure that $(\bm p, \bm q)$ solves~${\text{SP}}$$_{1}(\bm t')$ for all $\bm t' \in \mathcal{S}_1$ as well as~${\text{SP}}$$_{2}(\bm t')$ for all $\bm t' \in \mathcal{S}_{2}$. Our earlier reasoning also implies that $p_1(\bm t')=1$ and $q_1(\bm t')=0$ for all~$\bm t' \in \mathcal{S}_{1} \cup \mathcal{S}_{2}$. As any $\bm t \in \mathcal{S}_{3}$ is $2$-unilaterally reachable from $(\overline t_1, \underline t_2) \in \mathcal{S}_{2}$, and as~$p_1(\overline t_1, \underline t_2)=1$ and $p_2(\overline t_1, \underline t_2)=0$, Lemma~\ref{lem:deviationForFavored}(ii) implies that $p_2(\bm t)=q_2(\bm t)$. Thus, the objective function value of $(\bm p, \bm q)$ in ${\text{SP}}$$_{3}(\bm t)$ is bounded above by that of $(\bm p^\star, \bm q^\star)$ because
$$
\sum_{i \in \mathcal{I}} (p_i(\bm t)t_i- q_i(\bm t)c_i) \leq p_2(\bm t)(t_2-c_2) + p_1(\bm t)t_1 \leq t_1 = \sum_{i \in \mathcal{I}} (p_i^\star(\bm t)t_i- q_i^\star(\bm t)c_i),
$$
where the first inequality holds because~$c_1 \geq 0$. The second inequality exploits~\eqref{eq:FC} and the definition of~$\mathcal{S}_{3}$, which implies that $t_2-c_2 < \overline t_1 = t_1$. Thus,~$(\bm{p}^\star,\bm{q}^\star)$ solves~${\text{SP}}$$_{3}(\bm t)$. As $t_2-c_2 <\overline t_1 = t_1$, 
the mechanism $(\bm p, \bm q)$ can attain the optimal payoff $t_1$ only if~$p_1(\bm t)=1$ and $q_1(\bm t)=0$.

Next, set $k=4$, and consider any~$\bm{t} \in \mathcal{S}_{4}$ and $(\bm p, \bm q)$ feasible in ${\text{SP}}$$_{4}(\bm t)$. The constraints of~${\text{SP}}$$_{4}(\bm t)$ ensure that~$(\bm p, \bm q)$ solves~${\text{SP}}$$_{l}(\bm t')$ for all $\bm t' \in \mathcal{S}_l$ and~$l= 1,2,3$. The previous arguments imply that $p_1(\bm t')=1$ and $q_1(\bm t')=0$ for all $\bm t' \in \mathcal{S}_{1} \cup \mathcal{S}_{2} \cup \mathcal{S}_3$. 
%As $p_1(\bm t')=1$ and thus $p_2(\bm t')=0$ for all $\bm t' \in \mathcal{S}_{1} \cup \mathcal{S}_{2} \cup \mathcal{S}_3$, 
We can then use Lemma~\ref{lem:deviationForFavored}(i) and similar arguments as before to conclude that $p_1(\bm t)=1$. Thus, the objective function value of $(\bm p, \bm q)$ in ${\text{SP}}$$_{4}(\bm t)$ is bounded above by $t_1- q_1(\bm t)c_1 \leq t_1$, and~$(\bm{p}^\star,\bm{q}^\star)$ solves~${\text{SP}}$$_{4}(\bm t)$. As $c_1>0$, the mechanism $(\bm p, \bm q)$ can attain the same payoff only if $p_1(\bm t)=1$ and $q_1(\bm t)=0$.

Finally, set $k=5$, and consider any~$\bm{t} \in \mathcal{S}_{5}$ and~$(\bm p, \bm q)$ feasible in~${\text{SP}}$$_{5}(\bm t)$. The constraints of~${\text{SP}}$$_{5}(\bm t)$ ensure that $(\bm p, \bm q)$ solves~${\text{SP}}$$_{l}(\bm t')$ for all $\bm t' \in \mathcal{S}_l$ and~$l= 1,2,3,4$. Our earlier reasoning implies that $p_1(\bm t')=1$ and $q_1(\bm t')=0$ for all~$\bm t' \in \cup_{l=1}^4 \mathcal{S}_{l}$. As any $\bm t \in \mathcal{S}_{5}$ is $2$-unilaterally reachable from $( t_1, \underline t_2) \in \mathcal{S}_{1} \cup \mathcal{S}_{2}$, and as~$p_1( t_1, \underline t_2)=1$ and $p_2( t_1, \underline t_2)=0$, Lemma~\ref{lem:deviationForFavored}(ii) ensures that $p_2(\bm t)=q_2(\bm t)$. Thus, the objective function value of $(\bm p, \bm q)$ in ${\text{SP}}$$_{5}(\bm t)$ is bounded above by that of $(\bm p^\star, \bm q^\star)$ because
$$
\sum_{i \in \mathcal{I}} (p_i(\bm t)t_i- q_i(\bm t)c_i) \leq p_2(\bm t)(t_2-c_2) + p_1(\bm t)t_1 \leq t_2-c_2 = \sum_{i \in \mathcal{I}} (p_i^\star(\bm t)t_i- q_i^\star(\bm t)c_i),
$$
where the first inequality holds because~$p_2(\bm t)=q_2(\bm t)$ and~$c_1 \geq 0$. The second inequality exploits~\eqref{eq:FC} and the definition of~$\mathcal{S}_{5}$, which implies that $t_2-c_2 \geq \overline t_1 \geq t_1$. Thus,~$(\bm{p}^\star,\bm{q}^\star)$ solves~${\text{SP}}$$_{5}(\bm t)$.
In summary, we have thus verified that~$(\bm{p}^\star,\bm{q}^\star)$ satisfies condition~(ii) of Proposition~\ref{prop:unifyingPRO} for the partition~\eqref{eq: partition}.}

%Assume that $\arg \max_{i \in \mathcal{I}} \underline{\mu}_i = \{i^\star\}$ is a singleton. Then, any favored-agent mechanism with favored agent~$i^\star$ and threshold value~$\overline{t}_{i^\star}$ allocates the good to the favored agent without inspection in all scenarios $\bm{t} \in \mathcal{T}$ with~$\max_{i \neq i^\star} t_i-c_i<\overline{t}_{i^\star}$, that is, it satisfies the implications of Lemma~\ref{lem:PRO-revenue equivalence}. Furthermore, both the type~(i) and type~(ii) versions of this favored-agent mechanism satisfy the optimality condition of Lemma~\ref{lemma:markov-opt-condition1}. Thus, both versions of this mechanism are optimal. % when $\arg \max_{i \in \mathcal{I}} \underline{\mu}_i= \{i^\star\}$. 
%The next lemma implies that the type~(ii) version of this mechanism is also Pareto robustly optimal. 
% when $\arg \max_{i \in \mathcal{I}} \underline{\mu}_i= \{i^\star\}$ is a singleton. 

The following main theorem shows that \eqref{eq:MDP} admits a Pareto robustly optimal mechanism even if we abandon the simplifying assumption that ${\arg \max}_{i \in \mathcal{I}} \underline{\mu}_i$ is a singleton.

\begin{theorem}\label{thr:PRO-fav-agents-markov}
	If $\mathcal{P}$ is a Markov ambiguity set of the form \eqref{eq: Markov ambiguity set}, then any type~(ii)  favored-agent mechanism $(\bm{p}^\star,\bm{q}^\star)$ with favored agent ${i^\star} \in {\arg \max}_{i \in \mathcal{I}} \underline{\mu}_i$ and threshold value $\nu^\star = \overline{t}_{i^\star}$ is Pareto robustly optimal in \eqref{eq:MDP}.
\end{theorem}

The proof of Theorem \ref{thr:PRO-fav-agents-markov} constructs a perturbed ambiguity set $\mathcal{P}_{\varepsilon} \subseteq \mathcal{P}$, which is obtained by replacing $\underline{\mu}_{i^\star}$ with $\underline{\mu}_{i^\star}+\epsilon$ in the original Markov ambiguity set~\eqref{eq: Markov ambiguity set}. Here, we assume that $\varepsilon>0$ is sufficiently small for $\mathcal P_\varepsilon$ to remain nonempty. By construction, the expected type of agent~$i^\star$ has the largest lower bound in the perturbed ambiguity set~$\mathcal{P}_{\varepsilon}$, whereas the expected types of all other agents have strictly smaller lower bounds. %By Lemma~\ref{lemma:markov-revenue-equivalence}, the mechanism~$(\bm{p}^\star,\bm{q}^\star)$ is thus Pareto robustly optimal in problem~\eqref{eq:MDP} with ambiguity set~$\mathcal{P}_{\varepsilon}$.
As~$\mathcal{P}_{\varepsilon} \subseteq \mathcal{P}$, any mechanism $(\bm{p},\bm{q})$ that weakly Pareto robustly dominates $(\bm{p}^\star,\bm{q}^\star)$ with respect to~$\mathcal{P}$ must also weakly Pareto robustly dominate $(\bm{p}^\star,\bm{q}^\star)$ with respect to~$\mathcal{P}_{\varepsilon}$. By Lemma~\ref{lem:PRO-revenue equivalence} applied to $\mathcal{P}_{\varepsilon}$ instead of $\mathcal P$, we may thus conclude that the mechanisms~$(\bm{p},\bm{q})$ and $(\bm{p}^\star,\bm{q}^\star)$ generate the same payoff in every scenario $\bm{t} \in \mathcal{T}$. This in turn implies, however, that~%$(\bm{p},\bm{q})$ cannot Pareto robustly dominate~$(\bm{p}^\star,\bm{q}^\star)$ with respect to~$\mathcal P$, and thus~
$(\bm{p}^\star,\bm{q}^\star)$ must be Pareto robustly optimal in~\eqref{eq:MDP} with respect to~$\mathcal P$.

\section{Markov Ambiguity Sets with Independent Types}\label{sec: Markov Ambiguity Sets with Independent Types}

The Markov ambiguity set studied in Section~\ref{sec: Markov Ambiguity Sets} contains distributions under which the agents' types are dependent. Sometimes, however, the principal may have good reasons to assume that the agents' types are in fact {\em in}dependent. In this section we thus study a subset of the Markov ambiguity set~\eqref{eq: Markov ambiguity set} studied in Section~\ref{sec: Markov Ambiguity Sets}, which imposes the additional condition that the agents' types are mutually independent. Mathematically speaking, we thus investigate the Markov ambiguity set with independent types defined as
\begin{equation}\label{eq: Markov ambiguity set with independent types}
\begin{aligned}
\mathcal{P} = \left\{ \mathbb{P} \in \mathcal{P}_0(\mathcal{T}) ~:~ \begin{array}{l}
\mathbb{E}_{\mathbb{P}}[\tilde{t}_i] \in [\underline{\mu}_i,\overline{\mu}_i] \;\; \forall i \in \mathcal{I},\\
\tilde t_1, \dots, \tilde t_I \text{ are mutually independent under } \mathbb P \end{array}\right\}.
\end{aligned}
\end{equation}
By replacing the Markov ambiguity set~\eqref{eq: Markov ambiguity set} with its subset~\eqref{eq: Markov ambiguity set with independent types}, which contains only distributions under which the agents' types are independent, we can only increase but not decrease the optimal value of problem~\eqref{eq:MDP}. Proposition~\ref{prop:optimal-z-markov} thus implies that the optimal value of problem~\eqref{eq:MDP} with a Markov ambiguity set with independent types is bounded below by $\max_{i \in \mathcal I} \underline{\mu}_i$. The next proposition shows that this lower bound %is attained by a feasible mechanism and \textcolor{red}{-thus-(to be removed)} 
coincides in fact with the optimal value of problem~\eqref{eq:MDP}.

\begin{proposition}\label{prop:optimal-z-markov-independent}
If $\mathcal{P}$ is a Markov ambiguity set with independent types of the form \eqref{eq: Markov ambiguity set with independent types}, then problem \eqref{eq:MDP} is solvable, and $z^\star = \max_{i \in \mathcal{I}} \underline{\mu}_i$.
\end{proposition}

We do not provide a formal proof of Proposition~\ref{prop:optimal-z-markov-independent} because it is similar to that of Proposition~\ref{prop:optimal-z-markov}. However, the proof idea can be summarized as follows.  Proposition~\ref{prop:optimal-z-markov} implies that $\max_{i \in \mathcal{I}} \underline{\mu}_i$ provides a lower bound on~$z^\star$. Proposition~\ref{prop:optimal-z-markov-independent} thus follows if we can show that $\max_{i \in \mathcal{I}} \underline{\mu}_i$ provides also an upper bound on~$z^\star$. 
%which implies that~$\max_{i \in \mathcal{I}} \underline{\mu}_i$ provides a lower bound on problem~\eqref{eq:MDP}. 
This is indeed the case because the agents' types are independent under the Dirac distribution~$\delta_{\underline{\bm{\mu}}}$ that concentrates unit mass at $\underline{\bm{\mu}}$, which implies that~$\delta_{\underline{\bm{\mu}}}$ belongs to the Markov ambiguity set with independent types, and because the expected payoff of {\em any} feasible mechanism under~$\delta_{\underline{\bm{\mu}}}$ is bounded above by $\max_{i \in \mathcal I} \underline{\mu}_i$. 
%One readily verifies that this payoff is attained by the mechanism that allocates the good to an agent $i^\star \in \arg \max_{i \in \mathcal{I}} \underline{\mu}_i$ irrespective of~$\bm{t}\in\mathcal{T}$ and never inspects anyone's type. 
Propositions~\ref{prop:optimal-z-markov} and~\ref{prop:optimal-z-markov-independent} together suggest, perhaps surprisingly, that the principal does not benefit from knowing whether or not the agents' types are independent. At least, this information has no impact on the optimal worst-case expected payoff. 

%We will later see, however, that the principal may nevertheless benefit from the independence information because it allows her to increase her expected payoff in non-worst-case scenarios.

Theorem~\ref{thr:opt-fav-independent} below is reminiscent of Theorem~\ref{thr:opt-fav-markov} from Section~\ref{sec: Markov Ambiguity Sets} and shows that there is again a continuum of infinitely many optimal favored-agent mechanisms. 

\begin{theorem}\label{thr:opt-fav-independent}
If $\mathcal{P}$ is a Markov ambiguity set of the form \eqref{eq: Markov ambiguity set with independent types}, then any favored-agent mechanism with favored agent $i^\star \in {\arg \max}_{i \in \mathcal{I}} \underline{\mu}_i$ and threshold value $\nu^\star \geq {\max}_{i \in \mathcal{I}} \underline{\mu}_i$ is optimal in~\eqref{eq:MDP}.    
\end{theorem}

Comparing Theorems~\ref{thr:opt-fav-markov} and~\ref{thr:opt-fav-independent} reveals that information about the independence of the agents does not affect the choice of the optimal favored agent. However, it reduces the lowest optimal threshold value from $\overline{t}_{i^\star}$ to ${\max}_{i \in \mathcal{I}} \underline{\mu}_i$. Thus, the set of optimal favored-agent mechanisms {\em increases} if the principal learns that the agents are independent. This insight is unsurprising in view of %our earlier observation that the optimal payoff remains unchanged if the principal learns that the agents are independent. In fact, 
the impossibility to monetize such independence information (at least in the worst case), which implies that all mechanisms that were optimal under a Markov ambiguity set of the form~\eqref{eq: Markov ambiguity set} remain optimal under a Markov ambiguity set of the form~\eqref{eq: Markov ambiguity set with independent types}. However, the independence information allows the principal to choose an optimal threshold value that is independent of the favored agent.

%Particularly, the selection criteria of a favored agent remains the same, whereas the principal can select a lower threshold with the additional information of independence. Recall that by Remark \ref{rem: sub-optimal thresholds} any choice of threshold smaller than the highest possible type $\overline{t}_{i^\star}$ of the favored agent is suboptimal if the agents' types are not necessarily independent, $i.e.$, if the ambiguity set is given by \eqref{eq: Markov ambiguity set}.

\begin{remark}
\label{rem: sub-optimal thresholds independent types}
Theorem \ref{thr:opt-fav-independent} is sharp in the sense that there are problem instances for which any favored-agent mechanism with favored agent $i^\star \in \arg \max_{i \in \mathcal{I}} \underline{\mu}_i$ and threshold value $\nu < \max_{i \in \mathcal{I}} \underline{\mu}_i$ is strictly suboptimal.
%if $\mathcal{P}$ is a Markov ambiguity set of the form \eqref{eq: Markov ambiguity set with independent types}, and $\max_{i \neq i^\star} \overline{t}_i-c_i \geq \underline{\mu}_{i^\star}$. 
To see this, we revisit the instance of problem~\eqref{eq:MDP} described in Remark~\ref{rem: sub-optimal thresholds}, which involves $I=2$ agents with $c_1=c_2=2$ and a Markov ambiguity set of the form~\eqref{eq: Markov ambiguity set} with parameters $\mathcal{T}_1 =[1,6]$, $\mathcal{T}_2 =[0,10]$, $[\underline \mu_1,\overline\mu_1]=[4,5]$ and $[\underline \mu_2,\overline\mu_2]=[3,7]$. Now, however, we additionally assume that the agents' types are independent such that the ambiguity becomes an instance of~\eqref{eq: Markov ambiguity set with independent types}. Hence, by Proposition~\ref{prop:optimal-z-markov-independent}, the optimal value of problem \eqref{eq:MDP} still amounts to $\max_{i \in \mathcal{I}} \underline{\mu}_i=4$. Consider now any favored agent mechanism with favored agent $1 \in \arg \max_{i \in \mathcal{I}} \underline{\mu}_i$ and threshold value $\nu<\max_{i \in \mathcal{I}} \underline{\mu}_i=4$. In the following, we prove that this mechanism is suboptimal. To this end, assume first that $\nu<1$. If $\bm{t}=\underline{\bm{\mu}}=(4,3)$, then the mechanism allocates the good to agent 1 with verification and earns a payoff of $t_1-c_1=2$. As the discrete distribution that assigns unit probability to the scenario $\underline{\bm{\mu}}$ belongs to the Markov ambiguity set~\eqref{eq: Markov ambiguity set with independent types}, the worst-case expected payoff over all admissible distributions cannot exceed 2, which is strictly smaller than the optimal worst-case expected payoff. Assume next that $\nu \in [1,4)$, and 
%let $\varepsilon = 4-\nu$ that is strictly positive by definition. 
consider the discrete distribution $\mathbb{P}$ that assigns probability $\frac{1}{2}$ to the scenarios
%$(4,6-\varepsilon/4)$
$(4,5+\nu/4)$ and $(4,2)$ each. One readily verifies that $\tilde t_1$ is deterministic under~$\mathbb{P}$ and that~$\mathbb{P}$ belongs to the Markov ambiguity set~\eqref{eq: Markov ambiguity set with independent types}.
%This time, the following discrete distribution is feasible in the Markov ambiguity set \eqref{eq: Markov ambiguity set with independent types} for small enough $\varepsilon>0$. The agents' types are realized as $(4,6-\varepsilon/2)$ or $(4,2)$ with equal probability. 
In scenario $(4,5+\nu/4)$, the mechanism allocates the good to agent 2 with verification because $t_2-c_2 = (5+\nu/4)-2 > \nu$ and $t_2-c_2 > 2 = t_1 - c_1$. In scenario $(4,2)$, on the other hand, the mechanism allocates the good to agent 1 without verification. Consequently, the expected payoff of the mechanism under the distribution $\mathbb{P}$ amounts to $\frac{1}{2} (3+\nu/4)+ \frac{1}{2}4 = 3.5+\nu/8$, and the worst-case expected payoff over all admissible distributions cannot exceed $3.5+\nu/8$, which is strictly smaller than the optimal worst-case expected payoff. In summary, the mechanism is strictly suboptimal for all $\nu<4$.
\hfill$\square$
\end{remark}
%Figure \ref{figure:suboptimal-markov-independent} illustrates an example with two agents where $\underline{\mu}_1>\underline{\mu}_2$ and $\overline{t}_2-c_2 \geq \underline{\mu}_1$. Consider the favored agent mechanism with agent 1 as the favored and some $\nu<\underline{\mu}_1$ as the threshold. In this example, the nature may choose the following distribution from \eqref{eq: Markov ambiguity set with independent types}, under which the agents' expected types are $\underline{\mu}_i$: In both scenarios, agent 1 is of type $\underline{\mu}_1$. With some probability $\alpha \in (0,1)$, agent 2 has a type $t_2 \in (\nu +c_2,\underline{\mu}_1+c_2)$, and with probability $1-\alpha$, agent 2 has a type strictly less than $\nu +c_2$. Under this probability distribution with independent types, the aforementioned favored agent mechanism yields a payoff strictly less than $\underline{\mu}_1$:
% $$
% \alpha (t_2-c_2)+ (1- \alpha) \underline{\mu}_1 < \alpha \underline{\mu}_1+ (1- \alpha) \underline{\mu}_1 = \underline{\mu}_1.
% $$

% \begin{wrapfigure}{R}{0.45\textwidth}
% \centering
% \includegraphics[width=0.45\textwidth]{remark-markov-independent.png}
% \caption{A sub-optimal favored agent mechanism under \eqref{eq: Markov ambiguity set with independent types}. \label{figure:suboptimal-markov-independent}}
% \end{wrapfigure}
As in Sections~\ref{sec: support only} and~\ref{sec: Markov Ambiguity Sets}, we now show that among all optimal favored-agent mechanisms identified in Theorem~\ref{thr:opt-fav-independent}, there is always one that is Pareto robustly optimal; see Theorem~\ref{thr:PRO-fav-agents-independent} below. We first prove this result under the simplifying assumption that $\arg \max_{i \in \mathcal{I}} \underline{\mu}_i$ is a singleton, in which case the favored agent is uniquely determined. However, Theorem~\ref{thr:PRO-fav-agents-independent} will {\em not} depend on this assumption. 

%Assuming that $\arg \max_{i \in \mathcal{I}} \underline{\mu}_i = \{i^\star\}$, we first identify a distribution from $\mathcal{P}$ which has potential to be the worst-case distribution when the principal fails to choose a (Pareto robustly) optimal mechanism. Under this distribution, each agent has a two-point marginal distribution where at least one type of each agent $i \neq i^\star$ is guaranteed to be lower than s"lml"$\underline{\mu}_{i^\star}$, and the expected type of $i^\star$ is fixed to a given $\mu_{i^\star} \in [\underline{\mu}_{i^\star},\overline{\mu}_{i^\star}]$.

We first show that if $\arg \max_{i \in \mathcal{I}} \underline{\mu}_i = \{i^\star\}$, then any type profile $\bm t\in\mathcal T$ has a strictly positive probability under some discrete distribution in the Markov ambiguity set~\eqref{eq: Markov ambiguity set with independent types} with a prescribed expected value of~$\tilde t_{i^\star}$. It is further possible to require that, under this distribution, the type of each agent is supported on merely two points, the smaller of which falls strictly below~$\underline{\mu}_{i^\star}$ for every~$i \neq i^\star$.

\begin{lemma}\label{lem:findP-independent}
If $\mathcal{P}$ is a Markov ambiguity set of the form \eqref{eq: Markov ambiguity set with independent types} and $\arg \max_{i \in \mathcal{I}} \underline{\mu}_i = \{i^\star\}$ is a singleton, then, for any type profile $\bm{t} \in \mathcal{T}$ and any expected value $\mu_{i^\star} \in [\underline{\mu}_{i^\star},\overline{\mu}_{i^\star}]$ there exists a scenario $\hat{\bm{t}} \in \mathcal{T}$ with $\max_{i \neq i^\star}\hat{t}_i < \underline{\mu}_{i^\star}$ and a discrete distribution $\mathbb{P} \in \mathcal{P}$ such that
(i) $\mathbb{E}_{\mathbb{P}}[\tilde{t}_{i^\star}] =\mu_{i^\star}$,
(ii) $\mathbb{P}_{}(\tilde{{t}}_i \in \{{t}_i,\hat{{t}}_i \})=1$ for all $i \in \mathcal{I}$, and
(iii) $\mathbb{P}_{}(\tilde{\bm{t}} =\bm{t})>0$.
\end{lemma}

{  In the next lemma we show that if $\arg \max_{i \in \mathcal{I}} \underline{\mu}_i = \{i^\star\}$ is a singleton, then problem~\eqref{eq:MDP} admits a Pareto robustly optimal favored-agent mechanism~$(\bm{p}^\star,\bm{q}^\star)$. 
%The next lemma leverages Lemmas~\ref{lem:deviationForFavored} and \ref{lem:findP-independent} and Corollary \ref{Corollary of sufficiency proposition} of Proposition \ref{prop:unifyingPRO} to establish that if a feasible mechanism~$(\bm{p},\bm{q})$ generates the same or a higher {\em expected} payoff than $(\bm{p}^\star,\bm{q}^\star)$ under every {\em distribution} $\mathbb P\in \mathcal P$, then it must generate the same payoff as $(\bm{p}^\star,\bm{q}^\star)$ in every {\em scenario} $\bm t\in\mathcal T$. This readily implies that $(\bm{p}^\star,\bm{q}^\star)$ is Pareto robustly optimal in~\eqref{eq:MDP}.
}

\begin{lemma}\label{lem:independent-revenue-equivalence}
	Assume that $\mathcal{P}$ is a Markov ambiguity set of the form~\eqref{eq: Markov ambiguity set with independent types} and $\arg \max_{i \in \mathcal{I}} \underline{\mu}_i= \{i^\star\}$ is a singleton, and let $(\bm{p}^\star,\bm{q}^\star)$ be the type~(i) favored-agent mechanism with favored agent~${i^\star}$ and threshold value~$\nu^\star = \underline{\mu}_{i^\star}$. Then, any mechanism $(\bm{p},\bm{q}) \in \mathcal X$ that weakly Pareto robustly dominates $(\bm{p}^\star,\bm{q}^\star)$ must generate the same payoff as $(\bm{p}^\star,\bm{q}^\star)$ in every scenario $\bm{t} \in \mathcal{T}$, that is, we have
\begin{equation*}
	\begin{aligned}
	&\sum_{i \in \mathcal{I}} (p_i(\bm{t}) t_i - q_i(\bm{t}) c_i)  =  \sum_{i \in \mathcal{I}} (p_i^\star(\bm{t}) t_i - q_i^\star(\bm{t}) c_i)  &&\forall \bm{t} \in \mathcal{T}.
	\end{aligned}
\end{equation*}
\end{lemma}

Lemma~\ref{lem:independent-revenue-equivalence} is reminiscent of Lemma~\ref{lem:PRO-revenue equivalence}. {  The assumption that $\arg \max_{i \in \mathcal{I}} \underline{\mu}_i = \{i^\star\}$ is a singleton will be relaxed in Theorem~\ref{thr:PRO-fav-agents-independent} below.}
To gain some intuition into this result, we sketch the proof Lemma~\ref{lem:independent-revenue-equivalence} when there are only two agents with $\underline{\mu}_2<\underline{\mu}_1$, such that $\arg \max_{i \in \mathcal{I}} \underline{\mu}_i= \{1\}$ is a singleton, and where $\overline{t}_2 > c_2+\underline{\mu}_1$. These simplifying assumptions prevent tedious case distinctions. Our arguments will rely on the following partition of the type space~$\mathcal{T}$, which is illustrated in Figure~\ref{independent-figure}.
{ 
\begin{equation}\label{eq:partition sec 5}
\begin{aligned}
\mathcal{S}_1 &= \{\bm{t} \in  \mathcal{T} \,:\, t_2-c_2 \leq \underline{\mu}_1, \;  t_2 \leq \underline{\mu}_1 \; \text{and} \; t_1 \in (\underline{\mu}_1,\overline{\mu}_1] \}\\
\mathcal{S}_{2} &= \{\bm{t} \in  \mathcal{T} \,:\, t_2-c_2 \leq \underline{\mu}_1, \; t_2 > \underline{\mu}_1 \; \text{and} \; t_1 \in (\underline{\mu}_1,\overline{\mu}_1] \}\\
\mathcal{S}_{3} &= \{\bm{t} \in  \mathcal{T} \,:\, t_2-c_2 \leq \underline{\mu}_1, \; \text{and} \; t_1 \notin (\underline{\mu}_1,\overline{\mu}_1] \}\\
\mathcal{S}_4 &= \{\bm{t} \in  \mathcal{T} \,:\, t_2-c_2 > \underline{\mu}_1, \; \text{and} \; t_1= \underline \mu_1 \}\\
\mathcal{S}_5 &= \{\bm{t} \in  \mathcal{T} \,:\, t_2-c_2 > \underline{\mu}_1, \; \text{and} \; t_1 \neq \underline \mu_1 \}
\end{aligned}
\end{equation}
}
\tikzset{
pattern size/.store in=\mcSize, 
pattern size = 5pt,
pattern thickness/.store in=\mcThickness, 
pattern thickness = 0.3pt,
pattern radius/.store in=\mcRadius, 
pattern radius = 1pt}
\makeatletter
\pgfutil@ifundefined{pgf@pattern@name@_wgm5cv2xa}{
\pgfdeclarepatternformonly[\mcThickness,\mcSize]{_wgm5cv2xa}
{\pgfqpoint{0pt}{-\mcThickness}}
{\pgfpoint{\mcSize}{\mcSize}}
{\pgfpoint{\mcSize}{\mcSize}}
{
\pgfsetcolor{\tikz@pattern@color}
\pgfsetlinewidth{\mcThickness}
\pgfpathmoveto{\pgfqpoint{0pt}{\mcSize}}
\pgfpathlineto{\pgfpoint{\mcSize+\mcThickness}{-\mcThickness}}
\pgfusepath{stroke}
}}
\makeatother

% Pattern Info
 
\tikzset{
pattern size/.store in=\mcSize, 
pattern size = 5pt,
pattern thickness/.store in=\mcThickness, 
pattern thickness = 0.3pt,
pattern radius/.store in=\mcRadius, 
pattern radius = 1pt}
\makeatletter
\pgfutil@ifundefined{pgf@pattern@name@_9umps3dbp}{
\pgfdeclarepatternformonly[\mcThickness,\mcSize]{_9umps3dbp}
{\pgfqpoint{0pt}{0pt}}
{\pgfpoint{\mcSize+\mcThickness}{\mcSize+\mcThickness}}
{\pgfpoint{\mcSize}{\mcSize}}
{
\pgfsetcolor{\tikz@pattern@color}
\pgfsetlinewidth{\mcThickness}
\pgfpathmoveto{\pgfqpoint{0pt}{0pt}}
\pgfpathlineto{\pgfpoint{\mcSize+\mcThickness}{\mcSize+\mcThickness}}
\pgfusepath{stroke}
}}
\makeatother

% Pattern Info
 
\tikzset{
pattern size/.store in=\mcSize, 
pattern size = 5pt,
pattern thickness/.store in=\mcThickness, 
pattern thickness = 0.3pt,
pattern radius/.store in=\mcRadius, 
pattern radius = 1pt}
\makeatletter
\pgfutil@ifundefined{pgf@pattern@name@_ntabxexjz}{
\pgfdeclarepatternformonly[\mcThickness,\mcSize]{_ntabxexjz}
{\pgfqpoint{0pt}{0pt}}
{\pgfpoint{\mcSize+\mcThickness}{\mcSize+\mcThickness}}
{\pgfpoint{\mcSize}{\mcSize}}
{
\pgfsetcolor{\tikz@pattern@color}
\pgfsetlinewidth{\mcThickness}
\pgfpathmoveto{\pgfqpoint{0pt}{0pt}}
\pgfpathlineto{\pgfpoint{\mcSize+\mcThickness}{\mcSize+\mcThickness}}
\pgfusepath{stroke}
}}
\makeatother

% Pattern Info
 
\tikzset{
pattern size/.store in=\mcSize, 
pattern size = 5pt,
pattern thickness/.store in=\mcThickness, 
pattern thickness = 0.3pt,
pattern radius/.store in=\mcRadius, 
pattern radius = 1pt}
\makeatletter
\pgfutil@ifundefined{pgf@pattern@name@_6qql626ur}{
\pgfdeclarepatternformonly[\mcThickness,\mcSize]{_6qql626ur}
{\pgfqpoint{-\mcThickness}{-\mcThickness}}
{\pgfpoint{\mcSize}{\mcSize}}
{\pgfpoint{\mcSize}{\mcSize}}
{
\pgfsetcolor{\tikz@pattern@color}
\pgfsetlinewidth{\mcThickness}
\pgfpathmoveto{\pgfpointorigin}
\pgfpathlineto{\pgfpoint{0}{\mcSize}}
\pgfusepath{stroke}
}}
\makeatother

% Pattern Info
 
\tikzset{
pattern size/.store in=\mcSize, 
pattern size = 5pt,
pattern thickness/.store in=\mcThickness, 
pattern thickness = 0.3pt,
pattern radius/.store in=\mcRadius, 
pattern radius = 1pt}
\makeatletter
\pgfutil@ifundefined{pgf@pattern@name@_029dt5ijg lines}{
\pgfdeclarepatternformonly[\mcThickness,\mcSize]{_029dt5ijg}
{\pgfqpoint{0pt}{0pt}}
{\pgfpoint{\mcSize+\mcThickness}{\mcSize+\mcThickness}}
{\pgfpoint{\mcSize}{\mcSize}}
{\pgfsetcolor{\tikz@pattern@color}
\pgfsetlinewidth{\mcThickness}
\pgfpathmoveto{\pgfpointorigin}
\pgfpathlineto{\pgfpoint{\mcSize}{0}}
\pgfusepath{stroke}}}
\makeatother
\tikzset{every picture/.style={line width=0.75pt}} %set default line width to 0.75pt          

\begin{figure}
\centering

\begin{tikzpicture}[x=0.75pt,y=0.75pt,yscale=-1,xscale=1]

%Shape: Rectangle [id:dp26899129455409687] 
\draw  [draw opacity=0][fill={rgb, 255:red, 246; green, 220; blue, 138 }  ,fill opacity=0.9 ] (257.55,38.27) -- (437.2,38.27) -- (437.2,92.52) -- (257.55,92.52) -- cycle ;
%Shape: Rectangle [id:dp02738837564342056] 
\draw  [draw opacity=0][fill={rgb, 255:red, 252; green, 141; blue, 98 }  ,fill opacity=0.9 ] (328.5,38.43) -- (332.15,38.43) -- (332.15,92.68) -- (328.5,92.68) -- cycle ;
%Shape: Rectangle [id:dp6504926463492533] 
\draw  [draw opacity=0][pattern=_wgm5cv2xa,pattern size=8.25pt,pattern thickness=0.75pt,pattern radius=0pt, pattern color={rgb, 255:red, 255; green, 255; blue, 255}] (328.5,38.2) -- (332.5,38.2) -- (332.5,93.58) -- (328.5,93.58) -- cycle ;
%Shape: Rectangle [id:dp5718537620491368] 
\draw  [draw opacity=0][fill={rgb, 255:red, 141; green, 160; blue, 203 }  ,fill opacity=0.9 ] (258.33,92.52) -- (437.99,92.52) -- (437.99,210.56) -- (258.33,210.56) -- cycle ;
%Shape: Rectangle [id:dp01363327555427074] 
\draw  [draw opacity=0][pattern=_9umps3dbp,pattern size=8.25pt,pattern thickness=0.75pt,pattern radius=0pt, pattern color={rgb, 255:red, 255; green, 255; blue, 255}] (383.75,92.2) -- (437.85,92.2) -- (437.85,209.24) -- (383.75,209.24) -- cycle ;
%Shape: Rectangle [id:dp3494152799103656] 
\draw  [draw opacity=0][pattern=_ntabxexjz,pattern size=8.25pt,pattern thickness=0.75pt,pattern radius=0pt, pattern color={rgb, 255:red, 255; green, 255; blue, 255}] (257.89,93.68) -- (332.15,93.68) -- (332.15,210.72) -- (257.89,210.72) -- cycle ;
%Straight Lines [id:da581359827469236] 
\draw    (253.59,38.27) -- (257.55,38.27) ;
%Straight Lines [id:da4112754483228491] 
\draw    (438.08,214.63) -- (438.08,210.72) ;
%Shape: Rectangle [id:dp23491475066389866] 
\draw  [draw opacity=0] (257.55,93.27) -- (438.08,93.27) -- (438.08,210.72) -- (257.55,210.72) -- cycle ;
%Shape: Rectangle [id:dp2636004473959255] 
\draw  [draw opacity=0][fill={rgb, 255:red, 231; green, 138; blue, 195 }  ,fill opacity=0.9 ] (333.37,147.3) -- (383.35,147.3) -- (383.35,210.25) -- (333.37,210.25) -- cycle ;
%Shape: Rectangle [id:dp9969975646767084] 
\draw  [draw opacity=0][fill={rgb, 255:red, 102; green, 194; blue, 165 }  ,fill opacity=0.9 ] (333.23,92.58) -- (383.35,92.58) -- (383.35,147.3) -- (333.23,147.3) -- cycle ;
%Shape: Rectangle [id:dp24874037687055583] 
\draw  [draw opacity=0][pattern=_6qql626ur,pattern size=6pt,pattern thickness=1.1pt,pattern radius=0pt, pattern color={rgb, 255:red, 255; green, 255; blue, 255}] (332.5,148.08) -- (382.48,148.08) -- (382.48,211.03) -- (332.5,211.03) -- cycle ;
%Shape: Rectangle [id:dp9532659298839996] 
\draw  [draw opacity=0][pattern=_029dt5ijg,pattern size=6pt,pattern thickness=1.1pt,pattern radius=0pt, pattern color={rgb, 255:red, 255; green, 255; blue, 255}] (333.14,93.36) -- (383.27,93.36) -- (383.27,148.08) -- (333.14,148.08) -- cycle ;
%Straight Lines [id:da17196268730345543] 
\draw  [dash pattern={on 4.5pt off 4.5pt}]  (332.5,214.16) -- (332.5,92.4) ;
%Straight Lines [id:da030282135934512233] 
\draw  [dash pattern={on 4.5pt off 4.5pt}]  (383.27,214.33) -- (383.27,92.58) ;
%Straight Lines [id:da4370795805612717] 
\draw  [dash pattern={on 4.5pt off 4.5pt}]  (254.42,148.08) -- (432.34,148.08) -- (439.64,148.08) ;
%Straight Lines [id:da002109218706989102] 
\draw  [dash pattern={on 4.5pt off 4.5pt}]  (254.68,92.52) -- (432.6,92.52) -- (439.9,92.52) ;
%Straight Lines [id:da9839043116941806] 
\draw    (254.31,210.77) -- (450.93,210.77) ;
\draw [shift={(452.93,210.77)}, rotate = 180] [color={rgb, 255:red, 0; green, 0; blue, 0 }  ][line width=0.75]    (10.93,-3.29) .. controls (6.95,-1.4) and (3.31,-0.3) .. (0,0) .. controls (3.31,0.3) and (6.95,1.4) .. (10.93,3.29)   ;
%Straight Lines [id:da42740137409655676] 
\draw    (257.55,214.16) -- (257.55,25.91) ;
\draw [shift={(257.55,23.91)}, rotate = 90] [color={rgb, 255:red, 0; green, 0; blue, 0 }  ][line width=0.75]    (10.93,-3.29) .. controls (6.95,-1.4) and (3.31,-0.3) .. (0,0) .. controls (3.31,0.3) and (6.95,1.4) .. (10.93,3.29)   ;

% Text Node
\draw (251.17,9) node [anchor=north west][inner sep=0.75pt]    {$t_{2}$};
% Text Node
\draw (454.86,202) node [anchor=north west][inner sep=0.75pt]    {$t_{1}$};
% Text Node
\draw (348,167) node [anchor=north west][inner sep=0.75pt]  [font=\Large,color={rgb, 255:red, 0; green, 0; blue, 0 }  ,opacity=1 ] [align=left] {$\displaystyle \mathcal{S}_{1}$};
% Text Node
\draw (284,139) node [anchor=north west][inner sep=0.75pt]  [font=\Large,color={rgb, 255:red, 0; green, 0; blue, 0 }  ,opacity=1 ] [align=left] {$\displaystyle \mathcal{S}_{3}$};
% Text Node
\draw (348,109) node [anchor=north west][inner sep=0.75pt]  [font=\Large,color={rgb, 255:red, 0; green, 0; blue, 0 }  ,opacity=1 ] [align=left] {$\displaystyle \mathcal{S}_{2}$};
% Text Node
\draw (209,84) node [anchor=north west][inner sep=0.75pt]    {$c_{2} +\underline{\mu }_{1}$};
% Text Node
\draw (431.31,214.38) node [anchor=north west][inner sep=0.75pt]    {$\overline{t}_{1}$};
% Text Node
\draw (253.51,214) node [anchor=north west][inner sep=0.75pt]    {$\underline{t}_{1}$};
% Text Node
\draw (240.01,201) node [anchor=north west][inner sep=0.75pt]    {$\underline{t}_{2}$};
% Text Node
\draw (238.01,141) node [anchor=north west][inner sep=0.75pt]    {$\underline{\mu }_{1}$};
% Text Node
\draw (238,29.45) node [anchor=north west][inner sep=0.75pt]    {$\overline{t}_{2}$};
% Text Node
\draw (319.35,55) node [anchor=north west][inner sep=0.75pt]  [font=\Large,color={rgb, 255:red, 0; green, 0; blue, 0 }  ,opacity=1 ] [align=left] {$\displaystyle \mathcal{S}_{4}$};
% Text Node
\draw (324.93,215) node [anchor=north west][inner sep=0.75pt]    {$\underline{\mu }_{1}$};
% Text Node
\draw (374.75,214.6) node [anchor=north west][inner sep=0.75pt]    {$\overline{\mu }_{1}$};
% Text Node
\draw (377.35,55) node [anchor=north west][inner sep=0.75pt]  [font=\Large,color={rgb, 255:red, 0; green, 0; blue, 0 }  ,opacity=1 ] [align=left] {$\displaystyle \mathcal{S}_{5}$};

\end{tikzpicture}

\caption{Partition \eqref{eq:partition sec 5} of the type space~$\mathcal T$. \label{independent-figure}}
\end{figure}
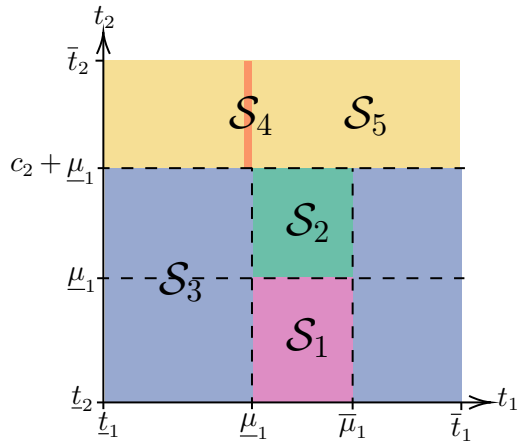

Under our simplifying assumptions about $\underline{\mu}_1$, $\underline{\mu}_2$, $\overline t_2$ and $c_2$, one can show that all of these sets are nonempty. We emphasize, however, that the formal proof of Lemma~\ref{lem:independent-revenue-equivalence} in the online appendix does not rely on any of the simplifying assumptions imposed here. 

{ 
Denote now by $(\bm{p}^\star,\bm{q}^\star)$ the type~(i) favored-agent mechanism with favored agent~${i^\star}$ and threshold value~$\nu^\star = \underline{\mu}_{i^\star}$. To prove Lemma~\ref{lem:independent-revenue-equivalence}, it suffices to show that the conditions~(i) and~(ii) of Proposition~\ref{prop:unifyingPRO} are satisfied for the partition~\eqref{eq:partition sec 5}. The claim then follows from Corollary~\ref{Corollary of sufficiency proposition}.
%we will leverage Corollary \ref{Corollary of sufficiency proposition} of Proposition \ref{prop:unifyingPRO}, which provides sufficient conditions for the claim made. In particular, to show that~any mechanism~$(\bm{p},\bm{q}) \in \mathcal X$ that weakly Pareto robustly dominates $(\bm{p}^\star,\bm{q}^\star)$ must generate the same payoff as $(\bm{p}^\star,\bm{q}^\star)$ in every scenario~$\bm t\in\mathcal T$, it suffices to verify the conditions~(i) and~(ii) in Proposition~\ref{prop:unifyingPRO} for the  partition~\eqref{eq:partition sec 5}.
We will exploit Lemma~\ref{lem:findP-independent} to verify condition~(i). The arguments needed to verify condition~(ii) closely parallel those used in the proof of Lemma~\ref{lem:PRO-revenue equivalence}. Thus, we omit this part of the proof for brevity.

We now show that condition~(i) is satisfied, that is, for any $k \in \{1, \dots, 5\}$ and $\bm t \in \mathcal S_k$ there exists $\mathbb P \in\mathcal P$ supported on~$\cup_{l=1}^k \mathcal{S}_l$ with $\mathbb P(\tilde{\bm t}= \bm t)>0$. By Lemma~\ref{lem:findP-independent}, given any $\bm t \in \mathcal{T}$ and $\mu_{1} \in [\underline \mu_{1},\overline{\mu}_{1}]$ there exists a scenario $\hat{\bm{t}} \in \mathcal{T}$ with $\hat t_2< \underline{\mu}_1$ 
and a discrete distribution $\mathbb{P} \in \mathcal{P}$ such that
$\mathbb{E}_{\mathbb{P}}[\tilde{t}_1] =\mu_1$, $\mathbb{P}(\tilde{{t}}_i \in \{{t}_i,\hat{{t}}_i \})=1$ for all $i \in \mathcal{I}$, and $\mathbb{P}(\tilde{\bm{t}} =\bm{t})>0$. We will show that~$\mathbb{P}$ is supported on~$\cup_{l=1}^k \mathcal{S}_l$ and therefore  satisfies condition~(i) for a judiciously chosen $\mu_{1} \in [\underline \mu_{1},\overline{\mu}_{1}]$. To this end, we first study some implications of Lemma~\ref{lem:findP-independent} on the support~$\mathbb S$ of~$\mathbb{P}$. %depends on $\bm t \in \mathcal T$ and $\mu_1 \in [\underline \mu_{1},\overline{\mu}_{1}]$. In the following, we denote by . We note that 
Clearly, we have $\mathbb S \subseteq \{\bm t' \in \mathcal{T} \,:\, t_i' \in \{t_i,\hat t_i\} \,\text{for all}\, i \in \mathcal{I}\}$, and as $\hat t_2< \underline{\mu}_1$, we also have
\begin{align}\label{eq:supportMarkovIndependentSketch1}
\mathbb S \subseteq \left\{\bm t' \in \mathcal{T} \,:\, t_2' \leq \max\{t_2,\underline \mu_1\} ~\text{and}~ t_2'-c_2 \leq \max\{t_2-c_2, \underline \mu_1\}\right\}.
\end{align}
In the special case when $t_1 = \mu_1$, then the valid relations $\mathbb{E}[\tilde{t}_1] =\mu_1$ and $\mathbb{P}(\tilde{{t}}_1 \in \{{t}_1,\hat{{t}}_1 \})=1$ imply that $\hat t_1 = t_1$. Hence, if $t_1 = \mu_1$, then we additionally have
\begin{align}\label{eq:supportMarkovIndependentSketch2}
\mathbb S \subseteq \left\{\bm t' \in \mathcal{T} \,:\, t_2' \leq \max\{t_2,\underline \mu_1\} ~\text{and}~ t_2'-c_2 \leq \max\{t_2-c_2, \underline \mu_1\} ~\text{and}~ t_{1}' = t_{1} \right\}.
\end{align}
To prove condition~(i), set first $k = 1$, and fix any $\bm t \in \mathcal{S}_1$. By the definition of~$\mathcal{S}_1$, we have $\max\{t_2,\underline \mu_1\} = \max\{t_2-c_2, \underline \mu_1\} = \underline{\mu}_1$ and $t_1 \in (\underline \mu_{1},\overline{\mu}_{1}]$. In this case, we choose~$\mu_1 = t_1$, in which case the support of~$\mathbb{P}$ satisfies~\eqref{eq:supportMarkovIndependentSketch2}. As $\max\{t_2,\underline \mu_1\} = \max\{t_2-c_2, \underline \mu_1\} = \underline \mu_1$, we obtain
\[
    \mathbb S \subseteq \left\{\bm t' \in \mathcal{T} \,:\, t_1' = t_{1} ~\text{and}~ t_2'-c_2 \leq \underline \mu_1 ~\text{and}~ t_2' \leq \underline \mu_1 \right\} \subseteq \mathcal S_1.
\]
This implies that $\mathbb{P}\in \mathcal{P}_0(\mathcal{S}_1)$. Hence, we have shown that condition~(i) holds for any~$\bm t \in \mathcal S_1$. 

Consider next~$k = 2$, and fix any~$\bm t \in \mathcal{S}_2$. By the definition of $\mathcal S_2$ we have $\max\{t_2-c_2, \underline \mu_1\} = \underline \mu_1$ and $t_1 \in (\underline \mu_{1},\overline{\mu}_{1}]$. In this case, we choose again~$\mu_1 = t_1$. As the support of~$\mathbb P$ satisfies~\eqref{eq:supportMarkovIndependentSketch2} and as $\max\{t_2-c_2, \underline \mu_1\} = \underline \mu_1$, we find $\mathbb S \subseteq  \{\bm t' \in \mathcal{T} \,:\, t_2'-c_2 \leq \underline \mu_1 ~\text{and}~t_1' = t_{1} \} \subseteq \mathcal S_1 \cup \mathcal S_2$, which implies that $\mathbb{P} \in\mathcal{P}_0(\mathcal{S}_1 \cup \mathcal S_2)$. Hence, we have shown that condition~(i) holds for any~$\bm t \in \mathcal S_2$. 

Next, set $k = 3$, and fix any $\bm t \in \mathcal{S}_3$. By the definition of $\mathcal S_3$, we again have $\max\{t_2-c_2, \underline \mu_1\} = \underline \mu_1$. In this case, we choose $\mu_1 \in [\underline \mu_{1},\overline{\mu}_{1}]$ arbitrarily. As the support of~$\mathbb P$ satisfies \eqref{eq:supportMarkovIndependentSketch1} and as $\max\{t_2-c_2, \underline \mu_1\} = \underline \mu_1$, we find $\mathbb S \subseteq  \{\bm t' \in \mathcal{T} \,:\, t_2'-c_2 \leq \underline \mu_1 \} \subseteq \mathcal S_1 \cup \mathcal S_2 \cup \mathcal S_3$, which implies that $\mathbb{P} \in \mathcal{P}_0(\mathcal{S}_1 \cup \mathcal S_2 \cup \mathcal S_3)$. Hence, we have shown that condition~(i) holds for any $\bm t \in \mathcal S_3$. 

Next, set $k = 4$, and fix any $\bm t \in \mathcal{S}_4$. By the definition of $\mathcal S_4$, we have $t_1 = \underline \mu_1 \in [\underline \mu_{1},\overline{\mu}_{1}]$. In this case, we choose $\mu_1 = t_1$. As the support of~$\mathbb P$ satisfies~\eqref{eq:supportMarkovIndependentSketch2}, we find $\mathbb S \subseteq  \{\bm t' \in \mathcal{T} \,:\, t_1' = t_{1} = \underline \mu_1 \}\subseteq \cup_{l=1}^4 \mathcal S_l$, which implies that $\mathbb{P} \in \mathcal{P}_0(\cup_{l=1}^4 \mathcal S_l)$. Hence, we have shown that condition~(i) holds for any $\bm t \in \mathcal S_4$. For~$k=5$, condition~(i) trivially holds for any~$\bm t \in \mathcal{S}_5$ thanks to Lemma~\ref{lem:findP-independent}.

In summary, we have verified that~$(\bm{p}^\star,\bm{q}^\star)$ satisfies condition~(i) of Proposition~\ref{prop:unifyingPRO} for the partition~\eqref{eq:partition sec 5}. To prove that~$(\bm{p}^\star,\bm{q}^\star)$ also satisfies condition~(ii), one can proceed as in the proof of Lemma~\ref{lem:PRO-revenue equivalence}. Details are omitted for brevity.}

The next theorem shows that the Pareto robust optimality result of Lemma~\ref{lem:independent-revenue-equivalence} remains valid even when ${\arg \max}_{i \in \mathcal{I}} \underline{\mu}_i$ is no longer guaranteed to be a singleton. 

%When ${\arg \max}_{i \in \mathcal{I}} \underline{\mu}_i=\{i^\star\}$, Pareto robust optimality of the type~(i) favored-agent mechanism $(\bm{p}^\star,\bm{q}^\star)$ with the favored agent $i^\star$ and threshold $\nu^\star=\underline{\mu}_{i^\star}$ follows from Proposition~\ref{lem:independent-revenue-equivalence}. That is, any mechanism $(\bm{p},\bm{q})$ that weakly Pareto robustly dominates $(\bm{p}^\star,\bm{q}^\star)$ generates the same expected payoff for the principal under every distribution $\mathbb{P} \in \mathcal{P}$. Hence, $(\bm{p}^\star,\bm{q}^\star)$ cannot be Pareto robustly dominated when ${\arg \max}_{i \in \mathcal{I}} \underline{\mu}_i=\{i^\star\}$. Moreover, as for the general Markov ambiguity sets, we can show that $(\bm{p}^\star,\bm{q}^\star)$ remains to be Pareto robust optimal even if ${\arg \max}_{i \in \mathcal{I}} \underline{\mu}_i$ is not a singleton. We can do this by revising the proof of Theorem \ref{thr:PRO-fav-agents-markov} in a way that only employs the distributions $\mathbb{P}$ under which the agents' types are independent.

\begin{theorem}\label{thr:PRO-fav-agents-independent}
	If $\mathcal{P}$ is a Markov ambiguity set of the form~\eqref{eq: Markov ambiguity set with independent types}, then any type~(i) favored-agent mechanism~$(\bm{p}^\star,\bm{q}^\star)$ with favored agent ${i^\star} \in {\arg \max}_{i \in \mathcal{I}} \underline{\mu}_i$ and threshold value $\nu^\star = \max_{i \in \mathcal{I}} \underline{\mu}_i$ is Pareto robustly optimal in~\eqref{eq:MDP}.
\end{theorem}

The proof of Theorem~\ref{thr:PRO-fav-agents-independent} is omitted because it widely parallels that of Theorem~\ref{thr:PRO-fav-agents-markov}. To close this section, we show that the favored-agent mechanism identified in Theorem~\ref{thr:PRO-fav-agents-markov} may cease to be Pareto robustly optimal when the Markov ambiguity set~\eqref{eq: Markov ambiguity set} is replaced with its subset~\eqref{eq: Markov ambiguity set with independent types} that imposes independence among the agents' types.

\begin{remark}\label{rem: on Pareto robust optimality}
%There are problem instances for which any favored-agent mechanism with favored agent $i^\star \in \arg \max_{i \in \mathcal{I}} \underline{\mu}_i$ and threshold value $\nu \geq \overline{t}_{i^\star}$ is not Pareto robustly optimal under Markov ambiguity sets \eqref{eq: Markov ambiguity set with independent types} with independent types. To see this, 
Consider an instance of the robust mechanism design problem~\eqref{eq:MDP} with $I=2$ agents, and assume that the input parameters satisfy $\underline{\mu}_1>\underline{\mu}_2$ and $\overline{t}_2-c_2 \geq \overline{t}_1 > \overline{\mu}_1$. In addition, let~$(\bm{p},\bm{q})$ be a type~(ii) favored-agent mechanism with favored agent $1 \in \arg \max_{i \in \mathcal{I}} \underline{\mu}_i$ and threshold value $\nu \geq \overline{t}_1$. In the special case where~$\nu=\overline t_1$, Theorem~\ref{thr:PRO-fav-agents-markov} implies that~$(\bm{p},\bm{q})$ is Pareto robustly optimal in~\eqref{eq:MDP} provided that~$\mathcal P$ is the Markov ambiguity set~\eqref{eq: Markov ambiguity set}. In the following, we prove that, for any~$\nu \geq \overline{t}_1$, $(\bm{p},\bm{q})$ is Pareto robustly dominated by another feasible mechanism when $\mathcal P$ is the Markov ambiguity set~\eqref{eq: Markov ambiguity set with independent types} with independent types.
%\footnote{One can use similar arguments to prove that the type~(i) variant of the same favored-agent mechanism is also Pareto dominated. Details are omitted for the sake of brevity.} 
To this end, consider an arbitrary distribution~$\mathbb{P}$ in the Markov ambiguity set~\eqref{eq: Markov ambiguity set with independent types}.
%under which the events $\tilde{t}_2-c_2 \leq \nu$, $\tilde{t}_2-c_2 > \nu$ and $\tilde{t}_2-c_2 \in (\overline{\mu}_1,\nu]$ have a strictly positive probability. 
%The latter condition ensures that conditioning on these events is well-defined under~$\mathbb P$. 
The expected payoff of~$(\bm{p},\bm{q})$ is then given by
\begin{align*}
    \mathbb{E}_{\mathbb P}\left[\sum_{i \in \{1,2\}} (p_i(\bm{t}) t_i - q_i(\bm{t}) c_i)\right] & = \left\{\begin{array}{l}
         \phantom{+}\mathbb{P}\left( \tilde{t}_2-c_2 \leq \overline{\mu}_1 \right) \mathbb{E}_{\mathbb{P}}[\tilde t_1 \,\vert\, \tilde{t}_2-c_2 \leq \overline{\mu}_1 ] \\[1ex]
         + \mathbb{P}\left( \tilde{t}_2-c_2 \in (\overline{\mu}_1,\nu) \right) \mathbb{E}_{\mathbb{P}}[\tilde t_1 \,\vert\, \tilde{t}_2-c_2 \in (\overline{\mu}_1,\nu) ]\\[1ex]
         + \mathbb{P}\left(\tilde{t}_2-c_2 \geq \nu \right) \mathbb{E}_{\mathbb{P}}[\textstyle \max_{i \in \{1,2\}} \tilde{t}_i-c_i \,\vert\, \tilde{t}_2-c_2 \geq \nu ]
    \end{array}\right.\\[1.5ex]
    & = \left\{\begin{array}{l}
         \phantom{+}\mathbb{P}\left( \tilde{t}_2-c_2 \leq \overline{\mu}_1 \right) \mathbb{E}_{\mathbb{P}}[\tilde t_1] \\[1ex]
         + \mathbb{P}\left( \tilde{t}_2-c_2 \in (\overline{\mu}_1,\nu) \right) \mathbb{E}_{\mathbb{P}}[\tilde t_1]\\[1ex]
         + \mathbb{P}\left(\tilde{t}_2-c_2 \geq \nu \right) \mathbb{E}_{\mathbb{P}}[\textstyle \max_{i \in \{1,2\}} \tilde{t}_i-c_i \,\vert\, \tilde{t}_2-c_2 \geq \nu ],
    \end{array}\right.
\end{align*}
%as the agents' types are independently distributed, and $\max_{i \in \{1,2\}} {t}_i-c_i \geq {t}_2-c_2= \nu^\star > \overline{\mu}_1 \geq \mathbb{E}_{\mathbb{P}}[t_1]$ in the case of tie ${t}_2-c_2 = \nu^\star$. 
where the second equality follows from the independence of the agents' types under~$\mathbb P$. Next, denote by $(\bm{p}',\bm{q}')$ the type~(i) favored-agent mechanism with favored agent~$1$ and threshold value~$\overline{\mu}_1$. By construction, the expected payoff of~$(\bm{p}',\bm{q}')$ under~$\mathbb P$ amounts to
\begin{equation*}
\begin{aligned}
    \mathbb{E}_{\mathbb P}\left[\sum_{i \in \{1,2\}} (p'_i(\bm{t}) t_i - q'_i(\bm{t}) c_i)\right] = \left\{\begin{array}{l}
         \phantom{+} \mathbb{P}\left( \tilde{t}_2-c_2 \leq \overline{\mu}_1 \right) \mathbb{E}_{\mathbb{P}}[\tilde t_1] \\[1ex]
         + \mathbb{P}\left(\tilde{t}_2-c_2 \in (\overline{\mu}_1,\nu) \right)\mathbb{E}_{\mathbb{P}}[\textstyle \max_{i \in \{1,2\}} \tilde{t}_i-c_i \,\vert\, \tilde{t}_2-c_2 \in (\overline{\mu}_1,\nu) ]\\[1ex]
         + \mathbb{P}\left( \tilde{t}_2-c_2 \geq \nu \right) \mathbb{E}_{\mathbb{P}}[\textstyle\max_{i \in \{1,2\}} \tilde{t}_i-c_i \,\vert\, \tilde{t}_2-c_2 \geq \nu ].
    \end{array} \right.
\end{aligned}
\end{equation*}
If $\mathbb{P}(\tilde{t}_2-c_2 \in (\overline{\mu}_1, \nu)) > 0$, then the expected payoff of~$(\bm{p}',\bm{q}')$ exceeds that of~$(\bm{p},\bm{q})$ under~$\mathbb P$ because 
\[
    \max_{i \in \{1,2\}} {t}_i-c_i \geq {t}_2-c_2 > \overline{\mu}_1 \geq \mathbb{E}_{\mathbb{P}}[\tilde t_1]
\]
for all $\bm t\in\mathcal T$ with ${t}_2-c_2 \in (\overline{\mu}_1,\nu)$.
%$\max_{i \in \{1,2\}} {t}_i-c_i \geq {t}_1-c_1 > {t}_1$ for any~$\bm t\in\mathcal T$. -- % The last inequality here does not hold
If $\mathbb{P}\left(\tilde{t}_2-c_2 \in (\overline{\mu}_1,\nu) \right) = 0$, on the other hand, then the expected payoffs of the two mechanisms coincide. In order to show that~$(\bm{p}',\bm{q}')$ Pareto robustly dominates~$(\bm{p},\bm{q})$ it thus suffices to construct a distribution~$\mathbb P^\star\in\mathcal P$ with $\mathbb{P}^\star(\tilde{t}_2-c_2 \in (\overline{\mu}_1,\nu) ) >0$. Such a distribution exists thanks to our assumption that $\overline{t}_2-c_2 \geq \overline{t}_1 > \overline{\mu}_1$. Indeed, we can define~$\mathbb P^\star$ as the two-point distribution that assigns probability~$\alpha = (\underline{\mu}_2 - \underline{t}_2)/((\overline{t}_1 + \overline{\mu}_1)/2 +c_2-\underline{t}_2)$ to scenario $(\underline{\mu}_1,(\overline{t}_1 + \overline{\mu}_1)/2+c_2)$ and probability $1-\alpha$ to scenario $(\underline{\mu}_1, \underline{t}_2)$. One readily verifies that this distribution belongs to the ambiguity set~\eqref{eq: Markov ambiguity set with independent types} and satisfies $\mathbb{P}^\star(\tilde{t}_2-c_2 \in (\overline{\mu}_1,\nu) )\geq\alpha >0$. Hence, the favored-agent mechanism~$(\bm{p},\bm{q})$ fails to be Pareto robustly optimal in problem~\eqref{eq:MDP} for any~$\nu\geq \overline t_1$ if~$\mathcal P$ is a Markov ambiguity set of the form~\eqref{eq: Markov ambiguity set with independent types} with independent types. \hfill $\square$
\end{remark}

In conjunction, Remarks~\ref{rem: sub-optimal thresholds} and~\ref{rem: on Pareto robust optimality} imply that for some instances of problem~\eqref{eq:MDP} there is {\em no} favored-agent mechanism that is Pareto robustly optimal simultaneously for {\em both} Markov ambiguity sets~\eqref{eq: Markov ambiguity set} and~\eqref{eq: Markov ambiguity set with independent types}. To see this, consider the instance of problem~\eqref{eq:MDP} described in Remark~\ref{rem: sub-optimal thresholds}, and note that this instance satisfies all assumptions of Remark~\ref{rem: on Pareto robust optimality}. One readily verifies that every favored-agent mechanism with favored agent~$i^\star \notin \arg \max_{i \in \mathcal{I}} \underline{\mu}_i$ is strictly suboptimal (and thus fails to be Pareto robustly optimal) for this problem instance under {\em both} ambiguity sets~\eqref{eq: Markov ambiguity set} and~\eqref{eq: Markov ambiguity set with independent types}. By Remark~\ref{rem: sub-optimal thresholds}, any favored-agent mechanism with favored agent~$i^\star \in \arg \max_{i \in \mathcal{I}} \underline{\mu}_i$ and threshold value~$\nu < \overline{t}_{i^\star}$ is strictly suboptimal (and thus fails to be Pareto robustly optimal) under the ambiguity set~\eqref{eq: Markov ambiguity set}. By Remark~\ref{rem: on Pareto robust optimality}, finally, any favored-agent mechanism with favored agent~$i^\star \in \arg \max_{i \in \mathcal{I}} \underline{\mu}_i$ and threshold value~$\nu \geq \overline{t}_{i^\star}$ fails to be Pareto robustly optimal under the ambiguity set~\eqref{eq: Markov ambiguity set with independent types}. This implies that it is crucial for the principal to know whether or not the agents' types are independent.

{ \section{Numerical Illustration}\label{sec: Numerical Illustration}

We now assess the Pareto robustly optimal mechanisms designed for different ambiguity sets against the optimal mechanism tailored to a crisp (but possibly misspecified) type distribution. For better comparability, we report the expected payoffs generated by different mechanisms relative to the maximum expected payoff achievable with full knowledge of the true type distribution. 

Throughout this section we assume that there are two agents with $\mathcal{T}_1 = [0.2, 2]$, $\mathcal{T}_2 = [0, 1.5]$, $[\underline{\mu}_1, \overline{\mu}_1] = [0.75, 1.25]$, $[\underline{\mu}_2, \overline{\mu}_2] = [0.5, 1]$, $c_1 = 0.1$, and $c_2 = 0.4$.
We also assume that the true type distribution is of the form $\mathbb{P}_\epsilon = (1 - \epsilon) \mathbb{P}^{\text{N}} + \epsilon \mathbb{P}^{\text{E}}$, where $\mathbb{P}^{\text{N}} \in \mathcal{P}_0(\mathcal{T})$ represents a nominal distribution that captures the seller's best guess for the unknown true distribution, and $\mathbb{P}^{\text{E}}\in \mathcal{P}_0(\mathcal{T})$ represents an extremal distribution to be specified later. The weight~$\epsilon \in [0, 1]$ determines the contamination level of~$\mathbb{P}^{\text{N}}$. Indeed, as $\epsilon$ increases, the true distribution $\mathbb{P}_\epsilon$ approaches~$\mathbb{P}^{\text{E}}$, thus becoming increasingly different from~$\mathbb{P}^{\text{N}}$.
In the following we use~$\mathcal N_{\mathcal T}(\bm \mu, \bm \Sigma)$ to denote the truncated normal distribution obtained by conditioning the (untruncated) normal distribution~$\mathcal N(\bm \mu, \bm \Sigma)$ on the event~$\tilde{\bm t}\in\mathcal T$. We then define $\mathbb{P}^{\text{N}} = \mathcal N_{\mathcal T}(\bm \mu_0, \bm \Sigma_0)$ as the truncated normal distribution with $\bm \mu_0=\overline{\bm t}/2$ 
and~$\bm \Sigma_0=\mathop{\rm diag}(\overline{\bm t})$.

We use three different models for the extremal distribution, that is, we set $\mathbb{P}_1^{\text{E}}= \mathcal N_{\mathcal T}(\underline{\bm{\mu}}, \bm \Sigma)$, $\mathbb{P}_2^{\text{E}}= \mathcal N_{\mathcal T}(\overline{\bm{\mu}}, \bm \Sigma)$ and $\mathbb{P}_3^{\text{E}}= \frac{1}{2} \mathcal N_{\mathcal T}(\underline{\bm t}, \bm \Sigma) + \frac{1}{2} \mathcal N_{\mathcal T}(\overline{\bm t}, \bm \Sigma) $ with $\bm\Sigma =\frac{1}{100}\bm I$. By construction, 
%Mean values of the truncated distributions $\mathbb{P}^{\underline{\bm{\mu}}}$ and $\mathbb{P}^{\overline{\bm{\mu}}}$ remain as $\underline{\bm{\mu}}$ and $\overline{\bm{\mu}}$, respectively, and the one of $\mathbb{P}^{(\frac{1}{2}\underline{\bm t}+\frac{1}{2}\overline{\bm t})}$ is equal to $[1.1, 0.75]$. Note that all of these mean values fall within $[\underline{\mu}_1, \overline{\mu}_1] \times [\underline{\mu}_2, \overline{\mu}_2]$. Distributions $\mathbb{P}^{\underline{\bm{\mu}}}$ and $\mathbb{P}^{\overline{\bm{\mu}}}$ 
$\mathbb{P}_1^{\text{E}}$ and~$\mathbb{P}_2^{\text{E}}$ concentrate all probability mass near the lower and upper bounds on the expected value of~$\tilde{\bm t}$, respectively, and~$\mathbb{P}_3^{\text{E}}$ concentrates half of the probability mass near~$\underline{\bm t}$ and the other half near~$\overline{\bm t}$. Note that~$\mathbb{P}_1^{\text{E}}$ represents a blurred version of the worst-case distribution that minimizes the optimal mechanism's expected payoff over the Markov ambiguity sets with or without independent types. In contrast, $\mathbb{P}_2^{\text{E}}$ represents a blurred version of the corresponding best-case distribution. Contaminating~$\mathbb{P}^{\text{N}}$ with~$\mathbb{P}_1^{\text{E}}$ or~$\mathbb{P}_2^{\text{E}}$ thus captures two complementary extreme scenarios. Lastly, note that the agent's types are perfectly correlated under~$\mathbb{P}_3^{\text{E}}$, in which case the independence assumption is in some sense maximally violated. This distribution facilitates a more nuanced comparison between the Pareto robustly optimal mechanisms for Markov ambiguity sets with and without independent types. %{\color{red} It is not clear to me why we do not define these extremal distributions as discrete distributions. What is the benefit of using Gaussians or mixtures of Gaussians? We should also discuss why we chose these three distributions. Are these (approximate) worst-case distributions for some ambiguity sets?}

For every~$\epsilon\in[0,1]$, we use the optimal mechanism under perfect distributional information as a baseline. This mechanism solves problem~\eqref{eq:MDP} in view of the singleton ambiguity set~\mbox{$\mathcal P= \{\mathbb P_\epsilon\}$}. We emphasize that this mechanism may {\em not} be a favored-agent mechanism because the types of the two agents fail to be independent under~$\mathbb P_\epsilon$. Thus, the infinite-dimensional linear program~\eqref{eq:MDP} cannot be solved exactly.
%in practice because often even a nominal distribution $\mathbb{P}^{\text{N}}$ is unknown due to limited realistically available data. Even if $\mathbb{P}^{\text{N}}$ is known, computing the nominal mechanism requires solving an infinite-dimensional optimization problem, specifically~\eqref{eq:MDP} with a known distribution. we compute the optimal mechanism approximately as follows. 
This prompts us to approximate~$\mathcal{T}$ by the grid~$\hat{\mathcal{T}} = \mathcal{T} \cap \delta \cdot \mathbb{Z}^I$ with~$\delta=0.05$ and to approximate $\mathbb{P}^{\text{N}}$ by a discrete distribution $\hat{\mathbb{P}}^{\text{N}}$ supported on~$\hat{\mathcal{T}}$. Specifically, the probabilities of~$\hat{\mathbb{P}}^{\text{N}}$ are obtained by evaluating the probability density function of the normal distribution~$\mathcal N(\bm \mu_0, \bm \Sigma_0)$ at all grid points in~$\hat{\mathcal{T}}$ and by normalizing them so that they sum to~$1$. Similarly, the three extremal distributions are approximated by discrete distributions~$\hat {\mathbb{P}}_1^{\text{E}}$, $\hat{\mathbb{P}}_2^{\text{E}}$ and~$\hat{\mathbb{P}}_3^{\text{E}}$. The optimal mechanism is then approximated by the solution of a discretized version of problem~\eqref{eq:MDP} with~$\mathcal P= \{\hat{\mathbb P}_\epsilon\}$, which enforces the robust constraints only on~$\hat{\mathcal{T}}$. Here, $\hat{\mathbb P}_\epsilon$ is defined in the obvious way as a mixture of the discretized nominal and extremal distributions. As~$\hat{\mathcal{T}} \subseteq \mathcal{T}$, the discretized mechanism design problem relaxes~\eqref{eq:MDP} and thus  overestimates the expected payoff that can be earned with the exact optimal mechanism. In addition, as~$\hat{\mathcal{T}}$ is finite, the discretized mechanism design problem constitutes a finite-dimensional linear program. All linear programs are implemented in MATLAB R2022a using the YALMIP interface and solved with GUROBI.
\begin{figure}
    \centering
    \begin{minipage}{0.32\textwidth}
        \includegraphics[width=\linewidth]{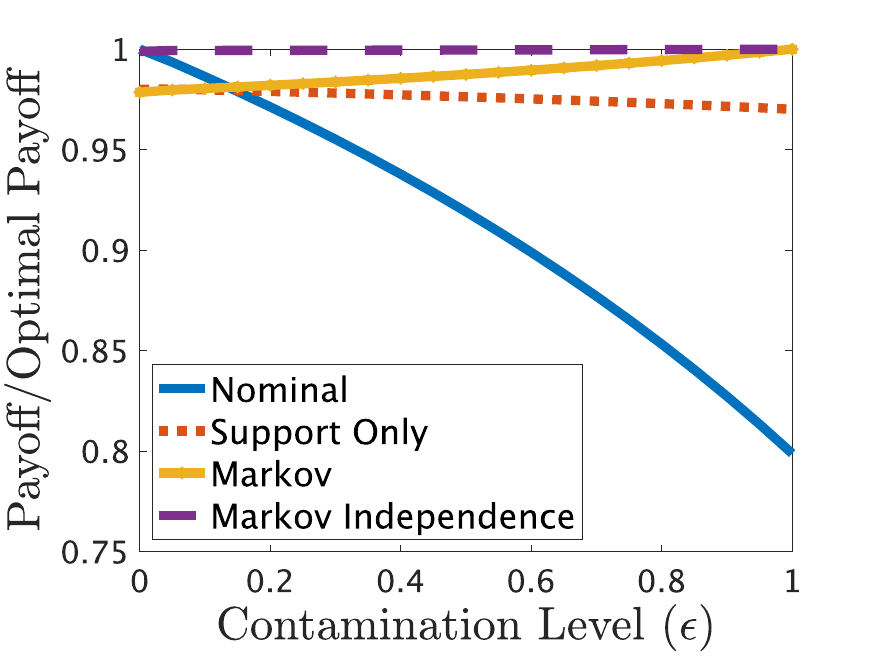}
        %\caption{Test 1}
    \end{minipage}
    \hfill
    \begin{minipage}{0.32\textwidth}
        \includegraphics[width=\linewidth]{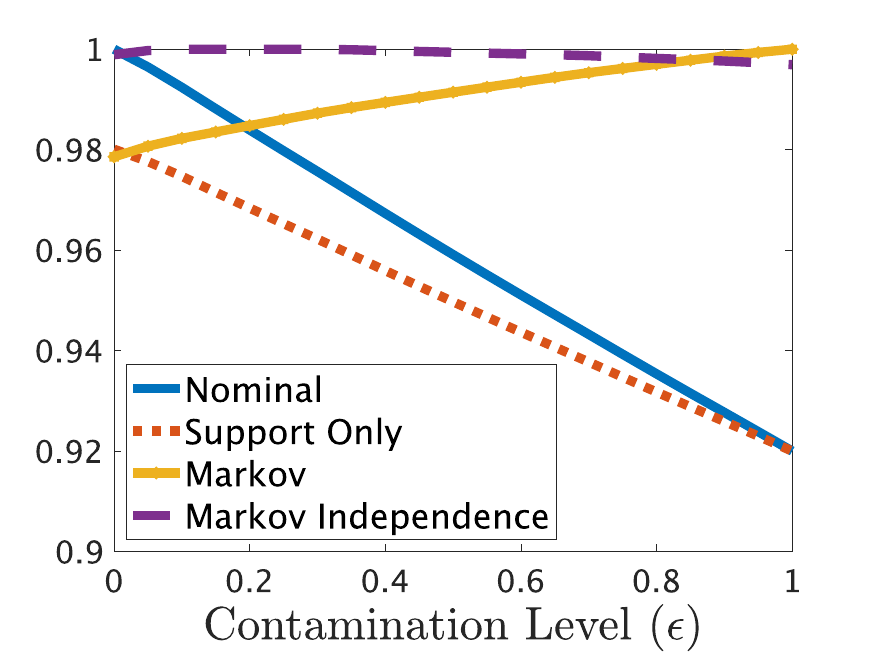} 
        %\caption{Test 2}
    \end{minipage}
    \hfill
    \begin{minipage}{0.32\textwidth}
        \includegraphics[width=\linewidth]{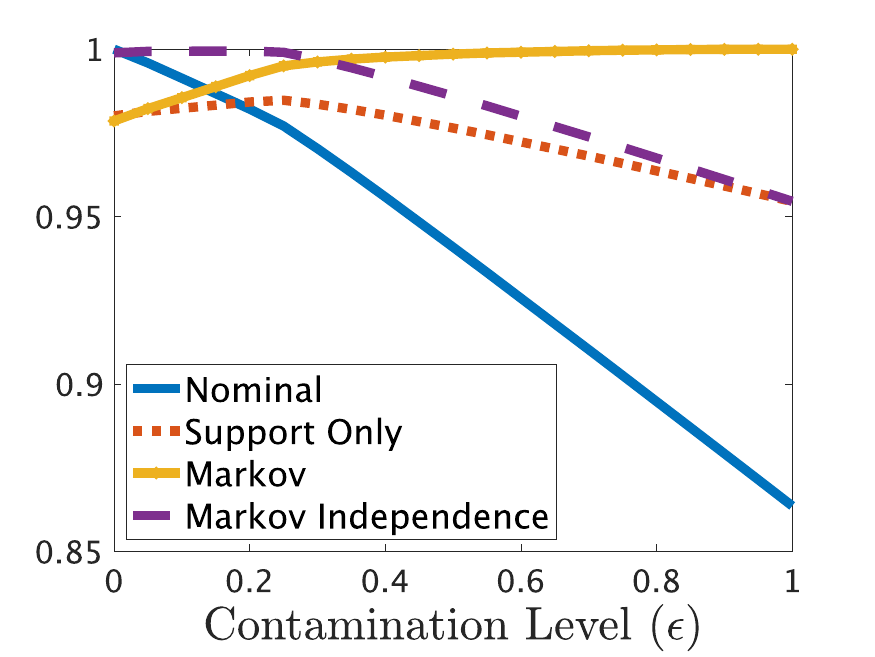} 
        %\caption{Test 2}
    \end{minipage}
\caption{Expected payoffs of the optimal mechanism tailored to $\hat{\mathbb P}^{\text{N}}$ and the proposed Pareto robustly optimal mechanisms for different contamination levels~$\epsilon$ and for~$\hat{\mathbb{P}}_1^{\text{E}}$ (left), $\hat{\mathbb{P}}_2^{\text{E}}$ (middle) and $\hat{\mathbb{P}}_3^{\text{E}}$ (right). All results are normalized by the expected payoff of the optimal mechanism with full distributional information.}
\label{fig:numerical illustration}
\end{figure}

In the first experiment we compare the different mechanisms in view of their {\em worst-case} payoffs across all scenarios in~$\bm t\in \hat{\mathcal{T}}$. While the worst-case payoff of the nominal mechanism tailored to~$\hat{\mathbb{P}}^{\text{N}}$ amounts to~$0$, the Pareto robustly optimal mechanisms tailored to support-only and Markov ambiguity sets with and without independent types all generate a worst-case payoff equal to~$0.2$. 
%The worst-case payoff $\min_{\bm{t} \in \hat{\mathcal{T}}} \; \sum_{i \in \mathcal{I}} (p_i(\bm{t}) t_i - q_i(\bm{t}) c_i)$ of the nominal mechanism across $\hat{\mathcal{T}}$ is zero, which is smaller than the worst-case payoff of our Pareto robustly optimal mechanisms, all of which achieve a worst-case payoff of $0.2$. 
When evaluating the worst case only over the scenarios $\bm t\in \hat{\mathcal{T}} \cap ([\underline{\mu}_1, \overline{\mu}_1] \times [\underline{\mu}_2, \overline{\mu}_2])$, 
%the worst-case payoff $\min_{\bm{t} \in \hat{\mathcal{T}} \cap ([\underline{\mu}_1, \overline{\mu}_1] \times [\underline{\mu}_2, \overline{\mu}_2])} \; \sum_{i \in \mathcal{I}} (p_i(\bm{t}) t_i - q_i(\bm{t}) c_i)$ of 
the nominal mechanism, which generates a worst-case payoff of~$0.5$, is still inferior to all Pareto robustly optimal mechanisms, which generate worst-case payoffs equal to~$0.65$ (for the support-only ambiguity set) and~$0.75$ (for the Markov ambiguity sets with or without independent types).

In the second experiment we compare the different mechanisms in view of their {\em expected} payoffs under~$\hat{\mathbb P}_\epsilon$ normalized by the maximum expected payoff achievable when~$\hat{\mathbb P}_\epsilon$ is known. Figure~\ref{fig:numerical illustration} visualizes the resulting relative payoffs as a function of the contamination level~$\epsilon \in[0,1]$ for the three different contaminating distributions~$\hat{\mathbb{P}}_1^{\text{E}}$, $\hat{\mathbb{P}}_2^{\text{E}}$ and~$\hat{\mathbb{P}}_3^{\text{E}}$. We observe that the Pareto robustly optimal mechanisms tailored to Markov ambiguity sets consistently outperform the nominal mechanism tailored to~$\hat{\mathbb{P}}^{\text{N}}$ even for small contamination levels. This outperformance becomes more pronounced as~$\epsilon$ increases. Notably, the Pareto robustly optimal mechanism for Markov ambiguity set with independent types is only marginally suboptimal at~$\epsilon = 0$ (in which case the nominal mechanism is optimal) and performs exceptionally well across all contamination levels  when the contaminating distribution is set to~$\hat{\mathbb{P}}_1^{\text{E}}$ or~$\hat{\mathbb{P}}_2^{\text{E}}$. This is perhaps expected because the agents' types are (almost) independent under these two distributions. If the contaminating distribution is set to~$\hat{\mathbb{P}}_3^{\text{E}}$, under which the types are highly correlated, then this mechanism's performance deteriorates for~$\epsilon\gtrsim 0.25$, and it is outperformed by
%and approaches the performance of 
the Pareto robustly optimal mechanism for Markov ambiguity sets {\em without} independent types. The Pareto robustly optimal mechanism for support-only ambiguity sets uses only minimal distributional information and thus displays a worse performance than the Pareto robustly optimal mechanisms for Markov ambiguity sets. Nevertheless, its performance is better or similar to that of the nominal mechanism. We thus conclude that the Pareto robustly optimal mechanisms not only provide the best possible worst-case guarantees but are also likely to outperform optimal mechanisms tailored to crisp distributions that are only marginally contaminated.

}

\section{Conclusions}
\label{sec:conclusions}
This paper studies optimal allocation problems with costly verification. Many allocation problems of this kind recur infrequently or never, and therefore it is unreasonable to assume that the principal has full knowledge of the distribution of the agents' types. This prompts us to formulate these allocation problems as distributionally robust mechanism design problems that explicitly account for (and hedge against) distributional ambiguity. We show that---like in the classical stochastic setting~\citep{ben2014optimal}---simple and interpretable mechanisms are optimal despite the extra layer of complexity introduced by the distributional ambiguity. Specifically, for three natural but increasingly restrictive ambiguity sets for the type distribution, we identify a large family of robustly optimal favored-agent mechanisms that maximize the principal's worst-case expected payoff. Moreover, for each of the three ambiguity sets, we identify a Pareto robustly optimal mechanism from within the family of all robustly optimal favored-agent mechanisms. These Pareto robustly optimal mechanisms not only maximize the worst-case expected payoff of the principal but also perform well when non-worst-case conditions prevail. In fact, these mechanisms strike an optimal trade-off between the expected payoffs under all distributions in the ambiguity set.
%Difference between the Pareto robustly optimal mechanisms \& when independence assumption makes sense...\\

The main results of this paper offer several insights of practical relevance. First, there is merit in acquiring information about the expected values of the agents' types. Indeed, at optimality, the principal's worst-case expected payoff is strictly higher under Markov ambiguity sets than under support-only ambiguity sets. For any given Markov ambiguity set, however, the principal does not benefit from knowing whether or not the agents' types are independent. At least, this information has no impact on the optimal worst-case expected payoff. 
%However, given identical information about the support sets and the expected values of the agents' types, the principal is indifferent between knowing that the agents' types are independent and not having access to any such information. Moreover, our results for Markov ambiguity sets reveal that 
Misrepresenting the agents' types as independent random variables may nevertheless have undesirable consequences, that is, it may mislead the principal into adopting a mechanism that fails to be Pareto robustly optimal and may even fail to be robustly optimal. We believe that the agents' types are unlikely to be independent in allocation problems with costly verification that arise naturally in reality. For example, a venture capitalist assigning seed funding to one of several start-up companies would be ill-advised to assume independence because innovations are often driven by societal trends, technical developments, or disruptive events ($e.g.$, the COVID-19 pandemic has led to a wave of supply chain start-ups), and therefore the potential gains from investing in these innovations cannot be independent. 

{  The allocation problem with costly verification addressed in this paper generalizes the stochastic model by \citet{ben2014optimal}, which posits that the type distribution is commonly known. Different variants of this stochastic model have been investigated in the recent literature. \citet{li2020mechanism} studies the impact of limited penalties by assuming that the principal can recover the good only {\em partially} when agents misreport their types. \citet{hu2024screening} assumes that the principal can only observe {\em signals} correlated with the types and that the cost of inspection increases with the accuracy of the signal, and \citet{chua2019optimal} assume that there are {\em multiple} homogeneous goods that need to be allocated to {\em different} agents. All of these problem variants are addressed with fundamentally different techniques than the baseline problem by \cite{ben2014optimal}. In addition, the corresponding optimal mechanisms are structurally different from favored-agent mechanisms studied here. More research is needed to study these problem variants under distributional ambiguity.}

\bibliographystyle{informs2014} % outcomment this and next line in Case 1
\bibliography{bibfile.bib} % if more than one, comma separated

% Appendix here
% Options are (1) APPENDIX (with or without general title) or 
%             (2) APPENDICES (if it has more than one unrelated sections)
% Outcomment the appropriate case if necessary
%
\newpage
\begin{APPENDIX}{ }

{ 
\section{Robustness Against the Agents’ Attitude Towards Ambiguity}\label{sec:Robustness Against the Agents’ Attitude Towards Ambiguity}
Incentive compatibility constraints are meant to ensure that all agents prefer to reveal their true types instead of lying. However, the probability that agent~$i$ receives the good, which amounts to $p_i(t_i, {\bm t}_{-i})$ under truthful reporting and to $p_i(t'_i, {\bm t}_{-i}) - q_i(t'_i, {\bm t}_{-i})$ under untruthful reporting, depends on the other agents' types and is thus uncertain. As agents may use different decision criteria to rank their choices under uncertainty, there is no canonical way of enforcing incentive compatibility. 
%We now demonstrate that  enforced in~\eqref{eq:MDP} protect the principal against uncertainty about the agents' attitudes towards ambiguity. Indeed,  Different agents' attitudes can lead to different types of incentive compatibility. 
For instance, if an agent believes that the type distribution is $\mathbb{P} \in \mathcal{P}_0(\mathcal{T})$, he might rank different choices by their expected utility under~$\mathbb P$. Alternatively, if an agent believes that the type distribution falls within an ambiguity set $\mathcal{P} \subseteq \mathcal{P}_0(\mathcal{T})$, he might rank different choices by their worst-case expected utility, their best-case expected utility, or a convex combination of these, in view of all distributions in~$\mathcal P$. Different decision criteria give rise to different incentive compatibility constraints. 
\begin{definition}
    A mechanism $(\bm p, \bm q)$ is called 
    \begin{itemize}
    \item \emph{Bayesian incentive compatible} with respect to a distribution $\mathbb P \in \mathcal P_0(\mathcal T)$ if for all $i \in \mathcal I$, $t_i, t_i' \in \mathcal T_i$,
    \begin{equation*}
        \begin{aligned}
            \mathbb E_{\mathbb P}[p_i(t_i, \tilde{\bm t}_{-i}) \mid \tilde{t_i} = t_i] \geq \mathbb E_{\mathbb P}[p_i(t'_i, \tilde{\bm t}_{-i}) - q_i(t'_i, \tilde{\bm t}_{-i}) \mid \tilde{t_i} = t_i],
        \end{aligned}
    \end{equation*} 
    \item \emph{Hurwicz incentive compatible} with respect to $\alpha \in [0,1]$ and $\mathcal P \subseteq \mathcal P_0(\mathcal T)$ if for all $i \in \mathcal I$, $t_i, t_i' \in \mathcal T_i$,
        \begin{equation*}
        \begin{aligned}
        &\alpha \inf_{\mathbb P \in \mathcal P} \mathbb E_{\mathbb P}[p_i(t_i, \tilde{\bm t}_{-i}) \mid \tilde{t_i} = t_i] + (1 - \alpha) \sup_{\mathbb P \in \mathcal P} \mathbb E_{\mathbb P}[p_i(t_i, \tilde{\bm t}_{-i}) \mid \tilde{t_i} = t_i] \\
        &\geq \alpha \inf_{\mathbb P \in \mathcal P} \mathbb E_{\mathbb P}[p_i(t'_i, \tilde{\bm t}_{-i}) - q_i(t'_i, \tilde{\bm t}_{-i}) \mid \tilde{t_i} = t_i] + (1-\alpha) \inf_{\mathbb P \in \mathcal P} \mathbb E_{\mathbb P}[p_i(t'_i, \tilde{\bm t}_{-i}) - q_i(t'_i, \tilde{\bm t}_{-i}) \mid \tilde{t_i} = t_i].
        \end{aligned}
    \end{equation*} 
    \end{itemize}
\end{definition}
%{\color{red}Arguments will hold also for monotonic utility functions of type $u(p - q)$... $u$ can be used to formulate different risk attitudes...}

The parameter $\alpha$ in the Hurwicz incentive compatibility constraints encodes the agent's risk aversion. Indeed, if $\alpha = 1$, the agent ranks outcomes based on their worst-case expected utility, which reflects risk-averse behavior. On the other hand, if $\alpha = 0$, the agent ranks outcomes based on the best-case expected payoff, which reflects risk-seeking behavior. The following proposition shows that the incentive compatibility constraints~\eqref{eq:IC} in problem~\eqref{eq:MDP} ensure that even agents with Bayesian or Hurwicz preferences have no incentive to misreport their true types. 
\begin{proposition}
    If $(\bm p, \bm q)$ satisfies~\eqref{eq:IC}, then it is Bayesian incentive compatible with respect to all $\mathbb P \in \mathcal P_0(\mathcal T)$ and Hurwicz incentive compatible with respect to all~$\alpha \in [0,1]$ and~$\mathcal P \subseteq \mathcal P_0(\mathcal T)$.
\end{proposition}
\proof{Proof.}
By taking expectations with respect to any $\mathbb P \in \mathcal P_0(\mathcal T)$ on both sides of the~\eqref{eq:IC} constraints, it is easy to see that~\eqref{eq:IC} implies Bayesian incentive compatibility. 
Next, consider any risk aversion parameter~$\alpha \in [0,1]$ and ambiguity set~$\mathcal P \subseteq \mathcal P_0(\mathcal T)$. We already know that Bayesian incentive compatibility holds for every distribution~$\mathbb{P} \in \mathcal{P}$. Maximizing or minimizing both sides of an inequality depending on~$\mathbb P$ over all $\mathbb P\in\mathcal P$ preserves the inequality. In addition, taking convex combinations of two valid inequalities (of the same direction) yields another valid inequality. Therefore, one readily verifies that $(\bm p, \bm q)$ satisfies Hurwicz incentive compatibility. 
\hfill \Halmos
\endproof

Proposition implies that the~\eqref{eq:IC} constraints enforced in problem~\eqref{eq:MDP} protect the principal against uncertainty about the agents' preferences.
}

\section{Proofs}

{ 
\proof{Proof of Lemma \ref{lem:deviationForFavored}.}
Fix any $i \in \mathcal I$, $\bm t \in \mathcal T$ and $(\bm p, \bm q) \in \mathcal X$, and suppose that $p_i(\bm t)-q_i(\bm t)=1$. We then have $1\geq p_i(t_i',\bm{t}_{-i}) \geq p_i(\bm{t})-q_i(\bm t)= 1$ for any $t'_i \in \mathcal T_i$, where the two inequalities follow from~\eqref{eq:FC} and~\eqref{eq:IC}, respectively. This implies that $p_i(t_i',\bm{t}_{-i}) = 1$, and thus assertion~(i) follows.

Next, fix any $i \in \mathcal I$, $\bm t \in \mathcal T$ and $(\bm p, \bm q) \in \mathcal X$, and suppose that $p_i(\bm t)=0$. We then have $0=p_i(\bm t) \geq p_i(t_i',\bm{t}_{-i})-q_i(t_i',\bm{t}_{-i})\geq 0$ for any $t'_i \in \mathcal T_i$, where the two inequalities follow from~\eqref{eq:IC} and~\eqref{eq:FC}, respectively.  This implies that $p_i(t_i',\bm t_{-i})=q_i(t_i',\bm t_{-i})$, and thus assertion~(ii) follows.
\hfill \halmos
\endproof

\proof{Proof of Proposition \ref{prop:unifyingPRO}.}
Fix any optimal mechanism $(\bm p^\star, \bm q^\star) \in \mathcal{X}$, and suppose that there exists a partition $\mathcal{S}_1, \ldots, \mathcal{S}_m$ of~$\mathcal T$ satisfying the conditions~(i) and~(ii) in the proposition statement. Consider now any feasible mechanism $(\bm p, \bm q) \in \mathcal{X}$ that Pareto robustly dominates $(\bm p^\star, \bm q^\star)$. In the following, we will use induction to  show that $(\bm p, \bm q)$ must generate the same payoff as $(\bm p^\star, \bm q^\star)$ in every scenario $\bm t \in \mathcal{T}$. Thus, $(\bm p, \bm q)$ cannot generate a strictly higher expected payoff than $(\bm p^\star, \bm q^\star)$ under any distribution, which contradicts our assumption that $(\bm p, \bm q)$ Pareto robustly dominates~$(\bm p^\star, \bm q^\star)$.

As for the base step of the induction, we aim to show that~$(\bm p, \bm q)$ generates the same payoff as~$(\bm p^\star, \bm q^\star)$ in every scenario $\bm t \in \mathcal{S}_1$. To this end, note first that~$(\bm p, \bm q)$ is feasible in~${\text{SP}}$$_{1}(\bm t)$. As~$(\bm p^\star, \bm q^\star)$ solves~${\text{SP}}$$_{1}(\bm t)$ thanks to assumption~(ii), the mechanism~$(\bm p, \bm q)$ must generate a weakly lower payoff than $(\bm p^\star, \bm q^\star)$ in scenario~$\bm t$. The same argument applies for every scenario~$\bm t \in \mathcal{S}_1$, and thus we have
\begin{align}\label{eq: boundByThePROmechanism}
    &\sum_{i \in \mathcal{I}} (p_i({\bm{t}}) {t}_i - q_i({\bm{t}}) c_i) \leq \sum_{i \in \mathcal{I}} (p_i^\star({\bm{t}}) {t}_i - q_i^\star({\bm{t}}) c_i) \quad\forall \bm t \in \mathcal{S}_1.
\end{align}
Assume now that the inequality in~\eqref{eq: boundByThePROmechanism} is strict for some~$\bm{t}' \in \mathcal{S}_1$.
%We claim that for $(\bm p, \bm q)$ to Pareto robustly dominate $(\bm p^\star, \bm q^\star)$, it should generate the same payoff as $(\bm p^\star, \bm q^\star)$ in any scenario $\bm t \in \mathcal{S}_1$. Suppose to the contrary that $(\bm p, \bm q)$ generates a strictly lower payoff than $(\bm p^\star, \bm q^\star)$ in some scenario $\bm{t}' \in \mathcal{S}_1$. 
By assumption~(i), there exists a distribution $\mathbb P\in \mathcal P\cap \mathcal{P}_0(\mathcal{S}_1)$ that assigns a strictly positive probability mass to scenario~$\bm{t}'$. %If $(\bm p, \bm q)$ generates a strictly lower payoff than $(\bm p^\star, \bm q^\star)$ in scenario $\bm{t}'$, then $(\bm p, \bm q)$ also generates a strictly lower expected payoff under $\mathbb{P}$, $i.e.$,
Taking expectations with respect to~$\mathbb P$ on both sides of~\eqref{eq: boundByThePROmechanism} thus yields
$$
\mathbb{E}_{\mathbb{P}}\left[\sum_{i \in \mathcal{I}} (p_i(\tilde{\bm{t}}) \tilde{t}_i - q_i(\tilde{\bm{t}}) c_i) \right]< \mathbb{E}_{\mathbb{P}}\left[\sum_{i \in \mathcal{I}} (p^\star_i(\tilde{\bm{t}}) \tilde{t}_i - q^\star_i(\tilde{\bm{t}}) c_i)  \right],
$$
which contradicts our initial assumption that $(\bm p, \bm q)$ Pareto robustly dominates $(\bm p^\star, \bm q^\star)$. Thus, the inequality in~\eqref{eq: boundByThePROmechanism} must be an equality for every~$\bm{t} \in \mathcal{S}_1$, that is, the mechanisms $(\bm p, \bm q)$ and $(\bm p^\star, \bm q^\star)$ must indeed generate the same payoff throughout~$\mathcal{S}_1$.

The induction step corresponding to~$k \geq 2$ is based on the hypothesis that $(\bm p, \bm q)$ and $(\bm p^\star, \bm q^\star)$ generate the same payoff on $\cup_{l=1}^{k-1} \mathcal{S}_l$. Our goal is to show that the two mechanisms also generate the same payoff in every scenario~$\bm t\in \mathcal{S}_k$.
%We now consider the conditions Proposition \ref{prop:unifyingPRO}(i) \& (ii) for the index value $k$. 
To this end, note first that $(\bm p, \bm q)$ is feasible in~\ref{eq:MDPk(t)} due to the induction hypothesis. As $(\bm p^\star, \bm q^\star)$ solves~\ref{eq:MDPk(t)} thanks to assumption~(ii), $(\bm p, \bm q)$ generates a weakly lower payoff than~$(\bm p^\star, \bm q^\star)$ in scenario~$\bm t$. As this is true for every scenario~$\bm t \in \mathcal{S}_k$, we have
\begin{align*}\label{eq: boundByThePROmechanism2}
    &\sum_{i \in \mathcal{I}} (p_i({\bm{t}}) {t}_i - q_i({\bm{t}}) c_i) \leq \sum_{i \in \mathcal{I}} (p_i^\star({\bm{t}}) {t}_i - q_i^\star({\bm{t}}) c_i) \quad \forall \bm t \in \mathcal{S}_k. %\cup_{l=1}^{k} \mathcal{S}_l.
\end{align*}
As in the base case, one can now show by contradiction that the inequality in the above expression is never strict. Hence, $(\bm p, \bm q)$ and $(\bm p^\star, \bm q^\star)$ generate the same payoff throughout~$\mathcal{S}_k$.
%suppose to the contrary that the above inequality is strict for some $\bm{t}' \in \mathcal{S}_k$. Proposition \ref{prop:unifyingPRO}(i) guarantees that there exists a distribution $\mathbb P \in \mathcal{P}$ that assigns strictly positive probability mass to the scenario $\bm{t}'$ and is supported on a subset of $\cup_{l=1}^k \mathcal{S}_l$. Recalling that the payoffs are equal on $\cup_{l=1}^{k-1} \mathcal{S}_l$ by induction hypothesis, by using similar arguments to base case, we can conclude that $(\bm p, \bm q)$ and $(\bm p^\star, \bm q^\star)$ should yield the same payoff under every $\bm t \in \mathcal{S}_k$.

By induction, the two mechanisms~$(\bm p, \bm q)$ and $(\bm p^\star, \bm q^\star)$ generate the same payoff on all of~$\mathcal T$ and thus the same {\em expected} payoff under every~$\mathbb P\in\mathcal P$. This contradicts the assumption that~$(\bm p, \bm q)$ Pareto robustly dominates~$(\bm p^\star, \bm q^\star)$. As $(\bm p^\star, \bm q^\star)$ is robustly optimal by assumption, and as~$(\bm p^\star, \bm q^\star)$ is {\em not} Pareto robustly dominated by any feasible mechanism, it is indeed Pareto robustly optimal.

Note that our arguments above also imply that any mechanism $(\bm p,\bm q) \in \mathcal{X}$ that weakly Pareto robustly dominates $(\bm p^\star,\bm q^\star)$ generates the same payoff as $(\bm p^\star,\bm q^\star)$ in every scenario $\bm t \in \mathcal{T}$. %In particular, our proof shows that $(\bm p,\bm q)$ cannot Pareto robustly dominate $(\bm p^\star,\bm q^\star)$ and can only weakly Pareto robustly dominate $(\bm p^\star,\bm q^\star)$, in which case their payoffs are the same for every $\bm t \in \mathcal T$.
\hfill \halmos
\endproof
}

\proof{Proof of Proposition \ref{prop:optimal-z-support}.}
%\begin{proof}[Proposition \ref{prop:optimal-z-support}]
Relaxing the incentive compatibility constraints and the first inequality in~\eqref{eq:FC} yields
%\noteck{$z^\star$ was not defined as a function before. It also shouldn't be a function here.}
\begin{equation*}
\begin{aligned}
z^\star &\leq& &\sup_{\bm{p}, \bm{q}} &&\inf_{\bm{t} \in \mathcal{T}} \; \sum_{i \in \mathcal{I}} (p_i(\bm{t}) t_i - q_i(\bm{t}) c_i)  \\
&&&\;\text{s.t.} &&%p_i : \mathcal{T} \rightarrow [0,1] \;\text{and}\; q_i : \mathcal{T} \rightarrow [0,1] \;\;\forall i \in \mathcal{I}
p_i, q_i \in \mathcal{L}(\mathcal{T}, [0,1]) \;\;\forall i \in \mathcal{I}\\
&&&&&\sum_{i \in \mathcal{I}} p_i(\bm{t}) \leq 1 \;\;\forall \bm{t} \in \mathcal{T}\\
&=& &\sup_{\bm{p}} &&\inf_{\bm{t} \in \mathcal{T}} \; \sum_{i \in \mathcal{I}} p_i(\bm{t}) t_i\\
&&&\;\text{s.t.} &&%p_i : \mathcal{T} \rightarrow [0,1] \;\;\forall i \in \mathcal{I}, 
p_i \in \mathcal{L}(\mathcal{T}, [0,1]) \;\;\forall i \in \mathcal{I}, \;\;\sum_{i \in \mathcal{I}} p_i(\bm{t}) \leq 1 \;\;\forall \bm{t} \in \mathcal{T},
%&\leq& &&&\inf_{\mathbb{P} \in \mathcal{P}} \; \mathbb{E}_{\mathbb{P}}\left[\max_{i \in \mathcal{I}} \tilde{t}_i \right]
\end{aligned}
\end{equation*}
where the equality holds because in the relaxed problem it is optimal to set~$q_i(\bm{t})=0$ for all $i \in \mathcal{I}$ and $\bm{t} \in \mathcal{T}$. As the resulting maximization problem over~$\bm{p}$ is separable with respect to $\bm{t} \in \mathcal{T}$, it is optimal to allocate the good in each scenario $\bm{t} \in \mathcal{T}$---with probability one---to an agent with maximal type. Therefore, $z^\star$ is bounded above by $\inf_{\bm{t} \in \mathcal{T}} \, \max_{i \in \mathcal{I}} {t}_i = \max_{i \in \mathcal{I}} \underline{t}_i$. However, this bound is attained by a mechanism that allocates the good to an agent $i' \in \arg \max_{i \in \mathcal{I}} \underline{t}_i$ irrespective of~$\bm{t}\in\mathcal{T}$ and never inspects anyone's type. Since this mechanism is feasible, the claim follows. \hfill \Halmos
%\end{proof}
\endproof

\vspace{2mm}

\proof{Proof of Theorem \ref{thr:opt-fav-support-only}.}
Select an arbitrary favored-agent mechanism with $i^\star \in {\arg \max}_{i \in \mathcal{I}} \underline{t}_i$ and $\nu^\star \geq \max_{i \in \mathcal{I}} \underline{t}_i$. Recall first that this mechanism is feasible in~\eqref{eq:RMDP}. Next, we will show that this mechanism attains a worst-case payoff that is at least as large as~$\max_{i \in \mathcal{I}} \underline{t}_i$, which implies via Proposition~\ref{prop:optimal-z-support} that it is in fact optimal in~\eqref{eq:RMDP}. To this end, fix an arbitrary type profile~$\bm{t}\in\mathcal{T}$. If~$\max_{i\neq i^\star} t_i-c_i< \nu^\star$, then condition~(i) in Definition~\ref{def:fav-agent-mechanism} implies that the principal's payoff amounts to $t_{i^\star} \geq \max_{i \in \mathcal{I}} \underline{t}_i$, where the inequality follows from the selection of~$i^\star$. If~$\max_{i\neq i^\star} t_i-c_i>\nu^\star$, then condition~(ii) in Definition~\ref{def:fav-agent-mechanism} implies that the principal's payoff amounts to~$\max_{i \in \mathcal{I}} t_i-c_i > \nu^\star \geq \max_{i \in \mathcal{I}} \underline{t}_i$, where the second inequality follows from the selection of~$\nu^\star$. If $\max_{i \neq i^\star} t_i-c_i = \nu^\star$, then the allocation functions are defined either as in condition~(i) or as in condition~(ii) of Definition~\ref{def:fav-agent-mechanism}. Thus, the principal's payoff amounts either to~$t_{i^\star}$ or to~$\max_{i \in \mathcal{I}} t_i-c_i \geq \nu^\star$, respectively, and is therefore again non-inferior to~$\max_{i \in \mathcal{I}} \underline{t}_i$. In summary, we have shown that the principal's payoff is non-inferior to~$z^\star = \max_{i \in \mathcal{I}} \underline{t}_i$ in all three cases. As scenario~$\bm{t}\in\mathcal{T}$ was chosen arbitrarily, this reasoning implies that the principal's worst-case payoff is also non-inferior to~$z^\star$. The favored-agent mechanism at hand is therefore optimal in~\eqref{eq:RMDP} by virtue of Proposition~\ref{prop:optimal-z-support}. \hfill \Halmos
\endproof

\vspace{2mm}
\proof{Proof of Lemma \ref{lem:randomization}.}
Assume first that $(\bm{p},\bm{q})$ is a favored-agent mechanism with favored agent $i^\star \in \mathcal{I}$ and threshold value~$\nu^\star \in \mathbb{R}$. Next, fix any agent~$i \in \mathcal{I}$ and any type profile~$\bm{t}_{-i} \in \mathcal{T}_{-i}$. If $i \neq i^\star$, then we have either $p_i(t_i,\bm{t}_{-i})=q_i(t_i,\bm{t}_{-i}) = 1$ or $p_i(t_i,\bm{t}_{-i})=q_i(t_i,\bm{t}_{-i}) = 0$ for all~$t_i\in\mathcal T_i$. This implies that $p_i(t_i,\bm{t}_{-i})-q_i(t_i,\bm{t}_{-i})=0$ is constant in $t_i \in \mathcal{T}_i$. If $i=i^\star$, then the fixed type profile~$\bm t_{-i^\star}$ uniquely determines whether the allocations are constructed as in case~(i) or as in case~(ii) of Definition~\ref{def:fav-agent-mechanism}. In case~(i) we have $p_{i^\star}(t_{i^\star},\bm{t}_{-{i^\star}})=1$ and $q_{i^\star}(t_{i^\star},\bm{t}_{-{i^\star}}) =0$ for all $t_{i^\star} \in \mathcal{T}_{i^\star}$, and thus $p_{i^\star}(t_{i^\star},\bm{t}_{-{i^\star}})-q_{i^\star}(t_{i^\star},\bm{t}_{-{i^\star}})=1$ is constant in~$t_{i^\star} \in \mathcal{T}_{i^\star}$. In case~(ii) we have either $p_{i^\star}(t_{i^\star},\bm{t}_{-{i^\star}})=1$ and $q_{i^\star}(t_{i^\star},\bm{t}_{-{i^\star}}) =1$ or $p_{i^\star}(t_{i^\star},\bm{t}_{-{i^\star}})=0$ and $q_{i^\star}(t_{i^\star},\bm{t}_{-{i^\star}}) =0$, and thus $p_{i^\star}(t_{i^\star},\bm{t}_{-{i^\star}})-q_{i^\star}(t_{i^\star},\bm{t}_{-{i^\star}})=0$ is again constant in~$t_{i^\star} \in \mathcal{T}_{i^\star}$. This establishes the claim for any favored-agent mechanism $(\bm{p},\bm{q})$.
% Either $p_{i^\star}(t_{i^\star},\bm{t}_{-{i^\star}})=1$ and $q_{i^\star}(t_{i^\star},\bm{t}_{-{i^\star}}) =0$ for all $t_{i^\star} \in \mathcal{T}_{i^\star}$ (case~(i) in Definition~\ref{def:fav-agent-mechanism}) or $p_{i^\star}(t_{i^\star},\bm{t}_{-{i^\star}})=1$ and $q_{i^\star}(t_{i^\star},\bm{t}_{-{i^\star}}) =1$ for all $t_{i^\star} \in \mathcal{T}_{i^\star}$ (case~(ii) in Definition~\ref{def:fav-agent-mechanism} when agent~$i^\star$ is the winner) or $p_{i^\star}(t_{i^\star},\bm{t}_{-{i^\star}})=0$ and $q_{i^\star}(t_{i^\star},\bm{t}_{-{i^\star}}) =0$ for all $t_{i^\star} \in \mathcal{T}_{i^\star}$ (case~(ii) in Definition~\ref{def:fav-agent-mechanism} when agent~$i^\star$ is {\em not} the winner). Which one of these situations will arise is uniquely determined by the fixed types~$\bm t_{-i^\star}$. In all three situations, $p_{i^\star}(t_{i^\star},\bm{t}_{-{i^\star}})-q_{i^\star}(t_{i^\star},\bm{t}_{-{i^\star}})$ is thus constant in~$t_{i^\star} \in \mathcal{T}_{i^\star}$, which establishes the claim when $(\bm{p},\bm{q})$ is a favored-agent mechanism.
Assume now that $(\bm{p},\bm{q})=\sum_{k\in\mathcal K} \pi_k (\bm p^k,\bm q^k)$ is a convex combination of favored-agent mechanisms~$(\bm p^k,\bm q^k)$, $k\in \mathcal K=\{1,\ldots, K\}$. Next, fix any~$i \in \mathcal{I}$ and~$\bm{t}_{-i} \in \mathcal{T}_{-i}$. From the first part of the proof we know that $p^k_i(t_i,\bm{t}_{-i})-q^k_i(t_i,\bm{t}_{-i})$ is constant in $t_i \in \mathcal{T}_i$ for each~$k\in\mathcal K$, and therefore $p_i(t_i,\bm{t}_{-i})-q_i(t_i,\bm{t}_{-i})$ is also constant in $t_i \in \mathcal{T}_i$. Similar arguments apply when $(\bm{p},\bm{q})$ represents a convex combination of infinitely many favored-agent mechanisms. \hfill \Halmos
\endproof

\vspace{2mm}
\proof{Proof of Theorem \ref{thr:PRO-fav-agents}.}
{  Throughout the proof, we denote by~$(\bm{p}^\star,\bm{q}^\star)$ the favored-agent mechanism of type~(i) with favored agent~${i^\star} \in {\arg \max}_{i \in \mathcal{I}} \underline{t}_i$ and threshold value~$\nu^\star= \max_{i \in \mathcal{I}} \underline{t}_i$. By construction, we thus have~$\nu^\star=\underline{t}_{i^\star}$. We also use the following partition of the type space~$\mathcal{T}$.
\begin{equation*}
\begin{aligned}
\mathcal{T}_{I} &= \{\bm{t} \in \mathcal{T} \,:\, \max_{i \neq i^\star}t_i-c_i \leq \underline{t}_{i^\star} \; \text{and} \; t_{i^\star} = \overline{t}_{i^\star}\}\\
\mathcal{T}_{II} &=  \{\bm{t} \in \mathcal{T} \,:\, \max_{i \neq i^\star}t_i-c_i \leq \underline{t}_{i^\star} \; \text{and} \; t_{i^\star} < \overline{t}_{i^\star}\}\\
\mathcal{T}_{III} &= \{\bm{t} \in \mathcal{T} \,:\, \max_{i \neq i^\star}t_i-c_i> \underline{t}_{i^\star} \}
\end{aligned}
\end{equation*}
The sets~$\mathcal{T}_{I}$ and $\mathcal{T}_{II}$ are nonempty and contain at least~$(\overline t_{i^\star},\underline{\bm t}_{-i^\star})$ and~$\underline{\bm t}$, respectively, because $\underline t_{i^\star} <\overline t_{i^\star}$ and $c_i > 0$ for all $i \in \mathcal{I}$.
However, $\mathcal{T}_{III}$ can be empty if $\underline t_{i^\star}$ or $c_i$, $i \neq i^\star$, are sufficiently large. }

{  To establish the Pareto robust optimality of $(\bm{p}^\star,\bm{q}^\star)$, we leverage Proposition~\ref{prop:unifyingPRO}. % and Lemma~\ref{lem:deviationForFavored}.
%Proposition~\ref{prop:unifyingPRO} provides sufficient conditions for the Pareto robust optimality of a mechanism that is robustly optimal. 
From Theorem~\ref{thr:opt-fav-support-only} we already know that $(\bm{p}^\star,\bm{q}^\star)$ is robustly optimal. To prove that~$(\bm{p}^\star,\bm{q}^\star)$ is also Pareto robustly optimal, it thus suffices to prove that the conditions~(i) and~(ii) in Proposition~\ref{prop:unifyingPRO} hold for some partition of the type space $\mathcal{T}$. Specifically, we will show that it holds for a refinement of the partition $\mathcal{T}_{I}, \mathcal{T}_{II}, \mathcal{T}_{III}$.
% That is, we need show that there exists a partition $\mathcal{T}_1, \mathcal{T}_2, ,\ldots, \mathcal{T}_m$ for some $m$ satisfying the following conditions for any $k=1,\ldots,m$, and $\bm{t} \in \mathcal{T}_k$:
%\begin{itemize}
%        \item[(i)] $\exists \mathbb P \in \mathcal{P}$ supported on a subset of $\cup_{l=1}^k \mathcal{T}_l$ and $\mathbb P(\tilde{\bm t}= \bm t)>0$,
%        \item[(ii)] $(\bm p^\star, \bm q^\star)$ solves the following optimization problem
%            \begin{equation}\label{eq:MDPk(t)} \tag{$\widehat{\text{MDP}}$$_{k,t}$}
%            \begin{aligned}
%                &\max_{(\bm{p}, \bm{q}) \in \mathcal{X}} &&\sum_{i \in \mathcal{I}} (p_i({\bm{t}}) {t}_i - q_i({\bm{t}}) c_i) \\
%                &\;\;\;\operatorname{s.t.} && \sum_{i \in \mathcal{I}} (p_i(\hat{\bm{t}}) \hat{t}_i - q_i(\hat{\bm{t}}) c_i) = \sum_{i \in \mathcal{I}} (p_i^\star(\hat{\bm{t}}) \hat{t}_i - q_i^\star(\hat{\bm{t}}) c_i) \;\;\forall \hat{\bm t} \in \cup_{l=1}^{k-1} \mathcal{T}_l.
%            \end{aligned}
%            \end{equation}
%    \end{itemize}
Condition~(i) trivially holds because the support-only ambiguity set contains all Dirac point distributions supported on any point~$\bm t \in \mathcal T$. It thus remains to verify condition~(ii).

The rest of the proof is divided into three steps focusing on the three subsets $\mathcal{T}_{I}$, $\mathcal{T}_{II}$ and $\mathcal{T}_{III}$. In this process, we will construct a refined partition $\mathcal S_1,\ldots,\mathcal S_m$ of~$\mathcal T$ with~$m=2I+1$, where $\mathcal S_1, \dots, \mathcal S_I$ forms a partition of~$\mathcal{T}_{I}$, $\mathcal S_{I+1}$ coincides with~$\mathcal{T}_{II}$, and $\mathcal S_{I+2}, \dots, \mathcal S_{2I+1}$ forms a partition of~$\mathcal{T}_{III}$. We will also exploit the non-locality of the incentive compatibility constraint~\eqref{eq:IC} via Lemma~\ref{lem:deviationForFavored} to show iteratively that~$(\bm p^\star, \bm q^\star)$ solves the scenario problem~\ref{eq:MDPk(t)} for all $\bm t \in \mathcal S_k$ and~$k=1,\ldots,m$.
%Moreover, in steps 1 and 2, we will also show that any mechanism that solves \eqref{eq:MDPk(t)} also has the same allocation probabilities as~$(\bm{p}^\star,\bm{q}^\star)$ for any $\bm t \in \mathcal{T}_I \cup \mathcal{T}_{II}$. This equivalence in allocation probabilities allows us to prove that $(\bm{p}^\star,\bm{q}^\star)$ also solves \eqref{eq:MDPk(t)} for the remaining scenarios. 

The subsequent reasoning critically relies on our assumption that~$(\bm{p}^\star,\bm{q}^\star)$ is a favored-agent mechanism of type~(i) with favored agent ${i^\star} \in {\arg \max}_{i \in \mathcal{I}} \underline{t}_i$ and threshold value $\nu^\star = \underline t_{i^\star}$. By Definition~\ref{def:fav-agent-mechanism}, this means that the principal's payoff in scenario~$\bm t$ amounts to~$t_{i^\star}$ when $\max_{i \neq i^\star} t_i - c_i \leq \underline{t}_{i^\star}$ ($i.e.$, when $\bm t \in \mathcal{T}_{I} \cup \mathcal{T}_{II}$) and to $\max_{i \in \mathcal{I}} t_i - c_i$ when  $\max_{i \neq i^\star} t_i - c_i > \underline{t}_{i^\star}$ ($i.e.$, when $\bm t \in \mathcal{T}_{III}$).}

\textbf{Step 1 ($\mathcal{T}_{I}$).} {  We partition~$\mathcal{T}_{I}$ into~$I$ subsets of the form
$$
\mathcal S_{k} = \left\{ \bm t \in \mathcal{T}_{I} \,:\, |\{i \in \mathcal{I} \,:\, t_i \geq t_{i^\star}\}| = k \right\}
$$
for $k=1,\ldots, I$. We will use induction on $k$ to prove that $(\bm{p}^\star,\bm{q}^\star)$ is optimal in~\ref{eq:MDPk(t)} and that any mechanism $(\bm p,\bm q)$ that solves~\ref{eq:MDPk(t)} must satisfy $\bm p(\bm{t})=\bm p^\star(\bm t)$ and $\bm q(\bm{t}) =\bm q^\star(\bm t)$ for all $\bm t\in\mathcal S_k$. % and , and that any optimal solution in all of \eqref{eq:MDPk(t)}, $k=1, \dots, I$, must have the same allocation probabilities as that of $(\bm{p}^\star,\bm{q}^\star)$ in any scenario $\bm t \in \mathcal T_{I}$. The case of $k=1$ serves as the base step.

As for the base step, %we have to show that $(\bm{p}^\star,\bm{q}^\star)$ solves the scenario problem~${\text{SP}}$$_{1}(\bm t)$ for any fixed~$\bm{t} \in \mathcal{S}_{1}$. To this end, 
note that any mechanism~$(\bm{p},\bm{q})\in\mathcal{X}$ is feasible in ${\text{SP}}$$_{1}(\bm t)$ for any~$\bm{t} \in \mathcal{S}_{1}$. In addition, the objective function value of $(\bm p, \bm q)$ is dominated by that of~$(\bm p^\star, \bm q^\star)$ in~${\text{SP}}$$_{1}(\bm t)$ because
\begin{equation*}
\begin{aligned}
\sum_{i \in \mathcal{I}} (p_i(\bm{t}) t_i - q_i(\bm{t}) c_i) \leq \sum_{i \in \mathcal{I}} p_i(\bm{t}) t_i \leq t_{i^\star} = \sum_{i \in \mathcal{I}} (p_i^\star(\bm{t}) t_i - q_i^\star(\bm{t}) c_i),
\end{aligned}
\end{equation*}
where the first inequality holds because $c_i>0$ for all $i \in \mathcal{I}$, and the second inequality follows from  the definition of $\mathcal{S}_{1}$, which implies that $t_i < t_{i^\star}$ for all $i \neq i^\star$. Thus, $(\bm{p}^\star,\bm{q}^\star)$ solves indeed~${\text{SP}}$$_{1}(\bm t)$. As $t_i < t_{i^\star}$ for all $i \neq i^\star$, the above arguments also reveal that any mechanism $(\bm{p}, \bm{q})$ that solves ${\text{SP}}$$_{1}(\bm t)$ must coincide with~$(\bm{p}^\star,\bm{q}^\star)$ on~$\mathcal T_1$, that is, $\bm p(\bm{t})=\bm p^\star(\bm t)$ and $\bm q(\bm{t}) =\bm q^\star(\bm t)$ for all $\bm t\in\mathcal T_1$. %$p_{i^\star}(\bm{t})=1$ and $q_{i^\star}(\bm{t}) =0$ for all $\bm t\in\mathcal T_1$. 

As for the induction step, assume that for all $l<k$ we know that $(\bm{p}^\star,\bm{q}^\star)$ solves~${\text{SP}}$$_{l}(\bm t)$ and that any mechanism $(\bm p,\bm q)$ that solves~${\text{SP}}$$_{l}(\bm t)$ must satisfy $\bm p(\bm{t})=\bm p^\star(\bm t)$ and $\bm q(\bm{t}) =\bm q^\star(\bm t)$ for all $\bm t\in\mathcal S_l$. Fix now any $\bm t \in \mathcal S_k$ and mechanism $(\bm p, \bm q)$ feasible in~\ref{eq:MDPk(t)}. The constraints in~\ref{eq:MDPk(t)} ensure that
%Our induction hypothesis is that any optimal solution of $(\widehat{\text{MDP}}$$_{n,t'})$ satisfy $p_{i^\star}(\bm{t}')=1$ and $q_{i^\star}(\bm{t}') =0$ for all $\bm t' \in \mathcal{T}_I^n$ and for all $n \leq k-1$ for some $k\in \{2,\ldots,\mathcal{I}\}$, and we fix an arbitrary $\bm t \in \mathcal{T}_I^{k}$. Now, the problem \eqref{eq:MDPk(t)} has an additional set of constraints except feasibility in $\mathcal{X}$:
$$
    \sum_{i \in \mathcal{I}} (p_i({\bm{t}'}) {t}_i' - q_i({\bm{t}'}) c_i) = \sum_{i \in \mathcal{I}} (p_i^\star({\bm{t}'}) {t}_i' - q_i^\star({\bm{t}'}) c_i) \quad \forall {\bm t'} \in \cup_{l=1}^{k-1}\mathcal{S}_l.
$$
As $(\bm{p}^\star,\bm{q}^\star)$ is optimal in~${\text{SP}}$$_{l}(\bm t')$ thanks to the induction hypothesis, this equality implies that~$(\bm p, \bm q)$ solves~${\text{SP}}$$_{l}(\bm t')$ for every~$\bm t'\in\mathcal S_l$ and~$l<k$. By the second part of the induction hypothesis, this in turn implies that $\bm p(\bm{t}')=\bm p^\star(\bm t')$ and $\bm q(\bm{t}') =\bm q^\star(\bm t')$ for all $\bm t'\in\mathcal S_l$ and for all $l<k$.
%By the induction hypothesis, we know that the allocation probabilities of $(\bm p, \bm q)$ should match those of $(\bm p^\star, \bm q^\star)$ on $\cup_{l=1}^{n-1}\mathcal{S}_l$
%
%, $i.e.$, $p_{i^\star}(\bm{t}')=1$ and $q_{i^\star}(\bm{t}') =0$ for all $\bm t' \in \cup_{l=1}^{k-1}\mathcal{S}_l$. 
%
By the definition of~$\mathcal S_k$, there are exactly $k-1$ agents~$i \neq i^\star$ with types~$t_i \geq {t}_{i^\star}$. 
%This feasibility condition together with our induction hypothesis imply that any feasible mechanism in \eqref{eq:MDPk(t)} should also satisfy $p_{i^\star}(\bm{t}')=1$ and $q_{i^\star}(\bm{t}') =0$ for all $\bm t' \in \mathcal{T}_I^n$ and for all $n \leq k-1$. This requirement restricts the allocation probabilities of the solutions of \eqref{eq:MDPk(t)} as follows. There exist exactly $k-1$ agents~$i \neq i^\star$ with types~$t_i \geq {t}_{i^\star}$. 
For any such agent~$i$, scenario $\bm t$ is $i$-unilaterally reachable from $(\underline t_i, \bm t_{-i})$. Note that $t_{i^\star} = \overline{t}_{i^\star} > \underline{t}_{i^\star} \geq \underline{t}_{i}$ for all $i \in \mathcal I$, where the equality follows from the definition of~$\mathcal T_I$, and the second inequality follows from the definition of~$i^\star$. This implies that  $(\underline t_i, \bm t_{-i}) \in \mathcal{S}_{k-1}$. Therefore, we know from the induction hypothesis that $p_i(\underline t_i, \bm t_{-i})=p^\star_i(\underline t_i, \bm t_{-i})=0$. % as $p_{i^\star}(\underline t_i, \bm t_{-i})=1$. 
By Lemma~\ref{lem:deviationForFavored}(ii), we then have $p_i(\bm{t}) = q_i(\bm{t})$ for all~$i \neq i^\star$ with~$t_i \geq {t}_{i^\star}$. Thus, the objective function value of $(\bm p, \bm q)$ in~\ref{eq:MDPk(t)} is bounded above by that of $(\bm p^\star, \bm q^\star)$ because
\begin{equation*}
\begin{aligned}
    \sum_{i \in \mathcal{I}} (p_i(\bm{t}) t_i - q_i(\bm{t}) c_i) \leq \sum_{\substack{i \in \mathcal I \setminus \{i^\star\} : \\ t_i \geq {t}_{i^\star}}} p_i(\bm{t}) (t_i- c_i) + \sum_{\substack{i \in \mathcal I : \\ t_i < {t}_{i^\star}}} p_i(\bm{t}) t_i + p_{i^\star}(\bm{t})t_{i^\star} \leq t_{i^\star} = \sum_{i \in \mathcal{I}} (p_i^\star(\bm{t}) t_i - q_i^\star(\bm{t}) c_i),
\end{aligned}
\end{equation*}
where the first inequality holds because $c_i >0$ for all $i \in \mathcal{I}$ and because $p_i(\bm{t}) = q_i(\bm{t})$ for all $i \neq i^\star$ with $t_i \geq {t}_{i^\star}$. The second inequality holds because $\bm t \in \mathcal T_I$, which implies that~$t_i-c_i\leq t_{i^\star}$ for all~$i \in \mathcal{I}$. Similarly, the equality holds because $\bm t \in \mathcal T_I$, in which case the payoff generated by $(\bm p^\star, \bm q^\star)$ amounts to~$t_{i^\star}$. Thus, $(\bm{p}^\star,\bm{q}^\star)$ solves~\ref{eq:MDPk(t)}. In addition, as ${t}_{i^\star}= \overline{t}_{i^\star}$ and $t_i-c_i \leq \underline t_{i^\star} < \overline{t}_{i^\star}$ for all $i \neq i^\star$, the two inequalities in the above expression can collapse to equalities only if~$p_i(\bm t)=0$ for every~$i\neq i^\star$, $p_{i^\star}(\bm t)=1$ and~$q_{i^\star}(\bm t)=0$. Put differently, $(\bm{p}, \bm{q})$ can only be optimal in~\ref{eq:MDPk(t)} if~$\bm p(\bm{t})=\bm p^\star(\bm t)$ and $\bm q(\bm{t}) =\bm q^\star(\bm t)$. As $\bm t\in\mathcal S_k$ was chosen arbitrarily, this observation completes the induction step. 
%In summary, the allocation probabilities of any optimal mechanism to \eqref{eq:MDPk(t)} must be equal to that of $(\bm{p}^\star,\bm{q}^\star)$.
}

\textbf{Step 2 ($\mathcal{T}_{II}$).} {  Define $\mathcal S_{I+1} = \mathcal{T}_{II}$, and set $k=I+1$. Next, fix an arbitrary scenario~$\bm{t} \in \mathcal S_k$ and mechanism~$(\bm{p}, \bm{q})$ feasible in~\ref{eq:MDPk(t)}. The constraints of~\ref{eq:MDPk(t)} ensure that~$(\bm{p}, \bm{q})$ solves~${\text{SP}}$$_{l}(\bm t')$ for every~$\bm t' \in \mathcal S_{l}$ and~$l<k$. From Step~1, we thus know that $p_{i^\star}(\bm{t}')=p^\star_{i^\star}(\bm{t}')=1$ and $q_{i^\star}(\bm{t}')=q^\star_{i^\star}(\bm{t}') =0$ for all $\bm t' \in \cup_{\ell=1}^{k-1} \mathcal S_\ell =  \mathcal T_I$. Clearly, $\bm t$ is $i^\star$-unilaterally reachable from $(\overline t_{i^\star}, \bm t_{-i^\star})$. As $(\overline t_{i^\star}, \bm t_{-i^\star}) \in \mathcal{T}_I$ and $p_{i^\star}(\overline t_{i^\star}, \bm t_{-i^\star}) -q_{i^\star}(\overline t_{i^\star}, \bm t_{-i^\star})  =1$, Lemma~\ref{lem:deviationForFavored}(i) implies that $p_{i^\star}(\bm{t}) = 1$. Thus, we have $t_{i^\star} - q_{i^\star}(\bm{t})c_{i^\star} \leq t_{i^\star}$, that is, the objective function value of~$(\bm p, \bm q)$ in~\ref{eq:MDPk(t)} is bounded above by that of~$(\bm p^\star,\bm q^\star)$, which implies that $(\bm{p}^\star,\bm{q}^\star)$ solves~\ref{eq:MDPk(t)}. Also, as $c_{i^\star} >0$, the mechanism~$(\bm{p}, \bm{q})$ can attain the optimal value~$t_{i^\star}$ of~\ref{eq:MDPk(t)} only if $p_{i^\star}(\bm{t})=1$ and $q_{i^\star}(\bm{t}) =0$, that is, only if~$\bm p(\bm{t})=\bm p^\star(\bm t)$ and~$\bm q(\bm{t}) =\bm q^\star(\bm t)$.
} 

\textbf{Step 3 ($\mathcal{T}_{III}$).}
{  We partition $\mathcal{T}_{III}$ into $I$ subsets of the form
\begin{equation*}
    \mathcal{S}_{k} = \{\bm t \in \mathcal{T}_{III} \,:\, |\{i \in \mathcal{I} \,:\, t_i > \underline t_{i^\star}\}| = k-I-1\},
\end{equation*}
% \begin{equation*}
%     \mathcal{S}_{I+1+n} = \{\bm t \in \mathcal{T}_{III} \,:\, |\{i \in \mathcal{I} \,:\, t_i > \underline t_{i^\star}\}| = n\},
% \end{equation*}
for $k=I+2, \dots, 2I+1$. We will use induction on~$k$ to show that, for any~$\bm t\in\mathcal S_k$, $(\bm{p}^\star,\bm{q}^\star)$ solves~\ref{eq:MDPk(t)} with optimal value $\max_{i' \in \mathcal{I}} t_{i'}-c_{i'}$ and that any mechanism~$(\bm{p}, \bm{q})$ that solves~\ref{eq:MDPk(t)} must satisfy
\begin{equation}\label{eq:allocation-to-max-net-payoff} 
    \sum_{i \in {\arg \max}_{i' \in \mathcal{I}} t_{i'}-c_{i'}} p_i(\bm t) = 1 \quad \text{and} \quad p_i(\bm{t})=q_i(\bm t) \quad \forall i \in {\arg \max}_{i' \in \mathcal{I}} t_{i'}-c_{i'}.
\end{equation}

As for the base step corresponding to~$k=I+2$, fix any~$\bm{t} \in \mathcal{S}_{I+2}$, and consider a mechanism~$(\bm p, \bm q)$ feasible in~\ref{eq:MDPk(t)}. The constraints of~\ref{eq:MDPk(t)} ensure that~$(\bm p, \bm q)$ solves~${\text{SP}}$$_{l}(\bm t')$ for every~$\bm t' \in \mathcal{S}_l$ and $l< I+2$. From Steps~1 and~2 we thus know that $p_{i^\star}(\bm{t}')=p^\star_{i^\star}(\bm{t}')=1$ and $q_{i^\star}(\bm{t}') =q^\star_{i^\star}(\bm{t}') =0$ for all $\bm t' \in \cup_{l=1}^{I+1}\mathcal{S}_l = \mathcal{T}_I \cup \mathcal{T}_{II}$. 
%Akin to Step 1, these requirements restrict the allocation probabilities of $(\widehat{\text{MDP}}$$_{\mathcal{I}+2,t})$ as follows. 
As $\bm{t} \in \mathcal{S}_{k}$ and~$k-I-1=1$, there exists exactly one agent $i^\circ \neq i^\star$ with~$t_{i^\circ} > \underline t_{i^\star}$. By the definition of~$\mathcal T_{III}$, agent~$i^\circ$ is the only agent whose adjusted type satisfies $t_{i^\circ} - c_{i^\circ}>\underline t_{i^\star}$. Note that~$\bm t$ is ${i^\circ}$-unilaterally reachable from~$(\underline t_{i^\circ}, \bm t_{-{i^\circ}})$. As~$\underline t_{i^\circ} -c_{i^\circ} \leq \underline t_{i^\star}$ by the definition of the favored agent~$i^\star$, we further have $(\underline t_{i^\circ}, \bm t_{-{i^\circ}}) \in \mathcal{T}_I \cup \mathcal{T}_{II}$. The reasoning in Steps~1 and~2 thus implies that~$p_{i^\circ}(\underline t_{i}, \bm t_{-{i}})=0$, which in turn implies via Lemma~\ref{lem:deviationForFavored}(ii) that $p_{i^\circ}(\bm t) =q_{i^\circ}(\bm t)$. Thus, the objective function value of $(\bm p, \bm q)$ in~\ref{eq:MDPk(t)} is bounded above by that of $(\bm p^\star, \bm q^\star)$. Indeed, we have
\begin{equation*}
\begin{aligned}
\sum_{i \in \mathcal{I}} (p_i(\bm{t}) t_i - q_i(\bm{t}) c_i)  \leq p_{i^\circ}(\bm t)(t_{i^\circ}-c_{i^\circ})+ \sum_{i \in \mathcal I \setminus \{i^\circ\}}  p_i(\bm{t}) t_i \leq \max_{i \in \mathcal{I}}t_i-c_i =\sum_{i \in \mathcal{I}} (p_i^\star(\bm{t}) t_i - q_i^\star(\bm{t}) c_i),
\end{aligned}
\end{equation*}
where the first inequality holds because $p_{i^\circ}(\bm t) =q_{i^\circ}(\bm t)$ and because $c_i >0$ for all $i \in \mathcal I$, whereas the second inequality holds because $t_{i^\circ}-c_{i^\circ}=\max_{i \neq i^\star}t_i-c_i > \underline t_{i^\star}$ and because $t_{i} \leq \underline t_{i^\star}$ for all $i \neq {i^\circ}$. Finally, the equality holds because $\bm t \in \mathcal T_{III}$, in which case $(\bm p^\star, \bm q^\star)$ generates a payoff of~$\max_{i\in\mathcal I} t_i - c_i$. Thus, $(\bm{p}^\star,\bm{q}^\star)$ solves~\ref{eq:MDPk(t)}. In addition, the objective function value of~$(\bm{p}, \bm{q})$ can equal $\max_{i \neq i^\star}t_i-c_i = t_{i^\circ}-c_{i^\circ}$ only if $p_{i^\circ}(\bm t) =q_{i^\circ}(\bm t)=1$. As $\bm t \in \mathcal S_k$ was chosen freely, this completes the base step.

As for the induction step, fix any $k\in\{I+3,\ldots,2I+1\}$. Assume that for all $l\in \{I+2,\ldots,k-1\}$ we know that $(\bm{p}^\star,\bm{q}^\star)$ solves ${\text{SP}}$$_l(\bm t)$ and that any mechanism $(\bm p,\bm q)$ that solves~${\text{SP}}$$_l(\bm t)$ satisfies~\eqref{eq:allocation-to-max-net-payoff} for all $\bm t \in \mathcal S_l$. Fix now any $\bm t \in \mathcal S_k$ and mechanism $(\bm p, \bm q)$ feasible in~\ref{eq:MDPk(t)}. The constraints in~\ref{eq:MDPk(t)} ensure that 
$(\bm p, \bm q)$ solves~${\text{SP}}$$_{l}(\bm t')$ for every $\bm t' \in \mathcal S_{l}$ and $l <k$. Thus, $(\bm p, \bm q)$ satisfies $p_{i^\star}(\bm{t}')=p^\star_{i^\star}(\bm{t}')=1$ and $q_{i^\star}(\bm{t}') =q^\star_{i^\star}(\bm{t}') =0$ for all $\bm t' \in \mathcal{T}_I \cup \mathcal{T}_{II}$, which follows from Steps~1 and~2, and it satisfies~\eqref{eq:allocation-to-max-net-payoff} for all $\bm t' \in \cup_{l=I+2}^{k-1} \mathcal S_l$, which follows from the induction hypothesis. As $\bm t\in\mathcal S_k$ and $k\in\{I+3,\ldots,2I+1\}$, there are exactly $k-I-1$ agents~$i \in \mathcal{I}$ with types~$t_i > \underline{t}_{i^\star}$. For any such agent $i$, scenario~$\bm t$ is $i$-unilaterally reachable from $(\underline t_i, \bm t_{-i})$. Note that either $(\underline t_i, \bm t_{-i}) \in \mathcal{T}_I \cup \mathcal{T}_{II}$ or $(\underline t_i, \bm t_{-i}) \in \cup_{l=I+2}^{k-1} \mathcal S_{l}$ because $\underline t_{i^\star} \geq \underline t_i$ for all $i \in \mathcal I$. We now show that $p_i(\underline t_i, \bm t_{-i})$ must vanish in both cases.
%Fix some $k \in \{\mathcal{I}+3,\ldots,2\mathcal{I}+1 \}$, and assume that any optimal solution of $(\widehat{\text{MDP}}$$_{n,t'})$ satisfies \eqref{eq:allocation-to-max-net-payoff} for all $\bm t' \in \mathcal{T}_{III}^n$ and for all $n = \mathcal{I}+1,\ldots, k-1$, and consider an arbitrary $\bm t \in \mathcal{T}_{III}^{k-(\mathcal{I}+1)}$. Similar to Step 1, the constraints of \eqref{eq:MDPk(t)} enforce that any feasible solution should optimally solve $(\widehat{\text{MDP}}$$_{n,t'})$ for all $\bm t'$ in the respective scenario sets and the respective $n \leq k-1$ that we considered prior to this step. 
%These requirements restrict the allocation probabilities of the solutions of \eqref{eq:MDPk(t)} as follows. There exists exactly $k-(\mathcal{I}+1)$ agents~$i \in \mathcal{I}$ with types~$t_i > {t}_{i^\star}$. For any such agent $i \in \mathcal{I}$, scenario $\bm t$ is $i-$unilaterally reachable from $(\underline t_i, \bm t_{-i}) \in \mathcal{T}_I \cup \mathcal{T}_{II} \cup \mathcal{T}_{III}^{k-(\mathcal{I}+1)-1}$. 
If $(\underline t_i, \bm t_{-i}) \in \mathcal{T}_I \cup \mathcal{T}_{II}$, then we have $i \neq i^\star$, in which case $p_i(\underline t_i, \bm t_{-i})=p^\star_i(\underline t_i, \bm t_{-i})=0$. If $(\underline t_i, \bm t_{-i}) \in \cup_{l=I+2}^{k-1} \mathcal S_{l}$, on the other hand, then the definition of~$i^\star$ implies that $\underline t_i-c_i \leq \underline t_{i^\star}$, and the definition of~$\mathcal T_{III}$ implies that $\max_{i' \in \mathcal I \setminus \{i^\star\}} t_{i'}-c_{i'}>\underline t_{i^\star}$. Hence, $i$ is no element of $\arg\max_{i' \in \mathcal I \setminus \{i^\star\}} t_{i'}-c_{i'}$, implying that $p_i(\underline t_i, \bm t_{-i})=0$ thanks to~\eqref{eq:allocation-to-max-net-payoff}. Lemma~\ref{lem:deviationForFavored}(ii) now ensures that $p_i(\bm{t}) = q_i(\bm{t})$ for all $i \in \mathcal{I}$ with $t_i > \underline{t}_{i^\star}$. Thus, the objective function value of $(\bm p, \bm q)$ in~\ref{eq:MDPk(t)} is bounded above by that of $(\bm p^\star, \bm q^\star)$ because
\begin{equation*}
    \sum_{i \in \mathcal{I}} (p_i(\bm{t}) t_i - q_i(\bm{t}) c_i) \leq \sum_{i \in \mathcal I : \, t_i > \underline{t}_{i^\star}} p_i(\bm{t}) (t_i- c_i) + \sum_{i \in \mathcal I : \, t_i \leq \underline{t}_{i^\star}} p_i(\bm{t}) t_i \leq \max_{i \in \mathcal{I}}t_i-c_i =\sum_{i \in \mathcal{I}} (p_i^\star(\bm{t}) t_i - q_i^\star(\bm{t}) c_i),
\end{equation*}
where the second inequality holds because $\max_{i \neq i^\star}t_i-c_i > \underline t_{i^\star}$. Thus, $(\bm{p}^\star,\bm{q}^\star)$ solves~\ref{eq:MDPk(t)}, and $(\bm p,\bm q)$ can solve~\ref{eq:MDPk(t)} only if it obeys~\eqref{eq:allocation-to-max-net-payoff}. This observation completes the induction step.} \hfill \Halmos
\endproof

\vspace{2mm}
\proof{Proof of Proposition \ref{prop:optimal-z-markov}.}
Relaxing the incentive compatibility constraints and the first inequality in \eqref{eq:FC} yields
\begin{equation*}
\begin{aligned}
z^\star &\leq& &\sup_{\bm{p}, \bm{q}} &&\inf_{\mathbb{P} \in \mathcal{P}} \; \mathbb{E}_{\mathbb{P}}\left[\sum_{i \in \mathcal{I}} (p_i(\tilde{\bm{t}}) \tilde{t}_i - q_i(\tilde{\bm{t}}) c_i) \right]\\
&&&\;\text{s.t.} &&p_i : \mathcal{T} \rightarrow [0,1] \;\text{and}\; q_i : \mathcal{T} \rightarrow [0,1] \;\;\forall i \in \mathcal{I}\\
&&&&&\sum_{i \in \mathcal{I}} p_i(\bm{t}) \leq 1 \;\;\forall \bm{t} \in \mathcal{T}\\
&=& &\sup_{\bm{p}} &&\inf_{\mathbb{P} \in \mathcal{P}} \; \mathbb{E}_{\mathbb{P}}\left[\sum_{i \in \mathcal{I}} p_i(\tilde{\bm{t}}) \tilde{t}_i \right]\\
&&&\;\text{s.t.} &&p_i : \mathcal{T} \rightarrow [0,1] \;\;\forall i \in \mathcal{I}, \;\;\sum_{i \in \mathcal{I}} p_i(\bm{t}) \leq 1 \;\;\forall \bm{t} \in \mathcal{T},
%&\leq& &&&\inf_{\mathbb{P} \in \mathcal{P}} \; \mathbb{E}_{\mathbb{P}}\left[\max_{i \in \mathcal{I}} \tilde{t}_i \right]
\end{aligned}
\end{equation*}
where the equality holds because it is optimal to set~$q_i(\bm{t})=0$ for all $i \in \mathcal{I}$ and $\bm{t} \in \mathcal{T}$ in the relaxed problem.
As $p_i \geq 0$ and $\sum_{i \in \mathcal{I}} p_i(\bm{t}) \leq 1$ for all $\bm{t} \in \mathcal{T}$, we further have
\begin{equation*}
\begin{aligned}
\sum_{i \in \mathcal{I}} p_i({\bm{t}}) {t}_i \leq \max_{i \in \mathcal{I}} {t}_i \;\;\forall \bm{t} \in \mathcal{T},
%z^\star &\leq& \inf_{\mathbb{P} \in \mathcal{P}} \; \mathbb{E}_{\mathbb{P}}\left[\max_{i \in \mathcal{I}} \tilde{t}_i \right].
\end{aligned}
\end{equation*}
which imply that $z^\star$ is bounded above by $\inf_{\mathbb{P} \in \mathcal{P}} \; \mathbb{E}_{\mathbb{P}}\left[\max_{i \in \mathcal{I}} \tilde{t}_i \right]$. Now, select an arbitrary $i^\star \in \arg \max_{i \in \mathcal{I}} \underline{\mu}_i$ and denote by $\delta_{\underline{\bm{\mu}}}$ the Dirac point mass at $\underline{\bm{\mu}}$. We have
\begin{equation*}\label{eq: kuhnBound}
\begin{aligned}
\mathbb{E}_{\delta_{\underline{\bm{\mu}}}}\left[\max_{i \in \mathcal{I}} \tilde{t}_i \right] \geq \inf_{\mathbb{P} \in \mathcal{P}} \; \mathbb{E}_{\mathbb{P}}\left[\max_{i \in \mathcal{I}} \tilde{t}_i \right] \geq \inf_{\mathbb{P} \in \mathcal{P}} \; \mathbb{E}_{\mathbb{P}}\left[\tilde{t}_{i^\star} \right] = \max_{i \in \mathcal{I}} \underline{\mu}_i,
\end{aligned}
\end{equation*}
where the first inequality holds because $\delta_{\underline{\bm{\mu}}} \in \mathcal{P}$, the second inequality holds because $\max_{i \in \mathcal{I}} {t}_i \geq {t}_{i^\star}$ for any $\bm{t} \in \mathcal{T}$, and the equality follows from the selection of $i^\star$ and the definition of the Markov ambiguity set~$\mathcal{P}$. As $\delta_{\underline{\bm{\mu}}}$ is the Dirac point mass at $\underline{\bm{\mu}}$, we also have $\mathbb{E}_{\delta_{\underline{\bm{\mu}}}}\left[\max_{i \in \mathcal{I}} \tilde{t}_i \right] = \max_{i \in \mathcal{I}} \underline{\mu}_i$ that implies $\inf_{\mathbb{P} \in \mathcal{P}} \; \mathbb{E}_{\mathbb{P}}\left[\max_{i \in \mathcal{I}} \tilde{t}_i \right] = \max_{i \in \mathcal{I}} \underline{\mu}_i$. Therefore, the optimal value $z^\star$ is bounded above by $\max_{i \in \mathcal{I}} \underline{\mu}_i$. However, this bound is attained by a mechanism that allocates the good to an agent $i^\star \in \arg \max_{i \in \mathcal{I}} \underline{\mu}_i$ irrespective of~$\bm{t}\in\mathcal{T}$ and never inspects anyone's type. Since this mechanism is feasible, the claim follows. \hfill \Halmos
\endproof

\vspace{2mm}
\proof{Proof of Theorem \ref{thr:opt-fav-markov}.}
Select an arbitrary favored-agent mechanism with $i^\star \in {\arg \max}_{i \in \mathcal{I}} \underline{\mu}_i$ and $\nu^\star \geq \overline{t}_{i^\star}$. Recall first that this mechanism is feasible in \eqref{eq:MDP}. Next, we will show that this mechanism attains a worst-case payoff that is at least as large as~$\max_{i \in \mathcal{I}} \underline{\mu}_i$, which implies via Proposition~\ref{prop:optimal-z-markov} that this mechanism is optimal in~\eqref{eq:MDP}. To this end, fix an arbitrary type profile $\bm t \in \mathcal{T}$.
If~$\max_{i\in\mathcal{I}} t_i-c_i< \nu^\star$, then condition~(i) in Definition~\ref{def:fav-agent-mechanism} implies that the principal's payoff amounts to $t_{i^\star}$. If~$\max_{i\in\mathcal{I}}t_i-c_i>\nu^\star$, then condition~(ii) in Definition~\ref{def:fav-agent-mechanism} implies that the principal's payoff amounts to~$\max_{i \in \mathcal{I}} t_i-c_i > \nu^\star \geq t_{i^\star}$, where the second inequality follows from the selection of~$\nu^\star$. If $\max_{i \neq i^\star} t_i-c_i = \nu^\star$, then the allocation functions are defined either as in condition~(i) or as in condition~(ii) of Definition~\ref{def:fav-agent-mechanism}. Thus, the principal's payoff amounts either to~$t_{i^\star}$ or to~$\max_{i \in \mathcal{I}} t_i-c_i \geq \nu^\star \geq t_{i^\star}$, respectively. In summary, we have shown that the principal's payoff is bigger than or equal to $t_{i^\star}$ in all three cases. As the type profile $\bm t$ was chosen arbitrarily, this implies that the principal's expected payoff under any distribution $\mathbb{P} \in \mathcal{P}$ is bounded below by $\mathbb{E}_{\mathbb{P}}\left[\tilde{t}_{i^\star} \right]$. By the definition of the Markov ambiguity set $\mathcal{P}$, the expectation $\mathbb{E}_{\mathbb{P}}\left[\tilde{t}_{i^\star} \right]$ cannot be lower than $z^\star = \max_{i \in \mathcal{I}} \underline{\mu}_i$ for any $\mathbb P \in \mathcal P$. Therefore, the principal's worst-case expected payoff under the favored-agent mechanism is bounded below by $z^\star$. The favored-agent mechanism at hand is therefore optimal in~\eqref{eq:RMDP} by virtue of Proposition~\ref{prop:optimal-z-markov}. \hfill \Halmos
\endproof

\vspace{2mm}
\proof{Proof of Lemma \ref{lem:findP}.}
For any $\bm{t} \in \mathcal{T}$, we will show that there exists a scenario $\hat{\bm{t}} \in \mathcal{T}$ that satisfies $\max_{i \neq i^\star} \hat{t}_i < {  \underline{\mu}_{i^\star}}$
%\hat{t}_{i^\star}$
and $\alpha \bm{t} + (1- \alpha)\hat{\bm{t}} = \underline{\bm{\mu}}$ for some $\alpha \in (0,1]$. This implies that the discrete distribution $\mathbb{P} = \alpha \delta_{\bm{t}} + (1 - \alpha) \delta_{\hat{\bm{t}}}$ belongs to the Markov ambiguity set $\mathcal{P}$ and moreover satisfies the properties (i)--(iii). 

To this end, consider any $\bm{t} \in \mathcal{T}$. If $\bm{t} = \underline{\bm{\mu}}$, set $\hat{\bm{t}} = \bm{t} = \underline{\bm{\mu}}$. As $\arg \max_{i \in \mathcal{I}} \underline{\mu}_i = \{i^\star\}$ is a singleton, scenario $\hat{\bm{t}}$ satisfies $\max_{i \neq i^\star} \hat{t}_i< {  \underline{\mu}_{i^\star}}$.
% \hat{t}_{i^\star}$. 
Moreover, note that $\alpha \bm{t} + (1- \alpha)\hat{\bm{t}} = \underline{\bm{\mu}}$ for any $\alpha \in (0,1]$. Similarly, for any $\alpha \in (0,1]$,  $\mathbb{P} = \alpha \delta_{\bm{t}} + (1 - \alpha) \delta_{\hat{\bm{t}}}= \delta_{\underline{\bm{\mu}}}$ is the Dirac point mass at $\underline{\bm{\mu}}$ and trivially satisfies the desired properties (i)--(iii).

If $\bm{t} \neq \underline{\bm{\mu}}$, define function $\hat{\bm{t}}(\alpha)$ through
$$
\hat{\bm{t}}(\alpha)=\frac{1}{1- \alpha}(\underline{\bm{\mu}}- \bm{t}) + \bm{t}.
$$
Note that, for any $\alpha \in [0,1)$, $\hat{\bm{t}}(\alpha)$ satisfies
\begin{equation*}
    \begin{aligned}
\alpha \bm{t} + (1- \alpha)\hat{\bm{t}}(\alpha) &= \alpha \bm{t} + (1- \alpha)\left(\frac{1}{1- \alpha}(\underline{\bm{\mu}}- \bm{t}) + \bm{t}\right)
%&= \alpha \bm{t} + (1- \alpha)\frac{\underline{\bm{\mu}}- \alpha\bm{t}}{1- \alpha} =\alpha \bm{t} + \underline{\bm{\mu}}- \alpha\bm{t}
= \underline{\bm{\mu}}.
    \end{aligned}
\end{equation*}
Thus, for any $\alpha \in [0,1)$, $\hat{\bm{t}} = \hat{\bm{t}}(\alpha)$ satisfies $\alpha \bm{t} + (1- \alpha)\hat{\bm{t}} = \underline{\bm{\mu}}$. We will next show that there exists an $\alpha \in (0,1)$ for which $\hat{\bm{t}} = \hat{\bm{t}}(\alpha)$ also satisfies $\max_{i \neq i^\star} \hat{t}_i < {  \underline{\mu}_{i^\star}}$.
%\hat{t}_{i^\star}$. 
To this end, first note that $\hat{\bm{t}}(\alpha)$ is a continuous function of $\alpha \in [0,1)$ and $\hat{\bm{t}}(0) = \underline{\bm{\mu}}$. Thus, for any $\varepsilon>0$, there exists $\alpha \in (0,1)$ such that $\hat{\bm{t}}(\alpha) \in \prod_{i \in \mathcal{I}} [\underline{\mu}_i -\varepsilon,\underline{\mu}_i +\varepsilon]$. We next show that any $\varepsilon>0$ that belongs to the set 
\begin{equation*}
    \begin{aligned}
    L = (0, \min_{i \in \mathcal{I}} \underline{\mu}_{i}-\underline{t}_{i}) \cap (0, \min_{i \in \mathcal{I}} \overline{t}_i-\underline{\mu}_i) \cap \big(0, (\underline{\mu}_{i^\star}-\max_{i \neq i^\star} \underline{\mu}_i)/2\big)
    \end{aligned}
\end{equation*}
ensures that $\prod_{i \in \mathcal{I}} [\underline{\mu}_i -\varepsilon,\underline{\mu}_i +\varepsilon] \subseteq \{\bm{t} \in  \mathcal{T} \,:\, \max_{i \neq i^\star} t_i < {  \underline{\mu}_{i^\star}} \}$.
%t_{i^\star}\}$. 
Note that set $L$ is non-empty because $\underline{t}_i < \underline{\mu}_i < \overline{\mu}_i < \overline{t}_i$ for all $i \in \mathcal{I}$ and $\arg \max_{i \in \mathcal{I}} \underline{\mu}_i = \{i^\star\}$ is a singleton. Consider any $\varepsilon \in L$.
As $\varepsilon<\min_{i \in \mathcal{I}} \underline{\mu}_{i}-\underline{t}_{i}$, any $\bm{t} \in \prod_{i \in \mathcal{I}} [\underline{\mu}_i -\varepsilon,\underline{\mu}_i +\varepsilon]$ satisfies 
$$
t_i \geq \underline{\mu}_i -\varepsilon > \underline{\mu}_i -(\min_{j \in \mathcal{I}} \underline{\mu}_{j}-\underline{t}_{j}) \geq \underline{\mu}_i-(\underline{\mu}_{i}-\underline{t}_{i}) = \underline{t}_{i} \;\;\;\forall i \in \mathcal{I}.
$$
Similarly, as $\varepsilon<\min_{i \in \mathcal{I}} \overline{t}_i-\underline{\mu}_i$, any $\bm{t} \in \prod_{i \in \mathcal{I}} [\underline{\mu}_i -\varepsilon,\underline{\mu}_i +\varepsilon]$ satisfies
$$
t_i \leq \underline{\mu}_i +\varepsilon < \underline{\mu}_i + (\min_{j \in \mathcal{I}} \overline{t}_j-\underline{\mu}_j) \leq \underline{\mu}_i +\overline{t}_i-\underline{\mu}_i =\overline{t}_i \;\;\;\forall i \in \mathcal{I}.
$$
Therefore, we have shown that $\prod_{i \in \mathcal{I}} [\underline{\mu}_i -\varepsilon,\underline{\mu}_i +\varepsilon] \subseteq \mathcal{T}$. Finally, any $\bm{t} \in \prod_{i \in \mathcal{I}} [\underline{\mu}_i -\varepsilon,\underline{\mu}_i +\varepsilon]$ satisfies
\begin{equation*}
    \begin{aligned}
%t_{i^\star}
{  \underline{\mu}_{i^\star}}&\geq \underline{\mu}_{i^\star}-\varepsilon>\underline{\mu}_{i^\star}-(\underline{\mu}_{i^\star}-\max_{j \neq i^\star} \underline{\mu}_j)/2= (\underline{\mu}_{i^\star}+\max_{j \neq i^\star} \underline{\mu}_j)/2\\
&=  \max_{j \neq i^\star} \underline{\mu}_j +(\underline{\mu}_{i^\star}-\max_{j \neq i^\star} \underline{\mu}_j)/2 > \max_{j \neq i^\star} \underline{\mu}_j+\varepsilon \geq \underline{\mu}_i+\varepsilon \geq t_i \quad \forall i \neq i^\star,
    \end{aligned}
\end{equation*}
where the second and third inequalities follow from $\varepsilon<(\underline{\mu}_{i^\star}-\max_{i \neq i^\star} \underline{\mu}_i)/2$. Thus, we have shown that $\prod_{i \in \mathcal{I}} [\underline{\mu}_i -\varepsilon,\underline{\mu}_i +\varepsilon] \subseteq \{\bm{t} \in  \mathcal{T} \,:\, \max_{i \neq i^\star} t_i < {  \underline{\mu}_{i^\star}} \}$
%t_{i^\star}\}$ 
for any $\varepsilon \in L$. As for any $\varepsilon \in L$ there exists $\alpha \in (0,1)$ such that $\hat{\bm{t}}(\alpha) \in \prod_{i \in \mathcal{I}} [\underline{\mu}_i -\varepsilon,\underline{\mu}_i +\varepsilon]$, the claim follows. \hfill \Halmos
\endproof

\vspace{2mm}
\proof{Proof of Lemma~\ref{lem:PRO-revenue equivalence}.}
{  Throughout the proof, we denote by~$(\bm{p}^\star,\bm{q}^\star)$ the favored-agent mechanism of type~(ii) with favored agent~${i^\star}$, where ${\arg \max}_{i \in \mathcal{I}} \underline{\mu}_i = \{i^\star\}$, and with threshold value~$\nu^\star= \overline{t}_{i^\star}$. We will use the following partition of the type space. 
\begin{equation}\label{eq: partition-proof} 
\begin{aligned}
\mathcal{T}_I &= \{\bm{t} \in \mathcal{T} \,:\, \max_{i \neq i^\star} t_i -c_i < \overline t_{i^\star} \; \text{and} \; \max_{i \neq i^\star} t_i < \underline \mu_{i^\star}  \; \text{and} \; t_{i^\star} = \underline \mu_{i^\star} \}\\
\mathcal{T}_{II} &=  \{\bm{t} \in \mathcal{T} \,:\, \max_{i \neq i^\star} t_i -c_i < \overline t_{i^\star} \; \text{and} \; \max_{i \neq i^\star} t_i < \underline \mu_{i^\star}  \; \text{and} \; t_{i^\star} \neq \underline \mu_{i^\star} \}\\
\mathcal{T}_{III} &= \{\bm{t} \in \mathcal{T} \,:\, \max_{i \neq i^\star} t_i -c_i < \overline t_{i^\star} \; \text{and} \; \max_{i \neq i^\star} t_i \geq \underline \mu_{i^\star} \; \text{and} \; t_{i^\star} = \overline t_{i^\star} \}\\
\mathcal{T}_{IV} &= \{\bm{t} \in \mathcal{T} \,:\, \max_{i \neq i^\star} t_i -c_i < \overline t_{i^\star} \; \text{and} \; \max_{i \neq i^\star} t_i \geq \underline \mu_{i^\star}  \; \text{and} \; t_{i^\star} < \overline t_{i^\star} \}\\
\mathcal{T}_V &= \{\bm{t} \in \mathcal{T} \,:\, \max_{i \neq i^\star} t_i-c_i \geq \overline{t}_{i^\star} \}
\end{aligned}
\end{equation}
Note that some of the conditions in the above definitions are redundant and are only introduced for ease of readability. 
The sets $\mathcal{T}_I$ and $\mathcal{T}_{II}$ are nonempty and contain at least $\underline{\bm \mu}$ and $(\overline{t}_{i^\star}, \underline{\bm \mu}_{-i^\star})$, respectively, because $\underline \mu_i < \underline \mu_{i^\star} < \overline{t}_{i^\star}$ for all $i \neq i^\star$. However, the sets $\mathcal{T}_{III}, \mathcal{T}_{IV}$ and $\mathcal{T}_V$ can be empty if $\underline \mu_{i^\star}$, $\overline{t}_{i^\star}$ or $c_i$, $i \neq i^\star$, are sufficiently large.

From Theorem~\ref{thr:opt-fav-markov} we already know that $(\bm{p}^\star,\bm{q}^\star)$ is robustly optimal. To prove Lemma \ref{lem:PRO-revenue equivalence}, we will leverage Corollary~\ref{Corollary of sufficiency proposition} of Proposition~\ref{prop:unifyingPRO}, which provides sufficient conditions for the claim made. In particular, to show that~any mechanism~$(\bm{p},\bm{q}) \in \mathcal X$ that weakly Pareto robustly dominates $(\bm{p}^\star,\bm{q}^\star)$ must generate the same payoff as $(\bm{p}^\star,\bm{q}^\star)$ in every scenario~$\bm t\in\mathcal T$, it suffices to verify the conditions~(i) and~(ii) in Proposition~\ref{prop:unifyingPRO} for some partition of the type space $\mathcal T$. We will show that they holds for a refinement of the partition $\mathcal{T}_{I}-\mathcal{T}_{V}$.
Specifically, we will construct a refined partition $\mathcal S_1,\ldots,\mathcal S_m$ of~$\mathcal T$ with~$m=2I+2$, where $\mathcal S_1$ coincides with~$\mathcal{T}_{I}$, $\mathcal S_{2}$ coincides with~$\mathcal{T}_{II}$, $\mathcal S_3, \dots, \mathcal S_{I+1}$ forms a partition of~$\mathcal{T}_{III}$, $\mathcal S_{I+2}$ coincides with~$\mathcal{T}_{IV}$ and $\mathcal S_{I+3}, \dots, \mathcal S_{2I+2}$ forms a partition of~$\mathcal{T}_{V}$. 

We first exploit Lemma~\ref{lem:findP} to show that condition~(i) holds. The subsequent arguments only depend on the definitions of $\mathcal S_1 = \mathcal{T}_{I}$ and $\mathcal S_2 = \mathcal{T}_{II}$. In contrast, the definitions of $\mathcal S_l$ for $l \in \{3, \dots, m\}$ do not play a role for proving condition~(i). Condition~(i) of Proposition \ref{prop:unifyingPRO} requires that, for any index $k \in \{1, \ldots,m\}$ and for any scenario $\bm t \in \mathcal{S}_k$, there exists $\mathbb{P} \in \mathcal{P} \cap \mathcal{P}_0(\cup_{l=1}^k \mathcal{S}_l)$ with $\mathbb{P}(\tilde{\bm{t}} =\bm{t})>0$.
Given any scenario $\bm t \in \mathcal{T}$, by Lemma~\ref{lem:findP}, there exists a scenario $\hat{\bm{t}} \in \mathcal{T}$ with $\max_{i\neq i^\star} \hat t_i< {  \underline{\mu}_{i^\star}}$
%\hat t_{i^\star}$ 
and a discrete distribution $\mathbb{P} \in \mathcal{P}$ such that
$\mathbb{E}_{\mathbb{P}}[\tilde{t}_i] =\underline{\mu}_i$ for all $i \in \mathcal{I}$, $\mathbb{P}(\tilde{\bm{t}} \in \{\bm{t},\hat{\bm{t}} \})=1$, and
$\mathbb{P}(\tilde{\bm{t}} =\bm{t})>0$.
Consider first any scenario $\bm t \in \mathcal{S}_1$. By the definition of $\mathcal S_1$, we have $t_{i^\star} = \underline \mu_{i^\star}$. Hence, the identity $\mathbb{E}_{\mathbb{P}}[\tilde{t}_{i^\star}] =\underline{\mu}_{i^\star}$ can be satisfied only if $\hat t_{i^\star} = \underline \mu_{i^\star}$. As $\max_{i\neq i^\star} \hat t_i< \underline{\mu}_{i^\star}$, this means that $\hat{\bm t} \in \mathcal S_1$ and $\mathbb P \in\mathcal P\cap \mathcal{P}_0(\mathcal{S}_1)$. Condition~(i) thus holds for $k=1$ and for any $\bm t \in \mathcal S_1$. For any $k \in \{2,\ldots,m\}$ and $\bm t \in \mathcal{S}_k$, condition~(i) can be easily verified thanks to Lemma \ref{lem:findP} and because $\hat{\bm t} \in \mathcal{S}_1 \cup \mathcal{S}_2 =\{\bm t \in \mathcal{T}: \; \max_{i \neq i^\star} t_i < \underline \mu_{i^\star}\}$, which implies that the discrete distribution $\mathbb P$ is supported on $\mathcal{S}_1 \cup \mathcal{S}_2 \cup \mathcal S_k$. 

Next, we prove condition~(ii). To this end, the rest of the proof is divided into five steps focusing on the five subsets $\mathcal{T}_{I}-\mathcal{T}_{V}$ and, in particular, on their respective refined partitions. We will exploit the non-locality of the incentive compatibility constraint~\eqref{eq:IC} via Lemma~\ref{lem:deviationForFavored} to show iteratively that~$(\bm p^\star, \bm q^\star)$ solves the scenario problem~\ref{eq:MDPk(t)} for all $\bm t \in \mathcal S_k$ and~$k=1,\ldots,m$. The subsequent reasoning critically relies on the definition of the mechanism~$(\bm{p}^\star,\bm{q}^\star)$, under which the principal's payoff in scenario~$\bm t$ amounts to~$t_{i^\star}$ when $\max_{i \neq i^\star} t_i - c_i < \overline{t}_{i^\star}$ ($i.e.$, when $\bm t \in \mathcal{T}_{I} \cup \mathcal{T}_{II}\cup \mathcal{T}_{III} \cup \mathcal{T}_{IV}$) and to $\max_{i \in \mathcal{I}} t_i - c_i$ when  $\max_{i \neq i^\star} t_i - c_i \geq \overline{t}_{i^\star}$ ($i.e.$, when $\bm t \in \mathcal{T}_{V}$).

\textbf{Step 1 ($\mathcal{T}_{I}$).} Define $\mathcal{S}_1 = \mathcal{T}_I$, and set $k=1$. Fix an arbitrary scenario $\bm t \in \mathcal{S}_1$ and mechanism $(\bm p, \bm q) \in \mathcal X$. Note that any $(\bm p, \bm q) \in \mathcal X$ is feasible in~${\text{SP}}$$_{1}(\bm t)$. In addition, the objective value of $(\bm p, \bm q)$ is dominated by that of $(\bm p^\star, \bm q^\star)$ because 
\begin{equation*}\label{eq:markov_contradiction_p<1}
\begin{aligned}
\sum_{i \in \mathcal{I}} (p_i(\bm{t}) t_i - q_i(\bm{t}) c_i) \leq \sum_{i \in \mathcal{I}} p_i(\bm{t}) t_i \leq t_{i^\star} = \sum_{i \in \mathcal{I}} (p_i^\star(\bm{t}) t_i - q_i^\star(\bm{t}) c_i),
\end{aligned}
\end{equation*}
where the first inequality holds because $c_i>0$ for all $i \in \mathcal{I}$, and the second inequality follows from the definition of $\mathcal{S}_1$, which implies that $t_i< t_{i^\star}$ for all $i \neq i^{\star}$. Thus, $(\bm p^\star, \bm q^\star)$ solves~${\text{SP}}$$_{1}(\bm t)$. As $t_i < t_{i^\star}$ for all $i \neq i^\star$, the above arguments also reveal that any mechanism $(\bm{p}, \bm{q})$ that solves ${\text{SP}}$$_{1}(\bm t)$ must satisfy $p_{i^\star}(\bm{t})=1$ and $q_{i^\star}(\bm{t})=0$ for any $\bm{t} \in \mathcal{S}_1 = \mathcal{T}_I$.

\textbf{Step 2 ($\mathcal{T}_{II}$).} Define $\mathcal{S}_2 = \mathcal{T}_{II}$. Fix an arbitrary scenario~$\bm t \in \mathcal{S}_2$ and mechanism $(\bm p, \bm q)$ feasible in~${\text{SP}}$$_{2}(\bm t)$. The constraints of~${\text{SP}}$$_{2}(\bm t)$ ensure that~$(\bm{p}, \bm{q})$ solves~${\text{SP}}$$_{1}(\bm t')$ for every~$\bm t' \in \mathcal S_{1} = \mathcal T_{I}$. From Step~1, we know that any such solution satisfies $p_{i^\star}(\bm{t})=1$ and $q_{i^\star}(\bm{t})=0$ for all $\bm t' \in \mathcal T_I$. Clearly,~$\bm t$ is $i^\star$-unilaterally reachable from $(\underline \mu_{i^\star}, \bm t_{-i^\star})$. As $(\underline \mu_{i^\star}, \bm t_{-i^\star}) \in \mathcal{T}_I$ and $p_{i^\star}(\underline \mu_{i^\star}, \bm t_{-i^\star}) -q_{i^\star}(\underline \mu_{i^\star}, \bm t_{-i^\star})  =1$, Lemma~\ref{lem:deviationForFavored}(i) implies that $p_{i^\star}(\bm{t}) = 1$. Thus, we have $t_{i^\star} - q_{i^\star}(\bm{t})c_{i^\star} \leq t_{i^\star}$, that is, the objective function value of~$(\bm p, \bm q)$ in~${\text{SP}}$$_{2}(\bm t)$ is bounded above by that of~$(\bm p^\star,\bm q^\star)$, which implies that $(\bm p^\star, \bm q^\star)$ solves~${\text{SP}}$$_{2}(\bm t)$.
Also, as $c_{i^\star} >0$, the mechanism $(\bm p, \bm q)$ can attain the optimal value~$t_{i^\star}$ only if $p_{i^\star}(\bm{t})=1$ and $q_{i^\star}(\bm{t}) =0$.

\textbf{Step 3 ($\mathcal{T}_{III}$).} We partition $\mathcal{T}_{III}$ into $I-1$ subsets of the form
\begin{equation*}
    \mathcal{S}_{k} = \left\{\bm t \in \mathcal{T}_{III} \,:\, |\{i \in \mathcal{I} \,:\, t_i \geq \underline \mu_{i^\star}\}| = k-1 \right\},
\end{equation*}
for $k=3, \dots, I+1$. Note that $|\{i \in \mathcal{I} : t_i \geq \underline \mu_{i^\star}\}| \geq 2$ for all $\bm{t} \in \mathcal{T}_{III}$ because $t_{i^\star}=\overline{t}_{i^\star}> \underline \mu_{i^\star}$ and $\max_{i \neq i^\star} t_i \geq \underline {\mu}_{i^\star}$ by the definition of $\mathcal{T}_{III}$. We will use induction on~$k$ to show that, for any $\bm t \in \mathcal{S}_{k}$, the mechanism $(\bm p^\star, \bm q^\star)$ solves~\ref{eq:MDPk(t)}, and any mechanism $(\bm p, \bm q)$ that solves~\ref{eq:MDPk(t)} must satisfy~$p_{i^\star}(\bm{t})=1$ and~$q_{i^\star}(\bm{t})=0$.

As for the base step corresponding to $k=3$, fix any~$\bm t \in \mathcal{S}_3$ and consider a mechanism $(\bm p, \bm q)$ feasible in~\ref{eq:MDPk(t)}. The constraints of~\ref{eq:MDPk(t)} ensure that~$(\bm p, \bm q)$ solves~${\text{SP}}$$_{l}(\bm t')$ for every~$\bm t' \in \mathcal{S}_l$ and $l< k$. From Steps~1 and~2 we thus know that $p_{i^\star}(\bm{t}')=1$ and $q_{i^\star}(\bm{t}') =0$ for all $\bm t' \in \cup_{l=1}^{2}\mathcal{S}_l = \mathcal{T}_I \cup \mathcal{T}_{II}$. 
As $\bm{t} \in \mathcal{S}_{k}$ and~$k-1=2$, there exists exactly one agent $i^\circ \neq i^\star$ with~$t_{i^\circ} \geq \underline \mu_{i^\star}$. Note that~$\bm t$ is ${i^\circ}$-unilaterally reachable from~$(\underline t_{i^\circ}, \bm t_{-{i^\circ}})$. As~$\underline t_{i^\circ} < \underline \mu_{i^\circ} < \underline \mu_{i^\star}$ by the definition of the favored agent~$i^\star$, we further have $(\underline t_{i^\circ}, \bm t_{-{i^\circ}}) \in \mathcal{T}_{II}$. The reasoning in Step~2 thus implies that~$p_{i^\circ}(\underline t_{i}, \bm t_{-{i}})=0$, which in turn implies via Lemma~\ref{lem:deviationForFavored}(ii) that $p_{i^\circ}(\bm t) =q_{i^\circ}(\bm t)$. Thus, the objective function value of $(\bm p, \bm q)$ in~\ref{eq:MDPk(t)} is bounded above by that of $(\bm p^\star, \bm q^\star)$. Indeed, we have
\begin{equation*}
\begin{aligned}
\sum_{i \in \mathcal{I}} (p_i(\bm{t}) t_i - q_i(\bm{t}) c_i)  \leq p_{i^\circ}(\bm t)(t_{i^\circ}-c_{i^\circ})+ \sum_{i \in \mathcal I \setminus \{i^\circ\}}  p_i(\bm{t}) t_i \leq t_{i^\star} = \sum_{i \in \mathcal{I}} (p_i^\star(\bm{t}) t_i - q_i^\star(\bm{t}) c_i),
\end{aligned}
\end{equation*}
where the first inequality holds because $p_{i^\circ}(\bm t) =q_{i^\circ}(\bm t)$ and because $c_i >0$ for all $i \in \mathcal I$, whereas the second inequality holds because $t_{i^\circ}-c_{i^\circ} < \overline t_{i^\star} = t_{i^\star}$ and because $t_{i} < \underline \mu_{i^\star}$ for all $i \neq {i^\circ}, i^\star$. Thus,  $(\bm p^\star, \bm q^\star)$ solves~\ref{eq:MDPk(t)}. In addition, the objective function value of~$(\bm{p}, \bm{q})$ can equal $t_{i^\star}$ only if $p_{i^\star}(\bm t)=1$ and $q_{i^\star}(\bm t)=0$. As $\bm t \in \mathcal S_k$ was chosen freely, this completes the base step.

As for the induction step, fix any $k\in\{4,\ldots,I+1\}$. Assume that for all $l\in \{3,\ldots,k-1\}$ we know that $(\bm p^\star, \bm q^\star)$ solves~${\text{SP}}$$_l(\bm t)$ and that any mechanism $(\bm p,\bm q)$ that solves~${\text{SP}}$$_l(\bm t)$ satisfies $p_{i^\star}(\bm t) =q_{i^\star}(\bm t)=1$ for all $\bm t \in \mathcal S_l$. Fix now any $\bm t \in \mathcal S_k$ and mechanism $(\bm p, \bm q)$ feasible in~\ref{eq:MDPk(t)}. The constraints in~\ref{eq:MDPk(t)} ensure that 
$(\bm p, \bm q)$ solves~${\text{SP}}$$_{l}(\bm t')$ for every $\bm t' \in \mathcal S_{l}$ and $l <k$. Thus, $(\bm p, \bm q)$ satisfies $p_{i^\star}(\bm{t}')=1$ and $q_{i^\star}(\bm{t}')=0$ for all $\bm t' \in \mathcal{T}_I \cup \mathcal{T}_{II} \cup (\cup_{l=3}^{k-1} \mathcal S_l)$, which follows from Steps~1 and~2 and from the induction hypothesis. As $\bm t\in\mathcal S_k$ and $k\in\{4,\ldots,I+1\}$, there are exactly $k-1$ agents~$i \in \mathcal{I}$ with types~$t_i \geq \underline{\mu}_{i^\star}$. For any such agent $i \neq i^\star$, scenario~$\bm t$ is $i$-unilaterally reachable from $(\underline t_i, \bm t_{-i})$. As $(\underline t_i, \bm t_{-i}) \in \mathcal S_{k-1}$, we have $p_i(\underline t_i, \bm t_{-i})=0$ thanks to our induction hypothesis. Lemma~\ref{lem:deviationForFavored}(ii) now ensures that $p_i(\bm{t}) = q_i(\bm{t})$ for all $i \in \mathcal{I}$ with $t_i \geq \underline{\mu}_{i^\star}$ and $i \neq i^\star$. Thus, the objective function value of any feasible $(\bm p, \bm q)$ in~\ref{eq:MDPk(t)} is bounded above by that of $(\bm p^\star, \bm q^\star)$, that is,
\begin{equation*}
    \sum_{i \in \mathcal{I}} (p_i(\bm{t}) t_i - q_i(\bm{t}) c_i) \leq \sum_{\substack{i \in \mathcal I \setminus \{i^\star\} :\\ \, t_i \geq \underline{\mu}_{i^\star}}} p_i(\bm{t}) (t_i- c_i) + \sum_{\substack{i \in \mathcal I : \\ t_i < \underline{\mu}_{i^\star}}} p_i(\bm{t}) t_i + p_{i^\star}(\bm{t}){t}_{i^\star}\leq \overline{t}_{i^\star} = \sum_{i \in \mathcal{I}} (p_i^\star(\bm{t}) t_i - q_i^\star(\bm{t}) c_i),
\end{equation*}
where the second inequality holds because $\max_{i \neq i^\star}t_i-c_i < \overline t_{i^\star}$. Thus, $(\bm p^\star, \bm q^\star)$ solves~\ref{eq:MDPk(t)}. Also, $(\bm p,\bm q)$ can solve~\ref{eq:MDPk(t)} only if it satisfies $p_{i^\star}(\bm t)=1$ and $q_{i^\star}(\bm t)=0$. This observation completes the induction step.

\textbf{Step 4 ($\mathcal{T}_{IV}$).} Define $\mathcal S_{I+2} = \mathcal{T}_{IV}$, and set $k=I+2$. Next, fix an arbitrary scenario~$\bm{t} \in \mathcal S_k$ and mechanism~$(\bm{p}, \bm{q})$ feasible in~\ref{eq:MDPk(t)}. The constraints of~\ref{eq:MDPk(t)} ensure that~$(\bm{p}, \bm{q})$ solves~${\text{SP}}$$_{l}(\bm t')$ for every~$\bm t' \in \mathcal S_{l}$ and~$l<k$. From Steps~1,~2, and~3, we thus know that $p_{i^\star}(\bm{t}')=1$ and $q_{i^\star}(\bm{t}') =0$ for all $\bm t' \in \cup_{\ell=1}^{k-1} \mathcal S_\ell =  \mathcal T_I \cup \mathcal T_{II} \cup \mathcal T_{III}$. Clearly, $\bm t$ is $i^\star$-unilaterally reachable from $(\overline t_{i^\star}, \bm t_{-i^\star}) \in \mathcal{T}_{III}$.
As $p_{i^\star}(\overline t_{i^\star}, \bm t_{-i^\star}) -q_{i^\star}(\overline t_{i^\star}, \bm t_{-i^\star})  =1$, by Lemma~\ref{lem:deviationForFavored}(i), we have $p_{i^\star}(\bm{t}) = 1$. Thus, the objective function value of~$(\bm p, \bm q)$ in~\ref{eq:MDPk(t)} is bounded above by $t_{i^\star} - q_{i^\star}(\bm{t})c_{i^\star} \leq t_{i^\star}$. Thus, $(\bm p^\star, \bm q^\star)$ solves~\ref{eq:MDPk(t)}. Also, as $c_{i^\star} >0$, the mechanism~$(\bm{p}, \bm{q})$ can attain the optimal value~$t_{i^\star}$ of~\ref{eq:MDPk(t)} only if $p_{i^\star}(\bm{t})=1$ and $q_{i^\star}(\bm{t}) =0$. 

\textbf{Step 5 ($\mathcal{T}_{V}$).} We partition $\mathcal{T}_{V}$ into $I$ subsets of the form
\begin{equation*}
    \mathcal{S}_{k} = \left\{\bm t \in \mathcal{T}_{V} \,:\, |\{i \in \mathcal{I} \,:\, t_i \geq \overline t_{i^\star}\}| = k-I-2 \right\},
\end{equation*}
for $k=I+3, \dots, 2I+2$. We will use induction on~$k$ to show that, for any~$\bm t\in\mathcal S_k$, $(\bm{p}^\star,\bm{q}^\star)$ solves~\ref{eq:MDPk(t)} with optimal value $\max_{i' \in \mathcal{I}} t_{i'}-c_{i'}$, and any mechanism~$(\bm{p}, \bm{q})$ that solves~\ref{eq:MDPk(t)} must satisfy
\begin{equation}\label{eq:allocation-to-max-net-payoff-same-as-before} 
    \sum_{i \in {\arg \max}_{i' \in \mathcal{I}} t_{i'}-c_{i'}} p_i(\bm t) = 1 \quad \text{and} \quad p_i(\bm{t})=q_i(\bm t) \quad \forall i \in {\arg \max}_{i' \in \mathcal{I}} t_{i'}-c_{i'}.
\end{equation}
As for the base step corresponding to~$k=I+3$, fix any~$\bm{t} \in \mathcal{S}_{I+3}$, and consider a mechanism~$(\bm p, \bm q)$ feasible in~\ref{eq:MDPk(t)}. The constraints of~\ref{eq:MDPk(t)} ensure that~$(\bm p, \bm q)$ solves~${\text{SP}}$$_{l}(\bm t')$ for every~$\bm t' \in \mathcal{S}_l$ and $l< I+3$. From Steps~1,~2,~3 and~4, we thus know that $p_{i^\star}(\bm{t}')=p^\star_{i^\star}(\bm{t}')=1$ and $q_{i^\star}(\bm{t}') =q^\star_{i^\star}(\bm{t}') =0$ for all $\bm t' \in \cup_{l=1}^{I+1}\mathcal{S}_l = \mathcal{T}_I \cup \mathcal{T}_{II} \cup \mathcal{T}_{III} \cup \mathcal{T}_{IV}$.
As $\bm{t} \in \mathcal{S}_{k}$ and~$k-I-2=1$, there exists exactly one agent~$i^\circ$ with~$t_{i^\circ} \geq \overline t_{i^\star}$. By the definition of~$\mathcal T_{V}$, agent~$i^\circ$ is the only agent whose adjusted type satisfies $t_{i^\circ} - c_{i^\circ}\geq \overline t_{i^\star}$ so that we must have $i^\circ \neq i^\star$. Note that~$\bm t$ is ${i^\circ}$-unilaterally reachable from~$(\underline t_{i^\circ}, \bm t_{-{i^\circ}})$. As~$\underline t_{i^\circ} -c_{i^\circ} < \underline \mu_{i^\circ} < \underline \mu_{i^\star}< \overline{t}_{i^\star}$ by the definition of the favored agent~$i^\star$, we further have $(\underline t_{i^\circ}, \bm t_{-{i^\circ}}) \in \mathcal{T}_I \cup \mathcal{T}_{II} \cup \mathcal{T}_{III} \cup \mathcal{T}_{IV}$. The reasoning in Steps~1,~2,~3 and~4 thus implies that~$p_{i^\circ}(\underline t_{i}, \bm t_{-{i}})=0$, which in turn implies via Lemma~\ref{lem:deviationForFavored}(ii) that $p_{i^\circ}(\bm t) =q_{i^\circ}(\bm t)$. Thus, the objective function value of $(\bm p, \bm q)$ in~\ref{eq:MDPk(t)} is bounded above by that of $(\bm p^\star, \bm q^\star)$. Indeed, we have
\begin{equation*}
\begin{aligned}
\sum_{i \in \mathcal{I}} (p_i(\bm{t}) t_i - q_i(\bm{t}) c_i)  \leq p_{i^\circ}(\bm t)(t_{i^\circ}-c_{i^\circ})+ \sum_{i \in \mathcal I \setminus \{i^\circ\}}  p_i(\bm{t}) t_i \leq \max_{i \in \mathcal{I}}t_i-c_i =\sum_{i \in \mathcal{I}} (p_i^\star(\bm{t}) t_i - q_i^\star(\bm{t}) c_i),
\end{aligned}
\end{equation*}
where the first inequality holds because $p_{i^\circ}(\bm t) =q_{i^\circ}(\bm t)$ and because $c_i >0$ for all $i \in \mathcal I$, whereas the second inequality holds because $t_{i^\circ}-c_{i^\circ}=\max_{i \neq i^\star}t_i-c_i \geq \overline t_{i^\star}$ and because $t_{i} - c_i < t_{i} < \overline t_{i^\star}$ for all $i \neq {i^\circ}$. Finally, the equality holds because $\bm t \in \mathcal T_{V}$, in which case $(\bm p^\star, \bm q^\star)$ generates a payoff of~$\max_{i\in\mathcal I} t_i - c_i$. Thus, $(\bm{p}^\star,\bm{q}^\star)$ solves~\ref{eq:MDPk(t)}. In addition, the objective function value of~$(\bm{p}, \bm{q})$ can equal $\max_{i \in \mathcal I}t_i-c_i = t_{i^\circ}-c_{i^\circ}$ only if $p_{i^\circ}(\bm t) =q_{i^\circ}(\bm t)=1$. As $\bm t \in \mathcal S_k$ was chosen freely, this completes the base step.

As for the induction step, fix any $k\in\{I+4,\ldots,2I+2\}$. Assume that for all $l\in \{I+3,\ldots,k-1\}$ we know that $(\bm{p}^\star,\bm{q}^\star)$ solves ${\text{SP}}$$_l(\bm t)$ and that any mechanism $(\bm p,\bm q)$ that solves~${\text{SP}}$$_l(\bm t)$ satisfies~\eqref{eq:allocation-to-max-net-payoff-same-as-before} for all $\bm t \in \mathcal S_l$. Fix now any $\bm t \in \mathcal S_k$ and mechanism $(\bm p, \bm q)$ feasible in~\ref{eq:MDPk(t)}. The constraints in~\ref{eq:MDPk(t)} ensure that 
$(\bm p, \bm q)$ solves~${\text{SP}}$$_{l}(\bm t')$ for every $\bm t' \in \mathcal S_{l}$ and $l <k$. Thus, $(\bm p, \bm q)$ satisfies $p_{i^\star}(\bm{t}')=p^\star_{i^\star}(\bm{t}')=1$ and $q_{i^\star}(\bm{t}') =q^\star_{i^\star}(\bm{t}') =0$ for all $\bm t' \in \mathcal{T}_I \cup \mathcal{T}_{II} \cup \mathcal{T}_{III} \cup \mathcal{T}_{IV}$, which follows from Steps~1,~2,~3 and~4, and it satisfies~\eqref{eq:allocation-to-max-net-payoff-same-as-before} for all $\bm t' \in \cup_{l=I+3}^{k-1} \mathcal S_l$, which follows from the induction hypothesis. As $\bm t\in\mathcal S_k$ and $k\in\{I+4,\ldots,2I+2\}$, there are exactly $k-I-2$ agents~$i \in \mathcal{I}$ with types~$t_i \geq \overline{t}_{i^\star}$. For any such agent~$i$, scenario~$\bm t$ is $i$-unilaterally reachable from $(\underline t_i, \bm t_{-i})$. Note that either $(\underline t_i, \bm t_{-i}) \in \mathcal{T}_I \cup \mathcal{T}_{II} \cup \mathcal{T}_{III} \cup \mathcal{T}_{IV}$ or $(\underline t_i, \bm t_{-i}) \in \cup_{l=I+3}^{k-1} \mathcal S_{l}$ because $\underline t_i <\underline \mu_i < \underline \mu_{i^\star}< \overline t_{i^\star}$ for all $i \in \mathcal I$. We now show that $p_i(\underline t_i, \bm t_{-i})$ must vanish in both cases.
If $(\underline t_i, \bm t_{-i}) \in \mathcal{T}_I \cup \mathcal{T}_{II}  \cup \mathcal{T}_{III} \cup \mathcal{T}_{IV}$, then we have $i \neq i^\star$, in which case $p_i(\underline t_i, \bm t_{-i})=p^\star_i(\underline t_i, \bm t_{-i})=0$. If $(\underline t_i, \bm t_{-i}) \in \cup_{l=I+3}^{k-1} \mathcal S_{l}$, on the other hand, then the definition of~$i^\star$ implies that $\underline t_i-c_i < \overline t_{i^\star}$, and the definition of~$\mathcal T_{V}$ implies that $\max_{i' \in \mathcal I \setminus \{i^\star\}} t_{i'}-c_{i'}\geq \overline t_{i^\star}$. Hence, $i$ is no element of $\arg\max_{i' \in \mathcal I \setminus \{i^\star\}} t_{i'}-c_{i'}$, implying that $p_i(\underline t_i, \bm t_{-i})=0$ thanks to~\eqref{eq:allocation-to-max-net-payoff-same-as-before}. Lemma~\ref{lem:deviationForFavored}(ii) now ensures that $p_i(\bm{t}) = q_i(\bm{t})$ for all $i \in \mathcal{I}$ with $t_i \geq \overline{t}_{i^\star}$. Thus, the objective function value of $(\bm p, \bm q)$ in~\ref{eq:MDPk(t)} is bounded above by that of $(\bm p^\star, \bm q^\star)$ because
\begin{equation*}
    \sum_{i \in \mathcal{I}} (p_i(\bm{t}) t_i - q_i(\bm{t}) c_i) \leq \sum_{i \in \mathcal I : \, t_i \geq \overline{t}_{i^\star}} p_i(\bm{t}) (t_i- c_i) + \sum_{i \in \mathcal I : \, t_i < \overline{t}_{i^\star}} p_i(\bm{t}) t_i \leq \max_{i \in \mathcal{I}}t_i-c_i =\sum_{i \in \mathcal{I}} (p_i^\star(\bm{t}) t_i - q_i^\star(\bm{t}) c_i),
\end{equation*}
where the second inequality holds because $\max_{i \neq i^\star}t_i-c_i \geq \overline t_{i^\star}$. Thus, $(\bm{p}^\star,\bm{q}^\star)$ solves~\ref{eq:MDPk(t)}, and $(\bm p,\bm q)$ can solve~\ref{eq:MDPk(t)} only if it obeys~\eqref{eq:allocation-to-max-net-payoff-same-as-before}. This observation completes the induction step.
\hfill \Halmos}
\endproof 

\vspace{2mm}
\proof{Proof of Theorem \ref{thr:PRO-fav-agents-markov}.}
Let $(\bm{p}^\star,\bm{q}^\star)$ denote the allocation probabilities of the favored-agent mechanism described in Theorem \ref{thr:PRO-fav-agents-markov}. We know that $(\bm{p}^\star,\bm{q}^\star)$ is optimal from Theorem \ref{thr:opt-fav-markov}. To show that it is also Pareto robustly optimal, fix a mechanism $(\bm{p},\bm{q}) \in \mathcal X$ and suppose that $(\bm{p},\bm{q})$ weakly Pareto robustly dominates $(\bm{p}^\star,\bm{q}^\star)$, $i.e.$, condition \eqref{eq:dominance} holds. We will show that $(\bm{p},\bm{q})$ cannot (strictly) Pareto robustly dominate $(\bm{p}^\star,\bm{q}^\star)$.

If ${\arg \max}_{i \in \mathcal{I}} \underline{\mu}_i=\{i^\star\}$ is a singleton, we know from Lemma~\ref{lem:PRO-revenue equivalence} that $(\bm{p},\bm{q})$ cannot generate strictly higher expected payoff under any $\mathbb P \in \mathcal P$, and $(\bm{p}^\star,\bm{q}^\star)$ is thus Pareto robustly optimal. 
Suppose now that ${\arg \max}_{i \in \mathcal{I}} \underline{\mu}_i$ is not a singleton. Select any $\varepsilon \in (0, \overline{\mu}_{i^\star}-\underline{\mu}_{i^\star})$ that exists because $\underline{\mu}_{i^\star}<\overline{\mu}_{i^\star}$, and define
\begin{equation*}
\begin{aligned}
\mathcal{P}_\varepsilon = \{ \mathbb{P} \in \mathcal{P} \,:\,  \mathbb{E}_{\mathbb{P}}[\tilde{t}_{i^\star}] \in [\underline{\mu}_{i^\star}+\varepsilon,\overline{\mu}_{i^\star}] \}.
\end{aligned}
\end{equation*}
Set $\mathcal{P}_\varepsilon$ represents another Markov ambiguity set where the lowest mean value $\underline{\mu}_{i^\star}$ of bidder $i^\star$ is shifted to $\underline{\mu}_{i^\star}+\varepsilon$. Note that agent $i^\star$ becomes the unique agent with the maximum lowest mean value under $\mathcal{P}_\varepsilon$.
As $\mathcal{P}_\varepsilon \subset \mathcal{P}$ by construction, we have
\begin{equation*}
\begin{aligned}
 &\mathbb{E}_{\mathbb{P}}\left[\sum_{i \in \mathcal{I}} (p_i(\tilde{\bm{t}}) \tilde{t}_i - q_i(\tilde{\bm{t}}) c_i) \right] \geq  \mathbb{E}_{\mathbb{P}}\left[\sum_{i \in \mathcal{I}} (p^\star_i(\tilde{\bm{t}}) \tilde{t}_i - q^\star_i(\tilde{\bm{t}}) c_i) \right] &&\forall \mathbb{P} \in \mathcal{P}_\varepsilon.
\end{aligned}
\end{equation*}
Thus, $(\bm{p},\bm{q})$ also weakly Pareto robustly dominates $(\bm{p}^\star,\bm{q}^\star)$ under the Markov ambiguity set $\mathcal{P}_\varepsilon$. By Lemma~\ref{lem:PRO-revenue equivalence}, we can now conclude that $(\bm{p},\bm{q})$ and $(\bm{p}^\star,\bm{q}^\star)$ generate the same payoff for the principal in any scenario $\bm{t} \in \mathcal{T}$. This implies that the expected payoff of $(\bm{p},\bm{q})$ cannot exceed the one of $(\bm{p}^\star,\bm{q}^\star)$ under any distribution $\mathbb P$ supported on $\mathcal T$. Thus, none of the inequalities in \eqref{eq:dominance} can be strict, and $(\bm{p},\bm{q})$ cannot Pareto robustly dominate $(\bm{p}^\star,\bm{q}^\star)$. The claim thus follows. \hfill \Halmos
\endproof

\vspace{2mm}
\proof{Proof of Theorem \ref{thr:opt-fav-independent}.}
Select any favored-agent mechanism with $i^\star \in {\arg \max}_{i \in \mathcal{I}} \underline{\mu}_i$ and $\nu^\star \geq {\max}_{i \in \mathcal{I}} \underline{\mu}_i$, denote by $(\bm{p},\bm{q})$ its allocation probabilities. Recall first that this mechanism is feasible in \eqref{eq:MDP}. We will prove that $(\bm{p},\bm{q})$ attains a worst-case expected payoff that is at least as large as $\max_{i \in \mathcal I} \underline{\mu}_i$, which implies via Proposition \ref{prop:optimal-z-markov-independent} that it is optimal in \eqref{eq:MDP}.

To this end, fix an arbitrary distribution $\mathbb{P} \in \mathcal{P}$ and suppose for ease of exposition that $\mathbb{P}\left(\max_{i \neq i^\star} \tilde{t}_i-c_i < \nu^\star \right)$, $\mathbb{P}\left(\max_{i \neq i^\star} \tilde{t}_i-c_i = \nu^\star\right)$ and $\mathbb{P}\left(\max_{i \neq i^\star} \tilde{t}_i-c_i > \nu^\star\right)$ are all strictly positive. We can write the principal's expected payoff from $(\bm{p},\bm{q})$ under $\mathbb{P}$ as
%\begin{equation}\label{eq: conditional expectations}
\begin{flalign}\label{eq: conditional expectations}
    \mathbb{E}_{\mathbb{P}}\left[\sum_{i \in \mathcal{I}} (p_i(\tilde{\bm{t}}) \tilde{t}_i - q_i(\tilde{\bm{t}}) c_i) \right]    = &\mathbb{P}\left(\max_{i \neq i^\star} \tilde{t}_i-c_i < \nu^\star \right) \mathbb{E}_{\mathbb{P}}\left[\sum_{i \in \mathcal{I}} (p_i(\tilde{\bm{t}}) \tilde{t}_i - q_i(\tilde{\bm{t}}) c_i) \,\bigg\vert\, \max_{i \neq i^\star} \tilde{t}_i-c_i < \nu^\star \right] &&\\ \nonumber
    &+ \mathbb{P}\left(\max_{i \neq i^\star} \tilde{t}_i-c_i = \nu^\star\right) \mathbb{E}_{\mathbb{P}}\left[\sum_{i \in \mathcal{I}} (p_i(\tilde{\bm{t}}) \tilde{t}_i - q_i(\tilde{\bm{t}}) c_i) \,\bigg\vert\, \max_{i \neq i^\star} \tilde{t}_i-c_i = \nu^\star \right] &&\\
    &+ \mathbb{P}\left(\max_{i \neq i^\star} \tilde{t}_i-c_i > \nu^\star\right) \mathbb{E}_{\mathbb{P}}\left[\sum_{i \in \mathcal{I}} (p_i(\tilde{\bm{t}}) \tilde{t}_i - q_i(\tilde{\bm{t}}) c_i) \,\bigg\vert\, \max_{i \neq i^\star} \tilde{t}_i-c_i > \nu^\star \right].&& \nonumber
\end{flalign}
%\end{equation}
If one or more of $\mathbb{P}\left(\max_{i \neq i^\star} \tilde{t}_i-c_i < \nu^\star \right)$, $\mathbb{P}\left(\max_{i \neq i^\star} \tilde{t}_i-c_i = \nu^\star\right)$ and $\mathbb{P}\left(\max_{i \neq i^\star} \tilde{t}_i-c_i > \nu^\star\right)$ are zero, the right-hand side of equation \eqref{eq: conditional expectations} can be adjusted by removing the respective terms, and the proof proceeds similarly.

In the following, we will show that all of the conditional expectations above, and therefore the principal's expected payoff under $\mathbb{P}$, are greater than or equal to $z^\star = \max_{i \in \mathcal I} \underline{\mu}_i$. 
If $\max_{i \neq i^\star} t_i-c_i < \nu^\star$, condition (i) in Definition \ref{def:fav-agent-mechanism} implies that the principal's payoff amounts to $t_{i^\star}$. This implies that
\begin{align*}
\mathbb{E}_{\mathbb{P}}\left[\sum_{i \in \mathcal{I}} (p_i(\tilde{\bm{t}}) \tilde{t}_i - q_i(\tilde{\bm{t}}) c_i) \,\bigg\vert\, \max_{i \neq i^\star} \tilde{t}_i-c_i < \nu^\star \right] &= \mathbb{E}_{\mathbb{P}}\left[ \tilde{t}_{i^\star} \,\bigg\vert\, \max_{i \neq i^\star} \tilde{t}_i-c_i < \nu^\star \right] \\
&= \mathbb{E}_{\mathbb{P}}\left[ \tilde{t}_{i^\star} \right] = \mu_{i^\star} = \max_{i \in \mathcal I} \underline{\mu}_i,
\end{align*}
where the second equality holds because the agents' types are independent.
If $\max_{i \neq i^\star} t_i-c_i > \nu^\star$, then condition (ii) in Definition \ref{def:fav-agent-mechanism} implies that 
the principal's payoff amounts to $\max_{i \in \mathcal{I}} t_i-c_i$. We thus have
\begin{align*}
\mathbb{E}_{\mathbb{P}}\left[\sum_{i \in \mathcal{I}} (p_i(\tilde{\bm{t}}) \tilde{t}_i - q_i(\tilde{\bm{t}}) c_i) \,\bigg\vert\, \max_{i \neq i^\star} \tilde{t}_i-c_i > \nu^\star \right] = \mathbb{E}_{\mathbb{P}}\left[ \max_{i \in \mathcal{I}} \tilde{t}_i-c_i \,\bigg\vert\, \max_{i \neq i^\star} \tilde{t}_i-c_i > \nu^\star \right] > \nu^\star \geq \max_{i \in \mathcal I} \underline{\mu}_i.
\end{align*}
If $\max_{i \neq i^\star} t_i-c_i = \nu^\star$, then the allocation functions are defined either as in condition (i) or as in condition (ii) of Definition \ref{def:fav-agent-mechanism}. If the allocation functions are defined as in condition (i), we have
\begin{align*}
\mathbb{E}_{\mathbb{P}}\left[\sum_{i \in \mathcal{I}} (p_i(\tilde{\bm{t}}) \tilde{t}_i - q_i(\tilde{\bm{t}}) c_i) \,\bigg\vert\, \max_{i \neq i^\star} \tilde{t}_i-c_i = \nu^\star \right] &= \mathbb{E}_{\mathbb{P}}\left[ \tilde{t}_{i^\star} \,\bigg\vert\, \max_{i \neq i^\star} \tilde{t}_i-c_i = \nu^\star \right] \\
&= \mathbb{E}_{\mathbb{P}}\left[ \tilde{t}_{i^\star} \right] = \mu_{i^\star} = \max_{i \in \mathcal I} \underline{\mu}_i,
\end{align*}
where the second equality again holds because the agents' types are independent. If the allocation functions are defined as in condition (ii), on the other hand, then 
\begin{align*}
\mathbb{E}_{\mathbb{P}}\left[\sum_{i \in \mathcal{I}} (p_i(\tilde{\bm{t}}) \tilde{t}_i - q_i(\tilde{\bm{t}}) c_i) \,\bigg\vert\, \max_{i \neq i^\star} \tilde{t}_i-c_i = \nu^\star \right] = \mathbb{E}_{\mathbb{P}}\left[ \max_{i \in \mathcal{I}} \tilde{t}_i-c_i \,\bigg\vert\, \max_{i \neq i^\star} \tilde{t}_i-c_i = \nu^\star \right] \geq \nu^\star \geq \max_{i \in \mathcal I} \underline{\mu}_i.
\end{align*}
In summary, we have shown that all of the conditional expectations in \eqref{eq: conditional expectations}, and therefore also the principal's expected payoff under $\mathbb P$, are non-inferior to $\max_{i \in \mathcal I} \underline{\mu}_i$. As the distribution $\mathbb P \in \mathcal P$ was chosen arbitrarily, this reasoning implies that the principal's worst-case expected payoff is also non-inferior to $\max_{i \in \mathcal I} \underline{\mu}_i$. The favored-agent mechanism at hand is therefore optimal in \eqref{eq:MDP} by virtue of Proposition~\ref{prop:optimal-z-markov-independent}. \hfill \Halmos
\endproof

\vspace{2mm}
\proof{Proof of Lemma \ref{lem:findP-independent}.}
Consider arbitrary $\bm{t} \in \mathcal{T}$ and $\mu_{i^\star} \in [\underline{\mu}_{i^\star},\overline{\mu}_{i^\star}]$. We will construct a scenario $\hat{\bm{t}} \in \mathcal{T}$, where $\max_{i \neq i^\star}\hat{t}_i < \underline{\mu}_{i^\star}$, and a discrete distribution $\mathbb{P} \in \mathcal{P}$ that satisfies (i)--(iii). 
To this end, we define $\hat{t}_i$ through
\begin{align*}
    &\hat{t}_i = \begin{cases}
    t_i &\text{if } t_i = \underline{\mu}_i\\
    \underline{t}_i &\text{if } t_i > \underline{\mu}_i\\
    \underline{\mu}_i+\varepsilon &\text{if } t_i < \underline{\mu}_i
    \end{cases} \;\;\;\forall i \in \mathcal I \setminus \{ i^\star \} \quad \text{and} \quad
    \hat{t}_{i^\star} = \begin{cases}
    t_{i^\star} &\text{if } t_{i^\star} = {\mu}_{i^\star}\\
    \underline{t}_{i^\star} &\text{if } t_{i^\star} > {\mu}_{i^\star}\\
    {\mu}_{i^\star}+\varepsilon &\text{if } t_{i^\star} < {\mu}_{i^\star},
    \end{cases}
\end{align*}
where $\varepsilon \in (0, \min_{i \in \mathcal{I}} \overline{t}_i-\overline{\mu}_i) \cap \big(0, (\underline{\mu}_{i^\star}-\max_{i \neq i^\star} \underline{\mu}_i)/2\big)$ is a fixed positive number. Note that there exists such  $\varepsilon > 0$ because $\underline{\mu}_i<\overline{\mu}_i<\overline{t}_i$ for all $i \in \mathcal{I}$ and $\arg \max_{i \in \mathcal{I}} \underline{\mu}_i = \{i^\star\}$ is a singleton.
We next show that $\hat{t}_i \in \mathcal T_i$ for all $i \in \mathcal{I}$ ($i.e.$, $\hat{\bm{t}} \in \mathcal T$) and $\max_{i \neq i^\star}\hat{t}_i < \underline{\mu}_{i^\star}$.
For any $i \in \mathcal I$, we have
$$
\hat{t}_{i} \leq \overline{\mu}_{i}+\varepsilon \leq \overline{\mu}_{i}+ \min_{j \in \mathcal{I}} (\overline{t}_j-\overline{\mu}_j) \leq \overline{\mu}_{i}+ \overline{t}_{i}-\overline{\mu}_{i} = \overline{t}_{i},
$$
where the first inequality follows from the definition of $\hat{t}_{i}$, and the second inequality holds because $\varepsilon < \min_{j \in \mathcal{I}} \overline{t}_j-\overline{\mu}_j$. The definition of $\hat{t}_{i}$ implies that we also have $\hat{t}_{i} \geq \underline{t}_i$. We thus showed that~$\hat{\bm{t}} \in \mathcal T$. 

For all $i \neq i^\star$, we moreover have
$$
\hat{t}_i \leq \underline{\mu}_i+\varepsilon \leq \underline{\mu}_i+ (\underline{\mu}_{i^\star}-\max_{j \neq i^\star} \underline{\mu}_j)/2 \leq \underline{\mu}_i+ (\underline{\mu}_{i^\star}- \underline{\mu}_i)/2 <\underline{\mu}_{i^\star},
$$
where the first inequality again follows from the definition of $\hat{t}_{i}$, the second inequality holds because $\varepsilon < (\underline{\mu}_{i^\star}-\max_{i \neq i^\star} \underline{\mu}_i)/2$, and the fourth inequality holds because $\arg \max_{i \in \mathcal{I}} \underline{\mu}_i = \{i^\star\}$ is a singleton. We thus showed that $\max_{i \neq i^\star}\hat{t}_i < \underline{\mu}_{i^\star}$.

Next, we will construct a discrete distribution $\mathbb{P}$ through the marginal distributions $\mathbb{P}_i = \alpha_i \delta_{t_i} + (1 - \alpha_i) \delta_{\hat{t}_i}$ of $\tilde{t}_i$'s, where $\alpha_i \in (0,1]$ for all $i \in \mathcal I$. We will then show that $\mathbb{P}$ belongs to the Markov ambiguity set $\mathcal{P}$ and moreover satisfies the properties (i)--(iii). To this end, we define $\alpha_i$ through
%When $t_i=\underline{\mu}_i$, we set $\alpha_i =1$. Otherwise, we rewrite the equality as $\alpha_i = (\underline{\mu}_i-\hat{t}_i)/(t_i-\hat{t}_i)$ and retrieve the value of $\alpha_i$:
\begin{align*}
    &\alpha_i= \begin{cases}
    1 &\text{if } t_i = \hat{t}_i,\\
    (\underline{\mu}_i-\hat{t}_i)/(t_i-\hat{t}_i) &\text{if } t_i \neq \hat{t}_i,
    \end{cases} \quad \forall i \in \mathcal I \setminus \{ i^\star \} 
    \end{align*}
    and
    \begin{align*}
    \alpha_{i^\star} = \begin{cases}
    1 &\text{if } t_{i^\star} = \hat{t}_{i^\star},\\
    ({\mu}_{i^\star} -\hat{t}_{i^\star})/(t_{i^\star}-\hat{t}_{i^\star}) &\text{if } t_{i^\star} \neq \hat{t}_{i^\star}.
    \end{cases}
\end{align*}
We first show that $\alpha_i \in (0,1]$ for all $i \in \mathcal{I}$. For any $i \in \mathcal{I}$, it is sufficient to show that the claim holds if $t_i \neq \hat{t}_i$. For any $i \neq i^\star$, if $t_i \neq \hat{t}_i$ and $t_i > \underline{\mu}_i$, we have
$$
\alpha_i =(\underline{\mu}_i-\hat{t}_i)/(t_i-\hat{t}_i)= (\underline{\mu}_i-\underline{t}_i)/(t_i-\underline{t}_i) \in (0,1),
$$
where the second equality follows from the definition of $\hat{t}_i$, and the inclusion holds because $t_i >\underline{\mu}_i>\underline{t}_i$. If $t_i \neq \hat{t}_i$ and $t_i < \underline{\mu}_i$, on the other hand, we have
$\alpha_i =-\varepsilon/(t_i-\underline{\mu}_i-\varepsilon) \in (0,1)$, where the equality again follows from the definition of $\hat{t}_i$, and the inclusion holds because $t_i<\underline{\mu}_i<\underline{\mu}_i+\varepsilon$. Note that if $t_i = \underline{\mu}_i$, then $\hat{t}_i = t_i$ by definition, and $\alpha_i = 1$. One can similarly show that $\alpha_{i^\star} \in (0,1]$ by replacing $\underline{\mu}_{i^\star}$ with $\mu_{i^\star}$ in the above arguments. Thus, $\alpha_i \in (0,1]$ for all $i \in \mathcal{I}$. 
We now define $\mathbb{P}$ through the marginal distributions $\mathbb{P}_i = \alpha_i \delta_{\bm{t}_i} + (1 - \alpha_i) \delta_{\hat{\bm{t}}_i}$, $i \in \mathcal I$, as follows.
\begin{align*}
    \mathbb{P}(\tilde{\bm{t}} =\bm{t}) &= \prod_{i \in \mathcal{I}} \mathbb{P}_i(\tilde{t}_i=t_i) \quad \forall \bm{t} \in \mathcal{T}
\end{align*}
By construction, $\tilde t_i$'s are mutually independent under $\mathbb P$. Hence, the expected type of each $i \in \mathcal{I}$ amounts to $\mathbb{E}_{\mathbb P}[\tilde t_i]=\alpha_i t_i + (1- \alpha_i)\hat{t}_i$.

We next show that $\mathbb{E}_{\mathbb P}[\tilde t_i] \in [\underline{\mu}_i,\overline{\mu}_i]$ for all $i \in \mathcal I$, which implies that $\mathbb P \in \mathcal P$. For any $i \neq i^\star$, if $t_i = \hat{t}_i$, then we have $t_i = \hat{t}_i = \underline{\mu}_i$ by definition of $\hat{t}_i$. The expected type therefore amounts to $\underline{\mu}_i$. If $t_i \neq \hat{t}_i$, on the other hand, we have
\begin{align*}
\mathbb{E}_{\mathbb P}[\tilde t_i]=\alpha_i t_i + (1- \alpha_i)\hat{t}_i = \alpha_i (t_i-\hat{t}_i) + \hat{t}_i = \frac{\underline{\mu}_i-\hat{t}_i}{t_i-\hat{t}_i} (t_i-\hat{t}_i) + \hat{t}_i = \underline{\mu}_i,
\end{align*}
where the third equality follows from the definition of $\alpha_i$. One can verify that $\mathbb{E}_{\mathbb P}[\tilde t_{i^\star}] = {\mu}_{i^\star}$ using similar arguments. We thus showed that $\mathbb{E}_{\mathbb P}[\tilde t_i] \in [\underline{\mu}_i,\overline{\mu}_i]$ for all $i \in \mathcal I$, and therefore $\mathbb P \in \mathcal P$.

It remains to show that $\mathbb P$ satisfies (i)--(iii). As we have $\mathbb{E}_{\mathbb P}[\tilde t_{i^\star}] = {\mu}_{i^\star}$, property (i) holds. The definition of $\mathbb P$ implies that (ii) and (iii) also hold. \hfill \Halmos
\endproof

\vspace{2mm}
\proof{Proof of Lemma \ref{lem:independent-revenue-equivalence}.}
{ 
Throughout the proof, we denote by~$(\bm{p}^\star,\bm{q}^\star)$ the favored-agent mechanism of type~(i) with favored agent~${i^\star}$, where ${\arg \max}_{i \in \mathcal{I}} \underline{\mu}_i = \{i^\star\}$, and threshold value~$\nu^\star= \underline{\mu}_{i^\star}$. We will use the following partition of $\mathcal{T}$.
\begin{equation*}
\begin{aligned}
\mathcal{T}_I &= \{\bm{t} \in \mathcal{T} \,:\, \max_{i \neq i^\star} t_i-c_i \leq \underline{\mu}_{i^\star} \; \text{and} \; \max_{i \neq i^\star} t_i \leq \underline{\mu}_{i^\star} \; \text{and} \; t_{i^\star} \in (\underline{\mu}_{i^\star},\overline{\mu}_{i^\star}]  \}\\
\mathcal{T}_{II} &=  \{\bm{t} \in \mathcal{T} \,:\, \max_{i \neq i^\star} t_i-c_i \leq \underline{\mu}_{i^\star}  \; \text{and} \; \max_{i \neq i^\star} t_i > \underline{\mu}_{i^\star} \; \text{and} \; t_{i^\star} \in (\underline{\mu}_{i^\star},\overline{\mu}_{i^\star}] \}\\
\mathcal{T}_{III} &= \{\bm{t} \in \mathcal{T} \,:\, \max_{i \neq i^\star} t_i-c_i \leq \underline{\mu}_{i^\star} \; \text{and} \; t_{i^\star} \notin (\underline{\mu}_{i^\star},\overline{\mu}_{i^\star}]  \}\\
\mathcal{T}_{IV} &= \{\bm{t} \in \mathcal{T} \,:\, \max_{i \neq i^\star} t_i-c_i > \underline{\mu}_{i^\star} \; \text{and} \; t_{i^\star} = \underline \mu_{i^\star} \}\\
\mathcal{T}_{V} &= \{\bm{t} \in \mathcal{T} \,:\, \max_{i \neq i^\star} t_i-c_i > \underline{\mu}_{i^\star} \; \text{and} \; t_{i^\star} \neq \underline \mu_{i^\star} \}
\end{aligned}
\end{equation*}
Note that $\mathcal{T}_I$ and $\mathcal{T}_{III}$ are nonempty because they contain at least $(\overline{\mu}_{i^\star}, \underline{\bm \mu}_{-i^\star})$ and $\underline{\bm \mu}$, respectively,
%as ${\arg \max}_{i \in \mathcal{I}} \underline{\mu}_i=\{{i^\star}\}$ and $[\underline{\mu}_i,\overline{\mu}_i] \in (\underline{t}_i,\overline{t}_i)$ for all $i \in \mathcal{I}$, 
but the sets $\mathcal{T}_{II}$, $\mathcal{T}_{IV}$ and $\mathcal{T}_{V}$ can be empty if $\underline{\mu}_{i^\star}$ or $c_i$ are sufficiently large for all $i \neq i^\star$.}

{  From Theorem~\ref{thr:opt-fav-independent} we know that $(\bm{p}^\star,\bm{q}^\star)$ is robustly optimal. To prove Lemma \ref{lem:independent-revenue-equivalence}, we will leverage Corollary \ref{Corollary of sufficiency proposition} of Proposition~\ref{prop:unifyingPRO}, which provides sufficient conditions for the claim made. In particular, to show that~any mechanism~$(\bm{p},\bm{q}) \in \mathcal X$ that weakly Pareto robustly dominates $(\bm{p}^\star,\bm{q}^\star)$ must generate the same payoff as $(\bm{p}^\star,\bm{q}^\star)$ in every scenario~$\bm t\in\mathcal T$, it suffices to verify the conditions~(i) and~(ii) in Proposition~\ref{prop:unifyingPRO} for some partition of the type space $\mathcal T$. 
We will show that this holds for a refinement of the partition $\mathcal{T}_{I}-\mathcal{T}_{V}$. Specifically, we will construct a refined partition $\mathcal S_1,\ldots,\mathcal S_m$ of~$\mathcal T$ with~$m=3I-1$, where $\mathcal S_1$ coincides with~$\mathcal{T}_{I}$, $\mathcal S_2, \dots, \mathcal S_{I}$ forms a partition of~$\mathcal{T}_{II}$, $\mathcal S_{I+1}$ coincides with~$\mathcal{T}_{III}$, $\mathcal S_{I+2}, \dots, \mathcal S_{2I}$ forms a partition of~$\mathcal{T}_{IV}$, and $\mathcal S_{2I+1}, \dots, \mathcal S_{3I-1}$ forms a partition of~$\mathcal{T}_{V}$. 

We will formally define the refined partition $\mathcal{S}_1,\ldots,\mathcal{S}_m$ while proving condition~(ii). We prove condition~(ii) in five steps, focusing on the five subsets $\mathcal{T}_{I},\ldots, \mathcal{T}_{V}$ and, in particular, on their respective refined partitions. We will exploit the non-locality of the incentive compatibility constraint~\eqref{eq:IC} via Lemma~\ref{lem:deviationForFavored} to show iteratively that~$(\bm p^\star, \bm q^\star)$ solves the scenario problem~\ref{eq:MDPk(t)} for all $\bm t \in \mathcal S_k$ and~$k=1,\ldots,m$. 
The subsequent reasoning critically relies on the definition of~$(\bm{p}^\star,\bm{q}^\star)$ under which the principal's payoff in scenario~$\bm t$ amounts to~$t_{i^\star}$ when $\max_{i \neq i^\star} t_i - c_i \leq \underline{\mu}_{i^\star}$ ($i.e.$, when $\bm t \in \mathcal{T}_{I} \cup \mathcal{T}_{II}\cup \mathcal{T}_{III}$) and to $\max_{i \in \mathcal{I}} t_i - c_i$ when  $\max_{i \neq i^\star} t_i - c_i > \underline{\mu}_{i^\star}$ ($i.e.$, when $\bm t \in \mathcal{T}_{IV} \cup \mathcal{T}_{V}$).}

\textbf{Step 1 ($\mathcal{T}_{I}$).} Define $\mathcal{S}_1 = \mathcal{T}_I$, and set $k=1$. Fix an arbitrary scenario $\bm t \in \mathcal{S}_1$ and mechanism $(\bm p, \bm q) \in \mathcal X$. Note that any $(\bm p, \bm q) \in \mathcal X$ is feasible in~${\text{SP}}$$_{1}(\bm t)$. In addition, the objective value of $(\bm p, \bm q)$ is dominated by that of $(\bm p^\star, \bm q^\star)$ because 
\begin{equation*}
\begin{aligned}
\sum_{i \in \mathcal{I}} (p_i(\bm{t}) t_i - q_i(\bm{t}) c_i) \leq \sum_{i \in \mathcal{I}} p_i(\bm{t}) t_i \leq t_{i^\star} = \sum_{i \in \mathcal{I}} (p_i^\star(\bm{t}) t_i - q_i^\star(\bm{t}) c_i),
\end{aligned}
\end{equation*}
where the first inequality holds because $c_i>0$ for all $i \in \mathcal{I}$, and the second inequality follows from the definition of $\mathcal{S}_1$, which implies that $t_i< t_{i^\star}$ for all $i \neq i^{\star}$. Thus, $(\bm p^\star, \bm q^\star)$ solves~${\text{SP}}$$_{1}(\bm t)$. As $t_i < t_{i^\star}$ for all $i \neq i^\star$, the above arguments also reveal that any mechanism $(\bm{p}, \bm{q})$ that solves ${\text{SP}}$$_{1}(\bm t)$ must satisfy $p_{i^\star}(\bm{t})=1$ and $q_{i^\star}(\bm{t})=0$ for any $\bm{t} \in \mathcal{S}_1 = \mathcal{T}_I$.

\textbf{Step 2 ($\mathcal{T}_{II}$).} We partition $\mathcal{T}_{II}$ into $I-1$ subsets of the form
\begin{equation*}
    \mathcal{S}_{k} = \left\{\bm t \in \mathcal{T}_{II} \,:\, |\{i \in \mathcal{I} \,:\, t_i > \underline \mu_{i^\star}\}| = k \right\},
\end{equation*}
for $k=2, \dots, I$. 
Note that $|\{i \in \mathcal{I} : t_i > \underline \mu_{i^\star}\}| \geq 2$ for all $\bm{t} \in \mathcal{T}_{II}$ because $t_{i^\star}\in (\underline \mu_{i^\star},\overline \mu_{i^\star}]$ and $\max_{i \neq i^\star} t_i > \underline {\mu}_{i^\star}$ by the definition of $\mathcal{T}_{II}$. We will use induction on~$k$ to show that, for any $\bm t \in \mathcal{S}_{k}$, $(\bm p^\star, \bm q^\star)$ solves~\ref{eq:MDPk(t)}, and any mechanism $(\bm p, \bm q)$ that solves~\ref{eq:MDPk(t)} must satisfy $p_{i^\star}(\bm{t})=1$ and~$q_{i^\star}(\bm{t})=0$.

As for the base step corresponding to $k=2$, fix any~$\bm t \in \mathcal{S}_2$ and consider a mechanism $(\bm p, \bm q)$ feasible in~\ref{eq:MDPk(t)}. The constraints of~\ref{eq:MDPk(t)} ensure that~$(\bm p, \bm q)$ solves~${\text{SP}}$$_{l}(\bm t')$ for every~$\bm t' \in \mathcal{S}_l$ and $l< k$. From Step~1 we thus know that $p_{i^\star}(\bm{t}')=1$ and $q_{i^\star}(\bm{t}') =0$ for all $\bm t' \in \mathcal{S}_1 = \mathcal{T}_I$. 
As $\bm{t} \in \mathcal{S}_{k}$ and~$k=2$, there exists exactly one agent $i^\circ \neq i^\star$ with~$t_{i^\circ} > \underline \mu_{i^\star}$. Note that~$\bm t$ is ${i^\circ}$-unilaterally reachable from~$(\underline t_{i^\circ}, \bm t_{-{i^\circ}})$. As~$\underline t_{i^\circ} < \underline \mu_{i^\circ} < \underline \mu_{i^\star}$ by the definition of the favored agent~$i^\star$, we further have $(\underline t_{i^\circ}, \bm t_{-{i^\circ}}) \in \mathcal{T}_{I}$. The reasoning in Step~1 thus implies that~$p_{i^\circ}(\underline t_{i}, \bm t_{-{i}})=0$, which in turn implies via Lemma~\ref{lem:deviationForFavored}(ii) that $p_{i^\circ}(\bm t) =q_{i^\circ}(\bm t)$. Thus, the objective function value of $(\bm p, \bm q)$ in~\ref{eq:MDPk(t)} is bounded above by that of $(\bm p^\star, \bm q^\star)$. Indeed, we have
\begin{equation*}
\begin{aligned}
\sum_{i \in \mathcal{I}} (p_i(\bm{t}) t_i - q_i(\bm{t}) c_i)  \leq p_{i^\circ}(\bm t)(t_{i^\circ}-c_{i^\circ})+ \sum_{i \in \mathcal I \setminus \{i^\circ\}}  p_i(\bm{t}) t_i \leq t_{i^\star} = \sum_{i \in \mathcal{I}} (p_i^\star(\bm{t}) t_i - q_i^\star(\bm{t}) c_i),
\end{aligned}
\end{equation*}
where the first inequality holds because $p_{i^\circ}(\bm t) =q_{i^\circ}(\bm t)$ and because $c_i >0$ for all $i \in \mathcal I$, whereas the second inequality holds because $t_{i^\circ}-c_{i^\circ} \leq \underline \mu_{i^\star} < t_{i^\star}$ and because $t_{i} \leq \underline \mu_{i^\star}$ for all $i \neq {i^\circ}, i^\star$. Thus,  $(\bm p^\star, \bm q^\star)$ solves~\ref{eq:MDPk(t)}. In addition, the objective function value of~$(\bm{p}, \bm{q})$ can equal $t_{i^\star}$ only if $p_{i^\star}(\bm t)=1$ and $q_{i^\star}(\bm t)=0$. As $\bm t \in \mathcal S_k$ was chosen freely, this completes the base step.
 
As for the induction step, fix any $k\in\{3,\ldots,I\}$. Assume that for all $l\in \{2,\ldots,k-1\}$ we know that $(\bm p^\star, \bm q^\star)$ solves~${\text{SP}}$$_l(\bm t)$ and that any mechanism $(\bm p,\bm q)$ that solves~${\text{SP}}$$_l(\bm t)$ satisfies $p_{i^\star}(\bm t)=1$ and $q_{i^\star}(\bm t)=0$ for all $\bm t \in \mathcal S_l$. Fix now any $\bm t \in \mathcal S_k$ and mechanism $(\bm p, \bm q)$ feasible in~\ref{eq:MDPk(t)}. The constraints in~\ref{eq:MDPk(t)} ensure that 
$(\bm p, \bm q)$ solves~${\text{SP}}$$_{l}(\bm t')$ for every $\bm t' \in \mathcal S_{l}$ and $l <k$. Thus, $(\bm p, \bm q)$ satisfies $p_{i^\star}(\bm{t}')=1$ and $q_{i^\star}(\bm{t}')=0$ for all $\bm t' \in \cup_{l=1}^{k-1} \mathcal S_l$, which follows from Step~1 and the induction hypothesis. As $\bm t\in\mathcal S_k$ and $k\in\{3,\ldots,I\}$, there are exactly $k$ agents~$i \in \mathcal{I}$ with types~$t_i > \underline{\mu}_{i^\star}$. For any such agent $i \neq i^\star$, scenario~$\bm t$ is $i$-unilaterally reachable from $(\underline t_i, \bm t_{-i})$. As $(\underline t_i, \bm t_{-i}) \in \mathcal S_{k-1}$, we have $p_i(\underline t_i, \bm t_{-i})=0$ thanks to our induction hypothesis. Lemma~\ref{lem:deviationForFavored}(ii) now ensures that $p_i(\bm{t}) = q_i(\bm{t})$ for all $i \in \mathcal{I}$ with $t_i \geq \underline{\mu}_{i^\star}$ and $i \neq i^\star$. Thus, the objective function value of any feasible $(\bm p, \bm q)$ in~\ref{eq:MDPk(t)} is bounded above by that of $(\bm p^\star, \bm q^\star)$, that is,
\begin{equation*}
    \sum_{i \in \mathcal{I}} (p_i(\bm{t}) t_i - q_i(\bm{t}) c_i) \leq \sum_{\substack{i \in \mathcal I \setminus \{i^\star\} :\\ \, t_i > \underline{\mu}_{i^\star}}} p_i(\bm{t}) (t_i- c_i) + \sum_{\substack{i \in \mathcal I : \\ t_i \leq \underline{\mu}_{i^\star}}} p_i(\bm{t}) t_i + p_{i^\star}(\bm{t}){t}_{i^\star}\leq {t}_{i^\star} = \sum_{i \in \mathcal{I}} (p_i^\star(\bm{t}) t_i - q_i^\star(\bm{t}) c_i),
\end{equation*}
where the second inequality holds because $\max_{i \neq i^\star}t_i-c_i < \underline \mu_{i^\star}$. Thus, $(\bm p^\star, \bm q^\star)$ solves~\ref{eq:MDPk(t)}. Also, $(\bm p,\bm q)$ can solve~\ref{eq:MDPk(t)} only if it satisfies $p_{i^\star}(\bm t)=1$ and $q_{i^\star}(\bm t)=0$. This observation completes the induction step.

% \textbf{Step 4 ($\mathcal{T}_{IV}$).} Define $\mathcal S_{I+2} = \mathcal{T}_{IV}$, and set $k=I+2$. Next, fix an arbitrary scenario~$\bm{t} \in \mathcal S_k$ and mechanism~$(\bm{p}, \bm{q})$ feasible in~\ref{eq:MDPk(t)}. The constraints of~\ref{eq:MDPk(t)} ensure that~$(\bm{p}, \bm{q})$ solves~${\text{SP}}$$_{l}(\bm t')$ for every~$\bm t' \in \mathcal S_{l}$ and~$l<k$. From Steps~1,~2, and~3 we thus know that $p_{i^\star}(\bm{t}')=1$ and $q_{i^\star}(\bm{t}') =0$ for all $\bm t' \in \cup_{\ell=1}^{k-1} \mathcal S_\ell =  \mathcal T_I \cup \mathcal T_{II} \cup \mathcal T_{III}$. Clearly, $\bm t$ is $i^\star$-unilaterally reachable from $(\overline \mu_{i^\star}, \bm t_{-i^\star})$.
% Note that we always have $(\overline \mu_{i^\star}, \bm t_{-i^\star}) \in \mathcal{T}_{III}$ because $\max_{i \neq i^\star} t_i > \underline{\mu}_{i^\star}$. Hence, we have $p_{i^\star}(\overline \mu_{i^\star}, \bm t_{-i^\star}) -q_{i^\star}(\overline \mu_{i^\star}, \bm t_{-i^\star})  =1$, which in turn implies via Lemma~\ref{lem:deviationForFavored}(i) that $p_{i^\star}(\bm{t}) = 1$. Thus, the objective function value of~$(\bm p, \bm q)$ in~\ref{eq:MDPk(t)} is bounded above by $t_{i^\star} - q_{i^\star}(\bm{t})c_{i^\star} \leq t_{i^\star}$. Hence, $(\bm p^\star, \bm q^\star)$ solves~\ref{eq:MDPk(t)}. Also, as $c_{i^\star} >0$, the mechanism~$(\bm{p}, \bm{q})$ can attain the optimal value~$t_{i^\star}$ of~\ref{eq:MDPk(t)} only if $p_{i^\star}(\bm{t})=1$ and $q_{i^\star}(\bm{t}) =0$. 
\textbf{Step 3 ($\mathcal{T}_{III}$).} Define $\mathcal S_{I+1} = \mathcal{T}_{III}$, and set $k=I+1$. Next, fix an arbitrary scenario~$\bm{t} \in \mathcal S_k$ and mechanism~$(\bm{p}, \bm{q})$ feasible in~\ref{eq:MDPk(t)}. The constraints of~\ref{eq:MDPk(t)} ensure that~$(\bm{p}, \bm{q})$ solves~${\text{SP}}$$_{l}(\bm t')$ for every~$\bm t' \in \mathcal S_{l}$ and~$l<k$. From Steps~1 and~2 we thus know that $p_{i^\star}(\bm{t}')=1$ and $q_{i^\star}(\bm{t}') =0$ for all $\bm t' \in \cup_{\ell=1}^{k-1} \mathcal S_\ell =  \mathcal T_I \cup \mathcal T_{II}$. Clearly, $\bm t$ is $i^\star$-unilaterally reachable from $(\overline \mu_{i^\star}, \bm t_{-i^\star}) \in \mathcal{T}_I \cup \mathcal{T}_{II}$.
Hence, we have $p_{i^\star}(\overline \mu_{i^\star}, \bm t_{-i^\star}) -q_{i^\star}(\overline \mu_{i^\star}, \bm t_{-i^\star})  =1$, which in turn implies via Lemma~\ref{lem:deviationForFavored}(i) that $p_{i^\star}(\bm{t}) = 1$. Thus, the objective function value of~$(\bm p, \bm q)$ in~\ref{eq:MDPk(t)} is bounded above by $t_{i^\star} - q_{i^\star}(\bm{t})c_{i^\star} \leq t_{i^\star}$. Hence, $(\bm p^\star, \bm q^\star)$ solves~\ref{eq:MDPk(t)}. Also, as $c_{i^\star} >0$, the mechanism~$(\bm{p}, \bm{q})$ can attain the optimal value~$t_{i^\star}$ of~\ref{eq:MDPk(t)} only if $p_{i^\star}(\bm{t})=1$ and $q_{i^\star}(\bm{t}) =0$. 

% \textbf{Step 5 ($\mathcal{T}_{V}$).} We partition $\mathcal{T}_{V}$ into $I$ subsets of the form
% \begin{equation*}
%     \mathcal{S}_{k} = \{\bm t \in \mathcal{T}_{V} \,:\, |\{i \in \mathcal{I} \,:\, t_i > \underline \mu_{i^\star}\}| = k-I-2\},
% \end{equation*}
% for $k=I+3, \dots, 2I+2$. We will use induction on~$k$ to show that, for any~$\bm t\in\mathcal S_k$, $(\bm{p}^\star,\bm{q}^\star)$ solves~\ref{eq:MDPk(t)} with optimal value $\max_{i' \in \mathcal{I}} t_{i'}-c_{i'}$ and that any mechanism~$(\bm{p}, \bm{q})$ that solves~\ref{eq:MDPk(t)} must satisfy
% \begin{equation}\tag{\ref{eq:allocation-to-max-net-payoff}} 
%     \sum_{i \in {\arg \max}_{i' \in \mathcal{I}} t_{i'}-c_{i'}} p_i(\bm t) = 1 \quad \text{and} \quad p_i(\bm{t})=q_i(\bm t) \quad \forall i \in {\arg \max}_{i' \in \mathcal{I}} t_{i'}-c_{i'}.
% \end{equation}
\textbf{Step 4 ($\mathcal{T}_{IV}$).} We partition $\mathcal{T}_{IV}$ into $I-1$ subsets of the form
\begin{equation*}
    \mathcal{S}_{k} = \left\{\bm t \in \mathcal{T}_{IV} \,:\, |\{i \in \mathcal{I} \,:\, t_i > \underline \mu_{i^\star}\}| = k-I-1 \right\},
\end{equation*}
for $k=I+2, \dots, 2I$. Note that, by the definition of $\mathcal{T}_{IV}$, we have $i^\star \notin \{i \in \mathcal{I} : t_i > \underline \mu_{i^\star}\}$ for any $\bm t \in \mathcal{T}_{IV}$ because $t_{i^\star}= \underline \mu_{i^\star}$.  We will use induction on~$k$ to show that, for any~$\bm t\in\mathcal S_k$, $(\bm{p}^\star,\bm{q}^\star)$ solves~\ref{eq:MDPk(t)} with optimal value $\max_{i' \in \mathcal{I}} t_{i'}-c_{i'}$ and that any mechanism~$(\bm{p}, \bm{q})$ that solves~\ref{eq:MDPk(t)} must satisfy
\begin{equation}\label{eq:allocation-to-max-net-payoff-same-as-before-2} 
    \sum_{i \in {\arg \max}_{i' \in \mathcal{I}} t_{i'}-c_{i'}} p_i(\bm t) = 1 \quad \text{and} \quad p_i(\bm{t})=q_i(\bm t) \quad \forall i \in {\arg \max}_{i' \in \mathcal{I}} t_{i'}-c_{i'}.
\end{equation}
As for the base step corresponding to~$k=I+2$, fix any~$\bm{t} \in \mathcal{S}_{I+2}$, and consider a mechanism~$(\bm p, \bm q)$ feasible in~\ref{eq:MDPk(t)}. The constraints of~\ref{eq:MDPk(t)} ensure that~$(\bm p, \bm q)$ solves~${\text{SP}}$$_{l}(\bm t')$ for every~$\bm t' \in \mathcal{S}_l$ and $l< I+2$. From Steps~1,~2 and~3, we thus know that $p_{i^\star}(\bm{t}')=p^\star_{i^\star}(\bm{t}')=1$ and $q_{i^\star}(\bm{t}') =q^\star_{i^\star}(\bm{t}') =0$ for all $\bm t' \in \cup_{l=1}^{I+1}\mathcal{S}_l = \mathcal{T}_I \cup \mathcal{T}_{II} \cup \mathcal{T}_{III}$.
As $\bm{t} \in \mathcal{S}_{k}$ and~$k-I-1=1$, there exists exactly one agent $i^\circ \neq i^\star$ with~$t_{i^\circ} > \underline \mu_{i^\star}$. 
Note that~$\bm t$ is ${i^\circ}$-unilaterally reachable from~$(\underline t_{i^\circ}, \bm t_{-{i^\circ}})$. As~$\underline t_{i^\circ} -c_{i^\circ} < \underline \mu_{i^\circ} < \underline \mu_{i^\star}$ by the definition of the favored agent~$i^\star$, we further have $(\underline t_{i^\circ}, \bm t_{-{i^\circ}}) \in \mathcal{T}_I \cup \mathcal{T}_{II} \cup \mathcal{T}_{III}$. The reasoning in Steps~1,~2 and~3 thus implies that~$p_{i^\circ}(\underline t_{i}, \bm t_{-{i}})=0$, which in turn implies via Lemma~\ref{lem:deviationForFavored}(ii) that $p_{i^\circ}(\bm t) =q_{i^\circ}(\bm t)$. Thus, the objective function value of $(\bm p, \bm q)$ in~\ref{eq:MDPk(t)} is bounded above by that of $(\bm p^\star, \bm q^\star)$. Indeed, we have
\begin{equation*}
\begin{aligned}
\sum_{i \in \mathcal{I}} (p_i(\bm{t}) t_i - q_i(\bm{t}) c_i)  \leq p_{i^\circ}(\bm t)(t_{i^\circ}-c_{i^\circ})+ \sum_{i \in \mathcal I \setminus \{i^\circ\}}  p_i(\bm{t}) t_i \leq \max_{i \in \mathcal{I}}t_i-c_i =\sum_{i \in \mathcal{I}} (p_i^\star(\bm{t}) t_i - q_i^\star(\bm{t}) c_i),
\end{aligned}
\end{equation*}
where the first inequality holds because $p_{i^\circ}(\bm t) =q_{i^\circ}(\bm t)$ and because $c_i >0$ for all $i \in \mathcal I$, whereas the second inequality holds because $t_{i^\circ}-c_{i^\circ}=\max_{i \neq i^\star}t_i-c_i > \underline \mu_{i^\star}$ and because $t_{i} \leq \underline \mu_{i^\star}$ for all $i \neq {i^\circ}$. Finally, the equality holds because $\bm t \in \mathcal T_{IV}$, in which case $(\bm p^\star, \bm q^\star)$ generates a payoff of~$\max_{i\in\mathcal I} t_i - c_i$. Thus, $(\bm{p}^\star,\bm{q}^\star)$ solves~\ref{eq:MDPk(t)}. In addition, the objective function value of~$(\bm{p}, \bm{q})$ can equal $\max_{i \neq i^\star}t_i-c_i = t_{i^\circ}-c_{i^\circ}$ only if $p_{i^\circ}(\bm t) =q_{i^\circ}(\bm t)=1$. As $\bm t \in \mathcal S_k$ was chosen freely, this completes the base step.

As for the induction step, fix any $k\in\{I+3,\ldots,2I\}$. Assume that for all $l\in \{I+2,\ldots,k-1\}$ we know that $(\bm{p}^\star,\bm{q}^\star)$ solves ${\text{SP}}$$_l(\bm t)$ and that any mechanism $(\bm p,\bm q)$ that solves~${\text{SP}}$$_l(\bm t)$ satisfies~\eqref{eq:allocation-to-max-net-payoff-same-as-before-2} for all $\bm t \in \mathcal S_l$. Fix now any $\bm t \in \mathcal S_k$ and mechanism $(\bm p, \bm q)$ feasible in~\ref{eq:MDPk(t)}. The constraints in~\ref{eq:MDPk(t)} ensure that 
$(\bm p, \bm q)$ solves~${\text{SP}}$$_{l}(\bm t')$ for every $\bm t' \in \mathcal S_{l}$ and $l <k$. Thus, $(\bm p, \bm q)$ satisfies $p_{i^\star}(\bm{t}')=p^\star_{i^\star}(\bm{t}')=1$ and $q_{i^\star}(\bm{t}') =q^\star_{i^\star}(\bm{t}') =0$ for all $\bm t' \in \mathcal{T}_I \cup \mathcal{T}_{II} \cup \mathcal{T}_{III}$, which follows from Steps~1,~2 and~3, and it satisfies~\eqref{eq:allocation-to-max-net-payoff-same-as-before-2} for all $\bm t' \in \cup_{l=I+2}^{k-1} \mathcal S_l$, which follows from the induction hypothesis. As $\bm t\in\mathcal S_k$ and $k\in\{I+3,\ldots,2I\}$, there are exactly $k-I-1$ agents~$i \in \mathcal{I} \setminus \{i^\star\}$ with types~$t_i > \underline \mu_{i^\star}$. For any such agent $i$, scenario~$\bm t$ is $i$-unilaterally reachable from $(\underline t_i, \bm t_{-i})$. Note that either $(\underline t_i, \bm t_{-i}) \in \mathcal{T}_I \cup \mathcal{T}_{II} \cup \mathcal{T}_{III}$ or $(\underline t_i, \bm t_{-i}) \in \cup_{l=I+2}^{k-1} \mathcal S_{l}$ because $\underline t_i <\underline \mu_i < \underline \mu_{i^\star}$ for all $i \in \mathcal I \setminus \{i^\star\}$. We now show that $p_i(\underline t_i, \bm t_{-i})$ must vanish in both cases.
If $(\underline t_i, \bm t_{-i}) \in \mathcal{T}_I \cup \mathcal{T}_{II}  \cup \mathcal{T}_{III}$, we have $p_i(\underline t_i, \bm t_{-i})=p^\star_i(\underline t_i, \bm t_{-i})=0$ because $i \neq i^\star$. If $(\underline t_i, \bm t_{-i}) \in \cup_{l=I+2}^{k-1} \mathcal S_{l}$, on the other hand, then the definition of~$i^\star$ implies that $\underline t_i-c_i < \underline \mu_{i^\star}$, and the definition of~$\mathcal T_{IV}$ implies that $\max_{i' \in \mathcal I \setminus \{i^\star\}} t_{i'}-c_{i'}> \underline \mu_{i^\star}$. Hence, $i$ is no element of $\arg\max_{i' \in \mathcal I \setminus \{i^\star\}} t_{i'}-c_{i'}$, implying that $p_i(\underline t_i, \bm t_{-i})=0$ thanks to~\eqref{eq:allocation-to-max-net-payoff-same-as-before-2}. Lemma~\ref{lem:deviationForFavored}(ii) now ensures that $p_i(\bm{t}) = q_i(\bm{t})$ for all $i \in \mathcal{I}$ with $t_i > \underline{\mu}_{i^\star}$. Thus, the objective function value of $(\bm p, \bm q)$ in~\ref{eq:MDPk(t)} is bounded above by that of $(\bm p^\star, \bm q^\star)$ because
\begin{equation*}
    \sum_{i \in \mathcal{I}} (p_i(\bm{t}) t_i - q_i(\bm{t}) c_i) \leq \sum_{i \in \mathcal I : \, t_i > \underline{\mu}_{i^\star}} p_i(\bm{t}) (t_i- c_i) + \sum_{i \in \mathcal I : \, t_i \leq \underline{\mu}_{i^\star}} p_i(\bm{t}) t_i \leq \max_{i \in \mathcal{I}}t_i-c_i =\sum_{i \in \mathcal{I}} (p_i^\star(\bm{t}) t_i - q_i^\star(\bm{t}) c_i),
\end{equation*}
where the second inequality holds because $\max_{i \neq i^\star}t_i-c_i > \underline \mu_{i^\star}$. Thus, $(\bm{p}^\star,\bm{q}^\star)$ solves~\ref{eq:MDPk(t)}, and $(\bm p,\bm q)$ can solve~\ref{eq:MDPk(t)} only if it obeys~\eqref{eq:allocation-to-max-net-payoff-same-as-before-2}. This observation completes the induction step.

\textbf{Step 5 ($\mathcal{T}_{V}$).} We partition $\mathcal{T}_{V}$ into $I-1$ subsets of the form
\begin{equation*}
    \mathcal{S}_{k} = \left\{\bm t \in \mathcal{T}_{V} \,:\, |\{i \in \mathcal{I} \setminus \{i^\star\} \,:\, t_i > \underline \mu_{i^\star}\}| = k-2I \right\},
\end{equation*}
for $k=2I+1, \dots, 3I-1$. We will use induction on~$k$ to show that, for any~$\bm t\in\mathcal S_k$, $(\bm{p}^\star,\bm{q}^\star)$ solves~\ref{eq:MDPk(t)} with optimal value $\max_{i' \in \mathcal{I}} t_{i'}-c_{i'}$, that any mechanism~$(\bm{p}, \bm{q})$ that solves~\ref{eq:MDPk(t)} must satisfy \eqref{eq:allocation-to-max-net-payoff-same-as-before-2}.

As for the base step corresponding to~$k=2I+1$, fix any~$\bm{t} \in \mathcal{S}_{2I+1}$, and consider a mechanism~$(\bm p, \bm q)$ feasible in~\ref{eq:MDPk(t)}. The constraints of~\ref{eq:MDPk(t)} ensure that~$(\bm p, \bm q)$ solves~${\text{SP}}$$_{l}(\bm t')$ for every~$\bm t' \in \mathcal{S}_l$ and $l< 2I+1$. From Steps~1,~2, and~3, we thus know that $p_{i^\star}(\bm{t}')=p^\star_{i^\star}(\bm{t}')=1$ and $q_{i^\star}(\bm{t}') =q^\star_{i^\star}(\bm{t}') =0$ for all $\bm t' \in \cup_{l=1}^{I+1}\mathcal{S}_l = \mathcal{T}_I \cup \mathcal{T}_{II} \cup \mathcal{T}_{III}$, and from Step~4, we know that~$(\bm p, \bm q)$ satisfies \eqref{eq:allocation-to-max-net-payoff-same-as-before-2} for all $\bm t' \in \cup_{l=I+2}^{2I}\mathcal{S}_l = \mathcal{T}_{IV}$.
As $\bm{t} \in \mathcal{S}_{k}$ and~$k-2I=1$, there exists exactly one agent $i^\circ \neq i^\star$ with~$t_{i^\circ} > \underline \mu_{i^\star}$. Now consider any agent $i \in \{i^\circ,i^\star\}$. Note that~$\bm t$ is ${i}$-unilaterally reachable from~$(\underline \mu_{i}, \bm t_{-{i}})$. If $i=i^\star$, then we have $(\underline \mu_{i^\circ}, \bm t_{-{i^\circ}}) \in \mathcal{T}_{IV}$. Our reasoning in Step~4 implies that $p_{i}(\underline \mu_{i}, \bm t_{-{i}})=0$ as $\underline \mu_{i^\star} < \max_{i' \in \mathcal{I}} t_i'-c_i'$. If $i\neq i^\star$, on the other hand, we must have $(\underline \mu_{i}, \bm t_{-{i}}) \in \mathcal{T}_I \cup \mathcal{T}_{II} \cup \mathcal{T}_{III}$, because, $\underline \mu_{i} -c_{i} < \underline \mu_{i^\star}$ by the definition of the favored agent~$i^\star$, and $i^\circ$ is the only agent in $\mathcal I \setminus \{i^\star\}$ with $t_{i^\circ}-c_{i^\circ}>\underline \mu_{i^\star}$. The reasoning in Steps~1,~2 and~3 thus implies that~$p_{i}(\underline \mu_{i}, \bm t_{-{i}})=0$. Then, for all $i \in \{i^\circ,i^\star\}$, Lemma~\ref{lem:deviationForFavored}(ii) implies that $p_{i}(\bm t) =q_{i}(\bm t)$. Thus, the objective function value of $(\bm p, \bm q)$ in~\ref{eq:MDPk(t)} is bounded above by that of $(\bm p^\star, \bm q^\star)$. Indeed, we have
\begin{equation*}
\begin{aligned}
\sum_{i \in \mathcal{I}} (p_i(\bm{t}) t_i - q_i(\bm{t}) c_i)  \leq \sum_{i \in \{i^\circ,i^\star\}}p_{i}(\bm t)(t_{i}-c_{i})+ \sum_{i \in \mathcal I \setminus \{i^\circ,i^\star\}}  p_i(\bm{t}) t_i \leq \max_{i \in \mathcal{I}}t_i-c_i =\sum_{i \in \mathcal{I}} (p_i^\star(\bm{t}) t_i - q_i^\star(\bm{t}) c_i),
\end{aligned}
\end{equation*}
where the first inequality holds because $p_{i}(\bm t) =q_{i}(\bm t)$ for all $i \in \{i^\circ,i^\star\}$ and because $c_i >0$ for all $i \in \mathcal I$, whereas the second inequality holds because $t_{i^\circ}-c_{i^\circ}=\max_{i \neq i^\star}t_i-c_i > \underline \mu_{i^\star}$ and because $t_{i} \leq \underline \mu_{i^\star}$ for all $i \notin \{i^\circ,i^\star\}$. Finally, the equality holds because $\bm t \in \mathcal T_{V}$, in which case $(\bm p^\star, \bm q^\star)$ generates a payoff of~$\max_{i\in\mathcal I} t_i - c_i$. Thus, $(\bm{p}^\star,\bm{q}^\star)$ solves~\ref{eq:MDPk(t)}. In addition, the objective function value of~$(\bm{p}, \bm{q})$ can equal $\max_{i \neq i^\star}t_i-c_i = t_{i^\circ}-c_{i^\circ}$ only if~$(\bm{p}, \bm{q})$ satisfies \eqref{eq:allocation-to-max-net-payoff-same-as-before-2}. As $\bm t \in \mathcal S_k$ was chosen freely, this completes the base step.

As for the induction step, fix any $k\in\{2I+2,\ldots,3I-1\}$. Assume that for all $l\in \{2I+1,\ldots,k-1\}$ we know that $(\bm{p}^\star,\bm{q}^\star)$ solves ${\text{SP}}$$_l(\bm t)$ and that any mechanism $(\bm p,\bm q)$ that solves~${\text{SP}}$$_l(\bm t)$ satisfies~\eqref{eq:allocation-to-max-net-payoff-same-as-before-2} for all $\bm t \in \mathcal S_l$. Fix now any $\bm t \in \mathcal S_k$ and mechanism $(\bm p, \bm q)$ feasible in~\ref{eq:MDPk(t)}. The constraints in~\ref{eq:MDPk(t)} ensure that 
$(\bm p, \bm q)$ solves~${\text{SP}}$$_{l}(\bm t')$ for every $\bm t' \in \mathcal S_{l}$ and $l <k$. Thus, $(\bm p, \bm q)$ satisfies $p_{i^\star}(\bm{t}')=p^\star_{i^\star}(\bm{t}')=1$ and $q_{i^\star}(\bm{t}') =q^\star_{i^\star}(\bm{t}') =0$ for all $\bm t' \in  \cup_{l=1}^{I+1} \mathcal S_l = \mathcal{T}_I \cup \mathcal{T}_{II} \cup \mathcal{T}_{III}$, which follows from Steps~1,~2 and~3, and it satisfies~\eqref{eq:allocation-to-max-net-payoff-same-as-before-2} for all $\bm t' \in \cup_{l=I+2}^{k-1} \mathcal S_l$, which follows from Step 4 and the induction hypothesis. As $\bm t\in\mathcal S_k$ and $k\in\{2I+2,\ldots,3I-1\}$, there are exactly $k-2I$ agents~$i \in \mathcal{I} \setminus \{i^\star\}$ with types~$t_i > \underline{\mu}_{i^\star}$. Let $i$ denote any such agent or agent $i^\star$. Then, scenario~$\bm t$ is $i$-unilaterally reachable from $(\underline \mu_i, \bm t_{-i})$. Note that we have $(\underline \mu_i, \bm t_{-i}) \in \mathcal{T}_{IV}$ if $i=i^\star$, and we have either $(\underline \mu_i, \bm t_{-i}) \in \mathcal{T}_I \cup \mathcal{T}_{II} \cup \mathcal{T}_{III}$ or $(\underline \mu_i, \bm t_{-i}) \in \cup_{l=2I+1}^{k-1} \mathcal S_{l}$ if $i \neq i^\star$, because $\underline \mu_i < \underline \mu_{i^\star}$ follows from the definition of $i^\star$.
We now show that $p_i(\underline \mu_i, \bm t_{-i})$ must vanish in any case.
If $(\underline \mu_i, \bm t_{-i}) \in \mathcal{T}_I \cup \mathcal{T}_{II}  \cup \mathcal{T}_{III}$, then we have $i \neq i^\star$, in which case $p_i(\underline t_i, \bm t_{-i})=p^\star_i(\underline t_i, \bm t_{-i})=0$. 
If $(\underline \mu_i, \bm t_{-i}) \in \mathcal{T}_{IV}$, then we have $i = i^\star$, in which case $p_{i^\star}(\underline t_{i^\star}, \bm t_{-{i^\star}})=0$ follows from~\eqref{eq:allocation-to-max-net-payoff-same-as-before-2} and $\max_{i \neq i^\star}t_i-c_i > \underline \mu_{i^\star}$. 
If $(\underline t_i, \bm t_{-i}) \in \cup_{l=I+2}^{k-1} \mathcal S_{l}$, on the other hand, then the definition of~$i^\star$ implies that $\underline t_i-c_i < \underline \mu_{i^\star}$, and the definition of~$\mathcal T_{V}$ implies that $\max_{i' \in \mathcal I \setminus \{i^\star\}} t_{i'}-c_{i'}> \underline \mu_{i^\star}$. Hence, $i$ is no element of $\arg\max_{i' \in \mathcal I \setminus \{i^\star\}} t_{i'}-c_{i'}$, implying that $p_i(\underline t_i, \bm t_{-i})=0$ thanks to~\eqref{eq:allocation-to-max-net-payoff-same-as-before-2}. Lemma~\ref{lem:deviationForFavored}(ii) now ensures that $p_i(\bm{t}) = q_i(\bm{t})$ for all $i \in \mathcal{I}$ with $t_i > \underline{\mu}_{i^\star}$ and for $i=i^\star$. Thus, the objective function value of $(\bm p, \bm q)$ in~\ref{eq:MDPk(t)} is bounded above by that of $(\bm p^\star, \bm q^\star)$ because
\begin{equation*}
\begin{aligned}
    \sum_{i \in \mathcal{I}} (p_i(\bm{t}) t_i - q_i(\bm{t}) c_i) \leq \sum_{i \in \mathcal I : \, t_i > \underline{\mu}_{i^\star}} p_i(\bm{t}) (t_i- c_i) +  p_{i^\star}(\bm{t}) (t_{i^\star}- c_{i^\star}) + \sum_{\substack{i \in \mathcal I \setminus \{i^\star\} :\\ \, t_i \leq \underline{\mu}_{i^\star} }} p_i(\bm{t}) t_i \\
    \leq \max_{i \in \mathcal{I}}t_i-c_i =\sum_{i \in \mathcal{I}} (p_i^\star(\bm{t}) t_i - q_i^\star(\bm{t}) c_i),
    \end{aligned}
\end{equation*}
where the second inequality holds because $\max_{i \neq i^\star}t_i-c_i > \underline \mu_{i^\star} \geq t_j$ for all $j \in \mathcal I \setminus \{i^\star\} :  t_j \leq \underline{\mu}_{i^\star}$. Thus, $(\bm{p}^\star,\bm{q}^\star)$ solves~\ref{eq:MDPk(t)}, and $(\bm p,\bm q)$ can solve~\ref{eq:MDPk(t)} only if it obeys~\eqref{eq:allocation-to-max-net-payoff-same-as-before-2}. This observation completes the induction step.

Next, we exploit Lemma~\ref{lem:findP-independent} to show that condition~(i) holds. Condition~(i) of Proposition \ref{prop:unifyingPRO} requires that, for any index $k \in \{1, \ldots,3I-1\}$ and for any scenario $\bm t \in \mathcal{S}_k$, there exists $\mathbb{P} \in \mathcal{P} \cap \mathcal{P}_0(\cup_{l=1}^k \mathcal{S}_l)$ with $\mathbb{P}(\tilde{\bm{t}} =\bm{t})>0$.
Given any scenario $\bm t \in \mathcal{T}$ and any $\mu_{i^\star} \in [\underline \mu_{i^\star},\overline{\mu}_{i^\star}]$, by Lemma~\ref{lem:findP-independent}, there exists a scenario $\hat{\bm{t}} \in \mathcal{T}$ with $\max_{i\neq i^\star} \hat t_i< \underline{\mu}_{i^\star}$
and a discrete distribution $\mathbb{P} \in \mathcal{P}$ such that
$\mathbb{E}_{\mathbb{P}}[\tilde{t}_{i^\star}] =\mu_{i^\star}$, $\mathbb{P}(\tilde{{t}}_i \in \{{t}_i,\hat{{t}}_i \})=1$ for all $i \in \mathcal{I}$, and $\mathbb{P}(\tilde{\bm{t}} =\bm{t})>0$.
We will show that, for any $k \in \{1, \ldots,3I-1\}$ and for any scenario $\bm t \in \mathcal{S}_k$, there exists $\mu_{i^\star} \in [\underline \mu_{i^\star},\overline{\mu}_{i^\star}]$ such that $\mathbb{P}$, as defined in Lemma~\ref{lem:findP-independent}, satisfies condition~(i). To this end, we first derive some useful implications of Lemma~\ref{lem:findP-independent} for arbitrary $\bm t \in \mathcal{T}$ and $\mu_{i^\star} \in [\underline \mu_{i^\star},\overline{\mu}_{i^\star}]$. In the following, we use~$\mathbb S$ to denote the support of the distribution~$\mathbb{P}$. Note that $\mathbb S \subseteq \{\bm t' \in \mathcal{T} : t_i' \in \{t_i,\hat t_i\} \; \text{for all} \; i \in \mathcal{I}\}$. Then, as $\max_{i\neq i^\star} \hat t_i< \underline{\mu}_{i^\star}$, we have $|\{i \in \mathcal{I} \setminus \{i^\star\}\,:\, t_{i}' > \underline \mu_{i^\star}\}| \leq |\{i \in \mathcal{I} \setminus \{i^\star\}\,:\, t_i > \underline \mu_{i^\star}\}|$ for any~$\bm t' \in \mathbb S$. For the same reason, we may also conclude that $\max_{i \neq i^\star} t_i'-c_i \leq \max\{\max_{i \neq i^\star} t_i-c_i, \underline \mu_{i^\star}\}$. We thus obtain 
\begin{align}\label{eq:supportMarkovIndependent1}
    \mathbb S
    &\subseteq
        \left\{\bm t' \in \mathcal{T} \,:\, f(\bm t') \leq f({\bm t}) \;\text{and}\; \max_{i \neq i^\star} t_i'-c_i \leq g({\bm t}) \right\},
\end{align}
where $f({\bm t}) = |\{i \in \mathcal{I} \setminus \{i^\star\}\,:\, t_i > \underline \mu_{i^\star}\}|$ and $g({\bm t})=\max\{\max_{i \neq i^\star} t_i-c_i,  \underline \mu_{i^\star}\}$. 

If $t_{i^\star} \in [\underline \mu_{i^\star},\overline \mu_{i^\star}]$ and $\mu_{i^\star}=t_{i^\star}$, then the identities $\mathbb{E}_{\mathbb{P}}[\tilde{t}_{i^\star}] ={\mu}_{i^\star}$ and $\mathbb{P}(\tilde{{t}}_{i^\star} \in \{{t}_{i^\star},\hat{{t}}_{i^\star} \})=1$ imply that $\hat t_{i^\star} = \mu_{i^\star}$. Together with \eqref{eq:supportMarkovIndependent1}, this in turn implies that
\begin{align}\label{eq:supportMarkovIndependent2}
    \mathbb S
    \subseteq
        \left\{\bm t' \in \mathcal{T} \,:\,  f(\bm t') \leq f({\bm t}) \text{ and } \max_{i \neq i^\star} t_i'-c_i \leq g({\bm t}) \text{ and } t_{i^\star}'= t_{i^\star} \right\}\quad
        \text{if}\quad t_{i^\star} = \mu_{i^\star}.
\end{align}
We are now ready to prove condition~(i).
Consider first $k=1$, and fix any $\bm t \in \mathcal{S}_1 = \mathcal T_I$. By the definition of $\mathcal T_I$, we have $f(\bm t) = 0$ and $g(\bm t) = \underline{\mu}_{i^\star}$. Observing that $t_{i^\star} \in (\underline{\mu}_{i^\star}, \overline{\mu}_{i^\star}]$, we may choose $\mu_{i^\star} = t_{i^\star}$ and define $\mathbb{P}$ as in Lemma~\ref{lem:findP-independent}. By~\eqref{eq:supportMarkovIndependent2} and as $f(\bm t) = 0$ and $g(\bm t) = \underline{\mu}_{i^\star}$, we thus obtain 
$$\mathbb S \subseteq \left\{\bm t' \in \mathcal{T} \,:\, t_{i^\star}'= t_{i^\star} \text{ and }  \max_{i \neq i^\star} t_i' -c_i \leq \underline \mu_{i^\star} \;\text{and}\; f(\bm t') \leq 0 \right\} \subseteq \mathcal{S}_1,$$
where the second inclusion holds because $\mathcal{S}_1 = \mathcal T_I$ and because of the definition of $\mathcal T_I$. We have therefore shown that $\mathbb{P} \in \mathcal{P}_0(\cup_{l=1}^k \mathcal{S}_l) = \mathcal{P}_0(\mathcal{S}_1)$. By Lemma~\ref{lem:findP-independent}, we further have $\mathbb{P} \in \mathcal P$ and $\mathbb{P}(\tilde{\bm{t}} =\bm{t})>0$, which completes the proof for $k=1$.

Next, consider any $k \in \{2, \dots, I\}$ and $\bm t \in \mathcal{S}_k \subseteq \mathcal T_{II}$. By the definitions of $\mathcal{S}_k$ and $\mathcal T_{II}$, we have $f(\bm t) = k-1$ and $g(\bm t) = \underline{\mu}_{i^\star}$. As $t_{i^\star} \in (\underline{\mu}_{i^\star}, \overline{\mu}_{i^\star}]$, we may again choose $\mu_{i^\star} = t_{i^\star}$ and define $\mathbb{P}$ as in Lemma \ref{lem:findP-independent}. By~\eqref{eq:supportMarkovIndependent2} and the identities $f(\bm t) = k-1$ and $g(\bm t) = \underline{\mu}_{i^\star}$, we then find
\begin{equation*}
    \begin{aligned}
        \mathbb S &\subseteq \left\{\bm t' \in \mathcal{T} \,:\, t_{i^\star}'= t_{i^\star} \;\text{and}\; \max_{i \neq i^\star} t_i' -c_i \leq \underline \mu_{i^\star} \;\text{and}\; f(\bm t') \leq k-1 \right\} \\
        &\subseteq \left\{\bm t' \in \mathcal{T}_I \cup  \mathcal{T}_{II} \,:\, f(\bm t') \leq k-1 \right\} \subseteq \cup_{l=1}^k \mathcal{S}_l,
    \end{aligned}
\end{equation*}
where the second and third inclusions follow from the definitions of $\mathcal{T}_I, \mathcal{T}_{II}$ and $\mathcal{S}_k$. Together with Lemma~\ref{lem:findP-independent}, this observation implies that $\mathbb{P} \in \mathcal P \cap \mathcal{P}_0(\cup_{l=1}^k \mathcal{S}_l)$ and $\mathbb{P}(\tilde{\bm{t}} =\bm{t})>0$.

Next, set $k=I+1$, and consider any $\bm t \in \mathcal{S}_{I+1} = \mathcal T_{III}$. In this case, we again have $g(\bm t) = \underline{\mu}_{i^\star}$. Consider any fixed ${\mu}_{i^\star} \in [\underline{\mu}_{i^\star}, \overline{\mu}_{i^\star}]$ and $\mathbb{P}$ as defined in Lemma \ref{lem:findP-independent}. By~\eqref{eq:supportMarkovIndependent1} and that $g(\bm t) = \underline{\mu}_{i^\star}$, we have 
$\mathbb S \subseteq \{\bm t' \in \mathcal{T} \,:\, \max_{i \neq i^\star} t_i' -c_i \leq \underline \mu_{i^\star}\} \subseteq \mathcal{T}_I \cup \mathcal{T}_{II}\cup \mathcal{T}_{III} = \cup_{l=1}^k \mathcal{S}_l$. Thus,  we again have $\mathbb{P} \in \mathcal P \cap \mathcal{P}_0(\cup_{l=1}^k \mathcal{S}_l)$ and $\mathbb{P}(\tilde{\bm{t}} =\bm{t})>0$.

Next, consider any $k \in \{I+2, \dots, 2I\}$ and $\bm t \in \mathcal S_k \subseteq \mathcal T_{IV}$. By the definition of $\mathcal S_k$ in Step~4, we have $f(\bm t) = k-I-1$. As $t_{i^\star} = \underline{\mu}_{i^\star}$ by the definition of $\mathcal T_{IV}$, we can choose $\mu_{i^\star} = t_{i^\star} = \underline{\mu}_{i^\star}$ and define~$\mathbb{P}$ as in Lemma \ref{lem:findP-independent}. We can now leverage~\eqref{eq:supportMarkovIndependent2} to conclude that 
$\mathbb S \subseteq \{\bm t' \in \mathcal{T} \,:\, t_{i^\star}'= \underline{\mu}_{i^\star} \;\text{and}\; f(\bm t') \leq k-I-1\}$, which is a subset of $\mathcal T_{III} \cup (\cup_{l=I+2}^k \mathcal{S}_l) \subseteq \cup_{l=1}^k \mathcal{S}_l$. This completes the proof for $k \in \{I+2, \dots, 2I\}$.

Finally, consider any $k \in \{2I+1, \dots, 3I-1\}$ and any $\bm t \in \mathcal S_k \subseteq \mathcal T_{V}$. Consider any fixed ${\mu}_{i^\star} \in [\underline{\mu}_{i^\star}, \overline{\mu}_{i^\star}]$, and define $\mathbb{P}$ as in Lemma \ref{lem:findP-independent}. By the definition of $\mathcal S_k$ in Step~5, we now have $f(\bm t) = k-2I$. Together with~\eqref{eq:supportMarkovIndependent1}, this implies that  $\mathbb S \subseteq \{\bm t' \in \mathcal{T} \,:\, f(\bm t') \leq k-2I\}$, which is a subset of $\cup_{l=1}^k \mathcal{S}_l$ by the definition of $\mathcal{S}_k$. The claim thus holds for $k \in \{2I+1, \dots, 3I-1\}$. This completes the proof of condition (i). Hence, the claim follows.
\hfill \Halmos
\endproof

\end{APPENDIX}
%
%   or 
%
% \begin{APPENDICES}
% \section{<Title of Section A>}
% \section{<Title of Section B>}
% etc
% \end{APPENDICES}

% Acknowledgments here
%\section*{Acknowledgments.}
% Enter the text of acknowledgments here

% References here (outcomment the appropriate case) 

% CASE 1: BiBTeX used to constantly update the references 
%   (while the paper is being written).

% CASE 2: BiBTeX used to generate mypaper.bbl (to be further fine tuned)
%\input{mypaper.bbl} % outcomment this line in Case 2

\end{document}